\documentclass{article}
\pdfoutput=1 

\usepackage{jheppub}
\usepackage{dashrule}
\usepackage[T1]{fontenc} 
\usepackage{braket}
\usepackage{float}
\usepackage{amsmath,amssymb}
\usepackage{graphicx}
\usepackage{textcomp}
\usepackage{mathtools}
\usepackage{eucal}
\usepackage{mathrsfs}
\usepackage{tikz}
\usepackage{makecell}
\usepackage{ytableau}
\usepackage[bbgreekl]{mathbbol}
\usepackage{stmaryrd}
\usepackage{makecell}
\usepackage{hyperref}
\DeclareMathAlphabet{\mathpzc}{OT1}{pzc}{m}{it}

\newcommand{\gO}{G^{\text{Out}}_\mathcal{N}}

\newcommand{\phN}{\varphi_{_\nn}}
\newcommand{\piN}{\pi_{\nn}}
\newcommand{\phNren}{\varphi_{_\nn\text{ren}}}
\newcommand{\JPb}{\multj_{\bar{P}}}
\newcommand{\JFb}{\multj_{\bar{F}}}

\newcommand{\ob}[1]{\ensuremath{\overline{\raisebox{0pt}[1.10\height]{#1}}}}
\newcommand{\overbar}[1]{\ob{${#1}$}}

\newcommand{\w}{\omega}

\newcommand{\nn}{\mathcal{N}}
\newcommand{\Dp}{D_+}
\newcommand{\Dm}{D_{-}}
\newcommand{\Gout}{G^{\text{Out}}}
\newcommand{\Gouts}{G^{\text{Out}\ast}}
\newcommand{\Kout}{K^{\text{Out}}}
\newcommand{\Kouts}{K^{\text{Out}\ast}}
\newcommand{\Khout}{\widehat{K}^{\text{Out}}}

\newcommand{\Gin}{G^{\text{In}}}

\newcommand{\ct}{\mathscr{C}}

\newcommand{\Nhh}{\mathcal{N}_{HH}}

\newcommand{\SCIP}{S_{\text{CIP}}}
\newcommand{\kO}{K_{\text{Out}}}
\newcommand{\kI}{K_{\text{In}}}
\newcommand{\khO}{\widehat{K}_{\text{Out}}}

\newcommand{\multj}{\mathcal{J}}

\newcommand{\Yellm}{\mathscr{Y}_{\ell\vec{m}}}
\newcommand{\Yellma}{\mathscr{Y}_{\ell m}}

\newcommand{\FC}{\overbar{\mathcal{C}}}
\newcommand{\FCf}{\mathcal{C}}
\newcommand{\gaugeV}{\overbar{\mathcal{V}}}
\newcommand{\gaugeVf}{\mathcal{V}}

\newcommand{\Erf}{E_r}
\newcommand{\Esf}{E_s}
\newcommand{\Evf}{E_v}

\newcommand{\Bvvf}{H_{vv}}
\newcommand{\Hsf}{H_s}
\newcommand{\Hvf}{H_v}

\newcommand{\Er}{\mathcal{E}_r}
\newcommand{\Es}{\mathcal{E}_s}
\newcommand{\Ev}{\mathcal{E}_v}
\newcommand{\Brs}{\mathcal{B}_{rs}}
\newcommand{\Brv}{\mathcal{B}_{rv}}
\newcommand{\Bvv}{\mathcal{H}_{vv}}
\newcommand{\Hs}{\mathcal{H}_s}
\newcommand{\Hv}{\mathcal{H}_v}

\newcommand{\VSH}{\mathbb{V}^{\alpha\ell\vec{m}}}

\title{Influence Phase of a dS Observer II: Electromagnetism}
\author[a]{R. Loganayagam,}
\author[a]{Omkar Shetye.}

\emailAdd{nayagam@icts.res.in}
\emailAdd{omkar.shetye@icts.res.in}

\affiliation[a]{ International Centre for Theoretical Sciences (ICTS-TIFR),
	Tata Institute of Fundamental Research,
	Shivakote, Hesaraghatta,
	Bangalore 560089, India.}

\abstract{We extend our proposal for static patch holography made in part I to the case of electromagnetism.
Using the on-shell action on de Sitter Schwinger-Keldysh (dS-SK) geometry, we derive the influence phase of an observer in dS$_{d+1}$ interacting with bulk electromagnetic fields. This influence phase, computed with appropriate boundary conditions and counterterms, encodes the physics of electromagnetic radiation reaction and corresponding dS Hawking radiation. The self-force is rendered finite through holographic renormalisation. In the short-time limit, we reproduce electromagnetic flat space radiation reaction ala Abraham-Lorentz-Dirac, along with cosmological corrections. We give a fully covariantised expression for this radiation reaction in several even spacetime dimensions. In this process, we also extend many existing results to cosmological spacetimes: multipole expansions including smearing effects, non-relativistic expansion for electromagnetism, and classical renormalisation in odd spacetime dimensions. Further, we review and extend many properties of vector spherical harmonics(VSHs) in arbitrary dimensions, explain the relation between spherical and cartesian VSH, and derive an addition theorem for VSHs. 
}

\begin{document}
\maketitle	
\section{Introduction }

The experimental detection of gravitational waves has brought renewed attention to the problem of radiation reaction and self-force in classical field theories. The standard puzzle is easily stated: given a point charge in arbitrary motion, the self-force due to its own electromagnetic fields seems naively infinite. Both experimental evidence as well as the momentum flux at infinity suggest that this conclusion is plainly wrong! There is a finite electromagnetic force on the particle due to a \emph{renormalised} field.

Over the past century, many successful proposals have been made to address the issue mentioned above. The main idea is twofold: we first solve for the field produced by a charge with \emph{outgoing} boundary conditions far away. Next, we identify and remove the divergent pieces from this field near the charge to get a finite answer. 

It is instructive to contrast this procedure against the holographic prescription in AdS/CFT to compute thermal CFT correlators. As described by Son and Starinets\cite{Son:2002sd}, the AdS/CFT computation proceeds again in two steps: first, we take a bulk black hole and impose \emph{infalling} boundary conditions at its horizon. Next, we look at this solution near the conformal boundary of AdS, put counter-terms and read off the renormalised CFT correlators\cite{Witten:1998qj,Balasubramanian:1999re,Skenderis:2002wp}. This similarity suggests that we can think of \emph{holography as a kind of radiation reaction}\footnote{See \cite{Klebanov:1997kc,Goldberger:2005cd} for how \emph{absorption} processes also have holographic features. }. Taking such a slogan seriously might give us a way to generalise holographic insights to spacetimes other than AdS.

In both the radiation reaction problem and in the AdS blackhole problem, the outgoing/infalling boundary conditions can only be an approximation. For example, a charge in de Sitter will see Hawking radiation from its horizon, whereas there is also Hawking radiation emerging out of AdS blackholes. Over the past few years, a geometric method has emerged to go beyond the strict retarded boundary conditions and take into account Hawking radiation. On the AdS side, this is the gravitational Schwinger-Keldysh construction\cite{Skenderis:2008dg,Skenderis:2008dh,vanRees:2009rw,deBoer:2018qqm,Glorioso:2018mmw,Chakrabarty:2019aeu,Jana:2020vyx}. In our previous work\cite{Loganayagam:2023pfb} which we refer to as Part I henceforth, we showed that a similar construction works for scalar radiation reaction in de Sitter. The idea here is to take holography seriously by imposing conditions only at the boundary and let the geometry dynamically impose partial retarded conditions at the horizon so as to admit Hawking effects. Formulated this way, the analogy between AdS holography and dS radiation reaction become quite clear.

In this work, we want to extend our scalar construction in Part I to the case of electromagnetism. The EM radiation reaction problem, of course, has a long and rich history\footnote{See, e.g., references \cite{Abraham1903,lorentz1909,Schott01011915,Dirac:1938nz,Wheeler1945,Coleman1982,deWitt1960,Hobbs1968} for the $3+1$ dimensional version. For higher dimensions, we refer the reader to \cite{Harte:2018iim,Birnholtz:2013ffa,Birnholtz:2013nta,Galtsov:2007zz,Galakhov:2007my,Mironov:2007mv,Mironov:2007nk,Kosyakov:1999np}.}, but we shall see that our `dS holographic perspective' adds new elements to this story. We will see how the counterterm procedure in the radiation reaction(RR) problem mirrors the one used in AdS holography. We would also like to point out how the RR problem in de Sitter is somewhat better behaved than one in flat spacetime.  In part I, we showed how almost all memory/tail effects in the scalar RR problem go away at cosmologically long times. This is true even in the case of odd-dimensional spacetimes, where the flat spacetime RR problem has serious memory/tail effects. We will see that the dS version of the EM RR problem is also better behaved in this sense.

In essence, the charge always forgets its past at cosmological time scales, and the long-time physics is always that of a Langevin particle executing  
Brownian motion within dS thermal bath. For fast motions/short time scales, the dS RR problem should approach the flat spacetime answers. Thus, de Sitter provides a nice infrared cut-off for the RR problem.\footnote{We are used to thinking of AdS as a good IR cutoff, but the fact that AdS is a confining spacetime makes the AdS RR problem non-Markovian (i.e., it is expected to have even worse memory problems than the flat spacetime version). Consequently, one does not expect a local description of the self-force at long times, unlike what happens in de Sitter.} This statement can be made mathematically precise at the level of Fuchsian ODEs that control free theory radial functions. In flat spacetime, these are Bessel-like functions with one regular singularity at $r=0$ and another irregular singularity at $r=\infty$.
Once we move to de Sitter, the irregular singularity at $r=\infty$ splits into two \emph{regular} singularities. The consequence is that radial functions in dS are hypergeometric functions with three regular singularities. 

We will now review the basic setup introduced in part I\cite{Loganayagam:2023pfb}. Our description here will be brief, and we will refer the reader to part I for a more extensive discussion. The main object of interest is the \emph{cosmological influence phase} (CIP), which encodes all physics as seen by an observer.\footnote{The idea of the influence phase was introduced by Feynman and Vernon\cite{FEYNMAN1963118} to capture the physics of open quantum systems. It is closely related to Schwinger-Keldysh\cite{Schwinger:1960qe,Keldysh:1964ud} path integrals. For a review, we will refer the reader to \cite{kamenev_2011,Sieberer:2015svu,breuer2002theory}. } The observer is specified by a twofold set of multipole moments that emit and absorb quanta of the ambient fields. In part I, we proposed this as the object analogous to the S matrix in flat spacetime or the generating function of CFT correlators in the case of AdS. 

In part I, we argued that the CIP can be computed by a de Sitter version of GKPW prescription\cite{Gubser:1998bc,Witten:1998qj}, i.e., by doing a quantum gravity path integral with appropriate boundary conditions for bulk fields at the observer's worldline. The main difference is that we are computing a real-time path integral, which evolves the density matrix, and hence, the data on the worldline is doubled. As in GKPW, we assume that the quantum gravity path integral is dominated by a semi-classical geometry called de Sitter Schwinger-Keldysh (dS-SK) geometry. We solve for the classical field equations in dS-SK with appropriate boundary conditions at the doubled worldline, i.e., by specifying the asymptotics of fields in terms of the observer's multipole moments. This
yields a combination of outgoing and dS-Hawking modes with relative coefficients having the right thermal factors.

Computing the bulk on-shell action for a scalar field with appropriate counterterms, we showed in part I that we get the cosmological influence phase that correctly encodes the physics of scalar self-force as well as thermal fluctuations in de Sitter. In that work, we also described how the high-frequency limit reproduces flat spacetime results and how the non-relativistic expansions can be resummed into dS-covariant self-force. This method can be thought of as the adaptation of real-time AdS/CFT techniques to the solipsistic picture of dS holography\cite{Anninos:2011af,Nakayama:2011qh}, which posits that the dual theory to a de Sitter static patch is a quantum mechanical system living on an observer's wordline at the centre of the static patch.

Our goal in this work is to generalise all these above statements to the case of electromagnetism. The first technical complication here is the appearance of vector spherical harmonics (VSHs) both in their spherical and cartesian avatars. We found that the existing literature on VSHs had many gaps that need to be addressed to solve our problem. We face this challenge head-on in our appendix \ref{app:VSH}, where the reader can find many new results about VSHs in arbitrary dimensions. They include a VSH version of the addition theorem, the relation between cartesian and spherical VSHs, and 
a set of toroidal operators in higher dimensions, which generalise the famous $-\vec{r}\times\vec{\nabla}$ operator in $\mathbb{R}^3$. Apart from these new results, we also review existing approaches to VSHs based on symmetric trace-free(STF) tensors as well as weight-shifting operators.

Once the technical machinery of VSHs is in place, the next step is to take the flat space EM multipole expansion and then see how it extends to de Sitter. Here we encounter our next obstacle. The existing literature on EM multipole expansions can be divided into two disjoint sets, one focused on spherical harmonic methods and the other using cartesian STF tensor methods. We found that the de Sitter problem requires an efficient mix of \emph{both} these methods. Even in flat spacetime, the conversion between these two methods is not clearly explained in the current literature. We address this lacuna in appendices \ref{app:FlatEMI} and \ref{app:FlatEMII}, which consolidate our knowledge about EM multipole expansions in flat spacetime.

In the main sections of the paper, we will restrict ourselves to explaining the physical ideas and relevant results of our calculations while relegating most of the technical details to the appendices. In section \S\ref{sec:EMindS}, we solve the electromagnetic field equations in the de Sitter static patch by mapping the electromagnetic fields to two scalar fields, analogous to the Hertz-Debye formalism in flat space. Section \S\ref{sec:renorm} conveys the main ideas of the paper: we outline a procedure to obtain the radiation reaction in a holographic prescription. This prescription is then justified by constructing the on-shell action in the dS-SK geometry, reproducing the same result. Finally, we obtain a further check on the radiation reaction by computing the Abraham-Lorentz-Dirac force in de Sitter as cosmological corrections to the flat space results in section \S\ref{sec:RR}. We end with a summary and discussion.  

\section{Electromagnetic Radiation in de Sitter}\label{sec:EMindS}

We will now summarise how to solve for radiative electromagnetic fields in the static patch of dS. Our analysis parallels the computation of \cite{Ishibashi:2004wx}, in that, we map the electromagnetic field equations to two scalar equations using spherical harmonic decomposition. The analogous scalars in flat space are referred to as Hertz-Debye potentials\footnote{ We review the flat space formalism in appendices \ref{app:FlatEMI} and \ref{app:FlatEMII} in our notation for the reader's convenience.}. Our notation will be chosen so that the formulae relate easily to their flat space analogues. We only convey essential formulae in this section, while a more extended analysis, including derivations of the formulae stated here, can be found in the appendix \ref{app:EMptsource}. 

We will work in the $d+1$ dimensional de Sitter static patch described by the following metric in the outgoing Eddington-Finklestein coordinate system:
\begin{equation}
    ds^2=-(1-r^2)du^2-2dudr+r^2d\Omega_{d-1}^2\ .
\end{equation}
This coordinate system covers the static patch of the south pole. The advantage of working with the outgoing coordinates is the ease of imposing retarded boundary conditions on our electromagnetic fields. We will exploit the spherical symmetry of the static patch to decompose the fields into harmonics on the sphere. We will also use the time-translation symmetry to work with the Fourier domain of $u$.

Given the spacetime, let us start with decomposing the Fourier transform of the electromagnetic field, $\mathcal{C}_{\mu\nu}(r,\w,\Omega)$, into spherical harmonics on the $S_{d-1}$. $\mathcal{C}_{\mu\nu}$ is an antisymmetric tensor field and hence admits the following decomposition: 
\begin{equation}
\begin{split}
    \FCf_{ru}(r,\w,\Omega)&\equiv \sum_{\ell\vec{m}}\Er(r,\w,\ell,\vec{m})\Yellm(\Omega) =  \FCf^{ur}(r,\w,\Omega)\ , \\
   \FCf_{rI}(r,\w,\Omega) &\equiv  \sum_{\ell\vec{m}} \Brs(r,\w,\ell,\vec{m}) \mathscr{D}_{I}\Yellm(\Omega)  + \sum_{\alpha\ell\vec{m}}\Brv(r,\w,\alpha,\ell,\vec{m})  \VSH_I(\Omega) = r^2\gamma_{IJ}\FCf^{Ju}(r,\w,\Omega) \ , \\
    \FCf_{Iu}(r,\w,\Omega) &= \sum_{\ell\vec{m}}\Es(r,\w,\ell,\vec{m}) \mathscr{D}_I \Yellm(\Omega) + \sum_{\alpha\ell\vec{m}} \Ev(r,\w,\alpha,\ell,\vec{m})   \VSH_I(\Omega)\ , \\
    \FCf_{IJ}(r,\w,\Omega) &\equiv \sum_{\alpha\ell\vec{m}} \Bvv(r,\w,\alpha,\ell,\vec{m}) \left[\mathscr{D}_I\VSH_J(\Omega)-\mathscr{D}_J \VSH_I(\Omega)\right] = r^4 \gamma_{IK} \gamma_{JL} \FCf^{KL}(r,\w,\alpha,\ell,\vec{m})\ ,
\end{split}
\end{equation}
where $\Yellm$ and $\mathbb{V}^{\alpha\ell\vec{m}}$ are scalar and divergenceless vector spherical harmonics\footnote{The $\VSH$ on $S_{d-1}$ are described in detail in appendix \ref{ssec:totoidalOp}. The appendix includes many explicit forms for these vector spherical harmonics: the generic expression in arbitrary dimensions can be found in eq.\eqref{eq:VSHdef}, whereas table \ref{tab:VSHexp} lists the explicit forms on $\mathbb{S}^2$ up to $\mathbb{S}^5$.} on $S_{d-1}$ respectively. The metric on the $S_{d-1}$ sphere is denoted by $\gamma_{IJ}$, and $\mathscr{D}_I$ is its corresponding covariant derivative. The $\mathscr{D}_I\Yellm$ are conventionally referred to as having `electric' parity while the $\VSH$ are regarded as `magnetic' parity\cite{Thorne:1980ru}. The difference in the parity eigenvalues forces the fields to decouple into electric and magnetic sectors, which do not mix. This decoupling will become further obvious as we solve the field equations. For convenience, we will also define time reversal covariant combinations of the $\mathcal{B}$'s and $\mathcal{E}$'s as follows to make the time-reversal symmetry of our equations explicit:
\begin{equation}
\begin{split}
    \Hs &\equiv (1-r^2) \Brs + \Es \ ,\quad 
    \Hv \equiv (1-r^2) \Brv + \Ev\ .
\end{split}
\end{equation}
To surmise, all the $\mathcal{E}$'s are time-reversal even while the $\mathcal{H}$'s are time-reversal odd, whereas $\{\Er,\Es,\Hs\}$ have electric parity and $\{\Ev,\Hv,\Bvv\}$ have magnetic parity.

In the absence of sources, the Maxwell equations can now be solved by mapping these electromagnetic fields to two scalars $\Phi_E$ and $\Phi_B$ (named as such due to their behaviour under parity). The mapping of the fields in terms of the scalars is as follows:  
\begin{equation}
\begin{split}
\Er(r,\w,\ell,\vec{m}) &= \frac{\ell(\ell+d-2)}{r^{d-1}}\Phi_E(r,\w,\ell,\vec{m}) \ ,\\
\Es(r,\w,\ell,\vec{m}) = \frac{1}{r^{d-3}}\Dp \Phi_E(r,\w,\ell,\vec{m}) \ &, \
\Ev(r,\w,\alpha,\ell,\vec{m}) = i\w \Phi_B(r,\w,\alpha,\ell,\vec{m})\ , \\
\Hs(r,\w,\ell,\vec{m}) = \frac{i\w }{r^{d-3}}\Phi_E(r,\w,\ell,\vec{m}) \ &, \ \Hv(r,\w,\alpha,\ell,\vec{m})  = \Dp \Phi_B(r,\w,\alpha,\ell,\vec{m}) \ ,\\ \Bvv(r,\w,\alpha,\ell,\vec{m})  &= \Phi_B(r,\w,\alpha,\ell,\vec{m}) \ .
%,\\ \Brs(r,\w,\ell,\vec{m}) = -\frac{1}{r^{d-3}}\partial_r \Phi_E(r,\w,\ell,\vec{m})\ &,\
%\Brv(r,\w,\alpha,\ell,\vec{m})  = \partial_r \Phi_B(r,\w,\alpha,\ell,\vec{m}) \ . \\
\end{split}
\end{equation}
Given such a mapping, one can easily check that the sourceless Maxwell equations are satisfied if the two scalars satisfy the following equations:
\begin{equation}
\begin{split}\label{eq:FreeDebyeEOMs}
    \frac{1}{r^{3-d}}D_+\left[r^{3-d}D_+\Phi_E\right]&+\omega^2\Phi_E-(1-r^2)\frac{\ell(\ell+d-2)}{r^2}\Phi_E=0\ ,\\
    \frac{1}{r^{d-3}}D_+\left[r^{d-3}D_+\Phi_B\right]&+\omega^2\Phi_B-(1-r^2)\frac{(\ell+1)(\ell+d-3)}{r^2}\Phi_B=0\ .
\end{split}
\end{equation}
This analysis maps the problem of understanding electromagnetism in de Sitter to that of two scalar fields satisfying the above equations of motion. In part I\cite{Loganayagam:2023pfb}, we studied a generic class of scalars parametrised by two numbers $\nn$ and $\mu$, which for the Debye potentials take specific values:$\{\nn=3-d,\mu=\frac{d}{2}-1\}$ for the electric sector and $\{\nn=d-3,\mu=\frac{d}{2}-2\}$ for the magnetic counterpart. Some of the analyses for these scalars follow similarly to part I but with two crucial differences. Part I analysed the scalars satisfying Dirichlet boundary conditions, whereas $\Phi_E$ satisfies Neumann boundary conditions. And even though $\Phi_B$ does satisfy Dirichlet boundary conditions, the fact that it appears along with vector spherical harmonics, rather than scalar ones, introduces new aspects to the multipole expansion as well as its appearance in the self-force. We will discuss these aspects further in the next section.

\section{Holographic renormalisation and self-force regularisation}\label{sec:renorm}

In this section, we will propose our prescription for obtaining the self-force in two different manners: first, we give a Son-Starinets-like prescription for computing the self-force as renormalised boundary correlators. Then, we show how one can obtain the same self-force as an on-shell action computed on a complex saddle geometry. One should contrast these two approaches analogously to the extrapolate/BDHM\cite{Banks:1998dd} vs differentiate/GKPW\cite{Gubser:1998bc} dictionaries of AdS/CFT\footnote{See\cite{Harlow:2011ke} for a comparison of the two dictionaries in both AdS as well as dS. Their dS analysis shows disagreement in the two approaches when applied to the case of dS-CFT correspondence\cite{Strominger:2001pn}, where the boundary dual lives at spacelike future infinity. }. The fact that both these approaches agree on the self-force acts as an additional check on our prescription.

We will first begin with describing the Son-Starinets prescription for de Sitter. For any localised source, one can think of the fields generated by the source as coming from electric and magnetic multipole moments $\multj^E(\w,\ell,\vec{m})$ and $\multj^B(\w,\alpha,\ell,\vec{m})$ specified at the origin. The multipole moments are defined by fixing the behaviour of the tangential electric field and the radial magnetic field at the origin $r=0$ of the static patch:
\begin{equation}\begin{split}
     \mathcal{J}^E(\w,\ell,\vec{m}) &\equiv  -\lim_{r\to 0}\ r^{\ell+d-2}\Es(r,\w,\ell,\vec{m}) \ ,\\
    \mathcal{J}^B(\w,\alpha,\ell,\vec{m}) &\equiv \lim_{r\to 0}\ r^{\ell+d-3}\Bvv(r,\w,\alpha,\ell,\vec{m}) \ .  
\end{split}\end{equation}

Once we have fixed these boundary conditions at $r=0$, one can solve for the Hertz-Debye scalars using the boundary to bulk propagators $\Gout_{E/B}$ such that \eqref{eq:FreeDebyeEOMs} are satisfied. The `out' superscript implies that we have imposed outgoing boundary conditions at the horizon. We write down the solutions in terms of these propagators as:  
\begin{equation}
\begin{split}
    \Phi_E(r,\w,\ell,\vec{m}) &=\frac{1}{\ell}\Gout_E(r,\w,\ell)\mathcal{J}^E(\w,\ell,\vec{m}) \ ,\\ 
    \Phi_B(r,\w,\alpha,\ell,\vec{m}) &=\Gout_B(r,\w,\ell)\mathcal{J}^B(\w,\alpha,\ell,\vec{m})\ .
\end{split}
\end{equation}
The $\Gout_{E/B}$ can be written down explicitly in terms of hypergeometric functions regular at the horizon:
\begin{equation}
	\begin{split}
  &G^{\text{Out}}_E(r,\w,\nu)=\frac{r^{\nu+\frac{d}{2}-1}(1+r)^{-i\w}}{\Gamma(1-i\w)\Gamma\left(\nu \right)}
		\\
		&\ \times \Gamma\left(\frac{\nu-
				\frac{d}{2}+3-i\w}{2}\right)\Gamma\left(\frac{\nu+\frac{d}{2}-1-i\w}{2}\right)\ {}_2F_1\left[\frac{\nu-
			\frac{d}{2}+3-i\w}{2},\frac{\nu+\frac{d}{2}-1-i\w}{2};1-i\w;1-r^2\right]\ ,\\
   	&G^{\text{Out}}_B(r,\w,\nu)=\frac{r^{\nu-\frac{d}{2}+2}(1+r)^{-i\w}}{\Gamma(1-i\w)\Gamma\left(\nu \right)}
		\\
		&\ \times \Gamma\left(\frac{\nu-
				\frac{d}{2}+2-i\w}{2}\right)\Gamma\left(\frac{\nu+\frac{d}{2}-i\w}{2}\right)\ {}_2F_1\left[\frac{\nu-
			\frac{d}{2}+2-i\w}{2},\frac{\nu+\frac{d}{2}-i\w}{2};1-i\w;1-r^2\right]\ .
	\end{split}
\end{equation}
where we have defined $\nu\equiv \ell+\frac{d}{2}-1$ for notational convenience. These functions play the same role as reverse Bessel polynomials in 3+1 dimensional flat spacetime\footnote{In appendix \ref{app:dSBesselPoly}, 
we give explicit polynomial expressions for $\Gout_{E/B}$(tables \ref{tab:GoutM1}, \ref{tab:GoutM2}, \ref{tab:GoutE1} and \ref{tab:GoutE2}) which reduce to reverse Bessel polynomial in the flat $d=3$ limit. }. We can now write the outgoing solutions for the electromagnetic fields as follows:
\begin{equation}
\begin{split}
\Er &= (\ell+d-2)\Gout_E(r,\w,\nu)\frac{\mathcal{J}^E}{r^{d-1}} \ ,\
\Ev =i\w \Gout_B(r,\w,\nu)\mathcal{J}^B\ ,  \\
\Es &= \left[\left\{\frac{(1-r^2)(2\nu+d-2)}{2r}+i\w r\right\}\Gout_E(r,\w,\nu)-2\nu(1-r)\Gout_E(r,\w+i,\nu+1)\right]\frac{\mathcal{J}^E}{\ell r^{d-3}}\ ,   \\
\Hs &= \Gout_{E}(r,\w,\nu)\frac{i\w \mathcal{J}^E}{\ell r^{d-3}} \ , \  \Bvv =\Gout_{B}(r,\w,\nu)\mathcal{J}^B\ ,\\
\Hv &=\left[\left\{\frac{(1-r^2)(2\nu-d+4)}{2r}+i\w r\right\}\Gout_E(r,\w,\nu)- 2\nu(1-r)\Gout_E(r,\w+i,\nu+1)\right]\mathcal{J}^B\ .
\end{split}
\end{equation}

Given these fields, it is now easy to extract out information about the radiation reaction. We claim that the boundary 2-point function that encodes the radiation reaction is obtained by renormalising the radial electric and tangential magnetic fields such that they have a finite behaviour as $r\to 0$. 
\begin{equation}\begin{split}
    \lim_{r\to 0}r^{1-\ell}\Er^{\text{ren}}=\lim_{r\to 0}r^{1-\ell}\left[\Er+\text{(counter-term proportional to $\Es$)}\right] &=\text{Radiation reaction on $\mathcal{J}^E$}\ ,\\
     \lim_{r\to 0}r^{1-\ell}\Hv^{\text{ren}}=\lim_{r\to 0}r^{-\ell}\left[\Hv+\text{(counter-term proportional to $\Bvv$)}\right] &=\text{Radiation reaction on $\mathcal{J}^B$}\ . 
\end{split}\end{equation}
This counterterming procedure is detailed in the appendix \ref{app:EMptsource}. In even spacetime dimensions, there are additional divergences which need to be countertermed further\footnote{See Eqns. \eqref{eq:EMfieldct} and \eqref{eq:EMcts} for counterterms required for all dimensions, whereas, Eqns. \eqref{eq:PhiActCtEven} and \eqref{eq:ctEvenH} specify counterterms specific to even dimensions.}. Once we have added appropriate counterterms to cancel the divergences in either odd or even dimensions, we obtain the boundary limits:
\begin{equation}\begin{split}
    \lim_{r\to 0}r^{1-\ell}\Er^{\text{ren}}&=-\frac{\ell+d-2}{\ell}\Kout_E(\w,\ell) \multj^E\ ,\\
     \lim_{r\to 0}r^{1-\ell}\Hv^{\text{ren}}&= -\Kout_B(\w,\ell) \multj^B\ . 
\end{split}\end{equation}
The functions $\Kout_{E/B}$ encode the radiation reaction kernel. This fact is made clear in further analysis, where we will show how it controls the decay time for multipole moments in the large time limit and also reproduces the Hubble corrections to the Abraham-Lorentz-Dirac force in a PN expansion. For now, we will quote the explicit expressions. In  dS$_{d+1}$ with even spacetime dimensions, we have
\begin{equation}\label{eq:KOutOddmain}
\begin{split}
	\Kout_E|_{\text{Odd d}} &
   =-e^{i\nu\pi}\frac{2\pi i}{\Gamma(\nu)^2} \frac{\Gamma\left(\frac{3-\frac{d}{2}+\nu-i\w}{2}\right)\Gamma\left(\frac{-1+\frac{d}{2}+\nu-i\w}{2}\right)}{\Gamma\left(\frac{3-\frac{d}{2}-\nu-i\w}{2}\right)\Gamma\left(\frac{-1+\frac{d}{2}-\nu-i\w}{2}\right)}\ , \\
   \Kout_B|_{\text{Odd d}} &= -e^{i\nu\pi}\frac{2\pi i}{\Gamma(\nu)^2} \frac{\Gamma\left(\frac{2-\frac{d}{2}+\nu-i\w}{2}\right)\Gamma\left(\frac{\frac{d}{2}+\nu-i\w}{2}\right)}{\Gamma\left(\frac{2-\frac{d}{2}-\nu-i\w}{2}\right)\Gamma\left(\frac{\frac{d}{2}-\nu-i\w}{2}\right)}\ .
	\end{split}
\end{equation}
The above expressions are polynomials in $\w$, which signifies the markovianity of the radiation reaction. We give explicit expressions of these polynomials for a few values of $d$ and $\ell$ in Table \ref{tab:KoutMark} in appendix \ref{app:EMptsource}.

The radiation reaction in odd-dimensional spacetimes is non-markovian. Explicitly, we have:
\begin{equation}\label{eq:KoutEvenmain}
	\begin{split}
		\Kout_E|_\text{Even $d$} &=\Delta\left(\nu,\frac{d}{2}-2,\w\right)\left[\psi^{(0)}\left(\frac{3-\frac{d}{2}+\nu-i\w}{2}\right)+\psi^{(0)}\left(\frac{-1+\frac{d}{2}+\nu-i\w}{2}\right)\right.\\ &\left.+\psi^{(0)}\left(\frac{3-\frac{d}{2}-\nu-i\w}{2}\right)+\psi^{(0)}\left(\frac{-1+\frac{d}{2}-\nu-i\w}{2}\right)-4\psi^{(0)}(\nu)\right]\ , \\
        \Kout_B|_\text{Even $d$} &=\Delta\left(\nu,\frac{d}{2}-1,\w\right)\left[\psi^{(0)}\left(\frac{2-\frac{d}{2}+\nu-i\w}{2}\right)+\psi^{(0)}\left(\frac{\frac{d}{2}+\nu-i\w}{2}\right)\right.\\ &\left.+\psi^{(0)}\left(\frac{2-\frac{d}{2}-\nu-i\w}{2}\right)+\psi^{(0)}\left(\frac{\frac{d}{2}-\nu-i\w}{2}\right)-4\psi^{(0)}(\nu)\right]\ ,
	\end{split}
\end{equation}
where,
\begin{equation}
\begin{split}
   \Delta(n,\mu,\w)&\equiv
  \frac{1}{\Gamma(n)^2}\prod\limits_{k=1}^{n}\left[\frac{\w^2}{4}+\frac{1}{4}(\mu-n+2k-1)^2\right]\ ,
\end{split}
\end{equation} 
and $\psi^{(0)}$ is digamma function. 
In this case, the radiation reaction is no longer markovian for generic $\w$. Nevertheless, one can take a long-time expansion of the above expression, which is well-defined, i.e. there are no poles at $\w=0$. This signifies the emergence of markovianity at cosmological time scales. There are however poles at some imaginary values of $\w$ coming from  digamma poles  not cancelled by the zeros of the $\Delta$ \footnote{These poles are sometimes called the `quasinormal modes' of the static patch. As we pointed out in part I, the poles corresponding to purely imaginary frequencies are more like Matsubara modes, and unlike quasinormal modes which have a non-zero real part that sets the frequency of ringdown.}. These poles of the radiation reaction kernels then indicate the breakdown of the long-time expansion at sub-horizon wavelengths and thus also the breakdown of markovianity. The non-markovian behaviour takes over at large frequencies, or equivalently in a flat space limit where, in odd-dimensional spacetime, one sees a non-local radiation reaction generically\cite{Birnholtz:2013ffa,Birnholtz:2013nta,Harte:2018iim}. In contrast, even dimensional flat spacetimes have a local electromagnetic radiation reaction. This is consistent with the behaviour in the flat limit:   
\begin{align}
\begin{split}
\lim_{\w\to \infty}\Kout_{E/B} = \left\{\begin{array}{cc} \frac{2\pi i}{\Gamma(\nu)^2}\left(\frac{\w}{2}\right)^{2\nu} & \text{for $d$  odd ,} \\
\frac{1}{ \Gamma(\nu)^2} \left(\frac{\w }{2}\right)^{2\nu} \ln \left(\frac{\w^4}{H^4}\right) & \text{for $d$  even .} \\
\end{array}\right.
\end{split}
\end{align}

What we have described till now is a Son-Starinets-like prescription\cite{Son:2002sd} to obtain the dS radiation reaction: we fix the gauge field $\gaugeV$ at the boundary($r=0$) and impose outgoing conditions at the horizon. The renormalised values of the conjugate fields $\mathcal{C}^{r\mu}$ then yield the radiation reaction kernels. Next, we will try to justify this prescription by explicitly computing the on-shell action on a doubled geometry, which does not involve any specific boundary conditions imposed in the bulk except for the continuity of solutions: a de Sitter version of the GKPW prescription.

Let us then solve the Debye scalar equations on the dS-SK geometry. Once the solutions are obtained, we can substitute them into the electromagnetic action and regularise it to obtain the open effective action of the observer. To begin with, we recall the dS-SK geometry described in part I: we extend the $r$ coordinate to a complex plane and define our geometry as a codimension 1 contour in the $r$ plane. This contour is parameterised by the `mock tortoise coordinate' $\zeta$ defined as:
\begin{equation}
	\zeta(r)=\frac{1}{i\pi}\int\limits^{0-i\epsilon}_r\frac{dr'}{1-r'^2}=\frac{1}{2\pi i}\ln\left(\frac{1-r}{1+r}\right)\ . \label{eq:zeta-def-main}
\end{equation}
$\zeta$ has logarithmic branch points at $r=\pm 1$ and is normalised to jump by $+1$ when you cross the branch cut from below to above.

	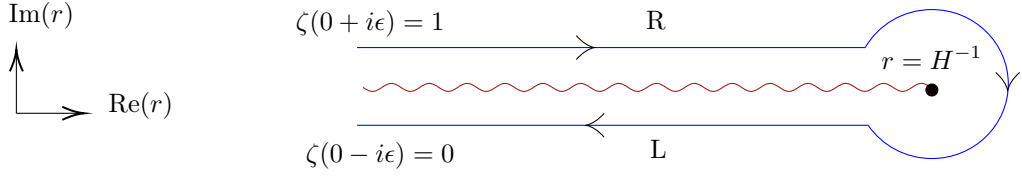
\begin{figure}[H]\label{fig:SKcontour}
		\centering
	\begin{tikzpicture}[x=0.75pt,y=0.75pt,yscale=-1,xscale=1]

%Straight Lines [id:da15508095782005948] 
\draw    (108,153) -- (108,123) ;
\draw [shift={(108,121)}, rotate = 90] [color={rgb, 255:red, 0; green, 0; blue, 0}  ][line width=0.75]    (10.93,-3.29) .. controls (6.95,-1.4) and (3.31,-0.3) .. (0,0) .. controls (3.31,0.3) and (6.95,1.4) .. (10.93,3.29)   ;
%Straight Lines [id:da773933967301495] 
\draw    (108,153) -- (140,153) ;
\draw [shift={(142,153)}, rotate = 180] [color={rgb, 255:red, 0; green, 0; blue, 0 }  ][line width=0.75]    (10.93,-3.29) .. controls (6.95,-1.4) and (3.31,-0.3) .. (0,0) .. controls (3.31,0.3) and (6.95,1.4) .. (10.93,3.29)   ;
%Shape: Wave [id:dp03668757509174492] 
\draw  [color={rgb, 255:red, 155; green, 25; blue, 25 }  ,draw opacity=1 ] (282,140) .. controls (283.55,141.02) and (285.03,142) .. (286.75,142) .. controls (288.47,142) and (289.95,141.02) .. (291.5,140) .. controls (293.05,138.98) and (294.53,138) .. (296.25,138) .. controls (297.97,138) and (299.45,138.98) .. (301,140) .. controls (302.55,141.02) and (304.03,142) .. (305.75,142) .. controls (307.47,142) and (308.95,141.02) .. (310.5,140) .. controls (312.05,138.98) and (313.53,138) .. (315.25,138) .. controls (316.97,138) and (318.45,138.98) .. (320,140) .. controls (321.55,141.02) and (323.03,142) .. (324.75,142) .. controls (326.47,142) and (327.95,141.02) .. (329.5,140) .. controls (331.05,138.98) and (332.53,138) .. (334.25,138) .. controls (335.97,138) and (337.45,138.98) .. (339,140) .. controls (340.55,141.02) and (342.03,142) .. (343.75,142) .. controls (345.47,142) and (346.95,141.02) .. (348.5,140) .. controls (350.05,138.98) and (351.53,138) .. (353.25,138) .. controls (354.97,138) and (356.45,138.98) .. (358,140) .. controls (359.55,141.02) and (361.03,142) .. (362.75,142) .. controls (364.47,142) and (365.95,141.02) .. (367.5,140) .. controls (369.05,138.98) and (370.53,138) .. (372.25,138) .. controls (373.97,138) and (375.45,138.98) .. (377,140) .. controls (378.55,141.02) and (380.03,142) .. (381.75,142) .. controls (383.47,142) and (384.95,141.02) .. (386.5,140) .. controls (388.05,138.98) and (389.53,138) .. (391.25,138) .. controls (392.97,138) and (394.45,138.98) .. (396,140) .. controls (397.55,141.02) and (399.03,142) .. (400.75,142) .. controls (402.47,142) and (403.95,141.02) .. (405.5,140) .. controls (407.05,138.98) and (408.53,138) .. (410.25,138) .. controls (411.97,138) and (413.45,138.98) .. (415,140) .. controls (416.55,141.02) and (418.03,142) .. (419.75,142) .. controls (421.47,142) and (422.95,141.02) .. (424.5,140) .. controls (426.05,138.98) and (427.53,138) .. (429.25,138) .. controls (430.97,138) and (432.45,138.98) .. (434,140) .. controls (435.55,141.02) and (437.03,142) .. (438.75,142) .. controls (440.47,142) and (441.95,141.02) .. (443.5,140) .. controls (445.05,138.98) and (446.53,138) .. (448.25,138) .. controls (449.97,138) and (451.45,138.98) .. (453,140) .. controls (454.55,141.02) and (456.03,142) .. (457.75,142) .. controls (459.47,142) and (460.95,141.02) .. (462.5,140) .. controls (464.05,138.98) and (465.53,138) .. (467.25,138) .. controls (468.97,138) and (470.45,138.98) .. (472,140) .. controls (473.55,141.02) and (475.03,142) .. (476.75,142) .. controls (478.47,142) and (479.95,141.02) .. (481.5,140) .. controls (483.05,138.98) and (484.53,138) .. (486.25,138) .. controls (487.97,138) and (489.45,138.98) .. (491,140) .. controls (492.55,141.02) and (494.03,142) .. (495.75,142) .. controls (497.47,142) and (498.95,141.02) .. (500.5,140) .. controls (502.05,138.98) and (503.53,138) .. (505.25,138) .. controls (506.97,138) and (508.45,138.98) .. (510,140) .. controls (511.55,141.02) and (513.03,142) .. (514.75,142) .. controls (516.47,142) and (517.95,141.02) .. (519.5,140) .. controls (521.05,138.98) and (522.53,138) .. (524.25,138) .. controls (525.97,138) and (527.45,138.98) .. (529,140) .. controls (530.55,141.02) and (532.03,142) .. (533.75,142) .. controls (535.47,142) and (536.95,141.02) .. (538.5,140) .. controls (540.05,138.98) and (541.53,138) .. (543.25,138) .. controls (544.97,138) and (546.45,138.98) .. (548,140) .. controls (549.55,141.02) and (551.03,142) .. (552.75,142) .. controls (554.47,142) and (555.95,141.02) .. (557.5,140) .. controls (559.05,138.98) and (560.53,138) .. (562.25,138) .. controls (563.97,138) and (565.45,138.98) .. (567,140) ;
%Straight Lines [id:da06962863897415716] 
\draw  [color={rgb, 255:red, 0; green, 0; blue, 255}  ]  (279,159.02) -- (535.51,159.02) ;
%Straight Lines [id:da7434676126846234] 
\draw   [color={rgb, 255:red, 0; green, 0; blue, 255}  ]  (279,120) -- (534,120) ;
%Shape: Arc [id:dp3995904585889173] 
\draw  [color={rgb, 255:red, 0; green, 0; blue, 255}  ] [draw opacity=0] (534,120) .. controls (540.57,108.6) and (553.1,100.91) .. (567.47,100.91) .. controls (588.66,100.91) and (605.83,117.65) .. (605.83,138.31) .. controls (605.83,158.96) and (588.66,175.71) .. (567.47,175.71) .. controls (554.13,175.71) and (542.39,169.08) .. (535.51,159.02) -- (567.47,138.31) -- cycle ; \draw  [color={rgb, 255:red, 0; green, 0; blue, 255}  ]  (534,120) .. controls (540.57,108.6) and (553.1,100.91) .. (567.47,100.91) .. controls (588.66,100.91) and (605.83,117.65) .. (605.83,138.31) .. controls (605.83,158.96) and (588.66,175.71) .. (567.47,175.71) .. controls (554.13,175.71) and (542.39,169.08) .. (535.51,159.02) ;  
%Shape: Circle [id:dp343518095513652] 
\draw  [fill={rgb, 255:red, 15; green, 1; blue, 1 }  ,fill opacity=1 ] (564.47,141.31) .. controls (564.47,139.65) and (565.81,138.31) .. (567.47,138.31) .. controls (569.12,138.31) and (570.47,139.65) .. (570.47,141.31) .. controls (570.47,142.96) and (569.12,144.31) .. (567.47,144.31) .. controls (565.81,144.31) and (564.47,142.96) .. (564.47,141.31) -- cycle ;
\draw   (403,165) .. controls (400,161.67) and (397,159.67) .. (394,159) .. controls (397,158.33) and (400,156.33) .. (403,153) ;
\draw   (389,114) .. controls (392,117.33) and (395,119.33) .. (398,120) .. controls (395,120.67) and (392,122.67) .. (389,126) ;
\draw   (611.5,132.5) .. controls (608.17,135.5) and (606.17,138.5) .. (605.5,141.5) .. controls (604.83,138.5) and (602.83,135.5) .. (599.5,132.5) ;

% Text Node
\draw (153,141) node [anchor=north west][inner sep=0.75pt]   [align=left] {Re($\displaystyle r$)};
% Text Node
\draw (103,96) node [anchor=north west][inner sep=0.75pt]   [align=left] {Im($\displaystyle r$)};
% Text Node
\draw (252,166) node [anchor=north west][inner sep=0.75pt]   [align=left] {$\displaystyle \zeta ( 0-i\epsilon ) =0$};
% Text Node
\draw (247,100) node [anchor=north west][inner sep=0.75pt]   [align=left] {$\displaystyle \zeta ( 0+i\epsilon ) =1$};
% Text Node
\draw (541,118) node [anchor=north west][inner sep=0.75pt]   [align=left] {$\displaystyle r=H^{-1}$};
% Text Node
\draw (422.83,100) node [anchor=north west][inner sep=0.75pt]   [align=left] {R};
% Text Node
\draw (424.83,165) node [anchor=north west][inner sep=0.75pt]   [align=left] {L};

\end{tikzpicture}

	\caption{The branch cut structure of $\zeta(r)$ in the complex $r$ plane at fixed $u$: branch-cut shown as a wiggly line. We also show the \emph{clockwise} dS-SK radial contour running from $\zeta=1$ to $\zeta=0$ (the blue curve in this figure and in Fig.\ref{fig:ds-sk}). The $\text{Im}\ r>0$ branch is the time-ordered/right branch, whereas the $\text{Im}\ r<0$ branch is the anti-time-ordered/left branch. }
	\end{figure}

        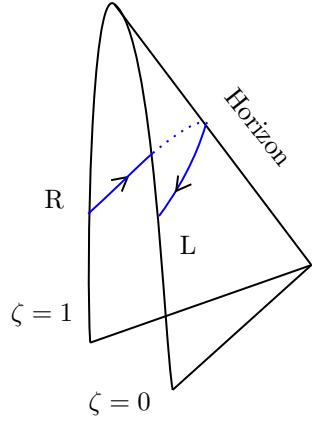
\begin{figure}[H]
\centering

\tikzset{every picture/.style={line width=0.75pt}} %set default line width to 0.75pt        

\begin{tikzpicture}[x=0.75pt,y=0.75pt,yscale=-1,xscale=1]

%Curve Lines [id:da5386802446200184] 
\draw    (315.84,42.81) .. controls (304.16,33.61) and (301.59,176.33) .. (303.78,212.86) ;
%Curve Lines [id:da21136090060260693] 
\draw    (315.84,42.81) .. controls (335.97,57.16) and (340.84,212.54) .. (345.17,236.57) ;
%Straight Lines [id:da8303619831252543] 
\draw    (414.79,173.88) -- (345.17,236.57) ;
%Straight Lines [id:da8514730172738603] 
\draw    (303.78,212.86) -- (414.79,173.88) ;
%Straight Lines [id:da7271233165258716] 
\draw    (315.84,42.81) -- (414.79,173.88) ;
%Curve Lines [id:da017569219223964372] 
\draw [color={rgb, 255:red, 0; green, 0; blue, 255 }  ,draw opacity=1 ]   (361.74,103.56) .. controls (359.65,110.82) and (353.52,128.79) .. (338.31,149.29) ;
\draw   (314.27,131.27) -- (322.57,129.69) -- (319.61,136.98) ;
\draw   (354.56,133.31) -- (346.93,136.67) -- (347.91,128.93) ;
%Curve Lines [id:da20387429499738996] 
\draw [color={rgb, 255:red, 0; green, 0; blue, 255 }  ,draw opacity=1 ]   (302.78,148.09) .. controls (321.32,131.21) and (327.35,124.41) .. (335.2,117.67) ;
%Curve Lines [id:da8101518038983148] 
\draw [color={rgb, 255:red, 0; green, 0; blue, 255 }  ,draw opacity=1 ] [dash pattern={on 0.84pt off 2.51pt}]  (335.2,117.67) .. controls (365.64,92.18) and (363.61,106.45) .. (361.74,103.56) ;

% Text Node
\draw (347.22,157.7) node [anchor=north west][inner sep=0.75pt]   [align=left] {L};
% Text Node
\draw (279.33,134.37) node [anchor=north west][inner sep=0.75pt]   [align=left] {R};
% Text Node
\draw (378.09,81.48) node [anchor=north west][inner sep=0.75pt]  [rotate=-51.47] [align=left] {Horizon};
% Text Node
\draw (301.41,236) node [anchor=north west][inner sep=0.75pt]   [align=left] {$\displaystyle \zeta =0$};
% Text Node
\draw (262.41,191) node [anchor=north west][inner sep=0.75pt]   [align=left] {$\displaystyle \zeta =1$};

\end{tikzpicture}
\caption{The two sheeted complex dS-SK geometry can be thought of as two static patches smoothly connected at the future horizon. The radial contour along an outgoing Eddington-Finkelstein slice (i.e., a constant $u$ slice) is shown in blue. The radial contour has an outgoing R branch and an incoming L branch.}
\label{fig:ds-sk}
\end{figure}

To solve for the fields on the dS-SK geometry, we require both sets of solutions: those which are outgoing at the horizon as well as those which are incoming. Given the outgoing propagators that we have already constructed, obtaining the incoming propagators is a matter of simple time reversal obtained by $\w\to-\w$ and $u\to 2\pi i \zeta-u$. This leads to the following expression for the incoming propagator: 
\begin{equation}
    \Gin_{E/B}=e^{-2\pi\w \zeta} \Gouts_{E/B}\ .
\end{equation}
The incoming waves have a branch cut inherited from $\zeta$. Hence, the incoming propagators pick up a factor of $e^{2\pi\w}$ when the argument goes from above the cut to below the cut on the $r$-plane. Since these modes coming in from the horizon correspond to Hawking radiation, the fact that the geometry encodes this monodromy is crucial to reproduce appropriate Bose-Einstein factors at the de Sitter temperature in our expressions.  

Given the outgoing/incoming waves, we can now impose appropriate boundary conditions on the scalars on either side of the dS-SK geometry. We have,
\begin{equation}\label{eq:phN_FPbasis}
\begin{split}
	\Phi_E(\zeta,\w,\ell,\vec{m})=\frac{1}{\ell}\left\{-\Gout_E(r,\w,\ell)\JFb^E(\w,\ell,\vec{m})+e^{2\pi\w(1-\zeta)}\Gouts_E(r,\w,\ell)\JPb^E(\w,\ell,\vec{m})\right\}\ ,\\
    \Phi_B(\zeta,\w,\alpha,\ell,\vec{m})=-\Gout_B(r,\w,\ell)\JFb^B(\w,\alpha,\ell,\vec{m})+e^{2\pi\w(1-\zeta)}\Gouts_B(r,\w,\ell)\JPb^B(\w,\alpha,\ell,\vec{m})\ ,
\end{split}
\end{equation}
where $\JFb^{E/B}$ and $\JPb^{E/B}$ are sources that radiate to the future and detect the fields from the past, respectively (see below for a definition). Even though the forms of $\Phi_E$ and $\Phi_B$  look similar, we should think of them as being obtained via different boundary conditions. We impose a double Neumann boundary condition on either side of the contour for the electric Debye potential, while a double Dirichlet boundary condition for the magnetic counterpart, i.e., we take
\begin{equation}
\begin{split}
    \lim_{\zeta\to 0}r^{\ell+1}\Dp\Phi_E = \multj^E_L\ ,& \qquad \lim_{\zeta\to 1}r^{\ell+1}\Dp\Phi_E = \multj^E_R \ ,\\
    \lim_{\zeta\to 0}r^{\ell+d-3}\Phi_B = \multj^B_L\ ,& \qquad \lim_{\zeta\to 1}r^{\ell+d-3}\Phi_B = \multj^B_R\ .
\end{split}
\end{equation}

These boundary conditions are equivalent to the following relations between the $F-P$ basis sources and $R-L$ sources on either side of the contour:
\begin{align}\label{eq:jPjF}
	\begin{split}
		\JFb(\w,\ell,\vec{m})&\equiv-\Bigl\{(1+n_\w)\multj_R(\w,\ell,\vec{m})-n_\w \multj_L(\w,\ell,\vec{m})\Bigr\}\\
        & =-\multj_A(\w,\ell,\vec{m})-\left(n_\w+\frac{1}{2}\right)\multj_D(\w,\ell,\vec{m})\ ,\\
		\JPb(\w,\ell,\vec{m})&\equiv-n_\w\Bigl\{\multj_R(\w,\ell,\vec{m})- \multj_L(\w,\ell,\vec{m})\Bigr\} \\ &=-n_\w\ \multj_D(\w,\ell,\vec{m})\ .
	\end{split}
\end{align}
Here we have introduced the average/difference sources $\multj_A\equiv \frac{1}{2}\multj_R+\frac{1}{2}\multj_L$ and $\multj_D\equiv\multj_R-\multj_L$, and $n_\w$ is the Bose-Einstein factor: 
	\begin{align}
	\begin{split}
		n_\w\equiv \frac{1}{e^{2\pi\w}-1} \ .
	\end{split}
\end{align}
This factor appears in the formulae due to the dS-SK geometry naturally incorporating detailed balance in the system, as noted in part I. 

Now that we have understood the nature of the solutions to electromagnetic equations on dS-SK, let us consider the evaluation of the on-shell action. The electromagnetic action after imposing the Maxwell equations becomes:
\begin{equation}
\begin{split}
 S =  \left[-\frac{1}{2}\int r^{d-1} dt \ d\Omega_{d-1} \FCf^{r\mu}\gaugeV_\mu\right]^{r_c-i\epsilon}_{r_c+i\epsilon} +S_{ct} \ .
\end{split}
\end{equation}
Here, $r_c$ acts as a cut-off regulator. Once we have added the appropriate counterterms\footnote{The reader is referred to Appendix \ref{app:EMptsource} for a detailed analysis of the gauge-invariant counterterms. We show the explicit gauge invariance of the counterterm action on the dSSK geometry in Eqn. \eqref{eq:GaugeInvCT}.} and obtained a finite action, we are left with the \emph{cosmological influence phase} for electromagnetically interacting observer:
\begin{equation}\label{eq:SCIPpt}
\begin{split}
	\SCIP&=-\sum_{\alpha\ell\vec{m}}\int\frac{d\w}{2\pi} \Kout_B(\w,\ell)\  \multj_D^{B*}\ \Bigl[\multj^B_A+\left(n_\w+\frac{1}{2}\right)\multj_D^B\Bigr] \\
    &\quad -\sum_{\ell\vec{m}}\frac{\ell+d-2}{\ell}\int\frac{d\w}{2\pi} \Kout_E(\w,\ell)\  \multj_D^{E*}\ \Bigl[\multj_A^E+\left(n_\w+\frac{1}{2}\right)\multj_D^E\Bigr]\ .
\end{split}
\end{equation}
This action passes a variety of checks: 
\begin{enumerate}
    \item The Schwinger-Keldysh collapse rule states that the action should evaluate to zero when the sources on either side are set to be the same, i.e. $\multj_R=\multj_L$ or equivalently $\multj_D=0$. This constraint comes from the unitarity of the dynamics of the open system and the environment put together\cite{kamenev_2011}. Given that we obtain no $\multj_A^\ast\multj_A$ terms in the action, our action satisfies this condition.
    \item The dissipative part of the action, which is encoded in the $\multj_D^\ast\multj_A$ terms of the action, is proportionate to the $\multj_D^\ast\multj_A$ terms, which encode the fluctuations. These fluctuations come with the appropriate proportionality factor to satisfy the fluctuation-dissipation theorem at the de Sitter Hawking temperature. 
\end{enumerate}
In the long time limit, one can expand the $\Kout_{E/B}$ to obtain a markovian open system where the multipole moments decay at the rates set by the coefficient of $-i\w$ in the low-frequency expansion. 
The decay time scales of different multipole moments are given as:
\begin{equation}
\tau^E_{dS}=i\frac{\ell+d-2}{\ell}\left[\frac{d}{d\w}\Kout_E\right]_{\w=0}\ ,\qquad \tau^B_{dS}=i\left[\frac{d}{d\w}\Kout_B\right]_{\w=0}\ .
\end{equation}
The extra $\ell$ and $d$ dependent factor appearing in the expression for the electric multipole comes from the on-shell action computation Eq.\eqref{eq:SCIPpt}. We give the values of $\tau_{dS}^B$ and $\tau_{dS}^E$ in tables \ref{tab:tau1} and \ref{tab:tau2} for the readers convenience. 

Here, we have derived the on-shell action for an observer defined by multipole moments at $r=0$. Given a model of the localised observer, we can construct these moments by appropriately smearing the observer's current density. We will see how to accomplish this in the next section. This analysis will also allow us to obtain the Abraham-Lorentz-Dirac force in de Sitter by modelling the observer as a single charged particle moving along an arbitrary trajectory close to the south pole in the static patch.

\begin{table}[H]
	\centering
	\caption{Decay time-scales for magnetic multipole moments in units of Hubble time($\tau^B_{dS}$).}\label{tab:tau1}
	\setlength{\extrarowheight}{2pt}
	\begin{tabular}{|c|c|c|c|c|c|}
		\hline
		$\mu=\frac{d}{2}-1$&$\ell=1$&$\ell=2$&$\ell=3$&$\ell=4$&$\ell=5$\\[0.5ex]\hline
		$d=3$ & $1$ & $\frac{4}{9}$ & $\frac{4}{25}$ & $\frac{64}{1225}$&$ \frac{64}{3969}$
		\\[0.5ex]
        $d=4$ & $1$ & $\frac{9\pi^2}{256}$ & $\frac{1}{9}$ & $\frac{225\pi^2}{65536}$&$ \frac{1}{100}$
		\\[0.5ex]
		$d=5$& $1$ &$ \frac{64}{225}$ & $\frac{4}{49}$ &
		$\frac{256}{11025} $& $\frac{64}{9801}$\\[0.5ex]
        $d=6$ & $1$ & $\frac{25\pi^2}{1024}$ & $\frac{1}{16}$ & $\frac{441\pi^2}{262144}$&$ \frac{1}{225}$
		\\[0.5ex]
		$d=7$& $1$ &$ \frac{256}{1225} $& $\frac{4}{81}$ &
		$\frac{16384}{1334025}$& $\frac{64}{20449}$ \\[0.5ex]
        $d=8$  & $1$ & $\frac{1225\pi^2}{65536}$ & $\frac{1}{25}$ & $\frac{3969\pi^2}{4194304}$&$ \frac{1}{441}$
		\\[0.5ex]
		$d=9$  & $1$ & $\frac{16384}{99225}$ &
		$\frac{4}{121}$ & $\frac{65536}{9018009}$ & $\frac{64}{38025}$ \\[0.5ex]
  
        $d=10$ & $1$ & $\frac{25\pi^2}{1024}$ & $\frac{1}{16}$ & $\frac{441\pi^2}{262144}$&$ \frac{1}{225}$
		\\[0.5ex]
		$d=11$ & $1$ &$\frac{65536}{480249} $ &
		$\frac{4}{169}$ & $\frac{1048576}{225450225}$ &$ \frac{64}{65025}$\\[0.5ex] \hline
	\end{tabular}
\end{table}

\begin{table}[H]
	\centering
	\caption{Decay time-scales for electric multipole moments in units of Hubble time($\tau^E_{dS}$).}\label{tab:tau2}
	\setlength{\extrarowheight}{2pt}
	\begin{tabular}{|c|c|c|c|c|c|}
		\hline
		$\mu=\frac{d}{2}-2$&$\ell=1$&$\ell=2$&$\ell=3$&$\ell=4$&$\ell=5$\\[0.5ex]\hline
		$d=3$ & $2$ & $\frac{2}{3}$ & $\frac{16}{75}$ & $\frac{16}{245}$&$ \frac{128}{6615}$
		\\[0.5ex]
        $d=4$ &  $\frac{3\pi^2}{16}$ & $\frac{1}{2}$ & $\frac{15\pi^2}{1024}$ & $\frac{1}{24}$ & $\frac{315\pi^2}{262144}$
		\\[0.5ex]
		$d=5$ & $\frac{16}{9}$ &$  \frac{2}{5}$ & $\frac{128}{1225}$ &
		$\frac{16}{567} $& $\frac{2048}{266805}$\\[0.5ex]
        $d=6$ &  $\frac{45\pi^2}{256}$ & $\frac{1}{3}$ & $\frac{525\pi^2}{65536}$ & $\frac{1}{50}$ & $\frac{2205\pi^2}{4194304}$
		\\[0.5ex]
		$d=7$  & $\frac{128}{75}$ &$\frac{2}{7} $& $\frac{2048}{33075}$ &
		$\frac{16}{1089}$& $\frac{32768}{9018009}$ \\[0.5ex]
        $d=8$  &  $\frac{175\pi^2}{1024}$ & $\frac{1}{4}$ & $\frac{1323\pi^2}{262144}$ & $\frac{1}{90}$ & $\frac{4455\pi^2}{16777216}$
		\\[0.5ex]
		$d=9$ & $\frac{2048}{1225}$ & $\frac{2}{9}$ &
		$\frac{32768}{800415}$ & $\frac{16}{1859}$ & $\frac{262144}{135270135}$ \\[0.5ex]
        $d=10$  &  $\frac{11025\pi^2}{65536}$ & $\frac{1}{5}$ & $\frac{14553\pi^2}{4194304}$ & $\frac{1}{147}$ & $\frac{637065\pi^2}{4294967296}$
		\\[0.5ex]
		$d=11$& $\frac{32768}{19845} $ &$\frac{2}{11}$ &
		$\frac{262144}{9018009}$ & $\frac{16}{2925}$ &$ \frac{2097152}{1861574715}$\\[0.5ex] \hline
	\end{tabular}
\end{table}
\section{Extended sources and electromagnetic radiation reaction in dS}\label{sec:RR}

In this section, we will obtain a generalisation of the Abraham-Lorentz-Dirac(ALD) force to a particle moving in $dS_{d+1}$ static patch. We will begin with constructing the multipole moments for an extended localised source near the south pole and then analyse the specific problem of a single particle trajectory.

To understand the electromagnetic fields in the presence of extended sources, we will decompose the current densities into harmonics on the sphere as we did with the fields and ask how the coefficients of the spherical harmonic expansion source the $\Phi_E$ and $\Phi_B$ consistently. One challenge is that the current density components are not all independent and are related by the conservation equation. We will parametrise the source decomposition such that the conservation equation is trivially satisfied. In particular, we choose the following expansion for the current density:
\begin{equation}
\begin{split}
    J^u &= \sum_{\ell \vec{m}} \left[\frac{1}{r^{d-1}}\partial_r\left\{r^{d-1}J_1(r,\w,\ell,\vec{m})\right\}-\frac{\ell(\ell+d-2)}{r^2}J_2(r,\w,\ell,\vec{m})\right] \mathscr{Y}_{\ell\vec{m}}(\hat{r})\ ,\\   
    J^r &=  \sum_{\ell \vec{m}}i\w J_1(r,\w,\ell,\vec{m}) \mathscr{Y}_{\ell\vec{m}}(\hat{r})\ , \\ 
    J^I &= \sum_{\ell \vec{m}} \frac{i\w}{r^2} J_2(r,\w,\ell,\vec{m})\mathscr{D}^I\mathscr{Y}_{\ell\vec{m}}(\hat{r})+\sum_{\alpha \ell \vec{m}} J_V(r,\w,\alpha,\ell,\vec{m}) \mathbb{V}^I_{\alpha \ell \vec{m}}(\hat{r}) \ .
\end{split}
\end{equation}
Given the orthogonality of the $\mathscr{D}_I\Yellm$ with the $\mathbb{V}_I^{\alpha\ell\vec{m}}$, only $J_1$ and $J_2$ source the $\Phi_E$ whereas only the $J_V$ sources the $\Phi_B$. Notice that the divergencelessness of the $\mathbb{V}_I^{\alpha\ell\vec{m}}$ makes this part of the current identically conserved, and hence, the conservation equation plays no role in the `magnetic' parity sector. 

Extended sources introduce modifications to our field expressions by source-local terms, i.e. terms present only where the source is non-zero. The modified expressions for the fields are then given by:
\begin{equation}
\begin{split}
    \Er = \frac{\ell(\ell+d-2)}{r^{d-1}}\Phi_E+J_1 &\ ,\quad
    \Es =r^{3-d}\Dp\Phi_E+(1-r^2)J_2 \ ,\\
    \Hs =r^{3-d}\Dp\Phi_E \ ,\quad 
    \Ev &= i\w\  \Phi_B \ ,\quad  \Hv = \Dp \Phi_B \ .
\end{split}
\end{equation}
Notice that the magnetic parity fields have no source modifications, which is a manifestation of their gauge invariance. 
The Hertz-Debye scalars, in this case, have to satisfy inhomogeneous differential equations: 
\begin{align}
\begin{split}
	\frac{1}{r^{3-d}}D_+\left[r^{3-d}D_+\Phi_E\right]+\omega^2\Phi_E-&(1-r^2)\frac{\ell(\ell+d-2)}{r^2}\Phi_E\\+
	r^{d-3}&\left[\Dp\left[(1-r^2)J_2\right]-(1-r^2)J_1\right]= 0\ ,\\
    \frac{1}{r^{d-3}}D_+ \left[r^{d-3}D_+\Phi_B\right] +  \omega^2\Phi_B-&(1-r^2)\frac{(\ell+1)(\ell+d-3)}{r^2}\Phi_B+r^2(1-r^2)J_V=0\ .
\end{split}
\end{align}
We can now solve the above differential equations using Green functions defined on the dS-SK geometry. Such bulk-to-bulk Green functions for scalar fields satisfying Dirichlet boundary conditions were discussed in part I. A similar analysis for the Neumann scalars is given in appendix \ref{app:NeuBlktoBlk}. Once we obtain the solutions for extended sources on dS-SK, the on-shell action again reduces to the same form as Eq.\eqref{eq:SCIPpt}, where we now have explicit expressions for the multipole moments in terms of the current density components:
\begin{equation}
\begin{split}
    \multj^E_R(r,\w,\ell,\vec{m}) & = \frac{1}{(\ell+d-2)}\int_R d^dx \ \Yellm^*(\hat{r})\frac{1}{r^{d-3}}\left(\frac{1-r}{1+r}\right)^{-\frac{i\w}{2}}\\
    &\qquad\qquad\qquad \times \Bigg[ \partial_r\Xi^E_n(r,\w,\ell) J_u(r,\w,\hat{r})+i \w\ \Xi^E_n(r,\w,\ell) J^r(r,\w,\hat{r})\Bigg]\ , \\
    \multj^E_L(r,\w,\ell,\vec{m}) & =  \frac{1}{(\ell+d-2)}\int_R d^dx \ \Yellm^*(\hat{r})\frac{1}{r^{d-3}}\left(\frac{1-r}{1+r}\right)^{-\frac{i\w}{2}}\\
    &\qquad\qquad\qquad \times \Bigg[ \partial_r\Xi^E_n(r,\w,\ell) J_u(r,\w,\hat{r})+i \w\ \Xi^E_n(r,\w,\ell) J^r(r,\w,\hat{r})\Bigg] \ ,\\
    \multj^B_R(\w,\alpha,\ell,\vec{m}) &= \int_R d^{d}r\left(\frac{1-r_0}{1+r_0}\right)^{-\frac{i\w }{2}}\Xi^B_n(r_0,\w,\ell)\VSH_I(\hat{r}_0) J^I(\vec{r}_0,\w)\ , \\
    \multj^B_L(\w,\alpha,\ell,\vec{m}) &= \int_L d^{d}r\left(\frac{1-r_0}{1+r_0}\right)^{-\frac{i\w }{2}}\Xi^B_n(r_0,\w,\ell)\VSH_I(\hat{r}_0) J^I(\vec{r}_0,\w)\ .   
\end{split}
\end{equation}
The functions $\Xi^E_n$ and $\Xi^B_n$ are the de Sitter analogues of the Bessel $J$ functions in flat space. They smear the current density appropriately such that the time delays from the extended nature of the source are correctly accounted for in the computations of the radiation. These smearing functions can be expressed in terms of hypergeometric functions as:
\begin{equation}
\begin{split}
    \Xi^B_n(r,\w,\ell)&\equiv\frac{1}{2\ell+d-2}r^{\ell+1}(1-r^2)^{-i\w/2} {}_2F_1\left[\frac{\ell+1-i\w}{2},\frac{\ell+d-1-i\w}{2};\ell+\frac{d}{2};r^2\right] \\
    \Xi^E_n(r,\w,\ell)&\equiv\frac{1}{2\ell+d-2}r^{\ell+d-2}(1-r^2)^{-\frac{i\w}{2}} {}_2F_1\left[\frac{\ell+2-i\w}{2},\frac{\ell+d-2-i\w}{2};\ell+\frac{d}{2};r^2\right] \ .
\end{split}
\end{equation}

Using these formulae for multipole moments, one can obtain the influence phase for any observer modelled as an extended current density source. 

Now that we understand the influence phase for extended sources, we move on to the problem of computing the radiation reaction(RR) force for a charged particle moving about the south pole. This particle also gets doubled in our dS-SK setup: a particle on each side of the doubled static patch. We want to obtain the radiative multipoles as $\multj^{E/B}_L$ and $\multj^{E/B}_R$ corresponding to the particles on the left and the right copies. Next, we will sketch the ideas and approximations used to obtain the RR force from our influence phase, while the detailed computations can be found in appendix \ref{app:dSPN}.

In the usual computation of the ALD force\cite{Jackson:1998nia,zangwill2013modern,wald2022advanced}, one uses a post-newtonian(PN) approximation\footnote{We are using here the gravitational terminology for the non-relativistic expansion. In the electromagnetism context, perhaps \emph{post-couloumbian expansion}\cite{Kunze_2001} is a more appropriate appellation.} to obtain the dipole radiation due to an accelerating charged particle. The particle consequently recoils due to momentum lost to the dipole radiation. This recoiling force also receives higher-order PN corrections. Such PN corrections to the RR have been computed in flat spacetimes in several even spacetime dimensions\cite{Galakhov:2007my, Birnholtz:2013nta}. We will take a similar approach to obtain the ALD force through a small velocity expansion. 

The extra scale $H$ modifies our post-newtonian scheme. The usual flat space PN scheme assumes the wavelength of the radiation to be much larger than the length scales traversed by the particle($\w r\ll 1$). This is consistent with the requirement of small velocities($v\ll c$). To obtain a sensible flat space result in the $H\to 0$ limit, we also require both the wavelength of the radiation and the scale of the particle trajectory to be much smaller than the cosmological scale($\w \gg H$, $rH\ll 1$). For the consistent validity of all these approximations, the particle needs to move slowly compared to the speed of light but rapidly compared to the cosmological time scales\footnote{This assumption is valid for most astrophysical and cosmological processes due to the smallness of $H$ relative to other frequency-scales in our universe.}. In our doubled geometry set-up, this implies a PN expansion of both the particles on either side of the geometry. The influence phase is then expressed in terms of the average and difference of the positions of the two particles.  

Till now, our multipole moments are defined in spherical polar coordinates, which are quite inconvenient for PN expansion. We can solve this by shifting to symmetric trace-free (STF) multipole moments in cartesian-like coordinates in de Sitter. Specifically, we work with appropriate coordinates that reduce to cartesian coordinates in the $H\to 0$ limit. We use a formalism developed in appendix \ref{app:VSH} to
convert the vector spherical harmonics on $\mathbb{S}^{d-1}$ to their cartesian counterparts in $\mathbb{R}^d$. To convert the influence phase evaluated in terms of spherical multipole moments to one in terms of STF moments, we invoke scalar/vector spherical harmonic addition theorems. The definition of STF moments, as well as the corresponding influence phase, can be found in appendix \ref{app:dSSKaction}\footnote{The STF multipole moments for electric and magnetic parity are given in Eqns. \eqref{eq:STFelectric} and \eqref{eq:STFmagnetic}, respectively. The action in terms of the STF multipole moments is provided in \eqref{eq:dSSKSTFaction}.}.

As noted in part I, the $\mathcal{J}_A\mathcal{J}^*_D$ part of the influence phase encodes dissipative physics, including the RR force. The $\Kout_{E/B}$ act as 2-point functions that capture the dissipative effects of one multipole moment on another. In even spacetime dimensions, this 2-point function is completely local: the multipole moment at a given instant is solely responsible for the RR force. The multipole moments are appropriately smeared distributions of the current density sourced by moving charged particles on either side of the geometry. The force is finally obtained by varying the Lagrangian with respect to the difference in the positions of the particles.    

Now that we have described the process of obtaining the RR force, we will quote some results. To begin with, we give the first Hubble correction to the $d+1$ dimensional ALD force:
\begin{align}
	\begin{split}
		F^i_\text{ALD}&=\frac{(-1)^{\frac{d+1}{2}}(d-1)}{|\mathbb{S}^{d-1}|d!!(d-2)!!}\Bigg\{ \partial^{d}_t x^i -H^2\frac{d}{6}\left(d^2-6d+11\right)\partial^{d-2}_t x^i\Bigg\}\ .
	\end{split}
\end{align}
As we argued in part I, the signs are consistent with the leading term in the force being dissipative. Due to dimensional considerations, each $H^2$ correction comes with two fewer derivatives. In the appendix, we give detailed calculations of the subleading PN corrections to this force in flat spacetime, along with $H^2$ and $H^4$ corrections.      

The PN terms of the RR force in flat space can be resummed to give Lorentz covariant answers, which are quoted for $d=3,5,7$\cite{Galakhov:2007my}. Similarly, we check that the de Sitter RR force resums to a de Sitter covariant result. Here, we will quote the de Sitter answers for $d=3,5,7$. In the appendix \ref{app:covRR}, the reader can find the corresponding results for $d=9,11$. We write the RR force in the form:

    \[ F^\mu_\text{RR}\equiv \frac{(-)^{\frac{d-1}{2}}}{|\mathbb{S}^{d-1}|(d-2)!!} f^\mu_d\]
where the values of $f^\mu_d$ are given as:
 \begin{equation}
\begin{split}
f^\mu_3 &\equiv \frac{P^{\mu\nu}}{3!!}\left\{-2a_\nu^{(1)}\right\}\ ,\\
f^\mu_5&\equiv \frac{ P^{\mu\nu}}{5!!}\left\{-4a_\nu^{(3)}+10\ (a\cdot a)\  a_\nu^{(1)}+30\ (a\cdot a^{(1)})\ a_\nu\right\}-H^2\frac{ P^{\mu\nu}}{5!!}\left\{16a_\nu^{(1)}\right\}\ ,\\
f^\mu_7&\equiv \frac{P^{\mu\nu}}{7!!}\left\{-6a_\nu^{(5)}+42\ (a\cdot a)\  a_\nu^{(3)}+210\ (a\cdot a^{(1)})\ a_\nu^{(2)}+224\ (a\cdot a^{(2)})\ a_\nu^{(1)}+\frac{574}{3}\ (a^{(1)}\cdot a^{(1)})\  a_\nu^{(1)}\right.\\
&\left.\qquad\qquad +126\ (a\cdot a^{(3)})\ a_\nu+280\ (a^{(1)}\cdot a^{(2)})\ a_\nu+O(a^5)\right\}\\
&\quad+H^2\frac{P^{\mu\nu}}{7!!}\left\{120 a_\nu^{(3)}-342\ (a\cdot a)\  a_\nu^{(1)}-978\ (a\cdot a^{(1)})\ a_\nu\right\} -H^4\frac{P^{\mu\nu}}{7!!}\left\{384a^{(1)}_\nu\right\}\ .\\
\end{split}
\end{equation}
where $P^{\mu\nu}=g^{\mu\nu}+v^\mu v^\nu$ such that $v^\mu=\frac{dx^\mu}{d\tau}$ is the proper velocity computed in the de Sitter background and $a^\mu=\frac{dv^\mu}{d\tau}$ is the corresponding dS covariant acceleration. $P$ acts as a projector transverse to the worldline of the particle.  We check our force in the flat limit against flat space results obtained in \cite{Galakhov:2007my} and \cite{Birnholtz:2013ffa} and find agreement for $d=3,5,7$. 

\section{Summary and Discussion}

In this paper, we extend the techniques developed in part I \cite{Loganayagam:2023pfb}  to 
include electromagnetic interactions. 
Our claim in part I was that the on-shell effective action evaluated on the dS-SK geometry yields the observer's influence phase. We have shown in this work how this works for
observers coupled to EM fields. Further, we have checked that our answers obey constraints imposed by bulk unitarity and the Kubo-Martin-Schwinger conditions. When the influence phase is expressed in terms of appropriate multipole moments, we show that the dissipative part correctly captures the radiation reaction felt by the observer. We have also checked that, in the $H\to 0$ limit, we recover the correct flat space limit.

We then take our observer to be a single charged particle moving along an arbitrary time-like trajectory in an even-dimensional spacetime. We compute the influence phase for such an observer in a non-relativistic approximation, where the Hubble corrections are also treated perturbatively. We then resum the corresponding self-force and show that they arrange themselves into a dS covariant vector. For odd-dimensional spacetimes, the radiation reaction on the observer is shown to renormalise the observer's multipole moments. On our way, we clarify and extend many results about vector spherical harmonics(VSHs)in arbitrary dimensions, as well as their use in EM multipole expansions (See appendices \ref{app:VSH},\ref{app:FlatEMI} and \ref{app:FlatEMII} for a detailed discussion).

\subsection*{Discussion}

In this work, we have focused on electromagnetic self-force. But we hope that many of the ideas here directly generalise to the gravitational case. In particular, we hope that the method of covariant counterterms introduced here can be extended to the gravitational case. This might be a useful alternative to existing methods to regularise the self-field\cite{Mino:1997,Quinn:1997,Quinn:2000wa,Detweiler:2002mi,Harte:2018iim}. We intend to explore some of this in our upcoming work on linearised gravity \cite{dSSKgravvec}.

 The more challenging analysis is considering gravitational non-linearities. We hope it will be possible to have a well-defined perturbation theory analogous to the multipolar-post-minkowskian (MPM) analysis in flat spacetime. A much simpler analysis is to consider scalar interactions in this setup which might provide insights into the more complicated problem of gravity.\footnote{In the non-gravitational case, for AdS black holes, there has been some progress in taking into account non-linearities perturbatively for contact as well as exchange real-time Witten diagrams\cite{Jana:2020vyx,Loganayagam:2022zmq,Loganayagam:2024mnj,Martin:2024mdm}.}

Given the enhanced electromagnetic dissipation any given radiative source experiences due to the presence of the cosmological constant, one could ask if there are any astrophysical/cosmological phenomena where such a dissipation would be relevant. The time(length) scales associated with such phenomena would be of the order of billions of (light-)years. This may rule out many sub-galactic-cluster scales but would contribute to intergalactic and large-scale structure dynamics. A more general analysis of FLRW spacetimes is required to address the problem in a more realistic setting.  

Even though our analysis focuses on the dissipative terms obtained in the effective action for the extended source, the conservative effects often dominate the dynamics of astrophysically/cosmologically relevant phenomena; e.g., the presence of a cosmological constant plays an important role in the relative dynamics of the local group with respect to the Virgo cluster\cite{Bisnovatyi-Kogan:2019iyz}. This motivates the study of the orbital dynamics of two interacting bodies in dS, which can be solved using the conservative pieces in our effective action for extended sources. The actual problem of galactic dynamics requires the gravitational effective action, but the scalar/electromagnetic counterpart serves as a simpler toy model to understand the binary problem. We hope to explore this avenue in upcoming work\footnote{A similar question has been addressed in the case of pure AdS in the newtonian limit\cite{Maxfield:2022hkd} where it can have interesting reflections upon the dual CFT.}.

\section*{Acknowledgements}
We thank Ofek Birnholtz, Kushal Chakraborty, Tuneer Chakraborty, Victor Godet, Sitender Kashyap, Alok Laddha, Godwin Martin, Shiraz Minwalla, Priyadarshi Paul, Harsh Rana, Mukund Rangamani, Arnab Rudra, Joseph Samuel, Shivam Sharma, Akhil Sivakumar, Sandip Trivedi and Amitabh Virmani for valuable discussions.  We acknowledge the support of the Department of Atomic Energy, Government of India, under project no. RTI4001, and would also like to acknowledge our debt to the people of India for their steady and generous support of research in the basic sciences.

\appendix

\section{Theory of vector spherical harmonics}\label{app:VSH}

In this subsection, we will give explicit expressions for the vector spherical harmonics(VSH) on $\mathbb{S}^{d-1}$. We will construct higher-dimensional analogues of the standard and well-known expressions for $\mathbb{S}^2$\cite{Morse:1953,Blatt:1952ije,Hill:1954}, which we refer to as orthonormal VSH. Past discussions of vector spherical harmonics on higher dimensional spheres appear in  \cite{Gerlach:1978gy,Rubin:1984tc,Higuchi:1986wu,BenAchour:2015aah,Lindblom:2017maa}. In particular, Higuchi\cite{Higuchi:1986wu} gives a recursive construction for an arbitrary tensor harmonic on $\mathbb{S}^{d-1}$ in terms of tensor harmonics on $\mathbb{S}^{d-2}$. In what follows, we will write explicit forms for the VSHs on $\mathbb{S}^{d-1}$, which agree with his recursion. 

Along with the usual orthonormal VSH expressed in spherical polar coordinates, one can also construct them in terms of symmetric trace-free(STF) tensors on the ambient $\mathbb{R}^d$. Their construction in cartesian coordinates helps in the post-newtonian expansions of multipole moments, and hence, they find their natural home in the literature on gravitational waves\cite{Thorne:1980ru,Damour:1990gj}. Their higher-dimensional analogues can be found in \cite{Henry:2021cek,Amalberti:2023ohj}. We will construct these cartesian STF VSH in all dimensions and show how they connect to previous constructions on $\mathbb{S}^2$ as well as for higher-dimensional spheres.

We begin with an explicit spherical coordinate system in $\mathbb{R}^d$ given by
\begin{equation}\begin{split}
		x_1&=r\ \sin\vartheta_{d-2}\ \sin\vartheta_{d-3}\ \ldots\ \sin\vartheta_2\ \sin\vartheta_1\ \cos\varphi\ ,\\
		x_2&=r\ \sin\vartheta_{d-2}\ \sin\vartheta_{d-3}\ \ldots\ \sin\vartheta_2\ \sin\vartheta_1\ \sin\varphi\ ,\\
		x_3&=r\ \sin\vartheta_{d-2}\ \sin\vartheta_{d-3}\ \ldots\ \sin\vartheta_2\ \cos\vartheta_1\ ,\\
		x_4&=r\ \sin\vartheta_{d-2}\ \sin\vartheta_{d-3}\ \ldots\ \cos\vartheta_2\  ,\\
		& \ldots\ ,\\
		x_{d-2}&=r\ \sin\vartheta_{d-2}\ \sin\vartheta_{d-3}\ \cos\vartheta_{d-4}\ ,\\
		x_{d-1}&=r\ \sin\vartheta_{d-2}\  \cos\vartheta_{d-3}\ ,\\
		x_{d}&=r\ \cos\vartheta_{d-2}\ .
\end{split}\end{equation}
Here the radius $r$ varies from $0$ to $\infty$ whereas the allowed values of  angles is $\vartheta_i\in[0,\pi]$ and $\varphi\in[0,2\pi)$. We will set $\vartheta_0\equiv\varphi$ and denote the coordinates on $\mathbb{S}^{d-1}$ as $\vartheta_I$ with $I=0,1,\ldots,d-2$. The sphere metric in these coordinates takes the form 
\begin{equation}\begin{split} d\Omega_{d-1}^2&\equiv \gamma_{IJ} d\vartheta_I d\vartheta_J = d\vartheta_{d-2}^2 
		+\sin^2\vartheta_{d-2}d\Omega_{d-2}^2=\ldots\\
		&= d\vartheta_{d-2}^2 
		+\sin^2\vartheta_{d-2} d\vartheta_{d-3}^2 
		 +\ldots+\prod_{K=J+1}^{d-2}\sin^2\vartheta_K\ d\vartheta_J^2+\ldots+\prod_{K=1}^{d-2}\sin^2\vartheta_K\ d\varphi^2
		.
\end{split}\end{equation}
In other words, the explicit metric coefficients are given by 
\begin{equation}\label{eq:sphMet}\begin{split}
\gamma_{IJ} =\left\{\begin{array}{cc} 
\prod_{K=I+1}^{d-2}\sin^2\vartheta_K & \text{when $I=J$\ ,}\\
0 & \text{otherwise.}\end{array}\right.
\end{split}\end{equation}
Since the metric is diagonal, its inverse is given by inverting the diagonal entries, i.e., $\gamma^{II}=\gamma_{II}^{-1}$. Another result we will need is the volume measure on the sphere $\sqrt{\gamma} =
\prod_{J=1}^{d-2}\sin^J\vartheta_J$ . By integrating this measure, we obtain the volume of $\mathbb{S}^{d-1}$ as
\begin{equation}|\mathbb{S}^{d-1}|\equiv \frac{2\pi^{\frac{d}{2}}}{\Gamma(\frac{d}{2})}\ ,
\end{equation}
We will denote the covariant derivative associated with the unit sphere metric as $\mathscr{D}_I$. For some of the conversions between partial derivatives in the spherical coordinates to cartesian coordinates, the following formula is useful:
\begin{equation}\label{eq:PushFsph}\begin{split}
\frac{\partial}{\partial\vartheta_I}&=-r\prod_{j=I}^{d-2}\sin\vartheta_j\ \frac{\partial}{\partial x_{_{I+2}}}+\frac{\cos\vartheta_I}{\sin\vartheta_I}\sum_{j=1}^{I+1}x_j\frac{\partial}{\partial x_{j}}\\
&=\frac{1}{\prod_{j=I}^{d-2}\sin\vartheta_j}\sum_{j=1}^{I+1}x_j\left\{x_{_{I+2}}\frac{\partial}{\partial x_{j}}-x_j\frac{\partial}{\partial x_{_{I+2}}}\right\}\ .
\end{split}\end{equation}
This is the push-forward of the coordinate basis vector fields on $\mathbb{S}^{d-1}$ to $\mathbb{R}^d$.

\subsection{SSHs and STF tensors}
We will begin our discussion by quickly reviewing the construction of scalar spherical harmonics (SSHs) on $\mathbb{S}^{d-1}$. Our goal
here is to list the key results needed for our purposes without any derivations or detailed justification.  We will refer the reader to Appendix A of \cite{Loganayagam:2023pfb} for a more detailed review of the expressions below and the rationale behind them.

SSHs  on $\mathbb{S}^{d-1}$ are labeled by a non-decreasing sequence of  non-negative integers
\begin{equation}\label{eq:mcondn}
\begin{split}
0\leq m_{_1}\leq m_2\ldots\leq m_{d-2}\leq m_{d-1}=\ell\ ,
\end{split}\end{equation}
and they form an orthonormal set of Laplace eigenfunctions on $\mathbb{S}^{d-1}$, i.e., we have
\begin{equation}
\begin{split}
\left[\mathscr{D}^2+\ell(\ell+d-2)\right]\mathscr{Y}_{\ell\vec{m}}= 0\ ,\quad 	
\int_{\mathbb{S}^{d-1}}
\mathscr{Y}_{\ell'\vec{m}'}^\ast\mathscr{Y}_{\ell\vec{m}} =\delta_{\ell'\ell}\delta_{\vec{m}'\vec{m}}\ .
\end{split}
\end{equation}
The explicit orthonormal basis of SSHs on $\mathbb{S}^{d-1}$ is well-known. We will write them down here following the notations of our previous work \cite{Loganayagam:2023pfb}:
\begin{equation}\label{eq:SSHdef}
\begin{split}
\mathscr{Y}_{\ell\vec{m}}(\hat{r})&\equiv C^S_{\ell\vec{m}}\ e^{\pm im_{_1}\varphi }\left[\prod_{k=1}^{d-2}(\sin\vartheta_k)^{m_k} P_{m_{k+1}-m_k}(k+2+2m_k,\cos\vartheta_k)\right]_{m_{d-1}=\ell}\ \text{with}\\
|C^S_{\ell\vec{m}}|^{-2}&\equiv 2\pi\prod_{i=1}^{d-2}\frac{|\mathbb{S}^{i+2m_i+1}|}{|\mathbb{S}^{i+2m_i}|\Nhh(i+2m_i+2,m_{i+1}-m_i)}\ .
\end{split}\end{equation}
The notation here is as follows: the symbol $P_\ell(d,z)$ denotes the generalised Legendre polynomials\footnote{These polynomials are related to the associated Legendre functions/Gegenbauer polynomials  via
\begin{equation}\begin{split}
		P_\lambda^{-\mu}(z)\equiv \frac{\left(\sqrt{1-z^2}\right)^{\mu}}{2^\mu \mu!}P_{\lambda-\mu}(2\mu+3;z)\ ,\quad C^{\frac{d}{2}-1}_\ell(z)\equiv (d-2)\frac{N_{HH}(d,\ell)}{2\ell+d-2}P_\ell(d,z)\ .
\end{split}\end{equation}}
\begin{equation}\label{Eq:LegGen}\begin{split}
\nn_{d,\ell}\Nhh(d,\ell)P_\ell(d,x)&=\sum_k \frac{\Gamma\left(\nu-k\right)}{2^{2k}k!\Gamma(\nu)}
\frac{(-)^k x^{\ell-2k}}{(\ell-2k)!} \ ,
\end{split}\end{equation}
where $\nu\equiv \frac{d}{2}+\ell-1$ and the RHS sum extends from $k=0$ until $k$ exceeds $\ell/2$. The symbol $ \Nhh(d,\ell)$ is an integer that counts the degeneracy $\ell^{th}$ SSHs on $\mathbb{S}^{d-1}$ (or equivalently, the number of homogeneous harmonic polynomials in $\mathbb{R}^d$: see below) and $\nn_{d,\ell}$ is an inverse integer given by
\begin{equation}\begin{split}  
\nn_{d,\ell}\equiv \frac{(d-2)!!}{(d+2\ell-2)!!}\ .
\end{split}\end{equation}
The inverse integer  $\nn_{d,\ell}$ is associated with the inner product of symmetric trace-free (STF) tensors. The generalised Legendre polynomials $P_\ell(d,z)$ defined above are normalised such that $P_\ell(d,z=1)=1$ with an inner product given by 
\begin{equation}\begin{split}
|\mathbb{S}^{d-2}|\int_0^\pi d\vartheta\ \sin^{d-2}\vartheta\  P_\ell(d,\cos\vartheta)P_{\ell'}(d,\cos\vartheta)
=\delta_{\ell\ell'}\frac{|\mathbb{S}^{d-1}|}{N_{HH}(d,\ell)}\ .
\end{split}\end{equation}
The normalisation appearing in Eq.\eqref{eq:SSHdef} follows from this inner product.

We will now obtain a formula for $ \Nhh(d,\ell)$ by an explicit counting of the number of $\vec{m}$ satisfying the condition given in Eq.\eqref{eq:mcondn}, i.e.,
\begin{equation}\label{eq:NHHsum}
\begin{split}
    \Nhh(d,m_{d-1})=&\sum\limits_{m_{d-2}=0}^{m_{d-1}}\sum\limits_{m_{d-3}=0}^{m_{d-2}}\dots \sum\limits_{m_2=0}^{m_3}\left\{1+\sum\limits_{m_1=1}^{m_2}2\right\}\ .
\end{split}
\end{equation}
As a check, setting $d=3$ yields the well-known result that there are $2m_2+1$ SSHs on $\mathbb{S}^2$ corresponding to the eigenvalue $-m_2(m_2+1)$.  We can perform this sum as follows: first, we note that the above identity implies a recursion relation of the form
\begin{equation}\label{eq:NHHRec}
\begin{split}
\Nhh(d,\ell)=\sum\limits_{m=0}^{\ell}\Nhh(d-1,m)\ .
\end{split}
\end{equation}
Using this relation, we can get the number of SSHs on $\mathbb{S}^{d-1}$ by starting from the count in $\mathbb{S}^2$ and then recursively summing the answer. An closed form expression for $\Nhh(d,\ell)$ satisfying the above recursion is given by
\begin{equation}\label{eq:NHH}\begin{split} \Nhh(d,\ell)\equiv 
		(2\ell+d-2) \frac{(\ell+d-3)!}{\ell! (d-2)!} \ .
\end{split}\end{equation}

The SSHs written in Eq.\eqref{eq:SSHdef} are simultaneous eigenfunctions of the laplacian on lower spheres $\mathbb{S}^{d-1},\mathbb{S}^{d-2},\ldots, \mathbb{S}^1$ respectively. The lower spheres are obtained by successively dropping the angles $\vartheta_{d-2},\vartheta_{d-3},\ldots$. In fact, the set of $m_i$'s in this construction
are indeed related to the lower sphere laplacians, viz.,
\begin{equation}\label{eq:mId}
\begin{split}
-\mathscr{D}^2_{\mathbb{S}^{I+1}}\mathscr{Y}_{\ell\vec{m}}(\hat{r})=-\gamma_{II}\sum\limits_{J=0}^{I}\frac{1}{\sqrt{\gamma}}\frac{\partial}{\partial \theta_J}\left\{\sqrt{\gamma}\ \gamma^{JJ}\frac{\partial}{\partial \theta_J}\mathscr{Y}_{\ell\vec{m}}(\hat{r})\right\}=  m_{I+1}(m_{I+1}+I)\ \mathscr{Y}_{\ell\vec{m}}(\hat{r})\ .
\end{split}
\end{equation}
This concludes our quick summary of the orthonormal SSHs on $\mathbb{S}^{d-1}$. We will now complement the discussion above with one on SSHs from the point of view of symmetric trace-free tensors and cartesian coordinates.

First, we observe that symmetric trace-free (STF) polynomials of degree $\ell$ in the radial unit vector, i.e., polynomials of the form 
\begin{equation}
\begin{split}
\hat{r}_{<i_1}\hat{r}_{i_2}\ldots \hat{r}_{i_\ell>}\ 
 \end{split}   
\end{equation}
constitute a basis of SSHs. Here, $\hat{r}_i\equiv \frac{x_i}{r}$, and we use angular brackets to denote STF projection, accomplished via a projector 
\begin{equation}
\begin{split}
\hat{r}_{<i_1}\ldots \hat{r}_{i_\ell>}\equiv (\Pi^S_{d,\ell})_{<i_1\ldots i_\ell>}^{<j_1\ldots j_\ell>} \hat{r}_{j_1}\ldots \hat{r}_{j_\ell}\ .
 \end{split}   
\end{equation}
We will give an explicit form for $\Pi^S$ shortly. For present purposes, we note that multiplying by $r^\ell$ makes them into homogeneous harmonic polynomials of cartesian coordinates in $\mathbb{R}^d$. Such polynomials, when restricted to the unit sphere $\mathbb{S}^{d-1}$, become scalar spherical harmonics (SSHs). 

The orthonormal SSHs of Eq.\eqref{eq:SSHdef} can all be written in the form
\begin{equation}\label{eq:OrthoToSTFssh}
\begin{split}
\mathscr{Y}_{\ell\vec{m}}(\hat{r})=\frac{1}{\ell !}\mathscr{Y}^{\ell\vec{m}}_{<i_1i_2\ldots i_\ell>}\hat{r}^{<i_1}\hat{r}^{i_2}\ldots \hat{r}^{i_\ell>}\ ,
\end{split}
\end{equation}
where $\mathscr{Y}^{(\ell\vec{m})}_{i_1i_2\ldots i_\ell}$ are STF tensors with constant cartesian components. One important example is the generalised Legendre polynomial of Eq.\eqref{Eq:LegGen}. Since $P_\ell\left(d,\hat{\kappa}\cdot\hat{r}\right)$ is a unique SSH invariant under $SO(d-1)$ rotations about  the $\hat{\kappa}$ axis, 
and since $\hat{\kappa}^{<i_1}\ldots \hat{\kappa}^{i_\ell>} $ is a unique $SO(d-1)$ invariant STF tensor, we should have that
\begin{equation}
\begin{split}
\nn_{d,\ell}\Nhh(d,\ell) P_\ell\left(d,\hat{\kappa}\cdot\hat{r}\right) =  \frac{1}{\ell!} \hat{\kappa}^{<i_1}\ldots \hat{\kappa}^{i_\ell>} \hat{r}_{<i_1}\ldots \hat{r}_{i_\ell>}= \frac{1}{\ell!} \hat{\kappa}^{i_1}\ldots \hat{\kappa}^{i_\ell}
(\Pi^S_{d,\ell})_{<i_1\ldots i_\ell>}^{<j_1\ldots j_\ell>} \hat{r}_{j_1}\ldots \hat{r}_{j_\ell}
\ .
 \end{split}   
\end{equation}
Here, we have fixed the relative normalisation by setting the coefficient of $(\hat{\kappa}\cdot\hat{r})^\ell$ to $\frac{1}{\ell!}$ on both sides. Since we already know the explicit expression for $P_\ell\left(d,\hat{\kappa}\cdot\hat{r}\right)$, the above expression can then be thought of as giving an explicit definition of $\Pi^S$.

Given any two vectors $\vec{r}$ and $\vec{\kappa}$, we define the following projected contraction
\begin{equation}\label{eq:STFLegSSH}
\begin{split}
\Pi^S_{d,\ell}(\vec{r}|\vec{\kappa})
&\equiv \Pi^S_{d,\ell}(\vec{\kappa}|\vec{r})\equiv \frac{1}{\ell!} \kappa^{i_1}\ldots \kappa^{i_\ell}
(\Pi^S_{d,\ell})_{<i_1\ldots i_\ell>}^{<j_1\ldots j_\ell>}\ r_{j_1}\ldots r_{j_\ell}=\nn_{d,\ell}\Nhh(d,\ell)(\kappa r)^\ell P_\ell\left(d,\hat{\kappa}\cdot\hat{r}\right)\\
&=  \left[\sum_{k=0}^{\lfloor\frac{\ell}{2}\rfloor}\ \frac{\Gamma\left(\nu-k\right)}{k!\ \Gamma\left(\nu\right)}  \left(-\frac{\kappa^2r^2}{4}\right)^{k}\frac{(\vec{\kappa}\cdot\vec{r})^{\ell-2k}}{(\ell-2k)!}\ \right]_{\nu=\frac{d}{2}+\ell-1}\ .
 \end{split}   
\end{equation}
This is an $\ell^{th}$ degree homogeneous polynomial in both $\vec{r}$ and $\vec{\kappa}$, and it is harmonic in both these variables, viz., 
\begin{equation}
\begin{split}
\nabla^2\Pi^S_{d,\ell}(\vec{r}|\vec{r}_0)&=\nabla^2_0\Pi^S_{d,\ell}(\vec{r}|\vec{r}_0)=0\ .
 \end{split}   
\end{equation}
It is, in fact, the unique polynomial which satisfies these properties up to an overall normalisation. The STF projector itself can then be obtained by differentiating this polynomial to strip off the $x_i$ and $\kappa_i$ factors. The STF projection also has a derivative operator representation which follows from the above formula:
\begin{equation}
\begin{split}
x^{<i_1}x^{i_2}\ldots x^{i_\ell>}&= \left[\sum_{k=0}^{\lfloor\frac{\ell}{2}\rfloor}\ \frac{\Gamma\left(\nu-k\right)}{k!\ \Gamma\left(\nu\right)}  \left(\frac{r}{2}\right)^{2k}(-\nabla^2)^k\right]_{\nu=\frac{d}{2}+\ell-1} x^{i_1}x^{i_2}\ldots x^{i_\ell}\ .
 \end{split}   
\end{equation}
Another representation of the STF projector, derived from the standard addition theorem for orthonormal SSHs, is
\begin{equation}\label{eq:SSHSTFAdd}\begin{split}
(\Pi^S_{d,\ell})^{<i_1i_2\ldots i_\ell>}_{ <j_1j_2\ldots j_\ell>}=\frac{\nn_{d,\ell}|\mathbb{S}^{d-1}|}{\ell!}\sum_{\vec{m}}
\mathscr{Y}_{\ell\vec{m}}^{\ast <i_1i_2\ldots i_\ell>} \mathscr{Y}^{\ell\vec{m}}_{<j_1j_2\ldots j_\ell>}\ ,
\end{split}\end{equation}
where $\mathscr{Y}^{(\ell\vec{m})}_{<i_1i_2\ldots i_\ell>} $ are the STF tensors which convert between the orthonormal basis and the STF basis. Equivalently, by contracting the STF indices with arbitrary vectors, we can write
\begin{equation}\label{eq:SSHaddnThm}\begin{split}
\Pi^S_{d,\ell}(\hat{r}_0|\hat{r})=\nn_{d,\ell}|\mathbb{S}^{d-1}|\sum_{\vec{m}}
\mathscr{Y}_{\ell\vec{m}}^{\ast}(\hat{r}_0) \mathscr{Y}^{\ell\vec{m}}(\hat{r})\ .
\end{split}\end{equation}

The above expression relates the STF projector to the standard inner product on SSHs: one gets an extra factor of $\nn_{d,\ell}|\mathbb{S}^{d-1}|$ relative to an orthonormal basis because of the overcompleteness of the STF basis. The same factor appears in the inner product computed in the STF basis:
\begin{equation}\label{Eq:SSHintSTF}\begin{split}
\int_{\mathbb{S}^{d-1}}
\left[\frac{1}{\ell !}\mathscr{Y}_{<i_1i_2\ldots i_\ell>}\hat{r}^{<i_1}\ldots \hat{r}^{i_\ell>}\right]
\left[\frac{1}{\ell !}\overline{\mathscr{Y}}_{<j_1j_2\ldots j_\ell>}\hat{r}^{<j_1}\ldots \hat{r}^{j_\ell>}\right]=
\frac{\nn_{d,\ell}|\mathbb{S}^{d-1}|}{\ell !}
\mathscr{Y}^{<i_1i_2\ldots i_\ell>} \overline{\mathscr{Y}}_{<i_1i_2\ldots i_\ell>}\ .
\end{split}\end{equation}
This is, in fact, necessary for the sum in Eq.\eqref{eq:SSHSTFAdd} to be a projector, i.e., for the idempotent property
\begin{equation}\begin{split}
(\Pi^S_{d,\ell})^{<i_1i_2\ldots i_\ell>}_{ <k_1k_2\ldots k_\ell>}(\Pi^S_{d,\ell})^{<k_1k_2\ldots k_\ell>}_{ <j_1j_2\ldots j_\ell>}=(\Pi^S_{d,\ell})^{<i_1i_2\ldots i_\ell>}_{ <j_1j_2\ldots j_\ell>}\ 
\end{split}\end{equation}
to hold. This concludes our brief overview of SSHs in the language of STF tensors. We will refer the reader to  Appendix (A.2) of \cite{Loganayagam:2023pfb} for a more detailed exposition with explicit expressions and derivations. We will now generalise these ideas to VSHs on $\mathbb{S}^{d-1}$.

\subsection{Toroidal derivatives and VSHs}\label{ssec:totoidalOp}
We will now move on to the subject of vector spherical harmonics (VSHs), i.e., divergence-free vector fields on $\mathbb{S}^{d-1}$, which are also eigenvectors of the sphere laplacian. From the cartesian viewpoint, these correspond to homogeneous, harmonic, divergence-free, polynomial vector fields on $\mathbb{R}^d$ that have no radial component. 

\subsubsection*{Harmonic vector fields in $\mathbb{R}^d$}
We will now construct such harmonic vector fields by applying an appropriate derivative operator on homogeneous harmonic polynomials $x^{<i_1}x^{i_2}\ldots x^{i_\ell>} $. Such a construction is well-known in $d=3$ where the toroidal operator $\vec{r}\times \vec{\nabla}$ will do the job.
Given that there is no cross-product for $d>3$, this statement does not generalize
as stated: \emph{there is, in fact, no one derivative operator that constructs all Vector polynomials from scalar polynomials in $d>3$}. However, we will now show that if we allow for two derivatives, we can indeed construct a full set of toroidal derivative operators for $d>3$. The standard $d=3$ construction will then be recovered as a degenerate special case. As far as we are aware, such a construction of toroidal operators for general dimensions has not appeared elsewhere and is entirely new.

Let $\mathbb{L}_{ij}\equiv x_i\partial_j-x_j\partial_i$ be the rotation  Killing vectors of $\mathbb{R}^d$ obeying  $SO(d)$ Lie-algebra
\begin{equation}\label{eq:LLcomm}
\begin{split}
[\mathbb{L}_{ij},\mathbb{L}_{kl}]=\delta_{ik}\mathbb{L}_{lj}-\delta_{jk}\mathbb{L}_{li}-\delta_{il}\mathbb{L}_{kj}+\delta_{jl}\mathbb{L}_{ki}\ .
\end{split}
\end{equation}
These operators obey relations of the form
\begin{equation}\label{eq:Lbianchi}
\begin{split}
x_k\mathbb{L}_{ij}+x_i\mathbb{L}_{jk}+x_j\mathbb{L}_{ki}=0=
\partial_k\mathbb{L}_{ij}+\partial_i\mathbb{L}_{jk}+\partial_j\mathbb{L}_{ki}\ .
\end{split}
\end{equation}
A useful corollary of the above relations is a sum of the form
\begin{equation}\label{eq:LbianchiRed}
\begin{split}
\sum_{ik}x_i\mathbb{L}_{jk}\mathbb{L}_{ik}=\frac{1}{2}x_j\sum_{ik}\mathbb{L}_{ik}\mathbb{L}_{ik}\ ,\quad 
\sum_{ik}\partial_i\mathbb{L}_{jk}\mathbb{L}_{ik}=\frac{1}{2}\partial_j \sum_{ik}\mathbb{L}_{ik}\mathbb{L}_{ik}\ ,
\end{split}
\end{equation}
where the sum over $i$ and $k$ are performed over the same subset of indices. These properties motivate the following definition of the toroidal operators 
\begin{equation}\label{eq:DeltaVdefCart}
\begin{split}
\mathbb{\Delta}_{i,\alpha+2}^{(\alpha)} f&\equiv \left\{\begin{array}{cc} 
\sum_{k=1}^{\alpha+1}\mathbb{L}_{k,\alpha+2}\mathbb{L}_{ki} f\ & \text{for} \ 1\leq i\leq \alpha+1\ ,\\
-\frac{1}{2}\sum_{j,k=1}^{\alpha+1}\mathbb{L}_{jk}\mathbb{L}_{jk}f\  & \text{for} \ i=\alpha+2 \ ,\\
  0 & \text{for} \ i>\alpha+2 \ ,
\end{array}\right.
\end{split}
\end{equation}
acting on an arbitrary function $f$ on $\mathbb{R}^d$. Here, $\alpha$ takes on values $\alpha=1,2,\ldots,(d-2)$, and the reason for our notation will become clear shortly. Equations Eq.\eqref{eq:LbianchiRed} imply that the vector field $\mathbb{\Delta}_{i,\alpha+2}^{(\alpha)} f$ is tangential to the sphere and is divergence-free for any $f$, viz.,
\begin{equation}
\begin{split}
\partial_i\mathbb{\Delta}_{i,\alpha+2}^{(\alpha)} f=0\ ,\  x^i\mathbb{\Delta}_{i,\alpha+2}^{(\alpha)} f=0\ .
\end{split}
\end{equation}
Further, since $\mathbb{L}_{ij}$s commute with the laplacian in $\mathbb{R}^d$, $\mathbb{\Delta}_{i,\alpha+2}^{(\alpha)}f$ is a harmonic vector field if $f$ is harmonic. We can then take $f$ to be any homogeneous harmonic polynomial in $\mathbb{R}^d$ to get a homogeneous harmonic vector field. Thus, an overcomplete basis of homogeneous harmonic vector fields of degree $\ell$ can be constructed by taking 
\begin{equation}
\begin{split}
\mathbb{\Delta}_{i,\alpha+2}^{(\alpha)} [x^{<i_1}x^{i_2}\ldots x^{i_\ell>}]\frac{\partial}{\partial x_i}
\end{split}
\end{equation}   
for $\alpha=1,2,\ldots,(d-2)$.
Such vector fields, when restricted to $\mathbb{S}^{d-1}$, yield a vector spherical harmonic (VSH). In the next subsection, we will construct an orthonormal basis for such VSHs.

The above set of toroidal operators can be generalised as follows. Say we are given a subspace $\mathbb{R}^{\alpha+2}\subseteq \mathbb{R}^d$. We can then define a toroidal operator corresponding to this subspace and a direction $j$ within that subspace via
\begin{equation}
\begin{split}
\mathbb{\Delta}^{(\alpha)}_{ij}f\equiv \Bigl\{\sum_{k\in \mathbb{R}^{\alpha+2}}\mathbb{L}_{kj}\mathbb{L}_{ki} -\frac{1}{2}\delta_{ij}\sum_{k,l\in \mathbb{R}^{\alpha+2}}\mathbb{L}_{kl}\mathbb{L}_{kl}\Bigr\}_{i,j\in \mathbb{R}^{\alpha+2}}f\ .
\end{split}
\end{equation}
This formula should be interpreted as follows: first of all, we get a non-zero answer only if $i,j$ directions are tangent to the subspace $\mathbb{R}^{\alpha+2}$ under question. Further sums indicated inside the bracket are over directions within $\mathbb{R}^{\alpha+2}$. If we take the subspace  $\mathbb{R}^{\alpha+2}$ spanned by the cartesian directions $\{x_1,x_2,\ldots,x_{\alpha+2}\}$ and choose $j$ to be equal along $x^{\alpha+2}$, and using
\begin{equation}
\begin{split}
\sum_{k=1}^{\alpha+2}\mathbb{L}_{k,\alpha+2}\mathbb{L}_{ki} &=
\sum_{k=1}^{\alpha+1}\mathbb{L}_{k,\alpha+2}\mathbb{L}_{ki} \  \text{for} \ 1\leq i\leq \alpha+1\ ,\\
\sum_{k=1}^{\alpha+2}\mathbb{L}_{k,\alpha+2}\mathbb{L}_{k,\alpha+2}-\frac{1}{2}\sum_{j,k=1}^{\alpha+2}\mathbb{L}_{jk}\mathbb{L}_{jk}&=-\frac{1}{2}\sum_{j,k=1}^{\alpha+1}\mathbb{L}_{jk}\mathbb{L}_{jk}\ ,
\end{split}
\end{equation}
we get back the toroidal operators defined before in Eq.\eqref{eq:DeltaVdefCart}.

A couple of remarks about the above form: first,  if the function $f$ is invariant under the $SO(\alpha+1)$ that rotates $\{x_1,x_2,\ldots\ ,x_{\alpha+1}\}$ , all the $\mathbb{L}_{ij}$s in 
Eq.\eqref{eq:DeltaVdefCart} annihilate $f$, and we get a vector field that is identically zero. Thus, \emph{the toroidal operators we have defined above have non-trivial kernels which become smaller as $\alpha$ increases.} Relatedly, at a given $\alpha$, the harmonic vector field should necessarily break $SO(\alpha+1)$. A second remark is that, for $\alpha=1$, the Eq.\eqref{eq:DeltaVdefCart} reduces to
\begin{equation}\label{eq:DeltaVdefCart3d}
\begin{split}
\mathbb{\Delta}_{i,3}^{(\alpha=1)} f&\equiv \left\{\begin{array}{cc} 
-\mathbb{L}_{23}\mathbb{L}_{12}f\ & \text{for} \ i=1\ ,\\
-\mathbb{L}_{31}\mathbb{L}_{12}f\ & \text{for} \ i=2\ ,\\
-\mathbb{L}_{12}^2f\  & \text{for} \ i=3 \ ,\\
  0 & \text{for} \ i>3 \ .
\end{array}\right.
\end{split}
\end{equation}
We recognise in RHS the familiar $3d$ toroidal operator $-\vec{r}\times\vec{\nabla}$ acting on $\mathbb{L}_{12}f$. As remarked above, the kernel of the above operator is the largest among all the toroidal operators: it is the set of  $SO(2)$ invariant functions, where the $SO(2)$ rotates the $12$ plane. But, in this special case of $\alpha=1$ (and only in this case), we can improve our toroidal operator by dropping an $\mathbb{L}_{12}$ and defining 
\begin{equation}\label{eq:DeltaVdefCart3dNew}
\begin{split}
\mathbb{\Delta}_{i,3}^{(\alpha=1)}|_{\text{New}} f&\equiv \left\{\begin{array}{cc} 
-\mathbb{L}_{23}f\ & \text{for} \ i=1\ ,\\
-\mathbb{L}_{31}f\ & \text{for} \ i=2\ ,\\
-\mathbb{L}_{12}f\  & \text{for} \ i=3 \ ,\\
  0 & \text{for} \ i>3 \ .
\end{array}\right.
\end{split}
\end{equation}
The kernel of this `improved' toroidal operator is smaller and is the set of  $SO(3)$ invariant functions, where the $SO(3)$ rotates the $x_1,x_2$ and $x_3$, i.e., the kernel is of the same size as the $\alpha=2$ toroidal operator. The usual 3d toroidal operator is improved in this sense.

\subsubsection*{Vector spherical harmonics on $\mathbb{S}^{d-1}$}
We now turn to a description in spherical coordinates.
The toroidal double-derivative operators on $\mathbb{S}^{d-1}$ take the following form:
\begin{equation}\label{eq:DeltaVdef}
\begin{split}
\mathbb{\Delta}_{I}^{\alpha} f&\equiv \sqrt{\gamma_{\alpha\alpha}} \times\left\{\begin{array}{cc} 
\frac{1}{\sin^{\alpha-2}\vartheta_\alpha}\frac{\partial}{\partial \vartheta_\alpha}\frac{\partial}{\partial \vartheta_I}\left[\sin^{\alpha-1}\vartheta_{\alpha}\ f\right]\ & \text{for} \ 0\leq I\leq \alpha-1\ ,\\
-\frac{1}{\sin\vartheta_\alpha} \mathscr{D}^2_{\mathbb{S}^\alpha}f\  & \text{for} \ I=\alpha \ ,\\
  0 & \text{for} \ I>\alpha \ .
\end{array}\right.
\end{split}
\end{equation}
Here, $f$ is an arbitrary function on the sphere,  the index $\alpha$ takes values from $1$ to $(d-2)$, thus defining $(d-2)$
different derivative operators. The index $I=0,1,\ldots,(d-2)$ denotes the vector directions on $\mathbb{S}^{d-1}$, $\mathscr{D}^2_{\mathbb{S}^{I+1}}$ is the lower sphere laplacian defined in Eq.\eqref{eq:mId}, and $\gamma_{IJ}$ are the sphere metric coefficients defined in Eq.\eqref{eq:sphMet}. The above  derivative operators exhibit the following useful properties, as can be established via direct computation:
\begin{itemize}
\item For an arbitrary  function $f$  on $\mathbb{S}^{d-1}$, the corresponding  vector field $\mathbb{\Delta}_{I}^{\alpha} f$  is divergence-free. 
\item The derivative operators $\mathbb{\Delta}_I^\alpha$ obey the following commutation relation with the sphere laplacian:
\begin{equation}\label{eq:DeltaEig}
\begin{split}
[\mathscr{D}^2,\mathbb{\Delta}_I^\alpha]f=\mathbb{\Delta}_I^\alpha f\ .
\end{split}
\end{equation}
If we distinguish between scalar and the vector laplacians on the sphere by subscripts $S$ and $V$ respectively, the relation above can also be stated as 
$\mathscr{D}_V^2\mathbb{\Delta}_I^\alpha f=\mathbb{\Delta}_I^\alpha (\mathscr{D}_S^2+1)f$ .
\item These vector fields are mutually orthogonal in the following sense: for any two functions $f$ and $g$ on $\mathbb{S}^{d-1}$, we have
\begin{equation}\label{eq:DeltaOrtho}
\begin{split}
\int_{\mathbb{S}^{d-1}}\gamma^{IJ}\ (\mathbb{\Delta}_{I}^{\alpha} f)\ (\mathbb{\Delta}_{J}^{\alpha'} g) =0\quad \text{for $\alpha\neq \alpha'$.}
\end{split}
\end{equation}
When $\alpha=\alpha'$, the same inner product evaluates to 
\begin{equation}\label{eq:DeltaNorm}
\begin{split}
\int_{\mathbb{S}^{d-1}}\gamma^{IJ}\ (\mathbb{\Delta}_{I}^{\alpha} f)\ (\mathbb{\Delta}_{J}^{\alpha} g) =\int_{\mathbb{S}^{d-1}}(\mathscr{D}^2_{\mathbb{S}^\alpha}f)\Bigl(\mathscr{D}^2_{\mathbb{S}^{\alpha+1}}g+(1-\alpha)g\Bigr)\ .
\end{split}
\end{equation}
\end{itemize}
Once these statements are established, the $\mathbb{\Delta}_{I}^{\alpha}$ operators can be used to give an explicit form for the vector spherical harmonics (VSHs). 

To this end, consider the vector fields defined\footnote{Our definitions here are consistent with the recursive construction by Higuchi\cite{Higuchi:1986wu}. See also appendix A.2 of \cite{Marolf:2008hg} where Higuchi's construction is reviewed).} by acting  $\mathbb{\Delta}_I^\alpha$'s on the orthonormal SSHs of Eq.\eqref{eq:SSHdef}:
\begin{equation}\label{eq:VSHdef}
\begin{split}
\mathbb{V}_{I}^{\alpha\ell\vec{m}}&\equiv C^V_{\alpha\ell\vec{m}} \mathbb{\Delta}_I^\alpha \mathscr{Y}_{\ell\vec{m}}(\hat{r})\\
&=C^V_{\alpha\ell\vec{m}} \sqrt{\gamma_{\alpha\alpha}} \times\left\{\begin{array}{cc} 
\frac{1}{\sin^{\alpha-2}\vartheta_\alpha}\frac{\partial}{\partial \vartheta_\alpha}\frac{\partial}{\partial \vartheta_I}\left[\sin^{\alpha-1}\vartheta_{\alpha}\ \mathscr{Y}_{\ell\vec{m}}(\hat{r})\right]\ & \text{for} \ 0\leq I\leq \alpha-1\ ,\\
 \frac{1}{\sin\vartheta_\alpha}m_\alpha(m_\alpha+\alpha-1)\ \mathscr{Y}_{\ell\vec{m}}(\hat{r})\  & \text{for} \ I=\alpha \ ,\\
  0 & \text{for} \ I>\alpha \ .
\end{array}\right.
\end{split}
\end{equation}
Here, we have simplified the $I=\alpha$ component by using Eq.\eqref{eq:mId},  and $C^V_{\alpha\ell\vec{m}}$ is a convenient normalization factor to be determined shortly. Using the first two properties $\mathbb{\Delta}_I^\alpha$ enumerated above, we conclude that $\mathbb{V}^I_{\alpha\ell\vec{m}}$ is  a divergence-free vector field  satisfying 
\begin{equation}
\begin{split}
\left[\mathscr{D}^2+\ell(\ell+d-2)-1\right] \mathbb{V}^I_{\alpha\ell\vec{m}}= 0\ .
\end{split}
\end{equation}
Eq.\eqref{eq:DeltaOrtho} then ensures the orthogonality 
of $\mathbb{V}^I_{\alpha\ell\vec{m}}$ and $\mathbb{V}^I_{\alpha'\ell'\vec{m}'}$
for $\alpha\neq \alpha'$. For $\alpha=\alpha'$, we use Eq.\eqref{eq:DeltaNorm} and Eq.\eqref{eq:mId} to get
\begin{equation}
\begin{split}
\int_{\mathbb{S}^{d-1}}&\gamma^{IJ}\mathbb{V}_{I}^{\alpha\ell\vec{m}}\mathbb{V}_{J}^{\alpha\ell'\vec{m}'\ast}\\
&=C^V_{\alpha\ell\vec{m}} C^{V\ast}_{\alpha\ell'\vec{m}'}   m_\alpha(m_\alpha+\alpha-1) (m'_{\alpha+1}+1)(m'_{\alpha+1}+\alpha-1)\int_{\mathbb{S}^{d-1}} \ \mathscr{Y}_{\ell\vec{m}}(\hat{r}) \mathscr{Y}_{\ell'\vec{m}'}^\ast(\hat{r})\ .
\end{split}
\end{equation}
From this, we conclude that $\mathbb{V}^I_{\alpha\ell\vec{m}}$ and $\mathbb{V}^I_{\alpha'\ell'\vec{m}'}$ are orthogonal unless $\alpha=\alpha',\ell=\ell'$ and $\vec{m}=\vec{m}'$. The above computation also determines the normalisation factor $C^V_{\alpha\ell\vec{m}}$ via
\begin{equation}\label{eq:VSHnorm}
\begin{split}
|C^V_{\alpha\ell\vec{m}}|^{-2} \equiv (m_{\alpha+1}+\alpha-1)(m_{\alpha+1}+1)m_{\alpha}(m_{\alpha}+\alpha-1)\ .
\end{split}
\end{equation}

We note that this normalization factor diverges if $m_\alpha = 0$, or if $\alpha=1$ and $m_{\alpha+1}=m_2=0$. This means that, unless the function multiplying 
this normalization factor in Eq.\eqref{eq:VSHdef} goes to zero in these cases, we have to discard the result for being non-normalisable. If instead, in any of these cases, \emph{all} the components in Eq.\eqref{eq:VSHdef} vanish, it is possible to get a finite result by taking a limit. We have to decide which case corresponds to which possibility. A careful analysis leads to the following  conclusions:
\begin{itemize}
\item For $\alpha>1$, the above expression yields a normalized VSH only if $m_\alpha > 0$. Thus, in this case, we should exclude the possibility that $m_\alpha=0$.
\item For $\alpha=1$, $m_\alpha=m_1$ can be taken to be zero and Eq.\eqref{eq:VSHdef} gives a finite answer when understood as a limit, provided 
$m_{\alpha+1}=m_2>0$.
\end{itemize}
As an example to illustrate the second point, consider the following  $\alpha=1$ VSH on $\mathbb{S}^4$:
\begin{equation}\begin{split}
\mathbb{V}_{I}^{1\ell\vec{m}}|_{\mathbb{S}^4}&\equiv \frac{\sin\vartheta_3\sin\vartheta_2}{m_1\sqrt{m_2(m_2+1)}} \left\{\begin{array}{cc} 
\sin\vartheta_1\frac{\partial}{\partial \vartheta_1}\frac{\partial}{\partial \varphi}\mathscr{Y}_{\ell\vec{m}}(\hat{r})\ & \text{for} \ I=0\ ,\\
 \frac{1}{\sin\vartheta_1}m_1^2\ \mathscr{Y}_{\ell\vec{m}}(\hat{r})\  & \text{for} \ I=1 \ ,\\
  0 & \text{for} \ I=2,3 \ .
\end{array}\right.
\end{split}\end{equation}
From Eq.\eqref{eq:SSHdef}, we can write $\frac{\partial}{\partial \varphi}\mathscr{Y}_{\ell\vec{m}}(\hat{r})=\pm im_1\ \mathscr{Y}_{\ell\vec{m}}(\hat{r})$. It is then clear that we get a finite result in the expression above as we take $m_1\to 0$ (provided $m_2>0$):
\begin{equation}\begin{split}
\lim_{m_1\to 0}(\mp i)\mathbb{V}_{I}^{1\ell\vec{m}}|_{\mathbb{S}^4}&\equiv \frac{\sin\vartheta_3\sin\vartheta_2 \sin\vartheta_1}{\sqrt{m_2(m_2+1)}} \left\{\begin{array}{cc} 
 \frac{\partial}{\partial \vartheta_1}\mathscr{Y}_{\ell\vec{m}}(\hat{r})|_{m_1=0}\ & \text{for} \ I=0\ ,\\
  0 & \text{for} \ I=1,2,3 \ .
\end{array}\right.
\end{split}\end{equation}
Alternately, we can avoid this subtlety for $\alpha=1$ altogether, by redefining the derivative operator $\mathbb{\Delta}_{I}^{\alpha=1}$ by stripping off a $\frac{\partial}{\partial \varphi}$ from it, i.e., we define
\begin{equation}
\begin{split}
\mathbb{\Delta}_{I}^{\alpha=1} f|_{\text{new}}&\equiv \sqrt{\gamma_{11}} \times\left\{\begin{array}{cc} 
-\sin\vartheta_1\frac{\partial f}{\partial \vartheta_1}\ & \text{for} \ I=0\ ,\\
\frac{1}{\sin\vartheta_1} \frac{\partial f}{\partial\varphi}\  & \text{for} \ I=1 \ ,\\
  0 & \text{for} \ I>1\ .
\end{array}\right.
\end{split}
\end{equation}
This corresponds exactly to the `improvement' of the $\alpha=1$ toroidal operator described before in the cartesian language.
Since $\frac{\partial}{\partial \varphi}$ is an isometry, this redefinition does not change any of the properties of $\mathbb{\Delta}_{I}^{\alpha=1}$ except for its overall normalisation (a factor of $m_1$ has to be dropped). The orthogonality Eq.\eqref{eq:DeltaOrtho} still holds, whereas Eq.\eqref{eq:DeltaNorm} becomes 
\begin{equation}\label{eq:DeltaNormNew}
\begin{split}
\int_{\mathbb{S}^{d-1}}\gamma^{IJ}\ (\mathbb{\Delta}_{I}^{\alpha=1} f)_\text{New}\ (\mathbb{\Delta}_{J}^{\alpha=1} g)_\text{New} =-\int_{\mathbb{S}^{d-1}}f\ \mathscr{D}^2_{\mathbb{S}^2}g\ .
\end{split}
\end{equation}
With this norm, the VSH quoted above then becomes 
\begin{equation}\begin{split}
\mathbb{V}_{I}^{1\ell\vec{m}}|_{\mathbb{S}^4,\text{New}}&\equiv \frac{\sin\vartheta_3\sin\vartheta_2}{\sqrt{m_2(m_2+1)}} \left\{\begin{array}{cc} 
-\sin\vartheta_1\frac{\partial}{\partial \vartheta_1}\mathscr{Y}_{\ell\vec{m}}(\hat{r})\ & \text{for} \ I=0\ ,\\
 \frac{1}{\sin\vartheta_1}\frac{\partial}{\partial \varphi}\ \mathscr{Y}_{\ell\vec{m}}(\hat{r})\  & \text{for} \ I=1 \ ,\\
  0 & \text{for} \ I=2,3 \ .
\end{array}\right.
\end{split}\end{equation}
The expression appearing here is, in fact, the standard VSH on $\mathbb{S}^2$ constructed via the toroidal operator $\vec{r}\times\vec{\nabla}$: rewritten in this form, no subtle limiting procedure is necessary to deal with the $m_1=0$ case. Adopting this new definition, we give in table~\ref{tab:VSHexp} the explicit form of VSHs in $\mathbb{S}^2,\mathbb{S}^3,\mathbb{S}^4$ and $\mathbb{S}^5$.

\begin{table}
\centering
\resizebox{0.95\columnwidth}{!}{
\begin{tabular}{ |c| } 
 \hline
 \textbf{VSHs on $\mathbb{S}^2$}\\
 \hline
 \\
\parbox{15cm}{\begin{equation*}\mathbb{V}_{I}^{1\ell\vec{m}}|_{\text{New}}\equiv \frac{1}{\sqrt{\ell(\ell+1)}}
\left\{\begin{array}{cc} 
-\sin\vartheta_1\frac{\partial}{\partial \vartheta_1}\mathscr{Y}_{\ell\vec{m}}(\hat{r})\ & \text{for} \ I=0\ ,\\
 \frac{1}{\sin\vartheta_1}\frac{\partial}{\partial \varphi}\ \mathscr{Y}_{\ell\vec{m}}(\hat{r})\  & \text{for} \ I=1  \ .
\end{array}\right.\end{equation*}}\\ 
 \hline
  \hline
 \textbf{VSHs on $\mathbb{S}^3$}\\
 \hline
 \\
\parbox{15cm}{\begin{equation*}
\begin{split}
\mathbb{V}_{I}^{2\ell\vec{m}}&\equiv\frac{1}{ (\ell+1)\sqrt{m_2(m_2+1)}}\left\{\begin{array}{cc} 
\frac{\partial}{\partial \vartheta_2}\frac{\partial}{\partial \vartheta_I}\left[\sin\vartheta_2\ \mathscr{Y}_{\ell\vec{m}}(\hat{r})\right]\ & \text{for} \ I=0,1\ ,\\
 \frac{1}{\sin\vartheta_2}m_2(m_2+1)\ \mathscr{Y}_{\ell\vec{m}}(\hat{r})\  & \text{for} \ I=2 \ .
\end{array}\right.\\
\mathbb{V}_{I}^{1\ell\vec{m}}|_{\text{New}}&\equiv \frac{\sin\vartheta_2}{\sqrt{m_2(m_2+1)}} \left\{\begin{array}{cc} 
-\sin\vartheta_1\frac{\partial}{\partial \vartheta_1}\mathscr{Y}_{\ell\vec{m}}(\hat{r})\ & \text{for} \ I=0\ ,\\
 \frac{1}{\sin\vartheta_1}\frac{\partial}{\partial \varphi}\ \mathscr{Y}_{\ell\vec{m}}(\hat{r})\  & \text{for} \ I=1 \ ,\\
  0 & \text{for} \ I=2 \ .
\end{array}\right.\end{split}\end{equation*}}\\ 
 \hline
  \hline
 \textbf{VSHs on $\mathbb{S}^4$}\\
 \hline
 \\
\parbox{15cm}{\begin{equation*}
\begin{split}
\mathbb{V}_{I}^{3\ell\vec{m}}&\equiv\frac{1}{ \sqrt{(\ell+1)(\ell+2)m_3(m_3+2)}}\left\{\begin{array}{cc} 
 \frac{1}{\sin\vartheta_3}\frac{\partial}{\partial \vartheta_3}\frac{\partial}{\partial \vartheta_I}\left[\sin^2\vartheta_3\ \mathscr{Y}_{\ell\vec{m}}(\hat{r})\right]\ & \text{for} \ I=0,1,2\ ,\\
 \frac{1}{\sin\vartheta_3}m_3(m_3+2)\ \mathscr{Y}_{\ell\vec{m}}(\hat{r})\  & \text{for} \ I=3 \ .
\end{array}\right.\\
\mathbb{V}_{I}^{2\ell\vec{m}}&\equiv\frac{\sin\vartheta_3}{ (m_3+1)\sqrt{m_2(m_2+1)}}\left\{\begin{array}{cc} 
\frac{\partial}{\partial \vartheta_2}\frac{\partial}{\partial \vartheta_I}\left[\sin\vartheta_2\ \mathscr{Y}_{\ell\vec{m}}(\hat{r})\right]\ & \text{for} \ I=0,1\ ,\\
 \frac{1}{\sin\vartheta_2}m_2(m_2+1)\ \mathscr{Y}_{\ell\vec{m}}(\hat{r})\  & \text{for} \ I=2 \ ,\\
  0 & \text{for} \ I=3 \ .
\end{array}\right.\\
\mathbb{V}_{I}^{1\ell\vec{m}}|_{\text{New}}&\equiv \frac{\sin\vartheta_3\sin\vartheta_2}{\sqrt{m_2(m_2+1)}} 
\left\{\begin{array}{cc} 
-\sin\vartheta_1\frac{\partial}{\partial \vartheta_1}\mathscr{Y}_{\ell\vec{m}}(\hat{r})\ & \text{for} \ I=0\ ,\\
 \frac{1}{\sin\vartheta_1}\frac{\partial}{\partial \varphi}\ \mathscr{Y}_{\ell\vec{m}}(\hat{r})\  & \text{for} \ I=1 \ ,\\
  0 & \text{for} \ I=2,3 \ .
\end{array}\right.
\end{split}\end{equation*}}\\ 
 \hline 
  \hline
 \textbf{VSHs on $\mathbb{S}^5$}\\
 \hline
 \\
\parbox{15cm}{\begin{equation*}
\begin{split}
\mathbb{V}_{I}^{4\ell\vec{m}}&\equiv\frac{1}{ \sqrt{(\ell+1)(\ell+3)m_4(m_4+3)}}\left\{\begin{array}{cc} 
 \frac{1}{\sin^2\vartheta_4}\frac{\partial}{\partial \vartheta_4}\frac{\partial}{\partial \vartheta_I}\left[\sin^3\vartheta_4\ \mathscr{Y}_{\ell\vec{m}}(\hat{r})\right]\ & \text{for} \ I=0,1,2,3\ ,\\
 \frac{1}{\sin\vartheta_4}m_4(m_4+3)\ \mathscr{Y}_{\ell\vec{m}}(\hat{r})\  & \text{for} \ I=4 \ .
\end{array}\right.\\
\mathbb{V}_{I}^{3\ell\vec{m}}&\equiv\frac{\sin\vartheta_4}{ \sqrt{(m_4+1)(m_4+2)m_3(m_3+2)}}\left\{\begin{array}{cc} 
 \frac{1}{\sin\vartheta_3}\frac{\partial}{\partial \vartheta_3}\frac{\partial}{\partial \vartheta_I}\left[\sin^2\vartheta_3\ \mathscr{Y}_{\ell\vec{m}}(\hat{r})\right]\ & \text{for} \ I=0,1,2\ ,\\
 \frac{1}{\sin\vartheta_3}m_3(m_3+2)\ \mathscr{Y}_{\ell\vec{m}}(\hat{r})\  & \text{for} \ I=3 \ ,\\
  0 & \text{for} \ I=4 \ .
\end{array}\right.\\
\mathbb{V}_{I}^{2\ell\vec{m}}&\equiv\frac{\sin\vartheta_4\sin\vartheta_3}{ (m_3+1)\sqrt{m_2(m_2+1)}}\left\{\begin{array}{cc} 
\frac{\partial}{\partial \vartheta_2}\frac{\partial}{\partial \vartheta_I}\left[\sin\vartheta_2\ \mathscr{Y}_{\ell\vec{m}}(\hat{r})\right]\ & \text{for} \ I=0,1\ ,\\
 \frac{1}{\sin\vartheta_2}m_2(m_2+1)\ \mathscr{Y}_{\ell\vec{m}}(\hat{r})\  & \text{for} \ I=2 \ ,\\
  0 & \text{for} \ I=3,4 \ .
\end{array}\right.\\
\mathbb{V}_{I}^{1\ell\vec{m}}|_{\text{New}}&\equiv \frac{\sin\vartheta_4\sin\vartheta_3\sin\vartheta_2}{\sqrt{m_2(m_2+1)}} \left\{\begin{array}{cc} 
-\sin\vartheta_1\frac{\partial}{\partial \vartheta_1}\mathscr{Y}_{\ell\vec{m}}(\hat{r})\ & \text{for} \ I=0\ ,\\
 \frac{1}{\sin\vartheta_1}\frac{\partial}{\partial \varphi}\ \mathscr{Y}_{\ell\vec{m}}(\hat{r})\  & \text{for} \ I=1 \ ,\\
  0 & \text{for} \ I=2,3,4 \ .
\end{array}\right.
\end{split}\end{equation*}}\\ 
 \hline  
\end{tabular}}
\caption{Explicit expressions for vector spherical harmonics (VSHs).}
\label{tab:VSHexp}
\end{table}

Before we conclude, it is often convenient to have a simple VSH for any given $\ell$ written down explicitly, on which computations can be done with ease. We will end this subsection by providing two such examples. The first example is the VSH corresponding to  
\begin{equation}\begin{split}
\alpha=1\ ,\ m_1=0\ ,\ m_2=m_3=\ldots=m_{d-2}=1\leq  m_{d-1}\equiv \ell\ .
\end{split}\end{equation}
The normalised SSH for this $\vec{m}$ is given by Eq.\eqref{eq:SSHdef} as
\begin{equation}\begin{split}
\mathscr{Y}_{\ell\vec{m}}(\hat{r})\ &=  \sqrt{2\pi\frac{\mathcal{N}_{HH}(d+2,\ell-1)}{|\mathbb{S}^{d+1}|}} P_{\ell-1}(d+2,\cos\vartheta_{d-2}) \cos\vartheta_1\prod_{J=2}^{d-2}\sin\vartheta_J
\end{split}\end{equation}
The corresponding VSH is given by 
\begin{equation}\begin{split}
\mathbb{V}_{I}&= \frac{\prod_{J=2}^{d-2}\sin\vartheta_J}{\sqrt{m_2(m_2+1)}|_{m_2=1}} \left\{\begin{array}{cc} 
-\sin\vartheta_1\frac{\partial}{\partial \vartheta_1}\mathscr{Y}_{\ell\vec{m}}(\hat{r})\ & \text{for} \ I=0\ ,\\
  0 & \text{for} \ I=1,2,3,\ldots, d-2 \ .
\end{array}\right.\\
&= \sqrt{\pi\frac{\mathcal{N}_{HH}(d+2,\ell-1)}{|\mathbb{S}^{d+1}|}} P_{\ell-1}(d+2,\cos\vartheta_{d-2})\left\{\begin{array}{cc} 
\prod_{J=1}^{d-2}\sin^2\vartheta_J \ & \text{for} \ I=0\ ,\\
  0 & \text{for} \ I=1,2,3,\ldots, d-2 \ .
\end{array}\right.
\end{split}\end{equation}
We can also present this as a vector field on $\mathbb{S}^{d-1}$ by raising the sphere index, viz., 
\begin{equation}\begin{split}
\mathbb{V}^{I}\frac{\partial}{\partial\vartheta_I}&= \sqrt{\pi\frac{\mathcal{N}_{HH}(d+2,\ell-1)}{|\mathbb{S}^{d+1}|}} P_{\ell-1}(d+2,\cos\vartheta_{d-2})\frac{\partial}{\partial\varphi}\ .
\end{split}\end{equation}
This vector-field can  be pushed-forward to $\mathbb{R}^d$ using Eq.\eqref{eq:PushFsph}: we then get a vector field which varies as
\begin{equation}\begin{split}
 P_{\ell-1}\left(d+2,\frac{x_d}{r}\right)\Bigl\{x_1\frac{\partial}{\partial x_2}-x_2\frac{\partial}{\partial x_2}\Bigr\}\ .
\end{split}\end{equation}
It is then evident that this vector field  is invariant under the 
$SO(2)$ rotations of $x_1-x_2$ plane as well as $SO(d-3)$ rotations of
$x_3,x_4,\ldots,x_{d-1}$. This is, hence, a simple VSH with a large group of symmetries, and we found it to be a convenient example to check our computations. 

The second example we discuss is a VSH with an even bigger symmetry of  $SO(d-2)$ that rotates $x_1,x_2,\ldots,x_{d-2}$. This is the most symmetric of all VSHs (for $d>4$) and will play an important role when we discuss the VSH addition theorem. The $SO(d-2)$-invariant VSH is obtained by taking
\begin{equation}\begin{split}
\alpha=d-2\ ,\ m_1=m_2=m_3=\ldots=m_{d-3}=0\ ,\ m_{d-2}=1\leq  m_{d-1}\equiv \ell\ .
\end{split}\end{equation}
The normalised SSH in this case is
\begin{equation}\begin{split}
\mathscr{Y}_{\ell\vec{m}}(\hat{r})\ &=  (d-1)\sqrt{\frac{1}{2\pi}\frac{\mathcal{N}_{HH}(d+2,\ell-1)}{|\mathbb{S}^{d+1}|}} \sin\vartheta_{d-2}P_{\ell-1}(d+2,\cos\vartheta_{d-2}) \cos\vartheta_{d-3}\ .
\end{split}\end{equation}
The corresponding VSH has the following form
\begin{equation}\begin{split}
\mathbb{V}^{I}\frac{\partial}{\partial\vartheta_I}&= \frac{(d-1)}{\sqrt{(d-2)(\ell+1)(\ell+d-3)}}\sqrt{\frac{1}{2\pi}\frac{\mathcal{N}_{HH}(d+2,\ell-1)}{|\mathbb{S}^{d+1}|}} \\
&\quad \times \Bigl\{(d-2)\cos\vartheta_{d-3} P_{\ell-1}(d+2,\cos\vartheta_{d-2})\frac{\partial}{\partial\vartheta_{d-2}}\Bigr.\\
&\Bigl.\qquad\quad -\frac{\sin\vartheta_{d-3}}{\sin^{d-2}\vartheta_{d-2}} \frac{d}{d\vartheta_{d-3}} [\sin^{d-2}\vartheta_{d-2}P_{\ell-1}(d+2,\cos\vartheta_{d-2})]\ \frac{\partial}{\partial\vartheta_{d-3}}\Bigr\}\ .
\end{split}\end{equation}
Stripping of the normalization pre-factor, its push-forward to $\mathbb{R}^d$ at radius $r$ (
computed via Eq.\eqref{eq:PushFsph}) yields
\begin{equation}\label{eq:LegVSH}
\begin{split}
&\frac{1}{d-1}r^{\ell-1}P_{\ell-1}\left(d+2,\frac{x_d}{r}\right)\left\{x_d\frac{\partial}{\partial x_{d-1}}-x_{d-1}\frac{\partial}{\partial x_{d}}\right\}\\
&\quad +\frac{(\ell-1)(\ell+d-1)}{(d+1)(d-1)(d-2)}r^{\ell-2}P_{\ell-2}\left(d+4,\frac{x_d}{r}\right)\sum_{j=1}^{d-2}x_j\left\{x_{d-1}\frac{\partial}{\partial x_j}-x_j\frac{\partial}{\partial x_{d-1}}\right\}\ .
\end{split}   
\end{equation}
Here, we have used the identity
\begin{equation}
\begin{split}
\frac{d}{dz} P_\ell(d,z)= \frac{\ell(\ell+d-2)}{d-1} P_{\ell-1}(d+2,z)
\end{split}   
\end{equation}
to compute the derivative of the generalised Legendre polynomials.

\subsubsection*{Addendum: Counting of VSHs}
Our explicit construction can be used to count the total number of VSHs for a given $\ell=m_{d-1}$: we will denote this by $\Nhh^V(d,\ell)=\Nhh^V(d,m_{d-1})$. To begin with, if there were no constraints on $\vec{m}$, all VSHs are obtained by acting $(d-2)$ derivative operators on $\Nhh(d,m_{d-1})$ number of SSHs and $\Nhh^V(d,m_{d-1})$ should just be  $(d-2)\Nhh(d,m_{d-1})$. But given the constraints on $\vec{m}$ described above, this is an overcounting, and a more careful counting is needed.

For a given $\alpha>1$, the number of VSHs we obtain is given by 
\begin{equation}
\begin{split}
\Nhh^{V,\alpha}(d,m_{d-1})&=\sum\limits_{m_{d-2}=1}^{m_{d-1}}\sum\limits_{m_{d-3}=1}^{m_{d-2}}\dots \sum\limits_{m_{\alpha}=1}^{m_{\alpha+1}}\sum\limits_{m_{\alpha-1}=0}^{m_{\alpha}}\sum\limits_{m_{\alpha-2}=0}^{m_{\alpha-1}}\dots \left\{1+\sum\limits_{m_1=1}^{m_2}2\right\}\\
&=\sum\limits_{m_{d-2}=1}^{m_{d-1}}\dots \sum\limits_{m_{\alpha}=1}^{m_{\alpha+1}}\Nhh(\alpha+1,m_{\alpha})\ .
\end{split}
\end{equation}
Here, we have imposed the constraint that $m_{d-1}\geq m_{d-2}\geq\ldots m_\alpha\geq 1$ and have used Eq.\eqref{eq:NHHsum} in the second line. In the next steps, we should systematically subtract out the forbidden $m_i$'s, e.g., the next few steps are given by 
\begin{equation}
\begin{split}
\Nhh^{V,\alpha}(d,m_{d-1})
&=\sum\limits_{m_{d-2}=1}^{m_{d-1}}\dots \sum\limits_{m_{\alpha+1}=1}^{m_{\alpha+2}}\left\{\Nhh(\alpha+2,m_{\alpha+1})-1\right\}\\
&=\sum\limits_{m_{d-2}=1}^{m_{d-1}}\dots \sum\limits_{m_{\alpha+2}=1}^{m_{\alpha+3}}\left\{\Nhh(\alpha+3,m_{\alpha+2})-1-m_{\alpha+2}\right\}\\
&=\sum\limits_{m_{d-2}=1}^{m_{d-1}}\dots \sum\limits_{m_{\alpha+3}=1}^{m_{\alpha+4}}\left\{\Nhh(\alpha+4,m_{\alpha+3})-1-\binom{m_{\alpha+3}}{1}-\binom{m_{\alpha+3}+1}{2}\right\}\ .
\end{split}
\end{equation}
We can show that, after $k$ steps, the above expression generalises to
\begin{equation}
\begin{split}
\Nhh^{V,\alpha}(d,m_{d-1})
&=\sum\limits_{m_{d-2}=1}^{m_{d-1}}\dots \sum\limits_{m_{\alpha+k}=1}^{m_{\alpha+k+1}}\left\{\Nhh(\alpha+k+1,m_{\alpha+k})-\sum\limits_{j=0}^{k-1}\binom{m_{\alpha+k}+j-1}{j}\right\}\ .
\end{split}
\end{equation}
This follows from the recursive use of the binomial identity
\begin{equation}
\sum\limits_{m=1}^{j}\binom{m+k-1}{k}=\binom{j+k}{k+1}\ .
\end{equation}
The recursion terminates when $k=d-1-\alpha$ and we get
\begin{equation}
\begin{split}
\Nhh^{V,\alpha>1}(d,m_{d-1})
&=\Nhh(d,m_{d-1})-\sum\limits_{j=0}^{d-2-\alpha}\binom{m_{d-1}+j-1}{j}.
\end{split}
\end{equation}
For $\alpha =1$,  we have to take $m_{d-1}\geq m_{d-2}\geq\ldots m_2\geq 1$, but $m_1$ is allowed to vanish. This is identical to the $\alpha=2$ case of the counting above. Thus, we get
\begin{equation}\begin{split}
    \Nhh^{V,\alpha=1}(d,m_{d-1})&= \Nhh^{V,\alpha=2}(d,m_{d-1})
    =\Nhh(d,m_{d-1})-\sum\limits_{j=0}^{d-4}\binom{m_{d-1}+j-1}{j}\ .
\end{split}\end{equation}
We can now sum over $\alpha$ to obtain the the total number of VSHs on $\mathbb{S}^{d-1}$ as
\begin{equation}
\begin{split}
\Nhh^{V}(d,m_{d-1})&=\sum\limits_{\alpha=1}^{d-2}\Nhh^{V,\alpha}(d,m_{d-1})= (d-2)\Nhh(d,m_{d-1})-\Nhh(d-2,m_{d-1}+1)
\end{split}
\end{equation}
As we shall describe in more detail later, $\Nhh^V(d,\ell)$ counts the number of transverse, divergence-free, homogeneous harmonic polynomial vector fields of degree $\ell$ in $\mathbb{R}^d$. We have the following explicit expression\cite{rubin1984eigenvalues,Rubin:1984tc}
\begin{equation}\label{eq:NHHVdef}\begin{split}
\Nhh^V(d,\ell)&=\frac{\ell(\ell+d-2)}{(\ell+1)(\ell+d-3)} (d-2)\Nhh(d,\ell) =(2\ell+d-2)\frac{\ell}{\ell+d-3}\binom{\ell+d-2}{\ell+1} \ .
\end{split}\end{equation}
The first expression shows that $(d-2)\Nhh(d,\ell)$ is actually a good approximation to $\Nhh^V(d,\ell)$ at large $\ell$. We also see that in $d=3$, the number of VSHs is identical to the number of SSHs. 

There is also a VSH analogue of the recursion relation Eq.\eqref{eq:NHHRec} given by
\begin{equation}\label{eq:NHHVRec}\begin{split}
\Nhh^V(d,\ell)&=\sum_{m=1}^\ell [\Nhh^V(d-1,m)+\Nhh(d-1,m)]\ ,
\end{split}\end{equation}
which can be shown using the above formula. Such a recursion relation is automatic in Higuchi's recursive construction of VSHs\cite{Higuchi:1986wu}, whereby VSHs in $\mathbb{S}^{d-1}$
are constructed from VSHs and SSHs in $\mathbb{S}^{d-2}$. 

\subsection{VSH projector and  addition theorem }\label{appS:VSHProj}
Till now, we have described vector spherical harmonics in terms of an orthonormal basis. While such a basis is ideal for defining a set of linearly independent multipole moments associated with an extended source distribution, it is often convenient to shift to a different basis based on symmetric trace-free (STF) tensors. While the set of harmonics defined this way is overcomplete, it is often easier to compute the multipole moments in this basis. This is especially so for moving particle sources, where convoluting the orthonormal VSHs against the world line would be a tedious exercise, resulting in unwieldy expressions. We will also prove in this subsection the addition theorem for vector spherical harmonics, a crucial tool in going back and forth between the spherical vs the cartesian description.\footnote{ VSH addition theorem for $d=3$ are discussed in \cite{freeden1998constructive,freeden2013special,freeden2022spherical}. This should not be confused with the conceptually completely different `translational' addition theorems\cite{stein1961addition,cruzan1962translational,danos1965multipole,XU1996285}.  }

We will begin by recasting our results on orthonormal VSHs in terms of STF tensors. We will proceed in analogy with our description of SSHs. The orthonormal VSHs defined in Eq.\eqref{eq:VSHdef} can be pushed-forward into $\mathbb{R}^d$ as follows:
\begin{equation}\label{eq:OrthoToSTFvsh}
\begin{split}
\frac{1}{r}\left(\frac{\partial x_i}{\partial \vartheta_J}\right)\mathbb{V}^{J}_{\alpha\ell\vec{m}}(\hat{r})\equiv \frac{1}{\ell !}\VSH_{i<i_1i_2\ldots i_\ell>}\hat{r}^{<i_1}\hat{r}^{i_2}\ldots \hat{r}^{i_\ell>}\ .
\end{split}
\end{equation}
Note that we have already indicated here the STF structure of the cartesian tensor in RHS. This can be justified as follows: we begin with the observation that the vector field $r^\ell\mathbb{V}^{I}_{\alpha\ell\vec{m}}(\hat{r})$
is a divergence-free harmonic vector field in $\mathbb{R}^d$. This means that 
each cartesian component should be harmonic separately: in fact, they should 
all be homogeneous harmonic polynomials of degree $\ell$. This means that the vector field
\begin{equation}
\begin{split}
\overrightarrow{\mathbb{V}}^{\alpha\ell\vec{m}}(\vec{r})\equiv \frac{1}{\ell !}\hat{e}_i\VSH_{i<i_1i_2\ldots i_\ell>}x^{<i_1}x^{i_2}\ldots x^{i_\ell>}\ ,
\end{split}
\end{equation}
defined using the tensor above should be harmonic in every cartesian component. This is possible if and only if the collection of indices $i_1i_2\ldots i_\ell$ is symmetric and trace-free as indicated.

Since this vector field $\overrightarrow{\mathbb{V}}^{\alpha\ell\vec{m}}$ is obtained by a push-forward of a divergence-free vector field on $\mathbb{S}^{d-1}$, we conclude that $\overrightarrow{\mathbb{V}}^{\alpha\ell\vec{m}}$ is a divergence-free vector field in $\mathbb{R}^d$, transverse to the radial direction, i.e.,
\begin{equation}
\begin{split}
\overrightarrow{\nabla}\cdot\overrightarrow{\mathbb{V}}^{\alpha\ell\vec{m}}=0\ ,\quad \overrightarrow{r}\cdot\overrightarrow{\mathbb{V}}^{\alpha\ell\vec{m}}=0\ .
\end{split}
\end{equation}
The first condition implies the vanishing of the following contraction 
\begin{equation}
\begin{split}
\delta^{ii_1}\VSH_{i<i_1i_2\ldots i_\ell>}=0\ ,
\end{split}
\end{equation}
whereas the second one indicates the vanishing of the full symmetrisation: 
\begin{equation}
\begin{split}
\VSH_{i<i_1i_2\ldots i_\ell>}x^i x^{i_1}x^{i_2}\ldots x^{i_\ell}=0\ .
\end{split}
\end{equation}
These properties imply that $\VSH_{i<i_1i_2\ldots i_\ell>}$ is a irreducible cartesian tensor of $SO(d)$ corresponding to the Young tableaux
\begin{center}
\ytableausetup{mathmode, boxframe=normal, boxsize=2em}
\begin{ytableau}
i_1 & i_2 & i_3 & \none[\dots]
& i_{\ell - 1} & i_\ell \\
i 
\end{ytableau}\quad  .
\end{center}
The theory of VSHs then becomes equivalent to the study of cartesian tensors with such symmetry.

Many of the results of SSHs directly generalise. For example, the inner product formula in Eq.\eqref{Eq:SSHintSTF} implies a similar formula for VSHs:
\begin{equation}\label{Eq:VSHintSTF}\begin{split}
\int_{\mathbb{S}^{d-1}}
\left[\frac{1}{\ell !}\mathbb{V}_{i<i_1i_2\ldots i_\ell>}\hat{r}^{<i_1}\ldots \hat{r}^{i_\ell>}\right]
\left[\frac{1}{\ell !}\overline{\mathbb{V}}_{i<j_1j_2\ldots j_\ell>}\hat{r}^{<j_1}\ldots \hat{r}^{j_\ell>}\right]=
\frac{\nn_{d,\ell}|\mathbb{S}^{d-1}|}{\ell !}
\mathbb{V}^{i<i_1i_2\ldots i_\ell>} \overline{\mathbb{V}}_{i<i_1i_2\ldots i_\ell>}\ ,
\end{split}\end{equation}
true for arbitrary $\mathbb{V}$ and $\overline{\mathbb{V}}$  with constant cartesian components. It then follows that the orthonormality of $\mathbb{V}^{I}_{\alpha\ell\vec{m}}$ can then be cast in terms of STF tensors as
\begin{equation}\begin{split}
\frac{\nn_{d,\ell}|\mathbb{S}^{d-1}|}{\ell !}
\mathbb{V}_{\alpha\ell\vec{m}'}^{\ast i<i_1i_2\ldots i_\ell>} \mathbb{V}^{\beta\ell\vec{m}}_{i<i_1i_2\ldots i_\ell>}=\delta^\alpha_\beta
\delta^{\vec{m}}_{\vec{m}'}\ .
\end{split}\end{equation}
Note the same extra factor of $\nn_{d,\ell}|\mathbb{S}^{d-1}|$ which appears in this inner product, as in the STF inner product for SSHs. With this extra factor,
we  can then define a vector STF projector analogous to the scalar STF projector defined in  Eq.\eqref{eq:SSHSTFAdd}:
\begin{equation}\label{eq:VSHSTFAdd}\begin{split}
(\Pi^V_{ij})^{<i_1i_2\ldots i_\ell>}_{ <j_1j_2\ldots j_\ell>}\equiv\frac{\nn_{d,\ell}|\mathbb{S}^{d-1}|}{\ell!}\sum_{\alpha\vec{m}}
\mathbb{V}_{\alpha\ell\vec{m}}^{\ast i<i_1i_2\ldots i_\ell>} \VSH_{j<j_1j_2\ldots j_\ell>}\ .
\end{split}\end{equation}
Given the above definition, the orthonormality relation then guarantees the idempotence of $\Pi^V$, i.e., we have
\begin{equation}\begin{split}
(\Pi^V_{ik})^{<i_1i_2\ldots i_\ell>}_{ <k_1k_2\ldots k_\ell>}(\Pi^V_{kj})^{<k_1k_2\ldots k_\ell>}_{ <j_1j_2\ldots j_\ell>}=(\Pi^V_{ij})^{<i_1i_2\ldots i_\ell>}_{ <j_1j_2\ldots j_\ell>}\ .
\end{split}\end{equation}
As we shall see later, this vector STF projector plays a crucial role in the vector multipole expansion for moving particle sources in dS. For this reason, in the rest of this subsection, we will provide a detailed treatment of this projector. Specifically, we seek explicit expressions to allow quick computations, as well as a catalogue of useful properties.

As we did for the scalar STF projector, it is convenient to define a projected contraction 
\begin{equation}
\begin{split}
\Pi^V_{ij}(\vec{r}|\vec{\kappa})
&\equiv \Pi^V_{ji}(\vec{\kappa}|\vec{r})\equiv \frac{1}{\ell!} \kappa^{i_1}\ldots \kappa^{i_\ell}
(\Pi^V_{ij})_{<i_1\ldots i_\ell>}^{<j_1\ldots j_\ell>}\ r_{j_1}\ldots r_{j_\ell}\ \ .
 \end{split}   
\end{equation}
Using Eq.\eqref{eq:VSHSTFAdd}, this projected contraction can equivalently be defined via an \emph{addition theorem} for VSHs
\begin{equation}\label{eq:VSHadd}\begin{split}
\nn_{d,\ell}|\mathbb{S}^{d-1}|\sum_{\alpha\vec{m}}\mathbb{V}_{i\ell\vec{m}}^{\ast\alpha}(\hat{r}_0)\mathbb{V}_{j\ell\vec{m}}^{\alpha}(\hat{r})=\Pi^V_{ij}(\hat{r}_0|\hat{r})_{d,\ell}=
\Pi^V_{ji}(\hat{r}|\hat{r}_0)_{d,\ell}\ . 
\end{split}\end{equation}
This is equivalent to the following addition theorem for orthonormal VSHs:
\begin{equation}\label{eq:VSHaddOrtho}\begin{split}
\nn_{d,\ell}|\mathbb{S}^{d-1}|\sum_{\alpha\vec{m}}\mathbb{V}_{I\ell\vec{m}}^{\ast\alpha}(\hat{r}_0)\mathbb{V}_{J\ell\vec{m}}^{\alpha}(\hat{r})=\Pi^V_{IJ}(\hat{r}_0|\hat{r})_{d,\ell}=
\Pi^V_{JI}(\hat{r}|\hat{r}_0)_{d,\ell}\ .
\end{split}\end{equation}
An appropriate push-forward relates these two formulae:
\begin{equation}\label{PiV_poltocart}
   \Pi^V_{ij}(\hat{r}_0|\hat{r})_{d,\ell}=\frac{1}{r r_0}\frac{\partial x_{0i}}{\partial \theta_0^I}\frac{\partial x_j}{\partial \theta^J}\Pi_V^{IJ}(\hat{r}_0|\hat{r})_{d,\ell}\ .
\end{equation}

The vector STF projector satisfies the following equations:
\begin{equation}\label{eq:PiVprop}
\begin{split}
\nabla^2\overleftrightarrow{\Pi}^V(\vec{r}|\vec{r}_0)&=\nabla^2_0\overleftrightarrow{\Pi}^V(\vec{r}|\vec{r}_0)=0\ ,\\
\overleftarrow{\nabla}_0\cdot\overleftrightarrow{\Pi}^V(\vec{r}|\vec{r}_0)&=\overrightarrow{\nabla}\cdot\overleftrightarrow{\Pi}^V(\vec{r}|\vec{r}_0)=0\ ,\\
\vec{r}\cdot\overleftrightarrow{\Pi}^V(\vec{r}|\vec{r}_0)&=\overleftrightarrow{\Pi}^V(\vec{r}|\vec{r}_0)\cdot\vec{r}_0=0\ , 
 \end{split}   
\end{equation}
where the cartesian tensor $\overleftrightarrow{\Pi}^V(\vec{r}|\vec{r}_0)$ is defined via
$\overleftrightarrow{\Pi}^V(\vec{r}|\vec{r}_0)\equiv \Pi^V_{ij}(\hat{r}|\hat{r}_0)\hat{e}_i\otimes\hat{e}_j$, and left arrow signifies vector dot product acting on the second index. Upto an overall normalisation, $\overleftrightarrow{\Pi}^V(\vec{r}|\vec{r}_0)$ is the unique tensor that is homogeneous of degree $\ell$ in both $\vec{r}$ and $\vec{r}_0$, as well as satisfying the above equations. We will now argue that the normalization of 
$\Pi^V$ can be fixed via the following relation:
\begin{equation}\label{eq:PiVnorm}
\begin{split}
\delta^{ij}\Pi^V_{ij}(\vec{r}|\vec{r}_0)
&=\frac{\Nhh^V(d,\ell)}{\Nhh(d,\ell)}\Pi^S(\vec{r}|\vec{r}_0)_{d,\ell}=\frac{\ell(\ell+d-2)}{(\ell+1)(\ell+d-3)} (d-2)\Pi^S(\vec{r}|\vec{r}_0)_{d,\ell}\ .
 \end{split}   
\end{equation}
First, for purely symmetry reasons, we should have $\delta^{ij}\Pi^V_{ij}\propto \Pi^S$. The reason is as follows: the combination $\delta^{ij}\Pi^V_{ij}$ is a $\ell^{th}$ degree harmonic polynomial in $\vec{r}$ and $\vec{r}_0$, invariant under simultaneous rotation of $\vec{r}$ and $\vec{r}_0$ and any such object should be proportional to $\Pi^S$. The constant of proportionality can then be fixed by comparing the integral
\begin{equation}
\begin{split}
\frac{1}{\nn_{d,\ell}|\mathbb{S}^{d-1}|}\int_{\mathbb{S}^{d-1}}\delta^{ij}\Pi^V_{ij}(\hat{r}|\hat{r})
&=\sum_{\alpha\vec{m}}\int_{\mathbb{S}^{d-1}}\mathbb{V}_{i\ell\vec{m}}^{\ast\alpha}(\hat{r})\mathbb{V}_{\ell\vec{m}}^{i\alpha}(\hat{r})=\Nhh^V(d,\ell)\ ,
 \end{split}   
\end{equation}
against the integral
\begin{equation}
\begin{split}
\frac{1}{\nn_{d,\ell}|\mathbb{S}^{d-1}|}\int_{\mathbb{S}^{d-1}}\Pi^S_{d,\ell}(\hat{r}|\hat{r})
&=\sum_{\vec{m}}\int_{\mathbb{S}^{d-1}}\mathscr{Y}_{\ell\vec{m}}^{\ast}(\hat{r})\mathscr{Y}_{\ell\vec{m}}(\hat{r})=\Nhh(d,\ell)\ .
 \end{split}   
\end{equation}
With this normalisation fixed, we have established Eqs.\eqref{eq:PiVprop} and \eqref{eq:PiVnorm}, which then serve to uniquely define $\Pi^V$. With this normalisation, we also have the following overcompleteness relation for STF VSHs:
\begin{equation}
\begin{split}
&\frac{1}{\nn_{d,\ell}|\mathbb{S}^{d-1}|}\int_{\hat{r}\in\mathbb{S}^{d-1}}\Pi^V_{ij}(\hat{r}_1|\hat{r})\Pi^V_{jk}(\hat{r}|\hat{r}_2)\\
&=\nn_{d,\ell}|\mathbb{S}^{d-1}|\sum_{\alpha_1\vec{m}_1}\sum_{\alpha_2\vec{m}_2}\mathbb{V}_{\ell\vec{m}_1}^{\ast i\alpha_1}(\hat{r}_1)\mathbb{V}_{\ell\vec{m}_2}^{ i\alpha_2}(\hat{r}_2)\int_{\hat{r}\in\mathbb{S}^{d-1}}\mathbb{V}_{i\ell\vec{m}_1}^{\alpha_1}(\hat{r})\mathbb{V}_{\ell\vec{m}_2}^{\ast i\alpha_2}(\hat{r})\\
&=\nn_{d,\ell}|\mathbb{S}^{d-1}|\sum_{\alpha\vec{m}}\mathbb{V}_{\ell\vec{m}}^{\ast i\alpha}(\hat{r}_1)\mathbb{V}_{\ell\vec{m}}^{ i\alpha}(\hat{r}_2)=\Pi^V_{ik}(\hat{r}_1|\hat{r}_2)\ .
 \end{split}   
\end{equation}

We will now use the above properties to explicitly construct $\Pi^V$. To this end, we remind the reader of our construction of the orthonormal VSHs via second-order differential operators on orthonormal SSHs. We will now employ a similar construction to derive the vector STF projector in terms of the scalar STF projector. We will start with the ansatz that $\Pi^V_{ij}(\vec{r}|\vec{r}_0)=\mathbb{\Delta}_{ij} \Pi^S(\vec{r}|\vec{r}_0)_{d,\ell}$ with $\mathbb{\Delta}_{ij}$ being a 2-derivative operator in $\vec{r}$. Since both 
$\Pi^S(\vec{r}|\vec{r}_0)_{d,\ell}$ as well as  $\Pi^V_{ij}(\vec{r}|\vec{r}_0)$
have the same homogeneity (i.e., they are both of degree $\ell$ in $\vec{r}$), the derivative should not change the number of $x$'s in each term. This leaves us with four possibilities:
\begin{equation}
\begin{split}
 \delta_{ij}\ ,\ x_j\partial_i\ ,\ x_i \partial_j\ ,\ x^2\partial_i \partial_j\ .
 \end{split}   
\end{equation}
Here, we have used the fact that when acting on a homogeneous polynomial, the operator $x_i\partial_i$ reduces to a number.

Next, we impose the constraint that $x^i\Pi^V_{ij}(\vec{r}|\vec{r}_0)=0$, which in turn implies that only the following two combinations can occur in $\mathbb{\Delta}_{ij}$:
\begin{equation}
\begin{split}
 \ell\delta_{ij}-x_j\partial_i\ ,\ (\ell-1)x_i \partial_j-x^2\partial_i \partial_j\ .
 \end{split}   
\end{equation}
Finally, imposing that $\Pi^V_{ij}(\vec{r}|\vec{\kappa})_{d,\ell}$ should be divergence-free in its first index, we conclude that only one combination is admissible, viz.,
\begin{equation}
\begin{split}
\mathbb{\Delta}_{ij} \propto \Bigl\{\ell\ \delta_{ij}-x_j\partial_i-\frac{1}{\ell+d-3}[(\ell-1)\ x_i-x^2\partial_i]\partial_j\Bigr\}\ .
 \end{split}   
\end{equation}
As a consistency check of our ansatz, one may also check that this yields a harmonic tensor in $\vec{r}$. Fixing the normalisation via Eq.\eqref{eq:PiVnorm}, we then obtain 
\begin{equation}\label{eq:PiVDform1}
\begin{split}
\Pi^V_{ij}(\vec{r}|\vec{r}_0)_{d,\ell} &\equiv  \mathbb{\Delta}_{ij} \Pi^S(\vec{r}|\vec{r}_0)_{d,\ell}\\
&\equiv \frac{1}{\ell+1}\Bigl\{\ell\ \delta_{ij}-x_j\partial_i-\frac{1}{\ell+d-3}[(\ell-1)\ x_i-x^2\partial_i]\partial_j\Bigr\}\Pi^S_{d,\ell}(\vec{r}|\vec{r}_0)\ .
 \end{split}   
\end{equation}
The $\mathbb{\Delta}_{ij} $  is also a toroidal operator similar to the ones described in \S\S\ref{ssec:totoidalOp}. Using the series form of $\Pi^S$, and performing the derivatives, we obtain the following form for the VSH projector:
\begin{equation}
\begin{split}
\Pi^V_{ij}(\vec{r}|\vec{r}_0)_{d,\ell}
&\equiv \frac{1}{(\ell+1)(\ell+d-3)}\sum_{k=0}^{\lfloor\frac{\ell}{2}\rfloor}\ \frac{\Gamma\left(\nu-k\right)}{k!\ \Gamma\left(\nu\right)}  \left(-\frac{r_0^2 r^2}{4}\right)^k\\
&\times\Bigl\{\delta_{ij} \left[\ell(\ell+d-2)-(\ell-2k)^2\right] \frac{(\vec{r}_0\cdot\vec{r})^{\ell-2k}}{(\ell-2k)!}-(d-2)\ x_{0i}x_{j}\frac{(\vec{r}_0\cdot\vec{r})^{\ell-2k-1}}{(\ell-2k-1)!}\Bigr.\\
&\Bigl.\qquad+\left[-(x_{i}x_{0j}+ x_{0i}x_j)(\vec{r}_0\cdot\vec{r})+\delta_{ij}(\vec{r}_0\cdot\vec{r})^2+r_0^2 x_{i}x_j + r^2 x_{0i}x_{0j}\right]\ \frac{(\vec{r}_0\cdot\vec{r})^{\ell-2k-2}}{(\ell-2k-2)!}\Bigr\}\ .
 \end{split}   
\end{equation}
A more useful form is obtained by grouping together the terms above to make transversality manifest:
\begin{equation}\label{eq:VSHPfinal}
\begin{split}
&\Pi^V_{ij}(\vec{r}|\vec{r}_0)_{d,\ell}V^j
=\frac{1}{\ell+1}\Biggl\{ \frac{(\vec{r}_0\cdot\vec{r})^{\ell-1}}{(\ell-1)!}[x\cdot(x_0\wedge V)]_i \Biggr.\\
&\ \Biggl. -  \frac{[r^2 V_{\perp i}-x_i(\vec{r}\cdot \vec{V}_\perp)]}{\ell+d-3}\sum_{k=0}^{\lfloor\frac{\ell}{2}\rfloor-1}\ \frac{\Gamma\left(\nu-k\right)}{k!\ \Gamma\left(\nu\right)}  \left(-\frac{r_0^2 r^2}{4}\right)^k  \frac{(\vec{r}_0\cdot\vec{r})^{\ell-2k-2}}{(\ell-2k-2)!}
\Biggr.\\
&\ \Biggl. -\frac{r^2}{4} \frac{[(\vec{r}_0\cdot\vec{r})V_{\perp i}-x_{0i}(\vec{r}\cdot \vec{V}_\perp )]}{(\ell+d-3)}\sum_{k=1}^{\lfloor\frac{\ell}{2}\rfloor}\ \frac{\Gamma\left(\nu-k\right)}{k!\ \Gamma\left(\nu\right)}  \left(-\frac{r_0^2 r^2}{4}\right)^{k-1}  \frac{(\vec{r}_0\cdot\vec{r})^{\ell-2k-1}}{(\ell-2k-1)!}(\ell+d-3-2k)\Biggr\}_{\nu=\ell+\frac{d}{2}-1} .
\end{split}
\end{equation}
Here, we have used the notation 
\begin{equation}\begin{split}
(x_0\wedge V)_{ij}&\equiv x_{0i}V_j-x_{0j}V_i\ ,\qquad
V_{\perp i}\equiv [x_0\cdot(x_0\wedge V)]_i=r_0^2V_i-(x_0\cdot V) x_{0i}\ 
\end{split}
\end{equation}
to simplify our expressions. The formula above is manifestly transverse to $\vec{r}$ in the first index and $\vec{r}_0$ in the second index. The simplest  case is where we take $\vec{r}_0=\hat{e}_d$ and $\vec{V}=\hat{e}_{d-1}$ where $\hat{e}_i$ are the cartesian basis vectors in $\mathbb{R}^d$. The above formula then simplifies to 
\begin{equation}
\begin{split}
\Pi^V_{i,d-1}(\vec{r}|\hat{e}_d)&\frac{\partial}{\partial x_i}= \Nhh^V(d,\ell)\ \nn_{d,\ell}\\
&\times \frac{1}{d-1}\Bigl[r^{\ell-1}P_{\ell-1}\left(d+2,\frac{x_d}{r}\right)\left\{x_d\frac{\partial}{\partial x_{d-1}}-x_{d-1}\frac{\partial}{\partial x_{d}}\right\}\Bigr.\\
&\quad\Bigl. +\frac{(\ell-1)(\ell+d-1)}{(d+1)(d-2)}r^{\ell-2}P_{\ell-2}\left(d+4,\frac{x_d}{r}\right)\sum_{j=1}^{d-2}x_j\left\{x_{d-1}\frac{\partial}{\partial x_j}-x_j\frac{\partial}{\partial x_{d-1}}\right\}\Bigr]\ ,
\end{split}   
\end{equation}
where $\Nhh^V(d,\ell)$ is the number of VSHs of degree $\ell$ on $\mathbb{S}^{d-1}$ (See Eq.\eqref{eq:NHHVdef}). We recognise here the appearance of the $SO(d-2)$ invariant harmonic vector field obtained by push-forward of the most symmetric VSH (see Eq.\eqref{eq:LegVSH}). This is then the vector analogue of the statement  that the scalar projector is proportional to the Legendre harmonic, i.e.,
\begin{equation}
\begin{split}
\Pi^S_{d,\ell}(\vec{r}|\hat{e}_d)
&=\nn_{d,\ell}\Nhh(d,\ell)r^\ell P_\ell\left(d,\frac{x_d}{r}\right)\ .
 \end{split}   
\end{equation}
This follows directly from Eq.\eqref{eq:STFLegSSH}.

\subsection{VSH projector using Young tableau methods}
In the last section, we remarked that VSHs correspond  to irreducible cartesian tensors of $SO(d)$ with the symmetry of their indices specified by the Young tableau
\begin{center}
\ytableausetup{mathmode, boxframe=normal, boxsize=2em}
\begin{ytableau}
i_1 & i_2 & i_3 & \none[\dots]
& i_{\ell - 1} & i_\ell \\
i 
\end{ytableau}\quad  .
\end{center}
For $SO(d)$, irreducible representations are obtained by Young tableaux with the number of rows less than $\frac{d}{2}$. This gives all irreducible tensor representations, except the $SO(2n)$ tensors with self-dual or anti-self-dual form indices (which correspond to reducible tableaux with exactly $\frac{d}{2}$ rows). 

We remind the reader how such a tableau should be interpreted in the context of $SO(d)$: the rows of the tableau indicate symmetrisation+trace-removal, and the columns indicate anti-symmetrisation.
These two steps, done sequentially, then give an irreducible tensor with an appropriate symmetry. Given a Young tableau, standard theorems in $SO(d)$ representation theory give formulae for dimension, character, Clebsh-Gordon decomposition, etc., of the corresponding irreducible representation. As an example, if $n_\alpha$ is the number of boxes in the $\alpha^{th}$ row, the dimension of the corresponding $SO(d)$ irrep is\cite{Samra_1979}(eqns. (3.1) and (3.2))
\begin{equation}
\prod_{\alpha>\beta | \alpha,\beta =1}^{\lfloor\frac{d}{2}\rfloor}  \frac{(\alpha-n_\alpha)-(\beta-n_\beta)}{\alpha-\beta}  \times \prod_{\alpha\geq\beta | \alpha,\beta =1}^{\lfloor\frac{d}{2}\rfloor}\frac{d-(\alpha-n_\alpha)-(\beta-n_\beta)}{d-\alpha-\beta}
\end{equation}
for odd $d$. For even $d$, a similar formula holds provided we drop all the $\alpha=\beta$ terms in the second product. The structure of these products is such that only non-empty rows contribute to $\beta$: if $\beta^{th}$ row is empty (i.e., if $n_\beta=0$) so is the $\alpha^{th}$ row (i.e., $n_\alpha=0$), and the contribution to the product is unity.

As an example, we can count the number of SSHs at a given $d$ and $\ell$ by taking  $n_1=\ell$ and $n_2=\ldots=n_{\lfloor\frac{d}{2}\rfloor}=0$. The two product factors become
\begin{equation}\begin{split}
&\frac{2-(1-\ell)}{1}\frac{3-(1-\ell)}{2}\ldots \frac{\lfloor\frac{d}{2}\rfloor-(1-\ell)}{\lfloor\frac{d}{2}\rfloor-1}=\binom{\ell+\lfloor\frac{d}{2}\rfloor-1}{\ell}\ ,\\
\end{split}\end{equation}
and 
\begin{equation}\begin{split}
&\textcolor{red}{\frac{d-2(1-\ell)}{d-2}}\frac{d-2-(1-\ell)}{d-2-1}\frac{d-3-(1-\ell)}{d-3-1}\ldots \frac{d-\lfloor\frac{d}{2}\rfloor-(1-\ell)}{d-\lfloor\frac{d}{2}\rfloor-1} \\
&\qquad=\frac{2\ell+d-2}{d-2}\binom{\ell+d-3}{\ell}\binom{\ell+\lfloor\frac{d}{2}\rfloor-1}{\ell}^{-1}\ .
\end{split}\end{equation}
Here, we have indicated by red the $\alpha=\beta$ factor present only in odd $d$. The net product then matches with the explicit count of SSHs given in Eq.\eqref{eq:NHH}. In a similar vein, we can count spin-s spherical harmonics by taking 
\begin{equation}
n_1=\ell,\ n_2=s,\   n_3=\ldots=n_{\lfloor\frac{d}{2}\rfloor}=0\ .
\end{equation}
A similar count as above, performed separately in odd vs even dimensions, gives the number of spin-s spherical harmonics as 
\begin{equation}\begin{split}
\Nhh^{(s)}(d,\ell)&\equiv  \frac{(2\ell+d-2)(2s+d-4)(\ell-s+1)(\ell+s+d-3)}{(d-2)(d-4)(\ell+1)(\ell+d-3)}\binom{\ell+d-3}{\ell}\binom{s+d-5}{s}\\
&=\frac{(2s+d-4)(\ell-s+1)(\ell+s+d-3)}{(d-4)(\ell+1)(\ell+d-3)}\binom{s+d-5}{s}\Nhh(d,\ell) \ .  
\end{split}\end{equation}
As for VSHs, we can set $s=1$ in the above formula and recover Eq.\eqref{eq:NHHVdef}.

The Young tableau methods can also be used to derive the VSH projector (up to an overall normalisation). 
In the rest of this subsection, we will review some of the Young tableau based methods existing in the literature. No new results are derived here however, so an uninterested reader may safely skip this subsection.

\subsubsection{Young-symmetriser}
We will follow the recent works by Henry-Faye-Blanchet(HFB)\cite{Henry:2021cek} as well as Amalberti-Larrouturou-Yang(ALY)\cite{Amalberti:2023ohj} to construct the projector corresponding to the above tableau. We begin with the following definition of the trace-free projector (see 
Eq.(A6) of  ALY or  2nd line of Eq.(A4) of HFB):
\begin{equation}\begin{split}
\mathbb{TF}[V^ax^{<i_1}x^{i_2}\ldots x^{i_\ell>}]&\equiv V^ax^{<i_1}x^{i_2}\ldots x^{i_\ell>}\\
&\quad-\frac{\ell(\nu-1)}{(\ell+d-3)\nu}
\text{STF}_{i_1\ldots i_\ell}[\delta^{ai_\ell}V_{b}x^{<i_1}x^{i_2}\ldots x^{i_{\ell-1}}x^{b>}]\\
&\equiv V^ax^{<i_1}x^{i_2}\ldots x^{i_\ell>}\\
&\quad-\frac{\ell(\nu-1)}{(\ell+d-3)\nu}
\text{Sym}_{i_1\ldots i_\ell}[\delta^{ai_\ell}V_{b}x^{<i_1}x^{i_2}\ldots x^{i_{\ell-1}}x^{b>}]\\
&\quad+\frac{1}{2}\frac{\ell(\ell-1)}{(\ell+d-3)\nu}
\text{Sym}_{i_1\ldots i_\ell}[\delta^{i_\ell i_{\ell-1}}V_{b}x^{<i_1}x^{i_2}\ldots x^{i_{\ell-2}}x^a x^{b>}]\ .
\end{split}\end{equation}
Here $\nu\equiv \ell+\frac{d}{2}-1$ and $\text{STF}/\text{Sym}$ denote the STF/symmetric projector onto its subscript indices, respectively.  This trace-free projection is the first step in the construction of the VSH projector. 

To understand the relation of this trace-free projector in our language, we define the following polynomial vector field
\begin{equation}\begin{split}
U^a&\equiv \frac{1}{\ell!}\kappa_{i_1}\ldots \kappa_{i_{\ell}}\mathbb{TF}[V^ax^{<i_1}x^{i_2}\ldots x^{i_\ell>}]\\
&\equiv V^a\Pi^S_{d,\ell}(\vec{\kappa}|\vec{r})\\
&\quad-\kappa^a\frac{\nu-1}{(\ell+d-3)\nu}
\times\frac{1}{(\ell-1)!}\kappa_{i_1}\ldots \kappa_{i_{\ell-1}}V_{b}x^{<i_1}x^{i_2}\ldots x^{i_{\ell-1}}x^{b>}\\
&\quad+\frac{\vec{\kappa}^2}{2}\frac{1}{(\ell+d-3)\nu}
\times\frac{1}{(\ell-2)!}\kappa_{i_1}\ldots \kappa_{i_{\ell-2}}V_{b}x^{<i_1}x^{i_2}\ldots x^{i_{\ell-2}}x^a x^{b>}\\
&=V^b\left\{\delta^a_b-\frac{\nu-1}{(\ell+d-3)\nu}\kappa^a\frac{\partial}{\partial \kappa_b} +\frac{\vec{\kappa}^2}{2}\frac{1}{(\ell+d-3)\nu}\frac{\partial^2}{\partial\kappa_a\partial \kappa_b}\right\}\Pi^S_{d,\ell}(\vec{\kappa}|\vec{r})\ .
\end{split}\end{equation}
We will now show that $\mathbb{TF}$ is indeed the trace-free projector as claimed by ALY, i.e., we have
\begin{equation}\begin{split}
\delta_{ai_\ell}\mathbb{TF}[V^ax^{<i_1}x^{i_2}\ldots x^{i_\ell>}]=0\ .
\end{split}\end{equation}
In our notation, this is equivalent to the assertion that $\frac{\partial U^a}{\partial \kappa_a}=0$. From the 
\begin{equation}\begin{split}
\frac{\partial U^a}{\partial \kappa_a}
&=V^b\frac{\partial }{\partial \kappa_a}\left\{\delta^a_b-\frac{\nu-1}{(d+\ell-3)\nu}\kappa^a\frac{\partial}{\partial \kappa_b} +\frac{\vec{\kappa}^2}{2}\frac{1}{(d+\ell-3)\nu}\frac{\partial^2}{\partial\kappa_a\partial \kappa_b}\right\}\Pi^S_{d,\ell}(\vec{\kappa}|\vec{r})\\
&=\left\{1-\frac{\nu-1}{(d+\ell-3)\nu}\left(\kappa^a\frac{\partial}{\partial \kappa_a}+d\right) +\frac{1}{(d+\ell-3)\nu}\left(\kappa^a\frac{\partial}{\partial \kappa_a}\right)\right\}V^b\frac{\partial}{\partial \kappa_b}\Pi^S_{d,\ell}(\vec{\kappa}|\vec{r})\ .
\end{split}\end{equation}
We have used the harmonicity of $\Pi^S_{d,\ell}(\vec{\kappa}|\vec{r})$ to simplify the last term in the last line. Since $V^b\frac{\partial}{\partial \kappa_b}\Pi^S_{d,\ell}$ is a homogeneous polynomial in $\kappa$ of degree $(\ell-1)$, we can use 
Euler's homogeneous function theorem to replace all $\kappa^a\frac{\partial}{\partial \kappa_a}$ above by $(\ell-1)$. With this replacement, the prefactor above vanishes, showing that $\mathbb{TF}$ is indeed the trace-free projector. 

A similar computation yields
\begin{equation}\begin{split}
\nabla_\kappa^2 U^a
&=V^b\left\{-2\frac{\nu-1}{(d+\ell-3)\nu}\frac{\partial^2}{\partial \kappa^a\partial \kappa_b} +\frac{1}{(\ell+d-3)\nu}\left(2\kappa^c\frac{\partial}{\partial \kappa_c}+d\right)\frac{\partial^2}{\partial\kappa_a\partial \kappa_b}\right\}\Pi^S_{d,\ell}(\vec{\kappa}|\vec{r})=0\ .
\end{split}\end{equation}
Thus, the vector field $U^a$ is a harmonic, divergence-free vector field in the $\kappa_i$ variables. This, in turn, implies that $U^a$ is a linear combination of the gradient of HHPs and toroidal vector fields. The gradient part has to be subtracted to get an irreducible tensor: as we shall see shortly, removing this gradient is equivalent to anti-symmetrisation imposed by the first column of the Young tableau.

The gradient part can be isolated by looking at the radial component
\begin{equation}\begin{split}
\kappa_a U^a
&=V^b\left\{\kappa_b-\frac{\vec{\kappa}^2}{2\nu}\frac{\partial}{\partial \kappa_b}\right\}\Pi^S_{d,\ell}(\vec{\kappa}|\vec{r})\ .
\end{split}\end{equation}
Removing this gradient then gives a toroidal vector field
\begin{equation}\begin{split}
U_\perp^a&\equiv  (\ell+1)U^a - \frac{\partial}{\partial \kappa_a}[\kappa_b U^b]\\
&=V^b\left\{\ell\delta^a_b-\kappa_b\frac{\partial}{\partial \kappa_a}-\frac{\ell-1}{\ell+d-3}\kappa^a\frac{\partial}{\partial \kappa_b} +\frac{\vec{\kappa}^2}{\ell+d-3}\frac{\partial^2}{\partial\kappa_a\partial \kappa_b}\right\}\Pi^S_{d,\ell}(\vec{\kappa}|\vec{r})\\
&=(\ell+1)\Pi^V_{ab}(\vec{\kappa}|\vec{r})V^b\ ,
\end{split}\end{equation}
where we have used our formula for $\Pi^V$ derived in Eq.\eqref{eq:PiVDform1}.  From the definition of $U^a$, we also have 
\begin{equation}\begin{split}
U_\perp^a&\equiv  (\ell+1)U^a - \frac{\partial}{\partial \kappa_a}[\kappa_b U^b]=\ell U^a-\kappa_b\frac{\partial U^b}{\partial \kappa_a}=\kappa_b\left(\frac{\partial U^a}{\partial \kappa_b}-\frac{\partial U^b}{\partial \kappa_a}\right)\\
&=\frac{1}{(\ell-1)!}\kappa_{i_1}\ldots \kappa_{i_{\ell-1}}\kappa_b\left(\delta^{bi_\ell}\mathbb{TF}[V^ax^{<i_1}x^{i_2}\ldots x^{i_\ell>}]-\delta^{ai_\ell}\mathbb{TF}[V^bx^{<i_1}x^{i_2}\ldots x^{i_\ell>}]\right)\\
&=\frac{1}{(\ell-1)!}\kappa_{i_1}\ldots \kappa_{i_{\ell-1}}\kappa_b\ \text{Anti}_{ab}\left[\mathbb{TF}[V^ax^{<i_1}x^{i_2}\ldots x^{i_{\ell-1}}x^{b>}]\right]\ .
\end{split}\end{equation}
Comparing, we get 
\begin{equation}\begin{split}
\frac{\ell+1}{\ell}\Pi^V_{ab}(\vec{\kappa}|\vec{r})V^b
&=\frac{1}{\ell !}\kappa_{i_1}\ldots \kappa_{i_{\ell}}\ \text{Anti}_{ai_\ell}\left[\mathbb{TF}[V^ax^{<i_1}x^{i_2}\ldots x^{i_{\ell}>}]\right]\ .
\end{split}\end{equation}
Stripping off the dummy $\kappa$ factors, we obtain the  following relation
\begin{equation}\label{eq:PiVAntiTF}\begin{split}
(\Pi^V)^{i<i_1i_2\ldots i_\ell>}_{j<j_1j_2\ldots j_\ell>}V^j x^{j_1} x^{j_2}\ldots x^{j_\ell}
&=\frac{1}{\ell+1} \text{Anti}_{ii_1}\left[\mathbb{TF}[V^ix^{<i_1}x^{i_2}\ldots x^{i_{\ell}>}]\right]\\
&\quad + \frac{1}{\ell+1} \text{Anti}_{ii_2}\left[\mathbb{TF}[V^ix^{<i_1}x^{i_2}\ldots x^{i_{\ell}>}]\right]\\
&\quad +\ldots+\frac{1}{\ell+1} \text{Anti}_{ii_\ell}\left[\mathbb{TF}[V^ix^{<i_1}x^{i_2}\ldots x^{i_{\ell}>}]\right]\ .
\end{split}\end{equation}
This is then the crucial relation we are after: it connects our vector projector to the sequential process of trace-removal followed by anti-symmetrisation. This relation can also be used to compare the results quoted in \cite{Henry:2021cek,Amalberti:2023ohj} against our expressions.

\subsubsection{VSH projector via weight shifting operators}
We will next describe a slightly different route to constructing the projector using Young tableau- the method of \emph{weight-shifting operators}
\cite{Karateev:2017jgd,Caron-Huot:2022jli}. We will first state here the general algorithm behind this method without proof, and then apply it to the special case of vector projector.\footnote{We would like to thank Arnab Rudra and Kushal Chakraborty for explaining this method to us and sharing with us their notes on this subject.} 

Say we need a projector for a representation corresponding to an arbitrary Young tableau with $h$ rows. Then we proceed as follows
\begin{itemize}
\item We first introduce $h$ number of cartesian positions: say we denote them by $x_{i,\alpha}$ corresponding to the $\alpha^{th}$ row. We will call $x_{i,\alpha}$ as the $\alpha^{th}$  row position.
\item Next, we construct a \emph{seed} polynomial, which is a product of factors, one factor for each column. The factor for a column is the completely anti-symmetric polynomial of the row positions, made out of all the rows which contribute to that column (see below for how this works for toroidal harmonic vector fields). The seed polynomial of a tableau with $n$ boxes is hence a polynomial of total homogeneity $n$. Further, the degree of homogeneity in each row position is the number of boxes in that row.
\item The third step is to apply a \emph{weight-shifting} differential operator on this seed polynomial. The  weight-shifting differential operator is given by the matrix product of  derivative operators, one each for every row. The $d\times d$  matrix of derivatives for $\alpha^{th}$ row is given by 
\begin{equation}\begin{split}
\left(\delta^j_k-\frac{x_{j,\alpha}}{N_\alpha}\frac{\partial}{\partial x_{k,\alpha}}\right)
\end{split}\end{equation}
where  $N_\alpha\equiv d-1-h+n_h-\alpha+n_\alpha$ with $n_\alpha$ denoting the number of boxes in $\alpha^{th}$ row. The only exception is the last row, for which we take a $d\times 1$ column matrix  of derivatives of the form
\begin{equation}\begin{split}
\left(\delta^j_p-\frac{x_{j,h}}{N_h-1}\frac{\partial}{\partial x_{p,h}}\right)\frac{\partial}{\partial x_{p,h}}\ .
\end{split}\end{equation}
The product of the square matrices with this column matrix then yields a $d\times 1$ column matrix of derivative operators.
\item The fourth step is to apply the projector corresponding to a Young tableau with one less box in the last row.
\end{itemize}
The claim is that the resultant polynomial gives a recursive construction for the required projector (up to some overall normalisation). We will refer the reader to \cite{Caron-Huot:2022jli} for a more detailed exposition of this algorithm with a variety of examples. Our interest is in applying this to the Young tableau
\begin{center}
\ytableausetup{mathmode, boxframe=normal, boxsize=2em}
\begin{ytableau}
i_1 & i_2 & i_3 & \none[\dots]
& i_{\ell - 1} & i_\ell \\
i 
\end{ytableau}\quad  
\end{center}
corresponding to VSHs. For this case, we have $h=2,n_h=1$ so that $d-1-h+n_h=d-2$ and the numbers $N_1=\ell+d-3$ and $N_2=d-3$. The projector applied after the weight-shifting operator is determined by the tableau with the $i$ box removed: 
\begin{center}
\ytableausetup{mathmode, boxframe=normal, boxsize=2em}
\begin{ytableau}
i_1 & i_2 & i_3 & \none[\dots]
& i_{\ell - 1} & i_\ell 
\end{ytableau}\quad  
\end{center}
This is the tableau for symmetric trace-free tensors and the final projector needed is just the SSH projector $\Pi^S$.

Let us begin by assigning the row positions $x_i$ and $y_i$ corresponding to the two rows of the tableaux above. The seed polynomial is then given by
\begin{equation}\begin{split}
x_{[i_1}y_{i]}x_{i_2}\ldots x_{i_\ell}=(x_{i_1}y_i-x_i y_{i_1})x_{i_2}\ldots x_{i_\ell}. 
\end{split}\end{equation}
Here, the first anti-symmetric factor corresponds to the first column, whereas the rest of the monomials are contributions from the rest of the columns.

The weight shifting differential operator is a column of derivative operators given by
\begin{equation}\begin{split}
\left(\delta^j_{k}-\frac{x^j}{\ell+d-3}\frac{\partial}{\partial x^k}\right)\left(\delta^k_{p}-\frac{y^k}{d-4}\frac{\partial}{\partial y^p}\right)\frac{\partial}{\partial y^p}. 
\end{split}\end{equation}
Since the seed polynomial in this case is linear in $y$, we can drop all the second derivatives in $y$ to simplify this to 
\begin{equation}\begin{split}
\left(\delta^j_{k}-\frac{x^j}{\ell+d-3}\frac{\partial}{\partial x^k}\right)\frac{\partial}{\partial y^k}. 
\end{split}\end{equation}
The fourth step involves the projector of a smaller Young tableau: in this case, this is just the scalar projector that can be realised via STF projecting differential operator on $x^i$. To summarise, the weight-shifting algorithm gives the following expression for the VSH projector:
\begin{equation}\begin{split}
(\Pi^V)&^{i<i_1\ldots i_\ell>}_{j<j_1\ldots j_\ell>}x^{j_1}x^{j_2}\ldots x^{j_\ell}\\
&\propto
\left[\sum_{k=0}^{\lfloor\frac{\ell}{2}\rfloor}\ \frac{\Gamma\left(\nu-k\right)}{k!\ \Gamma\left(\nu\right)}  \left(\frac{r}{2}\right)^{2k}(-\nabla^2)^k\right]_{\nu=\frac{d}{2}+\ell-1}\left(\delta^j_{p}-\frac{x^j}{\ell+d-3}\frac{\partial}{\partial x^p}\right)\frac{\partial}{\partial y^p}[x^{[i_1}y^{i]}x^{i_2}\ldots x^{i_\ell}]. 
\end{split}\end{equation}
We will now prove this relation and determine the proportionality constant. It is easier to work with the form obtained by contracting the free STF indices with a dummy variable and differentiating, viz.,
\begin{equation}\begin{split}
\frac{1}{\ell!}&
\left[\sum_{k=0}^{\lfloor\frac{\ell}{2}\rfloor}\ \frac{\Gamma\left(\nu-k\right)}{k!\ \Gamma\left(\nu\right)}  \left(\frac{r}{2}\right)^{2k}(-\nabla^2)^k\right]_{\nu=\frac{d}{2}+\ell-1}\left(\delta^j_{p}-\frac{x^j}{\ell+d-3}\frac{\partial}{\partial x^p}\right)[\delta^{i}_p(\vec{\kappa}\cdot\vec{r})^{\ell}-x^{i}\kappa^p(\vec{\kappa}\cdot\vec{r})^{\ell-1}] \\
 &=\frac{1}{\ell!}
\left[\sum_{k=0}^{\lfloor\frac{\ell}{2}\rfloor}\ \frac{\Gamma\left(\nu-k\right)}{k!\ \Gamma\left(\nu\right)}  \left(\frac{r}{2}\right)^{2k}(-\nabla^2)^k\right]_{\nu=\frac{d}{2}+\ell-1}\\
 &\qquad\left(\delta_{ij} (\vec{\kappa}\cdot\vec{r})^{\ell}-x^{i}\kappa^j(\vec{\kappa}\cdot\vec{r})^{\ell-1}-\frac{(\ell-1)\kappa^ix^j}{\ell+d-3}(\vec{\kappa}\cdot\vec{r})^{\ell-1}+(\ell-1)
\frac{\kappa^2x^i x^j}{\ell+d-3}(\vec{\kappa}\cdot\vec{r})^{\ell-2}\right) \ .
\end{split}\end{equation}
This form can be converted to a more familiar form by rewriting it in terms of $\kappa$ derivatives as 
\begin{equation}\begin{split}
&
\left[\sum_{k=0}^{\lfloor\frac{\ell}{2}\rfloor}\ \frac{\Gamma\left(\nu-k\right)}{k!\ \Gamma\left(\nu\right)}  \left(\frac{r}{2}\right)^{2k}(-\nabla^2)^k\right]_{\nu=\frac{d}{2}+\ell-1}\\
 &\qquad \frac{1}{\ell}\left(\ell \delta_{ij} -\kappa^j\frac{\partial}{\partial \kappa^i}-\frac{(\ell-1)}{\ell+d-3}\kappa^i\frac{\partial}{\partial \kappa^j}+
\frac{\kappa^2}{\ell+d-3}\frac{\partial^2}{\partial \kappa^i\partial \kappa^j}\right)\frac{(\vec{\kappa}\cdot\vec{r})^{\ell}}{\ell !}\\
&=\frac{1}{\ell}\left(\ell \delta_{ij} -\kappa^i\frac{\partial}{\partial \kappa^j}-\frac{(\ell-1)}{\ell+d-3}\kappa^j\frac{\partial}{\partial \kappa^i}+
\frac{\kappa^2}{\ell+d-3}\frac{\partial^2}{\partial \kappa^i\partial \kappa^j}\right)\Pi^S(\vec{\kappa}|\vec{r}) \ . 
\end{split}\end{equation}
We recognise here the derivative operator mapping SSH projector to VSH projector (see Eq.\eqref{eq:PiVDform1}).
We have thus proved that 
\begin{equation}\label{eq:wtShPiV}\begin{split}
\frac{\ell+1}{\ell}&\Pi^V_{ij}(\vec{\kappa}|\vec{r}) =\frac{1}{\ell!} \kappa_{i_1}\kappa_{i_2}\ldots \kappa_{i_\ell}\\
&\times \left[\sum_{k=0}^{\lfloor\frac{\ell}{2}\rfloor}\ \frac{\Gamma\left(\nu-k\right)}{k!\ \Gamma\left(\nu\right)}  \left(\frac{r}{2}\right)^{2k}(-\nabla^2)^k\right]_{\nu=\frac{d}{2}+\ell-1}\left(\delta^j_{p}-\frac{x^j}{\ell+d-3}\frac{\partial}{\partial x^p}\right)\frac{\partial}{\partial y^p}[x^{[i_1}y^{i]}x^{i_2}\ldots x^{i_\ell}] . 
\end{split}\end{equation}
This is an especially succinct formula for the VSH projector. The above derivation also shows the crucial difference between the weight-shifting method vs the method of a toroidal operator acting on $\Pi^S$: in the weight-shifting case,  the SSH projection is done at the very end. Such an exercise can hopefully be generalised to give a closed-form expression for spin-s projectors.

\section{Multipole expansion in flat space I}
\label{app:FlatEMI}
We will begin by describing the multipole expansion for Maxwell's theory in $\mathbb{R}^d$.
Our primary motivation is to have a benchmark to compare the $H\to 0$  limit of our dS expressions. We will emphasise two ways to think about multipole expansion: first in terms of orthonormal spherical harmonics, and second in terms of symmetric trace-free (STF) cartesian tensors. Both these formalisms have their own advantage, and both of them are necessary to compute radiation reaction in de Sitter.

The $d=3$ version of orthonormal multipoles is standard and is described in classic textbooks and articles\cite{Morse:1953,Blatt:1952ije,Jackson:1998nia,Barrera:1985a,Carrascal_1991, Hill:1954,Campbell:1977jf,Gray:1978a,Gray:1978b,zangwill2013modern}. The normalisations and conventions, however, differ from one reference to the other, and often even within the same textbook between statics and radiation. Discussion in most references is also incomplete in a variety of ways, e.g., they often do not describe how fields look like within sources. 

The STF multipoles in EM are discussed mainly in gravitational wave literature\cite{Damour:1990gj,Ross:2012fc} where STF tensors are used widely. We did not find any reference systematically describing the relation between these two kinds of multipoles, especially at the level of detail needed for our work.\footnote{The relation is straightforward in electrostatics, where only scalar spherical harmonics are involved: see Kip-Thorne's review \cite{Thorne:1980ru}, appendix A of \cite{Blanchet:1985sp},   the textbook by Poisson-Will\cite{poisson2014gravity}), or the book by Soffel-Han\cite{soffel2019applied}.The references \cite{Hartmann1994,Mathis:2007rk} discuss applications to celestial mechanics, and \cite{Meichsner:2015zma} describe the effects of gravitational vector moments.
What we did not find is a similar discussion of the conversion rules for magnetostatics and beyond. } Our goal here is to clearly describe the connection between the two kinds of multipoles in flat space EM: corresponding objects is dS can then be understood as a generalisation. For the convenience of the reader, we provide a detailed comparison between $d=3$ normalisations in  \S\ref{sec:MultNflat}. 

The discussion of multipoles for general dimensional EM can be found in \cite{Birnholtz:2013ffa,Bhattacharyya:2016nhn, Amalberti:2023ohj}. The main novelty in general $d$ is the fact that the magnetic field $B_{ij}$ is a 2-form and is no more a pseudo-vector field.  All references cited above emphasise the STF viewpoint.\footnote{The reference \cite{Bhattacharyya:2016nhn} takes a hybrid viewpoint, but its treatment of VSHs is closer to the cartesian STF approach. } We will show here that the results of the previous appendix \S\ref{app:VSH} can be used to give a description of both STF and orthonormal multipole moments in general $d$.

\subsection{Multipole expansion in statics I : toroidal currents}\label{appss:staticsB}
\subsubsection*{Magnetic fields due to toroidal currents}
Let us begin with the simpler setting of magnetostatics and then generalize to time-dependent situations involving magnetic multipole radiation. Consider the following problem in magnetostatics: imagine a steady \emph{toroidal} charge current, i.e., a time-independent, divergence-free current that flows everywhere tangentially to a thin spherical shell of radius $R$. Explicitly, we take a charge current density of the form
\begin{equation}
	\bar{J}^r=0\ ,\	\bar{J}^I(\vec{r})=\delta(r-R)\ \mathcal{K}^{I}(\hat{r})=\delta(r-R)\sum_{\alpha\ell\vec{m}}\mathcal{K}^{I}_{\alpha \ell \vec{m}}(\hat{r})\ ,
\end{equation}
where the index $I$ denotes the sphere directions and $\{\alpha,\ell, \vec{m}\}$ label an orthonormal basis of divergence-free vector fields 
on $\mathbb{S}^{d-1}$ denoted by $\mathbb{V}^{I}_{\alpha \ell \vec{m}}(\hat{r})$. We will find it convenient to take $\mathbb{V}^{I}_{\alpha \ell \vec{m}}(\hat{r})$ to be an orthonormal basis of \emph{Vector Spherical Harmonics}(VSHs) on $\mathbb{S}^{d-1}$, i.e., 
\begin{equation}
\begin{split}
\left[\mathscr{D}^2+\ell(\ell+d-2)-1\right] \mathbb{V}^I_{\alpha\ell\vec{m}}= 0\ ,\quad 		
\mathscr{D}_I \mathbb{V}^I_{\alpha\ell\vec{m}}= 0\  ,\quad
\int_{\mathbb{S}^{d-1}}\gamma_{IJ}
\mathbb{V}^{I\ast}_{\alpha'\ell'\vec{m}'}\mathbb{V}^J_{\alpha\ell\vec{m}} = \delta_{\alpha'\alpha}\delta_{\ell'\ell}\delta_{\vec{m}'\vec{m}}\ .
\end{split}
\end{equation}
Here $\gamma_{IJ}$ is the standard metric on $\mathbb{S}^{d-1}$ and  $\mathscr{D}_I$ is the corresponding covariant derivative. We will
also define VSH with lower indices as $\mathbb{V}_{I}^{\alpha\ell\vec{m}}\equiv   \gamma_{IJ}
\mathbb{V}^{J}_{\alpha\ell\vec{m}}$, i.e., in our conventions, the VSH indices will always be lowered using the unit sphere metric rather than the spacetime metric. To avoid confusion, all other raising and lowering of sphere indices will be written out explicitly using the unit sphere metric. We will also need the \emph{VSH addition theorem} (see the discussion around Eq.\eqref{eq:VSHaddOrtho})
\begin{equation}\begin{split}
\nn_{d,\ell}|\mathbb{S}^{d-1}|\sum_{\alpha\vec{m}}\mathbb{V}_{I}^{\ast\alpha\ell\vec{m}}(\hat{r}_0)\ \mathbb{V}_{J}^{\alpha\ell\vec{m}}(\hat{r})=\Pi^V_{IJ}(\hat{r}_0|\hat{r})=\Pi^V_{JI}(\hat{r}|\hat{r}_0)\ , 
\end{split}\end{equation}
where $|\mathbb{S}^{d-1}|$ is the volume of $\mathbb{S}^{d-1}$ and $\nn_{d,\ell}$ is an inverse integer given by
\begin{equation}|\mathbb{S}^{d-1}|\equiv \frac{2\pi^{\frac{d}{2}}}{\Gamma(\frac{d}{2})}\ ,
\quad
\nn_{d,\ell}\equiv \frac{(d-2)!!}{(d+2\ell-2)!!}\ .
\end{equation}

Further details about VSHs are explained in appendix \ref{app:VSH}, but they do not matter for present purposes. It suffices to note that we can decompose the surface current $\mathcal{K}^{I}$ into terms proportional to VSHs, i.e.,  we assume $\mathcal{K}^{I}_{\alpha \ell \vec{m}}(\hat{r})\propto \mathbb{V}^{I}_{\alpha \ell \vec{m}}(\hat{r})$. The coefficients in these decompositions can be determined using the orthonormality of VSHs:
\begin{equation}\label{eq:staticsVSHdecomp}
\begin{split}
\mathcal{K}^{I}_{\alpha \ell \vec{m}}(\hat{r})\equiv 
\mathbb{V}^{I}_{\alpha\ell\vec{m}}(\hat{r})\int_{\hat{r}_0\in\mathbb{S}^{d-1}}
[\VSH_{J}(\hat{r}_0)]^\ast\mathcal{K}^{J}(\hat{r}_0)\ .
\end{split}
\end{equation}
The advantage of such a decomposition is that we can solve for the magnetic field due to each component quite easily. The final answer is then obtained by superposition. Since the symmetry properties of each VSH under $SO(d)$ rotation is different, the vector potential $\gaugeV_\mu$ produced by $\mathcal{K}^{I}_{\alpha \ell \vec{m}}$ should be proportional to $\mathcal{K}^{I}_{\alpha \ell \vec{m}}$ upto an $r$ dependent pre-factor. 

Further, since the equation for this radial pre-factor can only depend on the eigenvalue of the spherical laplacian,
the $r$-dependent factor can only depend on $\ell$, i.e., we can take 
\begin{equation}
\begin{split}
\gaugeV_r=0\ ,\quad \gaugeV_I=\sum_{\ell}f_\ell(r)\sum_{\vec{m}\alpha} \gamma_{IJ}\overbar{\mathcal{K}}^{J}_{\alpha \ell \vec{m}}\ .
\end{split}
\end{equation}
This vector potential automatically satisfies the Coulomb gauge condition. Using the above ansatz, Maxwell equations can all be reduced to a single vector Poisson equation of the form\footnote{We work in SI units and set the Maxwell
coupling $g_{_\text{EM}}=\sqrt{\mu_0}=1$. While it is not relevant to the magnetostatics discussion, we will also set $c=1$ when we later discuss radiation.}
\begin{equation}\label{eq:VectorPoisson}
\begin{split}
-\frac{1}{r^{d-1}}\partial_r[r^{d-3}\partial_r\gaugeV_{I}]+\frac{1}{r^4}(-\mathscr{D}^2+d-2)\gaugeV_{I}=\gamma_{IJ} \bar{J}^J\ .
\end{split}
\end{equation}
For $\gaugeV_{I}$ varying as $\ell^{th}$ VSH, we can replace $-\mathscr{D}^2$ by $\ell(\ell+d-2)-1$.
Since
\[\ell(\ell+d-2)-1+d-2=(\ell+1)(\ell+d-3)\ ,\] 
we conclude that, away from the spherical shell, $f_\ell(r)$ should vary as $r^{\ell+1}$ or as $r^{-(\ell+d-3)}$. We should stitch together these two solutions continuously with an appropriate derivative discontinuity given by the current. We obtain the final answer
\begin{equation}
\gaugeV_r=0\ ,\quad    \gaugeV_I=\sum_{\ell\vec{m}\alpha}\frac{R^3\gamma_{IJ}\overbar{\mathcal{K}}^{J}_{\alpha \ell \vec{m}}}{2\ell+d-2}\left[\frac{r^{\ell+1}}{R^{\ell+1}}\Theta(r<R)+\frac{R^{\ell+d-3}}{r^{\ell+d-3}}\Theta(r>R)\right]\ .
\end{equation}
It is instructive to rewrite this answer in terms of the original data, i.e., the currents before decomposition into VSHs. This can be done by using Eq.\eqref{eq:staticsVSHdecomp}. We get
\begin{equation}
\begin{split}
\gaugeV_I&=\int_{\hat{r}_0\in\mathbb{S}^{d-1}}\sum_{\ell}\left[\frac{r^{\ell+1}}{R^{\ell+1}}\Theta(r<R)+\frac{R^{\ell+d-3}}{r^{\ell+d-3}}\Theta(r>R)\right] \frac{\sum_{\vec{m}\alpha}\mathbb{V}_I^{\alpha\ell\vec{m}}(\hat{r})
\mathbb{V}_J^{\alpha\ell\vec{m}\ast}(\hat{r}_0)}{2\ell+d-2}R^3\overbar{\mathcal{K}}^{J}(\hat{r}_0)\\
&=\sum_{\ell}\left[\frac{r^{\ell+1}}{R^{\ell+1}}\Theta(r<R)+\frac{R^{\ell+d-3}}{r^{\ell+d-3}}\Theta(r>R)\right]\frac{1}{\nn_{d,\ell}|\mathbb{S}^{d-1}|}\int_{\hat{r}_0\in\mathbb{S}^{d-1}} \frac{\Pi^V_{IJ}(\hat{r}|\hat{r}_0)_{d,\ell}}{2\ell+d-2}R^3\overbar{\mathcal{K}}^{J}(\hat{r}_0)\ .
\end{split}
\end{equation}
Here, we have used the VSH addition theorem at the second step.

Equivalently, we can rewrite this expression in terms of the full current density of the spherical shell, i.e., 
\begin{equation}\label{eq:ViStatics}
\begin{split}
\gaugeV_I
&=\int_{\vec{r}_0}\sum_{\ell} \frac{\mathbb{G}_B(r,r_0;\ell)}{\nn_{d,\ell}|\mathbb{S}^{d-1}|} \Pi^V_{IJ}(\hat{r}|\hat{r}_0)_{d,\ell}\bar{J}^{J}(\vec{r}_0)\quad \text{with}\\
\mathbb{G}_B(r,r_0;\ell)&\equiv \frac{1}{2\ell+d-2}\left\{\frac{r^{\ell+1}}{r_0^{\ell+d-3}}\Theta(r<r_0)+\frac{r_0^{\ell+1}}{r^{\ell+d-3}}\Theta(r>r_0)\right\}\ .
\end{split}
\end{equation}
This expression is, in fact, applicable  to a more general toroidal current distribution of the form
\begin{equation}
	\bar{J}^r=0\ ,\	\bar{J}^I=\sum_{\alpha\ell \vec{m}}\bar{J}_V(r,\alpha,\ell,\vec{m})\mathbb{V}^{I}_{\alpha \ell \vec{m}}(\hat{r})\ .
\end{equation}
As such, an arbitrary current distribution can be thought of as built from infinitely many thin spherical current sheets. The principle of superposition implies then that the resultant vector potential is still given by Eq.\eqref{eq:ViStatics}. The corresponding static magnetic fields are given by 
\begin{equation}
\begin{split}
\FC_{rI} &=\int_{\vec{r}_0}\sum_{\ell} \frac{\partial_r\mathbb{G}_B(r,r_0;\ell)}{\nn_{d,\ell}|\mathbb{S}^{d-1}|}\Pi^V_{IJ}(\hat{r}|\hat{r}_0)_{d,\ell}\bar{J}^{J}(\vec{r}_0) \ ,\\
\FC_{IJ}&=\int_{\vec{r}_0}\sum_{\ell}\frac{\mathbb{G}_B(r,r_0;\ell)}{\nn_{d,\ell}|\mathbb{S}^{d-1}|} \mathscr{D}_{[I}\Pi^V_{J]K}(\hat{r}|\hat{r}_0)_{d,\ell}
\bar{J}^{K}(\vec{r}_0)\ .
\end{split}
\end{equation}
Here we use the notation $A_{[I}B_{J]}\equiv A_I B_J-A_J B_I$. For future comparison, we will also write the decomposition into  orthonormal VSHs:
\begin{equation}\label{eq:MstatVSHexp}
\begin{split}
\gaugeV_r &=0\ ,\quad \gaugeV_I\equiv \sum_{\ell\vec{m}\alpha} \overline{\Phi}_{_B}(r,\alpha,\ell,\vec{m})\mathbb{V}_{I}^{\alpha\ell\vec{m}}(\hat{r})\ ,\\
\FC_{rI}&\equiv \sum_{\ell\vec{m}\alpha} \overline{H}_v (r,\alpha,\ell,\vec{m})\mathbb{V}_{I}^{\alpha\ell\vec{m}}(\hat{r}) \ ,\quad 
\FC_{IJ} \equiv \sum_{\ell\vec{m}\alpha} \overline{H}_{vv}(r,\alpha,\ell,\vec{m})\mathscr{D}_{[I}\mathbb{V}_{J]}^{\alpha\ell\vec{m}}(\hat{r})\ .
\end{split}
\end{equation}
Thus, knowing the scalar field $\overline{\Phi}_{_B}$ for every spherical mode is sufficient to characterise the magnetic field. We will call $\overline{\Phi}_{_B}$ as the \emph{magnetic Debye field}. The field strength can then be determined via
\begin{equation}
\begin{split}
\overline{H}_v=\partial_r \overline{\Phi}_{_B}\ ,\quad \overline{H}_{vv}= \overline{\Phi}_{_B} \ .
\end{split}
\end{equation}
The magnetic Debye field for the most general static current contribution can then be written as
\begin{equation}\label{eq:PhiBStatics}
\begin{split}
\overline{\Phi}_{_B}&\equiv
\int_{\vec{r}_0}\mathbb{G}_B(r,r_0;\ell)\mathbb{V}_I^{\alpha\ell\vec{m}\ast}(\hat{r}_0)\bar{J}^{I}(\vec{r}_0)=\int_0^\infty dr_0\ r_0^{d-1}\mathbb{G}_B(r,r_0;\ell) \bar{J}_V(r_0)\ ,
\end{split}
\end{equation}
where we have defined $\bar{J}_V(r_0)\equiv \int_{\hat{r}_0\in\mathbb{S}^{d-1}} \mathbb{V}_I^{\alpha\ell\vec{m}\ast}(\hat{r}_0)\bar{J}^{I}(\vec{r}_0)$.

\subsubsection*{Multipole expansion outside the sources}
The expressions above simplify considerably if we focus on the fields outside the currents. Defining  the spherical magnetic multipole moments via 
\begin{equation}\label{eq:jBstatic}
\begin{split}
\overbar{\multj}^B(\alpha,\ell,\vec{m}) &\equiv\frac{1}{2\ell+d-2}\int_{\vec{r}_0}r_0^{\ell+1}
\mathbb{V}_J^{\alpha\ell\vec{m}\ast}(\hat{r}_0)\bar{J}^{J}(\vec{r}_0)\ ,
\end{split}
\end{equation}
we can then write the magnetic Debye potential outside the currents as
\begin{equation}
\begin{split}
\overline{\Phi}_{_B}^{\text{Out}}(r,\alpha,\ell,\vec{m})=\frac{\overbar{\multj}^B(\alpha,\ell,\vec{m})}{r^{\ell+d-3}}\ ,
\end{split}
\end{equation}
and the corresponding magnetic field components are given by
\begin{equation}
\begin{split}
 \overline{H}_v^{\text{Out}} (r,\alpha,\ell,\vec{m}) =-(\ell+d-3)\frac{\overbar{\multj}^B(\alpha,\ell,\vec{m})}{r^{\ell+d-2}}\ ,\quad \
\overline{H}_{vv}^{\text{Out}}(r,\alpha,\ell,\vec{m}) =\frac{\overbar{\multj}^B(\alpha,\ell,\vec{m})}{r^{\ell+d-3}}\ .
\end{split}
\end{equation}
These expressions constitute the magnetostatic multipole expansion in $\mathbb{R}^d$. 
We will see later that the dS multipoles reduce to these expressions in an appropriate limit.

The magnetostatic multipole expansion can also be recast into a cartesian STF form. To do this, we will sum over the $\vec{m}\alpha$ indices, multiply by appropriate powers of $r$ and transform the spherical indices to get a harmonic vector field in $\mathbb{R}^d$. Each cartesian component of the vector field is then a $\ell^{th}$ degree homogeneous harmonic polynomial in cartesian coordinates. Such a vector field is of the form 
\begin{equation}
\begin{split}
\frac{1}{\ell!}\ {}^B\mathcal{Q}_{k<i_1 i_2\ldots i_\ell>}x^{i_1}\ldots x^{i_\ell}\ ,
\end{split}
\end{equation}
where ${}^B\mathcal{Q}^k_{<i_1 i_2\ldots i_\ell>}$ is a constant irreducible  tensor  corresponding to the Young tableaux
\begin{center}
\ytableausetup{mathmode, boxframe=normal, boxsize=2em}
\begin{ytableau}
i_1 & i_2 & i_3 & \none[\dots]
& i_{\ell - 1} & i_\ell \\
k 
\end{ytableau}\quad  .
\end{center}
The explicit relation between the  magnetic multipole moment tensor and the spherical magnetic moments is 
\begin{equation}\label{eq:jBtoQB}
\begin{split}
\frac{1}{\ell!}\ {}^B\overbar{\mathcal{Q}}_{k<i_1 i_2\ldots i_\ell>}x^{i_1}\ldots x^{i_\ell}\equiv \nn_{d,\ell-1}|\mathbb{S}^{d-1}|\left(\frac{\partial x_k}{\partial \vartheta_I}\right)\sum_{\vec{m}\alpha}\overbar{\multj}^B(\alpha,\ell,\vec{m}) r^{\ell-1}\mathbb{V}^I_{\alpha\ell\vec{m}}(\hat{r})\ .
\end{split}
\end{equation}
We have followed here the steps outlined above and included an additional normalisation factor of  $\nn_{d,\ell-1}|\mathbb{S}^{d-1}|=(2\ell+d-2)\nn_{d,\ell}|\mathbb{S}^{d-1}|$ for convenience. We will now convert everything to cartesian basis using \eqref{PiV_poltocart} and 
\begin{equation}
    r_0^2 \gamma_{JK}(\theta_0)\overbar{J}^K(\vec{r}_0)=\frac{\partial x_{0j}}{\partial \theta_{0}^J}\overbar{J}^j(\vec{r}_0)\ .
\end{equation}
Further, we  sum over $(\alpha,\vec{m})$  by invoking addition theorem, and  end up with
\begin{equation}
\begin{split}
\frac{1}{\ell!}\ {}^B\overbar{\mathcal{Q}}_{k<i_1 i_2\ldots i_\ell>}x^{i_1}\ldots x^{i_\ell}\equiv\int_{\vec{r}_0}\Pi^V_{kj}(\vec{r}|\vec{r}_0)\bar{J}^{j}(\vec{r}_0)\ ,
\end{split}
\end{equation}
or equivalently,
\begin{equation}
\begin{split}
{}^B\overbar{\mathcal{Q}}_{k<i_1 i_2\ldots i_\ell>}\equiv(\Pi^V_{kj})^{<i_1 i_2\ldots i_\ell>}_{<j_1 j_2\ldots j_\ell>}\int_{\vec{r}_0}x_0^{j_1}\ldots x_0^{j_\ell}\bar{J}^{j}(\vec{r}_0)\ .
\end{split}
\end{equation}
This equation gives a way to directly compute the cartesian moments from the current without going through orthonormal VSHs: we only need the vector STF projector constructed in Appendix\S\ref{appS:VSHProj}. We will give explicit expressions for cartesian moments below. 

The outside vector potential/magnetic field can be written in terms of the STF magnetic moment as\footnote{Here we use the notation $\text{Anti}_{jk}[T_{jk}]\equiv T_{jk}-T_{kj}$ for the anti-symmetrisation operator.}
\begin{equation}
\begin{split}
\gaugeV_k^{\text{Out}} &= \sum_{\ell} \frac{1}{\ell! \nn_{d,\ell-1}|\mathbb{S}^{d-1}|}{}^B\overbar{\mathcal{Q}}_{k<i_1 i_2\ldots i_\ell>}\frac{x^{i_1}\ldots x^{i_\ell}}{ r^{2\ell+d-2}}\ ,\\
\FC_{jk}^{\text{Out}}&= \text{Anti}_{jk}\sum_{\ell} \frac{1}{\ell! \nn_{d,\ell-1}|\mathbb{S}^{d-1}|}{}^B\overbar{\mathcal{Q}}_{j<i_1 i_2\ldots i_\ell>}[(2\ell+d-2)x^k x^{i_\ell}-r^2\ell\delta^{ki_\ell}]\frac{ x^{i_1}\ldots x^{i_{\ell-1}}}{ r^{2\ell+d}}\ .
\end{split}
\end{equation}
These expressions give the cartesian multipole expansion for magnetostatics in $\mathbb{R}^d$. It is conventional in STF literature to rewrite the potential as a derivative of the Newton-Coulomb potential using 
\begin{equation}
\begin{split}
\partial^{i_1}\partial^{i_2}\ldots \partial^{i_\ell}\Bigl\{\frac{1}{(d-2)|\mathbb{S}^{d-1}| r^{d-2}}\Bigr\}=\frac{(-)^\ell}{\nn_{d,\ell-1}|\mathbb{S}^{d-1}|}\frac{x^{<i_1}\ldots x^{i_\ell>}}{ r^{2\ell+d-2}}\ .
\end{split}
\end{equation}
This formula can be established by direct differentiation and using the harmonicity of the Newton-Coulomb potential away from the origin. Using this, we can write the vector potential as a series of derivatives acting on the Newton-Coulomb potential: 
\begin{equation}\label{eq:VstMagFlat}
\begin{split}
\gaugeV_k^{\text{Out}} &= \sum_{\ell} \frac{(-)^\ell}{\ell!}
\partial^{i_1}\partial^{i_2}\ldots \partial^{i_\ell}\Bigl\{\frac{{}^B\overbar{\mathcal{Q}}_{k<i_1 i_2\ldots i_\ell>}}{(d-2)|\mathbb{S}^{d-1}| r^{d-2}}\Bigr\}\ .
\end{split}
\end{equation}
Our derivation here follows closely the EM multipole expansion in $d=3$ described in \cite{Campbell:1977jf,Birnholtz:2013nta} and the discussion in general $d$ by \cite{Birnholtz:2013ffa}. The Cartesian multipole expansion in $d=3$ can be derived directly without using orthonormal VSHs as shown in \cite{Damour:1990gj,Ross:2012fc}( See \cite{Amalberti:2023ohj} for a recent generalisation to arbitrary dimensions). Such a method, however, does not readily generalise to the de Sitter static patch, as there are no static Cartesian coordinates in dS.

We will conclude our discussion with an explicit formula for STF magnetic moments ${}^B\overbar{\mathcal{Q}}_{k<i_1\ldots i_\ell> }$. Using the explicit form of the vector STF projector in Eq.\eqref{eq:VSHPfinal}, we can write
\begin{equation}
\begin{split}
\frac{1}{\ell !}&{}^B\overbar{\mathcal{Q}}_{j<i_1 \ldots i_\ell>}\kappa^{i_1}\ldots\kappa^{i_\ell}
=\frac{1}{\ell+1}\int_{\vec{r}}\Biggl\{ \frac{(\kappa\cdot r)^{\ell-1}}{(\ell-1)!}[\kappa\cdot(x\wedge \bar{J})]_j\Biggr.\\
&\ \Biggl. -  \frac{[\kappa^2\bar{I}_{j}-\kappa_j(\kappa\cdot \bar{I} )]}{\ell+d-3}\sum_{k=0}^{\lfloor\frac{\ell}{2}\rfloor-1}\ \frac{\Gamma\left(\nu-k\right)}{k!\ \Gamma\left(\nu\right)}  \left(-\frac{\kappa^2 r^2}{4}\right)^k  \frac{(\vec{\kappa}\cdot\vec{r})^{\ell-2k-2}}{(\ell-2k-2)!}
\Biggr.\\
&\ \Biggl. -\frac{\kappa^2}{4} \frac{[(\kappa\cdot r)\bar{I}_{j}-(\kappa\cdot \bar{I} )x_j]}{(\ell+d-3)}\sum_{k=1}^{\lfloor\frac{\ell}{2}\rfloor}\ \frac{\Gamma\left(\nu-k\right)}{k!\ \Gamma\left(\nu\right)}  \left(-\frac{\kappa^2 r^2}{4}\right)^{k-1}  \frac{(\vec{\kappa}\cdot\vec{r})^{\ell-2k-1}}{(\ell-2k-1)!}(\ell+d-3-2k)\Biggr\}_{\nu=\ell+\frac{d}{2}-1}\ .
\end{split}
\end{equation}
Here $\kappa^i$ is a dummy variable introduced to simplify our expressions, and we have used the notation 
\begin{equation}\begin{split}
(x\wedge\bar{J})_{ij}&\equiv x_i\bar{J}_j-x_j\bar{J}_i\ ,\qquad
\bar{I}_i\equiv [x\cdot(x\wedge\bar{J})]_i=r^2\bar{J}_i-(x\cdot\bar{J}) x_i\ 
\end{split}
\end{equation}
to simplify our expressions. For the first few $\ell$s, the above formula evaluates to
\begin{equation}\label{eq:QBexpcont1}
\begin{split}
{}^B\overbar{\mathcal{Q}}_{ki_1 }\kappa^{i_1} &=\frac{1}{2!}\int_{\vec{r}} [\kappa\cdot(x\wedge \bar{J})]^{k} \ ,\\
\frac{1}{2!}{}^B\overbar{\mathcal{Q}}_{k<i_1 i_2>}\kappa^{i_1}\kappa^{i_2}&=\frac{2}{3!}\int_{\vec{r}}\Biggl\{(\kappa\cdot r)[\kappa\cdot(x\wedge \bar{J})]^k
-\frac{1}{d-1}  [ \kappa^2\bar{I}^{k}-\kappa^k(\kappa\cdot \bar{I} )  ]\Biggr\}\ ,\\
\frac{1}{3!}{}^B\overbar{\mathcal{Q}}_{k<i_1 i_2 i_3>}\kappa^{i_1}\kappa^{i_2}\kappa^{i_3}&=\frac{3}{4!}\int_{\vec{r}}\Biggl\{
(\kappa\cdot r)^2[\kappa\cdot(x\wedge \bar{J})]^k\Biggr.\\
&\qquad\Biggl.
-\frac{2}{d} (\kappa\cdot r) [ \kappa^2\bar{I}^{k}-\kappa^k(\kappa\cdot \bar{I} )  ]-\frac{d-2}{d(d+2)}\kappa^2 [(\kappa\cdot r)\bar{I}^{k}-(\kappa\cdot \bar{I} )x^k] \Biggr\}\ ,\\
\end{split}
\end{equation}
\begin{equation}\label{eq:QBexpcont2}
\begin{split}
\frac{1}{4!}{}^B\overbar{\mathcal{Q}}_{k<i_1 i_2 i_3i_4>}\kappa^{i_1}\ldots\kappa^{i_4}&=\frac{4}{5!}\int_{\vec{r}}\Biggl\{ (\kappa\cdot r)^3[\kappa\cdot(x\wedge \bar{J})]^k\Biggr.\\
&\qquad\Biggl.-3\frac{(d+4)(\kappa\cdot r)^2-\kappa^2r^2}{(d+1)(d+4)} [ \kappa^2\bar{I}^{k}-\kappa^k(\kappa\cdot \bar{I} )  ]
\Biggr.\\
&\qquad\Biggl. -3\frac{d-1}{(d+1)(d+4)} \kappa^2(\kappa\cdot r)[(\kappa\cdot r)\bar{I}^{k}-(\kappa\cdot \bar{I} )x^k]\Biggr\}\ ,\\
\frac{1}{5!}{}^B\overbar{\mathcal{Q}}_{k<i_1 i_2 i_3i_4 i_5>}\kappa^{i_1}\ldots\kappa^{i_5}&=\frac{5}{6!}\int_{\vec{r}}\Biggl\{ (\kappa\cdot r)^4[\kappa\cdot(x\wedge \bar{J})]^k\Biggr.\\
&\qquad\Biggl. -4\frac{(d+6)(\kappa\cdot r)^3-\kappa^2r^2(\kappa\cdot r)}{(d+2)(d+6)} [ \kappa^2\bar{I}^{k}-\kappa^k(\kappa\cdot \bar{I} )  ]
\Biggr.\\
&\qquad\Biggl. -3\frac{2d(d+4)(\kappa\cdot r)^2-(d-2)\kappa^2r^2}{(d+2)(d+4)(d+6)} \kappa^2[(\kappa\cdot r)\bar{I}^{k}-(\kappa\cdot \bar{I} )x^k]\Biggr\}\ .
\end{split}
\end{equation}
Explicit tensor expressions can then be obtained by repeatedly differentiating the above formulae with respect to $\kappa^i$ to yield explicitly symmetrised expressions. For example, the STF magnetic dipole/quadrupole tensors are
\begin{equation}\label{eq:QBexp}
\begin{split}
&{}^B\overbar{\mathcal{Q}}_{ki_1 }\equiv\frac{1}{2}\int_{\vec{r}} (x\wedge\bar{J})^{i_1k} \ ,\\
&{}^B\overbar{\mathcal{Q}}_{k<i_1 i_2>}\equiv\frac{1}{3}\int_{\vec{r}}\Biggl\{
(x\wedge\bar{J})^{i_1k}x^{i_2}+(x\wedge\bar{J})^{i_2k}x^{i_1}
+\frac{1}{d-1} (\bar{I}^{i_1} \delta^{ki_2}+\bar{I}^{i_2} \delta^{ki_1}-2 \bar{I}^{k} \delta^{i_1i_2})\Biggr\}\ ,
\end{split}
\end{equation}
whereas the magnetic octopole tensor has the form
\begin{equation}
\begin{split}
&{}^B\overbar{\mathcal{Q}}_{k<i_1 i_2 i_3>}\equiv\frac{1}{4}\int_{\vec{r}}\Biggl\{
(x\wedge\bar{J})^{i_1k}x^{i_2}x^{i_3}+(x\wedge\bar{J})^{i_2k}x^{i_3}x^{i_1}+(x\wedge\bar{J})^{i_3k}x^{i_1}x^{i_2}\Biggr.\\
&\qquad\Biggl.
-\frac{3d+2}{d(d+2)} \bar{I}^{k}(x^{i_1}\delta^{i_2i_3}+x^{i_2}\delta^{i_3i_1}+x^{i_3}\delta^{i_1i_2})+\frac{d-2}{d(d+2)}x^k\left(\bar{I}^{i_1}\delta^{i_2i_3}+\bar{I}^{i_2}\delta^{i_3i_1}+\bar{I}^{i_3}\delta^{i_1i_2}\right)\Biggr.\\
&\qquad\Biggl.+\frac{\delta^{ki_1}}{d} (\bar{I}^{i_2}x^{i_3}+\bar{I}^{i_3}x^{i_2})+\frac{\delta^{ki_2}}{d}(\bar{I}^{i_3}x^{i_1}+\bar{I}^{i_1}x^{i_3})+\frac{\delta^{ki_3}}{d}(\bar{I}^{i_1}x^{i_2}+\bar{I}^{i_2}x^{i_1}) \Biggr\}\ .
\end{split}
\end{equation}

\subsection{Magnetic multipole radiation}\label{appss:radnB}
The expressions we found in our study of magnetostatics can be readily generalised to time-dependent toroidal currents. This is best done in the frequency domain.  The vector Poisson equation in \eqref{eq:VectorPoisson} generalises to the vector Helmholtz equation
\begin{equation}\label{eq:VectorHelm}
\begin{split}
-\frac{1}{r^{d-3}}\partial_r[r^{d-3}\partial_r\gaugeVf_{I}]-\w^2\gaugeVf_{I}+\frac{1}{r^2}(-\mathscr{D}^2+d-2)\gaugeVf_{I}=r^2\gamma_{_{IJ}} J^J\ ,
\end{split}
\end{equation}
where $\gaugeVf_I(\vec{r},\w)$ and $J^I(\vec{r},\w)$ are Fourier transforms of $\gaugeV_I(\vec{r},t)$ and $\bar{J}^I(\vec{r},t)$ respectively. More generally, we use overline for functions of time and remove them to denote the Fourier transforms.

\subsubsection*{Homogenous spherical waves}
Let us begin by finding the homogeneous solution for the above equation by separation of variables. We are interested in solutions whose angular dependence is given by the VSH $\mathbb{V}_I^{\alpha\ell\vec{m}}$. As we will briefly review, the radial dependencies are then controlled by Bessel-like functions.

For a given $(\alpha,\ell,\vec{m})$, there is a unique solution which is regular everywhere. It is given by:
\begin{equation}\label{eq:normBflat}
r^{\nu-\frac{d}{2}+2}{}_0F_1\left[1+\nu,-\frac{\w^2r^2}{4}\right] \mathbb{V}_I^{\alpha\ell\vec{m}}=\Gamma(1+\nu)\left(\frac{\w}{2}\right)^{-\nu}r^{2-\frac{d}{2}}J_{\nu}(\w r)\mathbb{V}_I^{\alpha\ell\vec{m}}\ ,
\end{equation}
where we have defined $\nu\equiv \ell+\frac{d}{2}-1$ to denote the rank of the Bessel function.
Here, we have given two forms of the solution: one in terms of ${}_0F_1$ and another in terms of the Bessel function. Although the Bessel form is standard among textbooks, we find the ${}_0F_1$ notation to be the most convenient. As we shall see below, the ${}_0F_1$ function above becomes the \emph{time-smearing function} used to define multipole moments for extended sources. In the gravitational wave literature, this ${}_0F_1$ function  often appears in the following integral form\cite{Damour:1990gj,Henry:2021cek}
\begin{equation}\label{eq:multipoleDelta}
{}_0F_1\left[1+\nu,-\frac{\w^2r^2}{4}\right] =\int_{-1}^1 dz\ \frac{\Gamma\left(1+\nu\right)}{\Gamma\left(\frac{1}{2}\right)\Gamma\left(\frac{1}{2}+\nu\right)}(1-z^2)^{\nu-\frac{1}{2}} e^{i\w z r}\equiv \int_{-1}^1 dz\ \delta_{\nu-\frac{1}{2}}(z)\  e^{i\w z r} \ .
\end{equation}
The `multipole delta function' $\delta_{\nu-\frac{1}{2}}(z)$ gives an even, positive, normalised measure on the interval $[-1,1]$. The above integral representation then interprets the ${}_0F_1$ function as a \emph{weighed superposition of time-delays}.

There is another solution which is regular everywhere except the origin:
\begin{align}
\begin{split}
\frac{1}{r^{\nu+\frac{d}{2}-2}}&\left\{{}_0F_1\left[1-\nu,-\frac{\w^2r^2}{4}\right]- \frac{\pi \cot \nu \pi}{\Gamma(\nu)\Gamma(1+\nu)}\left(\frac{\w r}{2}\right)^{2\nu}\  {}_0F_1\left[1+\nu,-\frac{\w^2r^2}{4}\right] \right\}  \mathbb{V}_I^{\alpha\ell\vec{m}}\\ &=-\left(\frac{\w}{2}\right)^\nu \frac{\pi\ r^{2-\frac{d}{2}}}{\Gamma(\nu)}Y_{\nu}(\w r)\mathbb{V}_I^{\alpha\ell\vec{m}}\ .
\end{split}
\end{align}
Here, $Y_\nu$ is the Neumann function: when $\nu\notin \mathbb{Z}$, this solution is obtained from the regular solution by the replacement $\nu\to-\nu$ and adding an appropriate amount of the regular solution. The expression in terms of ${}_0F_1$ should, however, be carefully interpreted whenever $\nu\equiv \ell+\frac{d}{2}-1$ is an integer (i.e., whenever the number of space dimensions $d$ is even). When $\nu$ is a positive integer, the hypergeometric series for ${}_0F_1[1-\nu,z]$ is divergent, and the  $\cot \nu\pi$ factor is also divergent. However, these two divergences cancel each other in the above expression, so the limit $\nu\to \text{Integer}$ exists and converges to the Neumann function. These statements should be contrasted against the case when $d$ is odd and $\nu$ is a half-integer: in this case, the hypergeometric series for ${}_0F_1[1-\nu,z]$ is convergent, and the factor $\cot \nu\pi$ evaluates to zero.

Alternately, one can  characterise the solutions according to their behaviours at $r=\infty$, i.e. either as outgoing or ingoing solutions:
\begin{align}
\begin{split}
\frac{1}{r^{\nu+\frac{d}{2}-2}}&\left\{{}_0F_1\left[1-\nu,-\frac{\w^2r^2}{4}\right]\pm (1\pm i\cot \nu \pi)\frac{2\pi i}{\Gamma(\nu)^2}\frac{1}{2\nu}\left(\frac{\w r}{2}\right)^{2\nu}\  {}_0F_1\left[1+\nu,-\frac{\w^2r^2}{4}\right]\right\}\mathbb{V}_I^{\alpha\ell\vec{m}}\\ & =\pm i\left(\frac{\w}{2}\right)^\nu \frac{\pi\ r^{2-\frac{d}{2}}}{\Gamma(\nu)}H^{\pm}_{\nu}(\w r)\mathbb{V}_I^{\alpha\ell\vec{m}}
\end{split}
\end{align}
Here, the solution with $H^+$ denotes the outgoing waves, while the $H^-$ denotes the incoming waves.
Our comments regarding the case where $d$ is even and $\nu$ is a positive integer still apply. Using the identity
\begin{align}
\begin{split}
\frac{ (+i) (1+ i\cot \nu \pi)}{ (-i) (1- i\cot \nu \pi)}=[e^{-i\pi}]^{2\nu}\ ,
\end{split}
\end{align}
it can be seen that the two
solutions here are related by time-reversal, i.e., under $\w\mapsto e^{i\pi}\w$, the outgoing wave is mapped to the incoming wave, and under $\w\mapsto e^{-i\pi}\w$ the incoming wave is mapped to the outgoing wave. For radiation reaction, we are mainly interested in outgoing waves whose radial part is given by 
\begin{align}
\begin{split}
G_{_B}^{\text{Out}}(r,\w,\ell)&\equiv \frac{1}{r^{\nu+\frac{d}{2}-2}}\\
&\times\left\{{}_0F_1\left[1-\nu,-\frac{\w^2r^2}{4}\right]+ (1+ i\cot \nu \pi)\frac{2\pi i}{\Gamma(\nu)^2}\frac{1}{2\nu}\left(\frac{\w r}{2}\right)^{2\nu}\  {}_0F_1\left[1+\nu,-\frac{\w^2r^2}{4}\right]\right\}\ .
\end{split}
\end{align}

When $\nu$ is a half-integer, the function $G_{_B}^{\text{Out}}$ can be greatly simplified by the use of \emph{reverse Bessel polynomials}. They are  defined via
\begin{equation}\label{eq:revBess}
\begin{split}
    \theta_{\nu-\frac{1}{2}}(z)&\equiv \sqrt{\frac{\pi}{2}}e^{z}z^\nu K_\nu(z)
    = \frac{2^{\nu}\Gamma(\nu)}{\sqrt{2\pi}}e^z\left\{\ {}_0F_1\left[1-\nu,\frac{z^2}{4}\right]+\frac{\Gamma(-\nu)}{\Gamma(\nu)}\left(\frac{z}{2}\right)^{2\nu}\ {}_0F_1\left[1+\nu,\frac{z^2}{4}\right]\right\}\\
    &=\sum_{n=0}^{\nu-\frac{1}{2}}\frac{z^{\nu-\frac{1}{2}-n}}{2^n n!}\frac{\left(\nu-\frac{1}{2}+n\right)!}{\left(\nu-\frac{1}{2}-n\right)!}\\
\end{split}
\end{equation}
Here $K_\nu(z)$ is the Macdonald function, and the second line shows that $\theta_{\nu-\frac{1}{2}}(z)$ is a polynomial of degree $\nu-\frac{1}{2}$ with positive integer coefficients,  the coefficient of $z^{\nu-\frac{1}{2}}$ being normalised to unity. Explicit forms of the first few reverse Bessel polynomials are tabulated below:

\begin{table}[H]
	\centering
		\caption{$\theta_{\nu-\frac{1}{2}}(z)$ for various values of $\nu$}
		\setlength{\extrarowheight}{2pt}
	\begin{tabular}{|c|c|}
		\hline\hline
		$\nu $&$\theta_{\nu-\frac{1}{2}}(z)$\\[0.5ex]\hline \hline
		$\frac{1}{2}$&  $1$  \\[0.5ex]\hline$\frac{3}{2}$  & $1+z$\\[0.5ex]\hline
		$\frac{5}{2}$&$3+3z+z^2+r^2z$\\[0.5ex]\hline
  $\frac{7}{2}$&$15+15z+6z^2+z^3$\\[0.5ex]\hline
  $\frac{9}{2}$&$105+105z+45z^2+10z^3+z^4$\\[0.5ex]\hline
  $\frac{11}{2}$&$945+945z+420z^2+105z^3+15z^4+z^5$\\[0.5ex]\hline
  $\frac{13}{2}$&$10395+10395z+4725z^2+1260z^3+210z^4+21z^5+z^6$\\[0.5ex]\hline
  \end{tabular}\label{tab:SmallTheta}
\end{table}
Another useful property of the reverse Bessel polynomials evident from the above table is the value of the constant term in these polynomials
\begin{equation}
\theta_{\nu-\frac{1}{2}}(0)=(2\nu-2)!!=\frac{(d-2)!!}{\nn_{d,\ell-1}} \ .    
\end{equation}
The radial part of the outgoing waves can then be written in the form
\begin{align}
\begin{split}
G_{_B}^{\text{Out}}(r,\w,\ell)= \frac{\theta_{\nu-\frac{1}{2}}(-i\w r)}{\theta_{\nu-\frac{1}{2}}(0)} \frac{e^{i\w r}}{r^{\nu+\frac{d}{2}-2}}=\nn_{d,\ell-1}\frac{\theta_{\nu-\frac{1}{2}}(-i\w r)}{(d-2)!!} \frac{e^{i\w r}}{r^{\nu+\frac{d}{2}-2}}
\end{split}
\end{align}
for $ \nu\in \mathbb{Z}+\frac{1}{2}$. 
The large $r$ asymptotics of this outgoing spherical wave are given by taking the largest power in the reverse Bessel polynomial:
\begin{align}\label{eq:GBasymp}
\begin{split}
G_{_B}^{\text{Out}}(r,\w,\ell)\to \nn_{d,\ell-1}\frac{(-i\w)^{\nu-\frac{1}{2}}}{(d-2)!!} \frac{e^{i\w r}}{r^{\frac{d-3}{2}}}\ \quad \text{as}\quad r\to\infty\ .
\end{split}
\end{align}
This asymptotic formula also holds for $d$ even and $\nu\in \mathbb{Z}$, provided the double factorial for even integers is defined recursively with the convention that 
$0!!\equiv \sqrt{\frac{2}{\pi}}$. For general $\nu$, we can justify this via the asymptotic expansion of Hankel functions (see \url{https://dlmf.nist.gov/10.17}):
\begin{align}
\begin{split}
\pm i\left(\frac{\w}{2}\right)^\nu \frac{\pi\ r^{2-\frac{d}{2}}}{\Gamma(\nu)}H^{\pm}_{\nu}(\w r)\to \left(\frac{\w\ e^{\mp i\frac{\pi}{2}}}{2}\right)^{\nu-\frac{1}{2}} \frac{\sqrt{\pi}}{\Gamma(\nu)}\frac{e^{\pm i\w r}}{r^{\frac{d-3}{2}}} \ \quad \text{as}\quad r\to\infty\ . 
\end{split}
\end{align}
We will conclude this discussion with some useful identities: the raising and lowering relations for Hankel functions lead to 
% \begin{equation}\label{eq:GBraising}
% \begin{split}
% \partial_rG_{_B}^{\text{Out}}(r,\w,\ell) +\frac{1}{r}\left(\frac{d}{2}-2-\nu\right)G_{_B}^{\text{Out}}(r,\w,\ell)&=-2\nu\ G_{_B}^{\text{Out}}(r,\w,\ell+1)\ ,\\
% \partial_rG_{_B}^{\text{Out}}(r,\w,\ell) +\frac{1}{r}\left(\frac{d}{2}-2+\nu\right)G_{_B}^{\text{Out}}(r,\w,\ell)&=\frac{\omega^2}{2(\nu-1)} G_{_B}^{\text{Out}}(r,\w,\ell-1)\  .
% \end{split}
% \end{equation}
\begin{equation}\label{eq:GBraising}
\begin{split}
-\frac{1}{r}\frac{\partial}{\partial r}\left[\frac{G_{_B}^{\text{Out}}(r,\w,\ell)}{\nn_{d,\ell-1}r^{\ell+1}}\right] &=\frac{G_{_B}^{\text{Out}}(r,\w,\ell+1)}{\nn_{d,\ell}r^{\ell+2}}\ ,\quad
\frac{1}{r}\frac{\partial}{\partial r}\left[\frac{G_{_B}^{\text{Out}}(r,\w,\ell)}{\nn_{d,\ell-1}\ r^{3-d-\ell}}\right] =\w^2 \frac{G_{_B}^{\text{Out}}(r,\w,\ell-1)}{\nn_{d,\ell-2}\ r^{4-d-\ell}}\ 
.
\end{split}
\end{equation}
These identities can be used to give the following formulae for the derivative of  $G_{_B}^{\text{Out}}$, i.e.,
\begin{equation}\label{eq:GBderiv}
\begin{split}
\partial_rG_{_B}^{\text{Out}}(r,\w,\ell) &=-\frac{\left(\ell+d-3\right)}{r} G_{_B}^{\text{Out}}(r,\w,\ell)\ +\frac{\omega^2}{2\ell+d-4} G_{_B}^{\text{Out}}(r,\w,\ell-1)\\
&=-\left(\ell+d-3\right)G_{_B}^{\text{Out}}(r,\w,\ell+1)+\frac{\ell+1}{2\ell+d-2}\frac{\omega^2}{2\ell+d-4} G_{_B}^{\text{Out}}(r,\w,\ell-1) \ .
\end{split}
\end{equation}

Another useful identity arises from the cartesian version of Eq.\eqref{eq:GBraising}:
\begin{equation}\label{eq:GBraisingC}
\begin{split}
-\partial_i\left[\frac{G_{_B}^{\text{Out}}(r,\w,\ell)}{\nn_{d,\ell-1}r^{\ell+1}}\right] &=x^i\frac{G_{_B}^{\text{Out}}(r,\w,\ell+1)}{\nn_{d,\ell}r^{\ell+2}}\ ,\quad
\partial_i\left[\frac{G_{_B}^{\text{Out}}(r,\w,\ell)}{\nn_{d,\ell-1}\ r^{3-d-\ell}}\right] =\w^2 x^i\frac{G_{_B}^{\text{Out}}(r,\w,\ell-1)}{\nn_{d,\ell-2}\ r^{4-d-\ell}}\ 
.
\end{split}
\end{equation}
By repeated application of the first identity, we get the following relation, which expresses the radial part of $\ell^{th}$ spherical wave as a derivative of the $\ell=0$ wave:
\begin{equation}\label{eq:dddGB}
\begin{split}
(-1)^\ell\partial^{<i_1}\partial^{i_2}\ldots \partial^{i_\ell>}\left\{\frac{G_{_B}^{\text{Out}}(r,\w,\ell=0) }{  (d-2) r}\right\}=\frac{G_{_B}^{\text{Out}}(r,\w,\ell)}{ \nn_{d,\ell-1}r^{\ell+1}}x^{<i_1}\ldots x^{i_\ell>}\ .
\end{split}
\end{equation}
Here we work with the convention that $\nn_{d,-1}=(d-2)$ which is the correct analytic extension of $\nn_{d,\ell}$ to negative $\ell$'s. In the above relation, the STF projection on the indices ensures that the derivatives acting on $x^i$'s always give zero at every step.
\subsubsection*{Green function for magnetic radiation}
Now that we understand the homogeneous solutions, we can solve the full inhomogeneous Helmholtz equation in Eq.\eqref{eq:VectorHelm} via Green functions. As we did in magnetostatics, we  will begin with an ansatz for the vector potential:
\begin{equation}
\begin{split}
\gaugeV_t(\Vec{r},t) &=\gaugeV_r(\Vec{r},t)=0\ ,\quad \gaugeV_I(\Vec{r},t)\equiv \sum_{\alpha\ell\vec{m}}\int_\w e^{-i\w t} \Phi_{_B}(r,\w,\alpha,\ell,\vec{m})\mathbb{V}_{I}^{\alpha\ell\vec{m}}(\hat{r})\ 
\end{split}
\end{equation}
where $\Phi_{_B}$ denotes the frequency domain magnetic  Debye field. Substituting this ansatz into Eq.\eqref{eq:VectorHelm}, we get a sourced Helmholtz equation for $\Phi_{_B}$:
\begin{equation}\label{eq:PhiBHelm}
\begin{split}
-\frac{1}{r^{d-3}}\partial_r[r^{d-3}\partial_r\Phi_{_B}]-\w^2\Phi_{_B}+\frac{1}{r^2}(\ell+1)(\ell+d-3)\Phi_{_B}=r^2J_V(r,\w)\ ,
\end{split}
\end{equation}
where the source appearing in the RHS is
\begin{equation}
\begin{split}
J_V(r,\w)\equiv \int_{\hat{r}\in\mathbb{S}^{d-1}} \mathbb{V}_I^{\alpha\ell\vec{m}\ast}(\hat{r})J^{I}(\vec{r},\w)\ .
\end{split}
\end{equation}
We will posit a solution of the form 
\begin{equation}\label{eq:PhiBdyn}
\begin{split}
\Phi_{_B}(r,\w,\alpha,\ell,\vec{m})&\equiv
\int_{\vec{r}_0}\mathbb{G}_B(r,r_0;\w,\ell)\mathbb{V}_J^{\alpha\ell\vec{m}\ast}(\hat{r}_0)J^{J}(\vec{r}_0,\w)\\
&=\int_0^\infty dr_0\ r_0^{d-1}\mathbb{G}_B(r,r_0;\w,\ell) J_V(r_0,\w)\ ,
\end{split}
\end{equation}
which generalises the static expression in Eq.\eqref{eq:PhiBStatics}. The Green function $\mathbb{G}_B$ obeys
\begin{equation}\label{eq:GBHelm}
\begin{split}
-\partial_r[r^{d-3}\partial_r\mathbb{G}_B]-\w^2r^{d-3}\mathbb{G}_B+(\ell+1)(\ell+d-3)r^{d-5}\mathbb{G}_B=\delta(r-r_0)\ ,
\end{split}
\end{equation}
and is built by stitching together the homogeneous solutions of the vector Helmholtz equation. Our normalisations here are such that $\mathbb{G}_B$ generalises the static Green function defined in Eq.\eqref{eq:ViStatics}. The conditions on the Green function $\mathbb{G}_B$ are that it should be continuous, its derivative should have an appropriate discontinuity, and it should match onto an outgoing wave far away from currents. These determine
\begin{equation}\label{eq:GB0F1}
\begin{split}
\mathbb{G}_{B}(r,r_0;\w,\ell)&=\frac{1}{2\nu}\frac{r_<^{\nu-\frac{d}{2}+2}}{r_{>}^{\nu+\frac{d}{2}-2}}{}_0F_1\left[1+\nu,-\frac{\w^2r_{<}^2}{4}\right]\\ &\times \left\{{}_0F_1\left[1-\nu,-\frac{\w^2r_{>}^2}{4}\right]+ (1+ i\cot \nu \pi)\frac{2\pi i}{\Gamma(\nu)^2}\frac{1}{2\nu}\left(\frac{\w r_{>}}{2}\right)^{2\nu}\  {}_0F_1\left[1+\nu,-\frac{\w^2r_{>}^2}{4}\right]\right\} \\
&=\frac{1}{2\nu}r_<^{\nu-\frac{d}{2}+2}{}_0F_1\left[1+\nu,-\frac{\w^2r_{<}^2}{4}\right] G_B^{\text{Out}}(r_{>},\w,\ell)\ ,
\end{split}
\end{equation}
where we have defined:
\begin{equation}
\begin{split}
r_>\equiv \text{Max}(r,r_0)\ ,\quad	
r_<\equiv \text{Min}(r,r_0)\ .
\end{split}
\end{equation}
The reader can check that the above expression reduces to  Eq.\eqref{eq:ViStatics} in the static (i.e., $\w\to 0$) limit. 
Once we have the vector potential, we can compute the 
the corresponding electric/magnetic field components. The VSH expansion of the field strengths takes the form
\begin{equation}
\begin{split}
     \FC_{rt}(\Vec{r},t)&=0\ ,\\
     \FC_{It}(\Vec{r},t)&=\sum_{\alpha\ell\vec{m}}\int_\w e^{-i\w t}\Evf(r,\w,\alpha,\ell,\vec{m}) \ \VSH_I(\vec{r})\ ,\\
     \FC_{rI}(\Vec{r},t)&=\sum_{\alpha\ell\vec{m}}\int_\w e^{-i\w t}\Hvf(r,\w,\alpha,\ell,\vec{m}) \ \VSH_I(\vec{r})\ ,\\
     \FC_{IJ}(\Vec{r},t)&=\sum_{\alpha\ell\vec{m}}\int_\w e^{-i\w t}\Bvvf (r,\w,\alpha,\ell,\vec{m})\ \left[\mathscr{D}_I\mathbb{V}_{J\alpha\ell\vec{m}}(\vec{r})-\mathscr{D}_J\mathbb{V}_{I\alpha\ell\vec{m}}(\vec{r})\right]\ ,
\end{split}
\end{equation}
where the components are given by
\begin{equation}\label{eq:MagneticFieldsFlat}
\begin{split}
\Hvf&=\partial_r\Phi_{_B}=\int_{\vec{r}_0}\partial_r\mathbb{G}_B(r,r_0;\w,\ell)\mathbb{V}_J^{\alpha\ell\vec{m}\ast}(\hat{r}_0)J^{J}(\vec{r}_0,\w)\ ,\\
\Bvvf &=\Phi_{_B}=\int_{\vec{r}_0}\mathbb{G}_B(r,r_0;\w,\ell)\mathbb{V}_J^{\alpha\ell\vec{m}\ast}(\hat{r}_0)J^{J}(\vec{r}_0,\w)\ ,\\
\Evf &= i\w\ \Phi_{_B}=i\w \int_{\vec{r}_0}\mathbb{G}_B(r,r_0;\w,\ell)\mathbb{V}_J^{\alpha\ell\vec{m}\ast}(\hat{r}_0)J^{J}(\vec{r}_0,\w) \ .
\end{split}
\end{equation}
Note that apart from the magnetic fields, we have an induced electric field when toroidal currents are time-dependent. These time-varying electric and magnetic fields sustain each other as they escape the source and propagate far away as radiation. 

\subsubsection*{Fields outside sources}
To get a handle on the structure of multipole radiation, we will now focus on fields outside the sources. We will generalise our definition of static magnetic multipole moment
in Eq.\eqref{eq:jBstatic} as follows:
\begin{equation}\label{eq:jBflat}
\begin{split}
\multj^B(\w,\alpha,\ell,\vec{m}) &\equiv\frac{1}{2\nu}\int_{\vec{r}_0}r_0^{\nu-\frac{d}{2}+2}{}_0F_1\left[1+\nu,-\frac{\w^2r_0^2}{4}\right]
\mathbb{V}_I^{\alpha\ell\vec{m}\ast}(\hat{r}_0)J^I(\vec{r}_0,\w)\ .
\end{split}
\end{equation}
It is clear from our general solution that this is the magnetic moment that determines the fields outside currents.
The magnetic Debye field outside the currents is given by 
\begin{equation}
\begin{split}
\Phi_{B}^{\text{Out}} &=G_{_B}^{\text{Out}}(r,\w,\ell)\ \multj^B(\w,\alpha,\ell,\vec{m})=\frac{\theta_{\nu-\frac{1}{2}}(-i\w r)}{\theta_{\nu-\frac{1}{2}}(0)} \frac{e^{i\w r}}{r^{\ell+d-3}}\multj^B(\w,\alpha,\ell,\vec{m}) \   .
\end{split}
\end{equation}
The values of field strength components are given explicitly in terms of $\multj^B$ by 
\begin{equation}\label{eq:HoutjB}
\begin{split}
\Bvvf^{\text{Out}} &=G_{_B}^{\text{Out}}(r,\w,\ell)\ \multj^B(\w,\alpha,\ell,\vec{m})=\frac{\theta_{\nu-\frac{1}{2}}(-i\w r)}{\theta_{\nu-\frac{1}{2}}(0)} \frac{e^{i\w r}}{r^{\ell+d-3}}\multj^B(\w,\alpha,\ell,\vec{m}) ,\\
\Hvf^{\text{Out}} 
&=-\frac{\left(\ell+d-3\right)}{r} G_{_B}^{\text{Out}}(r,\w,\ell)\ \multj^B(\w,\alpha,\ell,\vec{m})+\frac{\omega^2}{2\ell+d-4} G_{_B}^{\text{Out}}(r,\w,\ell-1)\ \multj^B(\w,\alpha,\ell,\vec{m})\\
&=-\left(\ell+d-3\right) \frac{\theta_{\nu-\frac{1}{2}}(-i\w r)}{\theta_{\nu-\frac{1}{2}}(0)} \frac{e^{i\w r}}{r^{\ell+d-2}}\ \multj^B(\w,\alpha,\ell,\vec{m})\\
&\qquad+\frac{\omega^2}{2\ell+d-4} \frac{\theta_{\nu-\frac{3}{2}}(-i\w r)}{\theta_{\nu-\frac{3}{2}}(0)} \frac{e^{i\w r}}{r^{\ell+d-4}}\ \multj^B(\w,\alpha,\ell,\vec{m}),\\
\Evf^{\text{Out}} &=i\w G_{_B}^{\text{Out}}(r,\w,\ell)\ \multj^B(\w,\alpha,\ell,\vec{m})=i\w\frac{\theta_{\nu-\frac{1}{2}}(-i\w r)}{\theta_{\nu-\frac{1}{2}}(0)} \frac{e^{i\w r}}{r^{\ell+d-3}}\multj^B(\w,\alpha,\ell,\vec{m})\   .
\end{split}
\end{equation}
Here we have used Eq.\eqref{eq:GBderiv} for evaluating $\partial_r G_{_B}^{\text{Out}}$. The large $r$ asymptotics can be worked out using Eq.\eqref{eq:GBasymp}:
\begin{equation}\label{eq:RadFieldsB}
\begin{split}
H_{vv}^{\text{Rad}} &=\Phi_{_B}^{\text{Rad}} =\frac{(-i\w)^{\nu-\frac{1}{2}}}{(2\nu-2)!!} \frac{e^{i\w r}}{r^{\frac{d-3}{2}}}\ \multj^B(\w,\alpha,\ell,\vec{m}) ,\\
H_v^{\text{Rad}} &=\partial_r\Phi_{_B}^{\text{Rad}}=-\frac{(-i\w)^{\nu+\frac{1}{2}}}{(2\nu-2)!!} \frac{e^{i\w r}}{r^{\frac{d-3}{2}}} \multj^B(\w,\alpha,\ell,\vec{m}),\\
E_{v}^{\text{Rad}} &=i\w\Phi_{_B}^{\text{Rad}}=-\frac{(-i\w)^{\nu+\frac{1}{2}}}{(2\nu-2)!!} \frac{e^{i\w r}}{r^{\frac{d-3}{2}}} \multj^B(\w,\alpha,\ell,\vec{m}) .
\end{split}
\end{equation}
We remind the reader that this holds even when $d$ is even, provided the double factorial for even integers is defined recursively with the convention that 
$0!!\equiv \sqrt{\frac{2}{\pi}}$. We will see later how these asymptotics get modified in dS  \eqref{eq:dShorizonFields}.

We next turn to the description in terms of cartesian STF tensors. The STF version of magnetic moment is still defined by Eq.\eqref{eq:jBtoQB}, but is now generalised to arbitrary frequency,i.e.,
\begin{equation}\label{eq:jBtoQB2}
\begin{split}
\frac{1}{\ell!}\ [{}^B\mathcal{Q}(\w)]_{k<i_1 i_2\ldots i_\ell>}\ x^{i_1}\ldots x^{i_\ell}\equiv \nn_{d,\ell-1}|\mathbb{S}^{d-1}|\left(\frac{\partial x_k}{\partial \vartheta_I}\right)\sum_{\vec{m}\alpha}\multj^B(\w,\alpha,\ell,\vec{m}) r^{\ell-1}\mathbb{V}^I_{\alpha\ell\vec{m}}(\hat{r})\ .
\end{split}
\end{equation}
Repeating the same logic as in magnetostatics, we can give a direct expression in terms of the vector STF projector: 
\begin{equation}\label{eq:QBgen}
\begin{split}
{}^B\mathcal{Q}_{k<i_1 i_2\ldots i_\ell>}(\w)\equiv(\Pi^V_{kj})^{<i_1 i_2\ldots i_\ell>}_{<j_1 j_2\ldots j_\ell>}\int_{\vec{r}_0}x_0^{j_1}\ldots x_0^{j_\ell}{}_0F_1\left[1+\nu,-\frac{\w^2r_0^2}{4}\right]J^{j}(\vec{r}_0,\w)\ .
\end{split}
\end{equation}
The tensor structure appearing here is exactly identical to that seen in statics (e.g., see \eqref{eq:QBexp}). The main difference in the time-dependent situation is the smearing due to time delays: using Eq.\eqref{eq:multipoleDelta}, we can write 
\begin{equation}
\begin{split}
{}^B\mathcal{Q}_{k<i_1 i_2\ldots i_\ell>}=(\Pi^V_{kj})^{<i_1 i_2\ldots i_\ell>}_{<j_1 j_2\ldots j_\ell>}\int_{-1}^1 dz\ \delta_{\nu-\frac{1}{2}}(z)\int_{\vec{r}_0}x_0^{j_1}\ldots x_0^{j_\ell}e^{i\w z r_0}J^{j}(\vec{r}_0,\w)\ .
\end{split}
\end{equation}
where 
\begin{equation}
\delta_{\nu-\frac{1}{2}}(z)\equiv\frac{\Gamma\left(1+\nu\right)}{\Gamma\left(\frac{1}{2}\right)\Gamma\left(\frac{1}{2}+\nu\right)}(1-z^2)^{\nu-\frac{1}{2}}=\frac{(2\nu)!!}{2^{\nu+\frac{1}{2}}\left(\nu-\frac{1}{2}\right)!}(1-z^2)^{\nu-\frac{1}{2}}  \ 
\end{equation}
with $\nu\equiv \ell+\frac{d}{2}-1$\footnote{For  even $d$, we work with the convention that  $0!!=\sqrt{\frac{2}{\pi}}$. }. The interpretation in terms of time delays is more transparent in the time domain where the above equation becomes
\begin{equation}
\begin{split}
{}^B\overline{\mathcal{Q}}_{k<i_1 i_2\ldots i_\ell>}(t)=(\Pi^V_{kj})^{<i_1 i_2\ldots i_\ell>}_{<j_1 j_2\ldots j_\ell>}\int_{\vec{r}_0}x_0^{j_1}\ldots x_0^{j_\ell}\int_{-1}^1 dz\ \delta_{\nu-\frac{1}{2}}(z)\bar{J}^{j}(\vec{r}_0,t-zr_0)\ .
\end{split}
\end{equation}
The time delay above is further compounded in fields by the standard retardation effect, i.e., the field depends on ${}^B\overline{\mathcal{Q}}_{k<i_1 i_2\ldots i_\ell>}(t-r)$ and hence on $\bar{J}^{j}(\vec{r}_0,t-r-zr_0)$.
Thus, for a source of size $R$ spread around the origin, we get a time delay seen by a detector ranging from $r-R$ (in the near end of the source) to $r+R$ (in the far end of the source). In an expanding universe, there is a further effect due to redshifts, which have to be correctly taken into account while defining multipole moments of cosmologically big sources.

With these comments, let us return to the task at hand: in terms of the magnetic multipole tensor, the vector potential/EM fields outside the sources can be written as\footnote{Here we use the notation $\text{Anti}_{jk}[T_{jk}]\equiv T_{jk}-T_{kj}$ for the anti-symmetrisation operator.}
\begin{equation}\label{eq:EMcartGenDB}
\begin{split}
\gaugeVf_k^{\text{Out}}(\vec{r},\w) &=\sum_{\ell} \frac{G_{_B}^{\text{Out}}(r,\w,\ell)}{\ell! \nn_{d,\ell-1}|\mathbb{S}^{d-1}|}{}^B\mathcal{Q}_{k<i_1 i_2\ldots i_\ell>}\frac{x^{i_1}\ldots x^{i_\ell}}{ r^{\ell+1}}\ ,\\
\FCf_{kt}^{\text{Out}}(\vec{r},\w) &=i\w \sum_{\ell} \frac{G_{_B}^{\text{Out}}(r,\w,\ell)}{\ell! \nn_{d,\ell-1}|\mathbb{S}^{d-1}|}{}^B\mathcal{Q}_{k<i_1 i_2\ldots i_\ell>}\frac{x^{i_1}\ldots x^{i_\ell}}{ r^{\ell+1}}\ ,\\
\FCf_{jk}^{\text{Out}}(\vec{r},\w)&= \text{Anti}_{jk}\sum_{\ell} \frac{G_{_B}^{\text{Out}}(r,\w,\ell)}{\ell! \nn_{d,\ell-1}|\mathbb{S}^{d-1}|}{}^B\mathcal{Q}_{j<i_1 i_2\ldots i_\ell>}[(2\ell+d-2)x^k x^{i_\ell}-r^2\ell\delta^{ki_\ell}]\frac{ x^{i_1}\ldots x^{i_{\ell-1}}}{ r^{\ell+3}}\\
&-\omega^2\text{Anti}_{jk}\sum_{\ell}\frac{G_{_B}^{\text{Out}}(r,\w,\ell-1)}{\ell! \nn_{d,\ell-2}|\mathbb{S}^{d-1}|}\ {}^B\mathcal{Q}_{j<i_1 i_2\ldots i_\ell>}\frac{x^kx^{i_1}\ldots x^{i_\ell}}{ r^{\ell+2}}\ .
\end{split}
\end{equation}
The field strengths here  can also be derived by direct cartesian differentiation (using Eq.\eqref{eq:GBraisingC} when needed).
The multipole vector potential given here can be rewritten  using Eq.\eqref{eq:dddGB} as
\begin{equation}
\begin{split}
\gaugeVf_k^{\text{Out}}(\vec{r},\w) 
&=\sum_{\ell} \frac{(-1)^\ell}{\ell!}\partial^{i_1}\partial^{i_2}\ldots \partial^{i_\ell}\left\{{}^B\mathcal{Q}_{k<i_1 i_2\ldots i_\ell>}\frac{G_{_B}^{\text{Out}}(r,\w,\ell=0) }{  (d-2)|\mathbb{S}^{d-1}| r}\right\}\ .
\end{split}
\end{equation}
For odd $d$, we can write
\begin{align}
\begin{split}
\frac{G_{_B}^{\text{Out}}(r,\w,\ell=0) }{  (d-2)|\mathbb{S}^{d-1}| r}=\frac{\theta_{\frac{d-3}{2}}(-i\w r)}{(d-2)!!} \frac{e^{i\w r}}{|\mathbb{S}^{d-1}|r^{d-2}}\ .
\end{split}
\end{align}
The $e^{i\w r}$ factor gives the standard retardation time-delay. Moving to time-domain, we then have a simple statement in $d=3$: the vector potential of a magnetic multipole is obtained by multiplying the retarded STF magnetic tensor with the Coulomb potential, followed by repeated differentiation. For odd $d>3$, we should apply an additional differential operator that depends on $r$ and with maximum $\frac{d-3}{2}$ time-derivatives
\begin{align}
\begin{split}
\frac{\theta_{\frac{d-3}{2}}(r\partial_t)}{(d-4)!!} \ ,
\end{split}
\end{align}
before the repeated differentiation\cite{Henry:2021cek}. Next, the large $r$ asymptotics of both the vector potential as well as the field strengths can be obtained via Eq.\eqref{eq:GBasymp}. We get
\begin{equation}\label{eq:RadBflat}
\begin{split}
\gaugeVf_k^{\text{Rad}}(\vec{r},\w) &=\frac{e^{i\w r}}{(d-2)!!|\mathbb{S}^{d-1}| r^{\frac{d-1}{2}} }\sum_{\ell} \frac{(-i\w)^{\nu-\frac{1}{2}}}{\ell! }{}^B\mathcal{Q}_{k<i_1 i_2\ldots i_\ell>}n^{i_1}\ldots n^{i_\ell}\ ,\\
\FCf_{kt}^{\text{Rad}}(\vec{r},\w) &=-\frac{e^{i\w r}}{(d-2)!!|\mathbb{S}^{d-1}| r^{\frac{d-1}{2}} }\sum_{\ell} \frac{(-i\w)^{\nu+\frac{1}{2}}}{\ell! }{}^B\mathcal{Q}_{k<i_1 i_2\ldots i_\ell>}n^{i_1}\ldots n^{i_\ell} \ ,\\
\FCf_{jk}^{\text{Rad}}(\vec{r},\w)&= \frac{e^{i\w r}}{(d-2)!!|\mathbb{S}^{d-1}| r^{\frac{d-1}{2}} }\times \text{Anti}_{jk}\Bigl\{ n^k\sum_{\ell}\frac{(-i\w)^{\nu+\frac{1}{2}}}{\ell!}\ {}^B\mathcal{Q}_{j<i_1 i_2\ldots i_\ell>}n^{i_1}\ldots n^{i_\ell}\ \Bigr\}\ ,
\end{split}
\end{equation}
where we have used the notation $n^i\equiv \frac{x^i}{r}$. The infinite sum appearing in the field strengths is the EM waveform or the \emph{light vector}.

\subsection{Multipole expansion in statics II : poloidal currents and charges}\label{appss:staticsE}
\subsubsection*{Magnetic fields due to poloidal currents}
We can go further and generalize to any time-independent, divergence-free current distribution. Such a current distribution, which \emph{not} toroidal (i.e., not purely tangential to the sphere directions), is said to be a \emph{poloidal} current distribution. A poloidal current can be expanded as 
\begin{equation}\begin{split}
	\bar{J}^r &=\sum_{\ell\vec{m}}\bar{J}_P(r,\ell,\vec{m})\mathscr{Y}_{\ell\vec{m}}(\hat{r})\ ,\quad 
 \bar{J}^I=\sum_{\ell\vec{m}}\bar{J}_Q(r,\ell,\vec{m})\gamma^{IJ}\mathscr{D}_J\mathscr{Y}_{\ell\vec{m}}(\hat{r})\ ,
\end{split}\end{equation}
where $\mathscr{Y}_{\ell\vec{m}}(\hat{r})$ are orthonormal scalar spherical harmonics (SSHs) on $\mathbb{S}^{d-1}$. They satisfy
\begin{equation}
\begin{split}
\left[\mathscr{D}^2+\ell(\ell+d-2)\right]\mathscr{Y}_{\ell\vec{m}}= 0\ ,\quad 	
\int_{\mathbb{S}^{d-1}}
\mathscr{Y}_{\ell'\vec{m}'}^\ast\mathscr{Y}_{\ell\vec{m}} =\delta_{\ell'\ell}\delta_{\vec{m}'\vec{m}}\ .
\end{split}
\end{equation}

We want to determine the vector potential and magnetic field due to such a poloidal current distribution. With some hindsight, we will abandon the Coulomb gauge and use instead a gauge where the vector potential is purely radial. The vector potential and magnetic field are then of the form
\begin{equation}
\begin{split}
\gaugeV_r
&\equiv -\sum_{\ell\vec{m}}\overline{H}_s(r,\ell,\vec{m})\mathscr{Y}_{\ell\vec{m}}(\hat{r})\ ,\quad \gaugeV_I=0\ ,\\
\FC_{rI} &= \sum_{\ell\vec{m}}\overline{H}_s(r,\ell,\vec{m})\mathscr{D}_I\mathscr{Y}_{\ell\vec{m}}(\hat{r})\ ,\quad
\FC_{IJ}=0\ .
\end{split}
\end{equation}
Using the sourced Maxwell equations\footnote{They are of the form
\begin{equation}
\begin{split}
-\gamma^{IJ}\mathscr{D}_I\FC_{Jr}=r^2\bar{J}^r\ ,&\quad 
\frac{1}{r^{d-1}}\partial_r[r^{d-3}\FC_{Ir}]+\frac{1}{r^4}\gamma^{JK}\mathscr{D}_K\FC_{IJ}= \gamma_{IJ} \bar{J}^J\ .
\end{split}
\end{equation}},
we obtain the following relation between the currents and the magnetic field:
\begin{equation}\label{eq:PolHsReln}\begin{split}
	\bar{J}^r &=-\sum_{\ell\vec{m}}\frac{\ell(\ell+d-2)}{r^2}\overline{H}_s(r,\ell,\vec{m})\mathscr{Y}_{\ell\vec{m}}(\hat{r})\ ,\\
\bar{J}^I&=-\sum_{\ell\vec{m}}\frac{1}{r^{d-1}}\partial_r[r^{d-3}\overline{H}_s(r,\ell,\vec{m})]\gamma^{IJ}\mathscr{D}_J\mathscr{Y}_{\ell\vec{m}}(\hat{r})\ ,
\end{split}\end{equation}
These two equations are not independent: they are related by  current conservation, viz.,
\begin{equation}\begin{split}
\frac{1}{r^{d-1}}\frac{\partial}{\partial r}(r^{d-1}\bar{J}^r)+\mathscr{D}_I\bar{J}^I=0\ .
\end{split}\end{equation}
Thus, it is enough to invert the first equation. Using the orthonormality of scalar spherical harmonics, we can write
\begin{equation}
\begin{split}
\overline{H}_s(r,\ell,\vec{m})&=-\frac{r^2}{\ell(\ell+d-2)}\int_{\hat{r}\in \mathbb{S}^{d-1}}\mathscr{Y}_{\ell\vec{m}}^\ast(\hat{r})\bar{J}^{r}(\vec{r})\\
&=-\frac{1}{\ell(\ell+d-2)}\frac{1}{r^{d-3}}\int_{\vec{r}_0}\mathscr{Y}_{\ell\vec{m}}^\ast(\hat{r}_0)\delta(r-r_0)\bar{J}^{r}(\vec{r}_0)\ .
\end{split}
\end{equation}
A few comments are in order: the result we have derived is valid if there is no $\ell=0$ component (i.e., there is no spherically symmetric component) in $\bar{J}^r$. A bit of thought shows that this has to be true:  a spherically symmetric radial current is inconsistent with charge conservation in the static limit. The next comment is about the locality in the radial direction: we see that if the poloidal current is confined within a radius $R$, its magnetic field also never extends beyond $R$. In particular, the far-field multipole expansion we derived before does not get corrected by poloidal currents.

When we move to time-varying currents, we will see that $\overline{H}_s$ part of the magnetic field can in fact escape the currents and travel out as EM radiation. This suggests that, in the full dynamical situation, $\overline{H}_s$ satisfies a wave equation with a source. Such an equation should reduce to a Poisson-like equation in the static limit. To see how this works, we combine the two equations of Eq.\eqref{eq:PolHsReln} into a Poisson-like equation:
\begin{equation}\label{eq:HsPoiss}\begin{split}
    r^{d-3}\partial_r\left(r^{3-d}\partial_r[r^{d-3}\overline{H}_s]\right)&-\frac{\ell(\ell+d-2)}{r^2}(r^{d-3}\overline{H}_s)\\
    &=\int_{\hat{r}\in \mathbb{S}^{d-1}}r^{d-3}\mathscr{Y}_{\ell\vec{m}}^\ast(\hat{r})\Bigl\{\bar{J}^r(\vec{r})+\frac{1}{\ell(\ell+d-2)}\partial_{r}\left[r^2\mathscr{D}_I\bar{J}^I(\vec{r})\right]\Bigr\}\ .
\end{split}\end{equation}
As we did for toroidal currents, we can solve this equation by thinking of the source as made of spherical shells and then integrating. we end up with 
\begin{equation}\label{eq:HsbStatics}
\begin{split}
r^{d-3}\overline{H}_s
&=-\int_{\vec{r}_0} \mathbb{G}_{E}(r,r_0,\ell)\
r_0^{1-d}\mathscr{Y}_{\ell\vec{m}}^\ast(\hat{r}_0)\Bigl\{\bar{J}^r(\vec{r}_0)+\frac{1}{\ell(\ell+d-2)}\partial_{r_0}\left[r_0^2\mathscr{D}_I\bar{J}^I(\vec{r}_0)\right]\Bigr\}\\ \text{with} &\quad 
\mathbb{G}_{E}(r,r_0,\ell)
\equiv \frac{1}{2\ell+d-2} \left\{\frac{r^{\ell+d-2}}{r_0^{\ell}}\Theta(r<r_0)+\frac{r_0^{\ell+d-2}}{r^{\ell}}\Theta(r>r_0)\right\}
\end{split}
\end{equation}
The locality in the radial direction can be seen by rewriting the source part of the integrand using the conservation of current:
\begin{equation}\begin{split}
&\bar{J}^r(\vec{r}_0)+\frac{1}{\ell(\ell+d-2)}\partial_{r_0}\left[r_0^2\mathscr{D}_I\bar{J}^I(\vec{r}_0)\right]\\
&\qquad = \bar{J}^r(\vec{r}_0)-\frac{1}{\ell(\ell+d-2)}\partial_{r_0}\left[r_0^{3-d}\partial_{r_0}(r_0^{d-1}\bar{J}^r(\vec{r}_0))\right]\ .
\end{split}
\end{equation}
The radial differential operator here is the same as the one in Eq.\eqref{eq:HsPoiss}, and after integration by parts, it acts on the Green function to give a delta function. We will see later how this conclusion changes if the poloidal currents become time-dependent. 

\subsubsection*{Electric fields due to static charges}
We can repeat our magnetostatic analysis for electrostatics. Let us begin with a time-independent
surface charge density $\bar{\sigma}(\hat{r})$ spread out on a thin spherical shell of radius $R$. Explicitly, we take a charge current density of the form
\begin{equation}
	\bar{J}^t=\delta(r-R)\ \bar{\sigma}(\hat{r})=\delta(r-R)\sum_{\ell\vec{m}}\bar{\sigma}_{\ell \vec{m}}(\hat{r})\ ,
\end{equation}
where we have expanded out the charge density in terms of orthonormal Scalar Spherical Harmonics (SSHs) on $\mathbb{S}^{d-1}$ labelled by  $\{\ell, \vec{m}\}$. Using such a decomposition, we can solve for the electric field due to each component and then add it up to get the final answer. Since the symmetry properties of each SSH under $SO(d)$ rotation is different, the scalar potential $\gaugeV_t$ produced by $\bar{\sigma}_{\ell \vec{m}}$ should be proportional to $\bar{\sigma}_{\ell \vec{m}}$. We take an ansatz of the form 
\begin{equation}
\begin{split}
\gaugeV_t=\sum_{\ell}f_\ell(r)\sum_{\vec{m}} \bar{\sigma}_{\ell \vec{m}}(\hat{r})\ ,
\end{split}
\end{equation}
and impose the scalar Poisson equation
\begin{equation}\label{eq:ScalarPoisson}
\begin{split}
\frac{1}{r^{d-1}}\partial_r[r^{d-1}\FC_{rt}]+\frac{1}{r^2}\gamma^{JK}\mathscr{D}_K\FC_{It}=\frac{1}{r^{d-1}}\partial_r[r^{d-1}\partial_r\gaugeV_{t}]+\frac{1}{r^2}\mathscr{D}^2\gaugeV_{t}= \bar{J}^t\ .
\end{split}
\end{equation}
Replacing  $-\mathscr{D}^2$ by $\ell(\ell+d-2)$,
we conclude that, away from the spherical shell, $f_\ell(r)$ should vary as $r^{\ell}$ or as $r^{-(\ell+d-2)}$. We should stitch together these two solutions continuously with an appropriate derivative discontinuity given by the charge density. We obtain the final answer
\begin{equation}
\gaugeV_t=-\sum_{\ell\vec{m}}\frac{R\bar{\sigma}_{\ell \vec{m}}(\hat{r})}{2\ell+d-2}\left[\frac{r^{\ell}}{R^{\ell}}\Theta(r<R)+\frac{R^{\ell+d-2}}{r^{\ell+d-2}}\Theta(r>R)\right]\ .
\end{equation}
In terms of the original data, we have
\begin{equation}
\begin{split}
\gaugeV_t&=-\int_{\hat{r}_0\in\mathbb{S}^{d-1}}\sum_{\ell}\left[\frac{r^{\ell}}{R^{\ell}}\Theta(r<R)+\frac{R^{\ell+d-2}}{r^{\ell+d-2}}\Theta(r>R)\right] \frac{\sum_{\vec{m}}\mathscr{Y}_{\ell\vec{m}}(\hat{r})
\mathscr{Y}_{\ell\vec{m}}^\ast(\hat{r}_0)}{2\ell+d-2}R\bar{\sigma}_{\ell \vec{m}}(\hat{r}_0)\\
&=-\sum_{\ell}\left[\frac{r^{\ell}}{R^{\ell}}\Theta(r<R)+\frac{R^{\ell+d-2}}{r^{\ell+d-2}}\Theta(r>R)\right]\frac{1}{\nn_{d,\ell}|\mathbb{S}^{d-1}|}\int_{\hat{r}_0\in\mathbb{S}^{d-1}} \frac{\Pi^S_{d,\ell}(\hat{r}|\hat{r}_0)}{2\ell+d-2}R\bar{\sigma}_{\ell \vec{m}}(\hat{r}_0)\ \\
&=-\int_{\vec{r}_0}\sum_{\ell}\left\{\frac{r^{\ell}}{r_0^{\ell+d-2}}\Theta(r<r_0)+\frac{r_0^{\ell}}{r^{\ell+d-2}}\Theta(r>r_0)\right\} \frac{1}{\nn_{d,\ell}|\mathbb{S}^{d-1}|} \frac{\Pi^S_{d,\ell}(\hat{r}|\hat{r}_0)}{2\ell+d-2}\bar{J}^t(\vec{r}_0)
\end{split}
\end{equation}
Here, we have used the SSH addition theorem in the second step. As we did in magnetostatics, we can expand the electrostatic potential as well as the electric fields in terms of spherical harmonics:
\begin{equation}
\begin{split}
\gaugeV_t &=\sum_{\ell\vec{m}} \overline{E}_{s}(r,\ell,\vec{m})\mathscr{Y}_{\ell\vec{m}}(\hat{r})\ ,\\
\FC_{rt}&\equiv \sum_{\ell\vec{m}} \overline{E}_r (r,\ell,\vec{m})\mathscr{Y}_{\ell\vec{m}}(\hat{r}) \ ,\quad 
\FC_{It} \equiv \sum_{\ell\vec{m}} \overline{E}_{s}(r,\ell,\vec{m})\mathscr{D}_{I}\mathscr{Y}_{\ell\vec{m}}(\hat{r})\ ,
\end{split}
\end{equation}
with $\overline{E}_r=\partial_r\overline{E}_{s}$. The potential function then has a Green-function expression
\begin{equation}
\begin{split}
\overline{E}_{s}
&=-\frac{1}{2\ell+d-2}\int_{\vec{r}_0}\left\{\frac{r^{\ell}}{r_0^{\ell+d-2}}\Theta(r<r_0)+\frac{r_0^{\ell}}{r^{\ell+d-2}}\Theta(r>r_0)\right\}  \mathscr{Y}_{\ell\vec{m}}^\ast(\hat{r}_0)\bar{J}^t(\vec{r}_0)\ .
\end{split}
\end{equation}

\subsubsection*{Multipole expansion outside the electric sources}
Unlike the poloidal currents, the static charge distributions do give rise to fields outside them. The electric field outside the charges is then given by
\begin{equation}\label{eq:staticEflat}
\begin{split}
 \overline{E}_s^{\text{Out}} (r,\ell,\vec{m}) =-\frac{\overbar{\multj}^E(\ell,\vec{m})}{r^{\ell+d-2}}\ ,\quad \
\overline{E}_{r}^{\text{Out}}(r,\ell,\vec{m}) =(\ell+d-2)\frac{\overbar{\multj}^E(\ell,\vec{m})}{r^{\ell+d-1}}\ ,
\end{split}
\end{equation}
where $\overbar{\multj}^E$ denotes the spherical electric multipole moments defined via 
\begin{equation}\label{eq:jEstat}
\begin{split}
\overbar{\multj}^E(\ell,\vec{m}) &\equiv\frac{1}{2\ell+d-2}\int_{\vec{r}_0}r_0^{\ell}
\mathscr{Y}_{\ell\vec{m}}^\ast(\hat{r}_0)\bar{J}^t(\vec{r}_0)\ .
\end{split}
\end{equation}
The corresponding cartesian multipole moment can be  defined from the spherical moments via SSH addition theorem, viz.,
\begin{equation}\label{eq:jEtoQE}
\begin{split}
\frac{1}{\ell!}\ {}^E\overbar{\mathcal{Q}}_{<i_1 i_2\ldots i_\ell>}x^{i_1}\ldots x^{i_\ell}&\equiv \nn_{d,\ell-1}|\mathbb{S}^{d-1}|\sum_{\vec{m}}\overbar{\multj}^E(\ell,\vec{m}) r^{\ell}\mathscr{Y}_{\ell\vec{m}}(\hat{r})=\int_{\vec{r}_0}\Pi^S(\vec{r}|\vec{r}_0)\bar{J}^{t}(\vec{r}_0)\ .
\end{split}
\end{equation}
or equivalently, we have
\begin{equation}
\begin{split}
{}^E\overbar{\mathcal{Q}}_{<i_1 i_2\ldots i_\ell>}\equiv (\Pi^S)^{<i_1 i_2\ldots i_\ell>}_{<j_1 j_2\ldots j_\ell>}\int_{\vec{r}_0}x_0^{j_1}\ldots x_0^{j_\ell}\bar{J}^{t}(\vec{r}_0)\ .
\end{split}
\end{equation}
The scalar potential and the cartesian components of the electric field take the form
\begin{equation}\label{eq:EstaicSTF}
\begin{split}
\gaugeV_t^{\text{Out}} &= -\sum_{\ell} \frac{1}{\ell! \nn_{d,\ell-1}|\mathbb{S}^{d-1}|}{}^E\overbar{\mathcal{Q}}_{<i_1 i_2\ldots i_\ell>}\frac{x^{i_1}\ldots x^{i_\ell}}{ r^{2\ell+d-2}}=-\sum_{\ell} \frac{(-)^\ell}{\ell!}
\partial^{i_1}\partial^{i_2}\ldots \partial^{i_\ell}\Bigl\{\frac{{}^E\overbar{\mathcal{Q}}_{<i_1 i_2\ldots i_\ell>}}{(d-2)|\mathbb{S}^{d-1}| r^{d-2}}\Bigr\}\ ,\\
\FC_{jt}^{\text{Out}}&= \sum_{\ell} \frac{1}{\ell! \nn_{d,\ell-1}|\mathbb{S}^{d-1}|}{}^E\overbar{\mathcal{Q}}_{<i_1 i_2\ldots i_\ell>}[(2\ell+d-2)x^j x^{i_\ell}-r^2\ell\delta^{ji_\ell}]\frac{ x^{i_1}\ldots x^{i_{\ell-1}}}{ r^{2\ell+d}}\ .
\end{split}
\end{equation}
All this is completely analogous to our discussion of magnetostatics in STF language. As we did for magnetic STF moments, explicit expressions for the first few electric STF moments can be written down by contracting against a dummy variable:
\begin{equation}
\begin{split}
{}^E\overbar{\mathcal{Q}} & = \int_{\vec{r}}\bar{J}^{t}(\vec{r})\ ,\\
{}^E\overbar{\mathcal{Q}}_{<i_1>}\kappa^{i_1} & = \int_{\vec{r}}(\kappa\cdot r)\ \bar{J}^{t}(\vec{r})\ ,\\
\frac{1}{2!}{}^E\overbar{\mathcal{Q}}_{<i_1 i_2>}\kappa^{i_1}\kappa^{i_2} & = \int_{\vec{r}}\left[\frac{(\kappa\cdot r)^2}{2!}-\frac{\kappa^2r^2}{2d}\right]\bar{J}^{t}(\vec{r})\ , \\
\frac{1}{3!}{}^E\overbar{\mathcal{Q}}_{<i_1 i_2i_3>}\kappa^{i_1}\kappa^{i_2}\kappa^{i_3} & = \int_{\vec{r}}\left[\frac{(\kappa\cdot r)^3}{3!}-\frac{\kappa^2 r^2}{2(d+2)} (\kappa\cdot r)\right]\bar{J}^{t}(\vec{r})\ , \\
\frac{1}{4!}{}^E\overbar{\mathcal{Q}}_{<i_1 i_2 i_3 i_4>}\kappa^{i_1}\kappa^{i_2}\kappa^{i_3} \kappa^{i_4} & = \int_{\vec{r}}\Bigg[ \frac{(\kappa\cdot r)^4}{4!}-\frac{\kappa^2r^2}{2(d+4)}\frac{(\kappa\cdot r)^2}{2!}+\frac{\kappa^4 r^4}{8(d+4)(d+2)}\Bigg]\bar{J}^{t}(\vec{r})\ ,\\
\frac{1}{5!}{}^E\overbar{\mathcal{Q}}_{<i_1 i_2 i_3 i_4 i_5>}\kappa^{i_1}\kappa^{i_2}\kappa^{i_3} \kappa^{i_4}\kappa^{i_5}&=\int_{\vec{r}}\Bigg[\frac{(\kappa\cdot r)^5}{5!}-\frac{\kappa^2 r^2}{2(d+6)}\frac{(\kappa\cdot r)^3}{3!} +\frac{\kappa^4 r^4}{8(d+6)(d+4)}(\kappa\cdot r)\Bigg]\bar{J}^{t}(\vec{r})\ .
\end{split}
\end{equation}
The corresponding STF tensors can be obtained by differentiating with respect to $\kappa_i$. Till $\ell=4$, they are given as
\begin{equation}
\begin{split}
{}^E\overbar{\mathcal{Q}} & = \int_{\vec{r}}\bar{J}^{t}(\vec{r})\ ,\\
{}^E\overbar{\mathcal{Q}}_{<i_1>} & = \int_{\vec{r}}x^{i_1}\bar{J}^{t}(\vec{r})\ ,\\
{}^E\overbar{\mathcal{Q}}_{<i_1 i_2>} & = \int_{\vec{r}}\left[x^{i_1}x^{i_2}-\frac{r^2}{d}\delta^{i_1i_2}\right]\bar{J}^{t}(\vec{r})\ , \\
{}^E\overbar{\mathcal{Q}}_{<i_1 i_2i_3>} & = \int_{\vec{r}}\left[x^{i_1}x^{i_2}x^{i_3}-\frac{r^2}{d+2} \left(x^{i_1}\delta^{i_2i_3}+x^{i_2}\delta^{i_1i_3}+x^{i_3}\delta^{i_1i_2}\right)\right]\bar{J}^{t}(\vec{r})\ , \\
\end{split}
\end{equation}
and
\begin{equation}
\begin{split}
{}^E\overbar{\mathcal{Q}}_{<i_1 i_2 i_3 i_4>} & = \int_{\vec{r}}\Bigg[ x^{i_1}x^{i_2}x^{i_3}x^{i_4}\\
&\qquad-\frac{r^2}{d+4}\Big(x^{i_1}x^{i_2}\delta^{i_3i_4}+x^{i_1}x^{i_3}\delta^{i_2i_4}+x^{i_1}x^{i_4}\delta^{i_2i_3}\\
&\qquad \qquad \qquad \qquad\qquad \qquad + x^{i_2}x^{i_3}\delta^{i_1i_4}+x^{i_2}x^{i_4}\delta^{i_1i_3}+x^{i_3}x^{i_4}\delta^{i_1i_2}\Big)\\
&\qquad +\frac{r^4}{(d+4)(d+2)}\left(\delta^{i_1i_2}\delta^{i_3i_4}+\delta^{i_1i_3}\delta^{i_2i_4}+\delta^{i_1i_4}\delta^{i_2i_3}\right)\Bigg]\bar{J}^{t}(\vec{r})\ .
\end{split}
\end{equation}
% \begin{equation}
% \begin{split}
% {}^E\overbar{\mathcal{Q}}_{<i_1 i_2 i_3 i_4 i_5>}&=\int_{\vec{r}}\Bigg[x^{i_1}x^{i_2}x^{i_3}x^{i_4}x^{i_5}-\frac{x^2}{d+6}\Bigl(x^{i_1}x^{i_2}x^{i_3}\delta^{i_4i_5}+x^{i_1}x^{i_2}x^{i_4}\delta^{i_3i_5}+x^{i_1}x^{i_3}x^{i_4}\delta^{i_2i_5}\Bigr.\\ &\quad \Bigl. +x^{i_2}x^{i_3}x^{i_4}\delta^{i_1i_5}+x^{i_5}x^{i_1}x^{i_2}\delta^{i_3i_4}+x^{i_5}x^{i_1}x^{i_3}\delta^{i_2i_4}+x^{i_5}x^{i_1}x^{i_4}\delta^{i_2i_3}+x^{i_2}x^{i_4}x^{i_5}\delta^{i_1i_3}\Bigr.\\ &\quad \Bigl.+x^{i_5}x^{i_2}x^{i_3}\delta^{i_1i_4}+x^{i_5}x^{i_3}x^{i_4}\delta^{i_1i_2}\Bigr) +\frac{x^4}{(d+6)(d+4)}\Bigl(x^{i_1}\delta^{i_2i_3}\delta^{i_4i_5}+x^{i_1}\delta^{i_2i_4}\delta^{i_3i_5}\Bigr.\\
% &\ \Bigl.+x^{i_1}\delta^{i_2i_5}\delta^{i_3i_4}+ x^{i_2}\delta^{i_1i_3}\delta^{i_4i_5}+x^{i_2}\delta^{i_1i_4}\delta^{i_3i_5}+x^{i_2}\delta^{i_1i_5}\delta^{i_3i_4}\Bigr.\\
% &\
% \Bigl.+x^{i_3}\delta^{i_1i_2}\delta^{i_4i_5}+x^{i_3}\delta^{i_1i_4}\delta^{i_2i_5}+x^{i_3}\delta^{i_1i_5}\delta^{i_2i_4}+x^{i_4}\delta^{i_1i_2}\delta^{i_3i_5}+x^{i_4}\delta^{i_1i_3}\delta^{i_2i_5}\Bigr.\\
% &\ \Bigl. +x^{i_4}\delta^{i_1i_5}\delta^{i_2i_3}+x^{i_5}\delta^{i_1i_2}\delta^{i_3i_4}+x^{i_5}\delta^{i_1i_3}\delta^{i_2i_4}+x^{i_5}\delta^{i_1i_4}\delta^{i_2i_3}\Bigr)\Bigg]\bar{J}^{t}(\vec{r})
% \end{split}
% \end{equation}

\subsection{Electric multipole radiation }\label{appss:radnE}
We will now turn to the problem of radiation from time-varying charge distributions and poloidal currents. Over and above the issues discussed above for the toroidal currents, there are some new subtleties which show up in this case.

First, in the dynamical setting, poloidal currents inevitably accompany changes in charge configurations, and the spacetime charge current $\bar{J}^\mu(\vec{r},t)$ should satisfy the  conservation equation 
\begin{equation}\begin{split}
\partial_t\bar{J}^t+\frac{1}{r^{d-1}}\frac{\partial}{\partial r}(r^{d-1}\bar{J}^r)+\mathscr{D}_I\bar{J}^I=0\ .
\end{split}\end{equation}
Maxwell equations are mathematically consistent only if the sources obey this constraint. This means that to solve for the EM fields, we should characterise the class of currents consistent with charge conservation. Second, we need to deal with Gauss law constraint in this sector, i.e., one of the Maxwell equations serves to constrain the initial data of EM fields. Both these facts are intimately tied to gauge invariance in electromagnetism: since the gauge parameter is a scalar function, it is expandable into scalar spherical harmonics (SSHs), and its effect is visible in the scalar sector.

We will address the problem of conservation by imagining that the charge flow is described well by a time-varying  electric polarisation field, i.e., we take
\begin{align}
\label{eq:JfromP}
\begin{split}
    \bar{J}^t (\vec{r},t) &=-\frac{1}{r^{d-1}}\frac{\partial}{\partial r}[r^{d-1}\bar{P}^r(\vec{r},t)]-\mathscr{D}_I\bar{P}^I(\vec{r},t)  \ ,\\
    \bar{J}^r (\vec{r},t) &= \partial_t \bar{P}^r(\vec{r},t)\ ,\quad
    \bar{J}^I (\vec{r},t) =\partial_t \bar{P}^I(\vec{r},t)\ ,
\end{split}
\end{align}
which automatically satisfies the conservation equation. Such a polarisation field can always be defined by time integrating the  current densities, i.e.,
\begin{align}
\begin{split}
\bar{P}^r(\vec{r},t)\equiv \int dt\ 
\bar{J}^r (\vec{r},t) \ ,\quad
    \bar{P}^I(\vec{r},t) \equiv \int dt\ \bar{J}^I (\vec{r},t)\ .
\end{split}
\end{align}
Such polarisation fields also help simplify Gauss law inside charged matter: it becomes the statement of $\vec{E}+\vec{P}$ being divergence-free. This parametrisation is not without its subtleties, as we shall discuss in \S\S\ref{appss:Elimit}. But, for now, we will take such time-varying polarisation fields are given and proceed.

We will solve the Maxwell equations by passing to the frequency domain and expanding all fields in terms of scalar spherical harmonics (SSHs). The expansion of the polarisation fields is 
\begin{equation}\label{eq:PfJ1J2}
\begin{split}
\bar{P}^r(\vec{r},t) &\equiv \sum_{\ell\vec{m}} \int_\w e^{-i\w t}J_1(r,\w,\ell,\vec{m})\mathscr{Y}_{\ell\vec{m}}(\hat{r})\ ,\\
r^2\gamma_{_{IJ}} \bar{P}^J(\vec{r},t)
&\equiv \sum_{\ell\vec{m}}\int_\w e^{-i\w t}J_2(r,\w,\ell,\vec{m})\mathscr{D}_I\mathscr{Y}_{\ell\vec{m}}(\hat{r})\ .
\end{split}
\end{equation}
Using the same notation/gauge as in statics, we take the scalar/vector potential to be
\begin{equation}\label{eq:VexpE}
\begin{split}
\gaugeV_t(\vec{r},t) &\equiv \sum_{\ell\vec{m}} \int_\w e^{-i\w t}\Esf(r,\w,\ell,\vec{m})\mathscr{Y}_{\ell\vec{m}}(\hat{r})\ ,\\
\gaugeV_r(\vec{r},t)
&\equiv -\sum_{\ell\vec{m}}\int_\w e^{-i\w t}\Hsf(r,\w,\ell,\vec{m})\mathscr{Y}_{\ell\vec{m}}(\hat{r})\ ,\\
 \gaugeV_I(\vec{r},t) &=0\ .
\end{split}
\end{equation}
We aim to solve for $\Esf,\Hsf$ in terms of $J_1$ and $J_2$. To this end, we first compute the EM fields corresponding to the above potential:
\begin{equation}
\begin{split}
     \FC_{rt}(\Vec{r},t)&=\sum_{\ell\vec{m}}\int_\w e^{-i\w t}\Erf(r,\w,\ell,\vec{m})\  \Yellm(\hat{r})\ ,\\
     \FC_{It}(\Vec{r},t)&=\sum_{\ell\vec{m}}\int_\w e^{-i\w t}\Esf(r,\w,\ell,\vec{m})\ \mathscr{D}_I\Yellm(\hat{r})\ ,\\
     \FC_{rI}(\Vec{r},t)&=\sum_{\ell\vec{m}}\int_\w e^{-i\w t}\Hsf(r,\w,\ell,\vec{m}) \ \mathscr{D}_I\Yellm(\hat{r})\ , \\
     \FC_{IJ}(\Vec{r},t)&=0\ ,
\end{split}
\end{equation}
where $\Erf=\partial_r\Esf-i\w \Hsf$ . This relation giving  $\Erf$ in terms of $\Esf$ and $\Hsf$ can also be directly obtained from the Bianchi identity (or the unsourced Maxwell equations). 

Next, we write down the sourced Maxwell equations:
\begin{equation}
\begin{split}
 \frac{1}{r^{d-1}}\partial_r\left(r^{d-1}[\Erf+J_1] \right)-\frac{\ell(\ell+d-2)}{r^2}[\Esf+J_2]  &= 0 \quad\quad \text{(t-Eqn)}\ ,\\
-\frac{\ell(\ell+d-2)}{r^2}\Hsf+  i\w[\Erf+J_1]  &= 0\ \quad\quad \text{(r-Eqn)}\ ,\\
-\frac{1}{r^{d-3}}\partial_r\left(r^{d-3}\Hsf\right)+i\w [\Esf+J_2] &=0 \quad\quad \text{(I-Eqn)}\ .
\end{split}
\end{equation}
We note how the electric field always shows up in the $\vec{E}+\vec{P}$ combination. The above set of coupled ODEs can be solved by introducing the \emph{electric Debye field} $\Phi_E(r,\w,\ell,\vec{m})$ such that 
\begin{equation}
\begin{split}
    \Erf+J_1 &= \frac{\ell(\ell+d-2)}{r^{d-1}}\Phi_E\ ,\quad \Esf+J_2= \frac{1}{r^{d-3}}\partial_r\Phi_E\ ,\quad \Hsf = \frac{i\w}{r^{d-3}}\Phi_E\ .
\end{split}
\end{equation}
The relation $\Erf=\partial_r\Esf-i\w \Hsf$  then becomes the follow inhomogeneous Helmholtz equation for  $\Phi_E$:
\begin{equation}\label{eq:PhiEHelm}
    -\frac{1}{r^{3-d}}\partial_r\left(r^{3-d}\partial_r\Phi_E\right)-\w^2 \Phi_E+\frac{\ell(\ell+d-2)}{r^2}\Phi_E=r^{d-3}\left[J_1-\partial_rJ_2\right]\ .
\end{equation}
Hereon, the procedure here is similar to the one adopted for the magnetic Debye field $\Phi_B$. The above equation is solved by finding an appropriate Green function $\mathbb{G}_E$ such that 
\begin{equation}\label{eq:PhiEformGE}
    \Phi_E(r,\w,\ell,\vec{m}) =\int_0^\infty dr_0 \ \mathbb{G}_E(r,r_0;\w,\ell) \left[J_1(r_0,\w,\ell,\vec{m})-\partial_{r_0} J_2(r_0,\w,\ell,\vec{m})\right]
\end{equation}
The  Green function $\mathbb{G}_E$ then obeys
\begin{equation}\label{eq:GEHelm}
\begin{split}
-\partial_r[r^{3-d}\partial_r\mathbb{G}_E]-\w^2r^{3-d}\mathbb{G}_E+\ell(\ell+d-2)r^{1-d}\mathbb{G}_E=\delta(r-r_0)\ ,
\end{split}
\end{equation}
The magnetic Helmholtz equation in Eq.\eqref{eq:GBHelm} can be mapped to the electric Helmholtz equation through the replacements
\begin{equation}
d\mapsto 6-d\ ,\quad \ell\mapsto \ell+d-3\ ,\quad 
\nu\mapsto \nu
\end{equation}
where $\nu\equiv \ell+\frac{d}{2}-1$. Through such a replacement, the results we derived in the magnetic case can be recycled here. Given this, we will be content in stating just the final results in what follows. 

\subsubsection*{Homogeneous spherical waves}
The homogenous solutions are as follows: the solution regular everywhere is
\begin{equation}
r^{\nu+\frac{d}{2}-1}{}_0F_1\left[1+\nu,-\frac{\w^2r^2}{4}\right] =\Gamma(1+\nu)\left(\frac{\w}{2}\right)^{-\nu}r^{\frac{d}{2}-1}J_{\nu}(\w r)\ ,
\end{equation}
and this solution plays the role of the time-delay smearing function in the electric case. The  solution that is regular everywhere except the origin is given by
\begin{align}
\begin{split}
\frac{1}{r^{\nu-\frac{d}{2}+1}}&\left\{{}_0F_1\left[1-\nu,-\frac{\w^2r^2}{4}\right]- \frac{\pi \cot \nu \pi}{\Gamma(\nu)\Gamma(1+\nu)}\left(\frac{\w r}{2}\right)^{2\nu}\  {}_0F_1\left[1+\nu,-\frac{\w^2r^2}{4}\right] \right\}  \\ &=-\left(\frac{\w}{2}\right)^\nu \frac{\pi\ r^{\frac{d}{2}-1}}{\Gamma(\nu)}Y_{\nu}(\w r)\ .
\end{split}
\end{align}
The outgoing/ingoing solutions are
\begin{align}
\begin{split}
\frac{1}{r^{\nu-\frac{d}{2}+1}}&\left\{{}_0F_1\left[1-\nu,-\frac{\w^2r^2}{4}\right]\pm (1\pm i\cot \nu \pi)\frac{2\pi i}{\Gamma(\nu)^2}\frac{1}{2\nu}\left(\frac{\w r}{2}\right)^{2\nu}\  {}_0F_1\left[1+\nu,-\frac{\w^2r^2}{4}\right]\right\}\mathbb{V}_I^{\alpha\ell\vec{m}}\\ & =\pm i\left(\frac{\w}{2}\right)^\nu \frac{\pi\ r^{\frac{d}{2}-1}}{\Gamma(\nu)}H^{\pm}_{\nu}(\w r)\mathbb{V}_I^{\alpha\ell\vec{m}}
\end{split}
\end{align}
As in the magnetic case, when $d$ is even and $\nu$ is a positive integer, these expressions become indeterminate, and we should take a limit.  The radial part of the outgoing waves is 
\begin{align}
\begin{split}
G_{_E}^{\text{Out}}(r,\w,\ell)&\equiv \frac{1}{r^{\nu-\frac{d}{2}+1}}\\
&\times\left\{{}_0F_1\left[1-\nu,-\frac{\w^2r^2}{4}\right]+ (1+ i\cot \nu \pi)\frac{2\pi i}{\Gamma(\nu)^2}\frac{1}{2\nu}\left(\frac{\w r}{2}\right)^{2\nu}\  {}_0F_1\left[1+\nu,-\frac{\w^2r^2}{4}\right]\right\}\ .
\end{split}
\end{align}

When $\nu$ is a half-integer, the function $G_{_E}^{\text{Out}}$ can be expressed in terms of reverse Bessel polynomials as
\begin{align}
\begin{split}
G_{_E}^{\text{Out}}(r,\w,\ell)= \frac{\theta_{\nu-\frac{1}{2}}(-i\w r)}{\theta_{\nu-\frac{1}{2}}(0)} \frac{e^{i\w r}}{r^{\nu-\frac{d}{2}+1}}\ \quad \text{for}\quad \nu\in \mathbb{Z}+\frac{1}{2}\ ,
\end{split}
\end{align}
whose large $r$ asymptotic are given by 
\begin{align}\label{eq:GEasymp}
\begin{split}
G_{_E}^{\text{Out}}(r,\w,\ell)\to \nn_{d,\ell-1}\frac{(-i\w)^{\nu-\frac{1}{2}}}{(d-2)!!} \frac{e^{i\w r}}{r^{\frac{3-d}{2}}}\ \quad \text{as}\quad r\to\infty\ .
\end{split}
\end{align}

The raising and lowering relations become
% \begin{equation}\label{eq:GEraising}
% \begin{split}
% \partial_rG_{_E}^{\text{Out}}(r,\w,\ell) +\frac{1}{r}\left(-\frac{d}{2}+1-\nu\right)G_{_E}^{\text{Out}}(r,\w,\ell)&=-2\nu\ G_{_E}^{\text{Out}}(r,\w,\ell+1)\ ,\\
% \partial_rG_{_E}^{\text{Out}}(r,\w,\ell) +\frac{1}{r}\left(-\frac{d}{2}+1+\nu\right)G_{_E}^{\text{Out}}(r,\w,\ell)&=\frac{\omega^2}{2(\nu-1)} G_{_E}^{\text{Out}}(r,\w,\ell-1)\  ,
% \end{split}
% \end{equation}
\begin{equation}\label{eq:GEraising}
\begin{split}
-\frac{1}{r}\frac{\partial}{\partial r}\left[\frac{G_{_E}^{\text{Out}}(r,\w,\ell)}{\nn_{d,\ell-1}r^{\ell+d-2}}\right] &=\frac{G_{_E}^{\text{Out}}(r,\w,\ell+1)}{\nn_{d,\ell}r^{\ell+d-1}}\ ,\quad
\frac{1}{r}\frac{\partial}{\partial r}\left[\frac{G_{_E}^{\text{Out}}(r,\w,\ell)}{\nn_{d,\ell-1}\ r^{-\ell}}\right] =\w^2 \frac{G_{_E}^{\text{Out}}(r,\w,\ell-1)}{\nn_{d,\ell-2}\ r^{1-\ell}}\ ,
\end{split}
\end{equation}
and the derivative of  $G_{_E}^{\text{Out}}$ is 
\begin{equation}\label{eq:GEderiv}
\begin{split}
\partial_rG_{_E}^{\text{Out}}(r,\w,\ell) &=-\frac{\ell}{r} G_{_E}^{\text{Out}}(r,\w,\ell)\ +\frac{\omega^2}{2\ell+d-4} G_{_E}^{\text{Out}}(r,\w,\ell-1)\\
&=-\ell\ G_{_E}^{\text{Out}}(r,\w,\ell+1)+\frac{\ell+d-2}{2\ell+d-2}\frac{\omega^2}{2\ell+d-4} G_{_E}^{\text{Out}}(r,\w,\ell-1) \ .
\end{split}
\end{equation}
The cartesian version of raising/lowering relations is
\begin{equation}\label{eq:GEraisingC}
\begin{split}
-\partial_i\left[\frac{G_{_E}^{\text{Out}}(r,\w,\ell)}{\nn_{d,\ell-1}r^{\ell+d-2}}\right] &=x^i\frac{G_{_E}^{\text{Out}}(r,\w,\ell+1)}{\nn_{d,\ell}r^{\ell+d-1}}\ ,\quad
\partial_i\left[\frac{G_{_E}^{\text{Out}}(r,\w,\ell)}{\nn_{d,\ell-1}\ r^{-\ell}}\right] =\w^2 x^i\frac{G_{_@}^{\text{Out}}(r,\w,\ell-1)}{\nn_{d,\ell-2}\ r^{1-\ell}}\ ,
\end{split}
\end{equation}
and the electric analog of Eq.\eqref{eq:dddGB} is given by 
\begin{equation}\label{eq:dddGE}
\begin{split}
(-1)^\ell\partial^{<i_1}\partial^{i_2}\ldots \partial^{i_\ell>}\left\{\frac{G_{_E}^{\text{Out}}(r,\w,\ell=0) }{  (d-2) r^{d-2}}\right\}=\frac{G_{_E}^{\text{Out}}(r,\w,\ell)}{ \nn_{d,\ell-1}r^{\ell+d-2}}x^{<i_1}\ldots x^{i_\ell>}\ .
\end{split}
\end{equation}

\subsubsection*{Green function for electric radiation}
The Green function $\mathbb{G}_E$ can be constructed by stitching together the homogeneous solutions continuously but with an appropriate discontinuity in its derivative. We have
\begin{equation}\label{eq:GE0F1}
\begin{split}
\mathbb{G}_{E}(r,r_0;\w,\ell)&=\frac{1}{2\nu}\frac{r_<^{\nu+\frac{d}{2}-1}}{r_{>}^{\nu-\frac{d}{2}+1}}{}_0F_1\left[1+\nu,-\frac{\w^2r_{<}^2}{4}\right]\\ &\times \left\{{}_0F_1\left[1-\nu,-\frac{\w^2r_{>}^2}{4}\right]+ (1+ i\cot \nu \pi)\frac{2\pi i}{\Gamma(\nu)^2}\frac{1}{2\nu}\left(\frac{\w r_{>}}{2}\right)^{2\nu}\  {}_0F_1\left[1+\nu,-\frac{\w^2r_{>}^2}{4}\right]\right\} \\
&=\frac{1}{2\nu}r_<^{\nu+\frac{d}{2}-1}{}_0F_1\left[1+\nu,-\frac{\w^2r_{<}^2}{4}\right] G_E^{\text{Out}}(r_{>},\w,\ell)\ ,
\end{split}
\end{equation}
where we have defined:
\begin{equation}
\begin{split}
r_>\equiv \text{Max}(r,r_0)\ ,\quad	
r_<\equiv \text{Min}(r,r_0)\ .
\end{split}
\end{equation}
We can compute the 
the corresponding electric/magnetic field components as
\begin{equation}\label{eq:EMfJ1J2}
\begin{split}
\Esf+J_2&= \frac{1}{r^{d-3}}\partial_r\Phi_E\\
&=\frac{1}{r^{d-3}}\int_0^\infty dr_0 \ \partial_r\mathbb{G}_E(r,r_0;\w,\ell) \left[J_1(r_0,\w,\ell,\vec{m})-\partial_{r_0} J_2(r_0,\w,\ell,\vec{m})\right]\ ,\\
\Erf+J_1 &=\frac{\ell(\ell+d-2)}{r^{d-1}}\Phi_E\\
&=\frac{\ell(\ell+d-2)}{r^{d-1}}\int_0^\infty dr_0 \ \mathbb{G}_E(r,r_0;\w,\ell) \left[J_1(r_0,\w,\ell,\vec{m})-\partial_{r_0} J_2(r_0,\w,\ell,\vec{m})\right]\ ,\\
\Hsf &= \frac{i\w}{r^{d-3}}\Phi_E\\
&=\frac{i\w}{r^{d-3}} \int_0^\infty dr_0 \ \mathbb{G}_E(r,r_0;\w,\ell) \left[J_1(r_0,\w,\ell,\vec{m})-\partial_{r_0} J_2(r_0,\w,\ell,\vec{m})\right] \ .
\end{split}
\end{equation}
These time-varying electric and magnetic fields sustain each other and propagate outwards as electric multipole radiation. 

\subsubsection*{Fields outside sources}
We will now turn to fields outside the sources, determined by electric multipole moments. For $\ell\neq 0$, we will define a dynamic version of the electric multipole  moment via
\begin{equation}\label{eq:JEfromJ1J2}
\begin{split}
\multj^E(\w,\ell,\vec{m}) &\equiv\frac{\ell}{2\nu}\int_0^\infty dr_0\  r_0^{\nu+\frac{d}{2}-1}{}_0F_1\left[1+\nu,-\frac{\w^2r_0^2}{4}\right]
\Bigl\{J_1(r_0,\w,\ell,\vec{m})-\partial_{r_0} J_2(r_0,\w,\ell,\vec{m})\Bigr\}\ .
\end{split}
\end{equation}
We will justify this definition and generalise it to the $\ell=0$ case later. While this looks very different from the electric multipole moments defined in the static case, we will later see that it reduces to them in the appropriate limit.

The electric Debye field outside the sources is given by
\begin{equation}
\begin{split}
\Phi_{E}^{\text{Out}} &=\frac{1}{\ell}G_{_E}^{\text{Out}}(r,\w,\ell)\ \multj^E(\w,\ell,\vec{m})=\frac{1}{\ell}\frac{\theta_{\nu-\frac{1}{2}}(-i\w r)}{\theta_{\nu-\frac{1}{2}}(0)} \frac{e^{i\w r}}{r^{\ell}}\multj^E(\w,\ell,\vec{m}) \   .
\end{split}
\end{equation}
The corresponding  field strength components are 
\begin{equation}\label{eq:EoutjE}
\begin{split}
\Erf^{\text{Out}} &=\frac{\ell+d-2}{r^{d-1}}G_{_E}^{\text{Out}}(r,\w,\ell)\ \multj^E(\w,\ell,\vec{m})=(\ell+d-2)\frac{\theta_{\nu-\frac{1}{2}}(-i\w r)}{\theta_{\nu-\frac{1}{2}}(0)} \frac{e^{i\w r}}{r^{\ell+d-1}} \multj^E(\w,\ell,\vec{m}),\\
\Esf^{\text{Out}} &=-\frac{1}{ r^{d-2}} G_{_E}^{\text{Out}}(r,\w,\ell)\ \multj^E(\w,\ell,\vec{m})+\frac{1}{r^{d-3}}\frac{\omega^2}{\ell(2\ell+d-4)} G_{_E}^{\text{Out}}(r,\w,\ell-1)\ \multj^E(\w,\ell,\vec{m})\\
&=-\frac{\theta_{\nu-\frac{1}{2}}(-i\w r)}{\theta_{\nu-\frac{1}{2}}(0)} \frac{e^{i\w r}}{r^{\ell+d-2}} \multj^E(\w,\ell,\vec{m})+\frac{\omega^2}{\ell(2\ell+d-4)} \frac{\theta_{\nu-\frac{3}{2}}(-i\w r)}{\theta_{\nu-\frac{3}{2}}(0)} \frac{e^{i\w r}}{r^{\ell+d-4}} \multj^E(\w,\ell,\vec{m}),\\
\Hsf^{\text{Out}} &=\frac{i\w}{r^{d-3}} \frac{1}{\ell}G_{_E}^{\text{Out}}(r,\w,\ell)\ \multj^E(\w,\ell,\vec{m})=\frac{i\w}{\ell}\frac{\theta_{\nu-\frac{1}{2}}(-i\w r)}{\theta_{\nu-\frac{1}{2}}(0)} \frac{e^{i\w r}}{r^{\ell+d-3}} \multj^E(\w,\ell,\vec{m}) .
\end{split}
\end{equation}
Here we have used Eq.\eqref{eq:GEderiv} for evaluating $\partial_r G_{_E}^{\text{Out}}$. One check of our multipole moment definition is that it reduces to the correct static expressions in $\w \to 0$ limit, i.e., we get
\begin{equation}
\begin{split}
 \overline{E}_s^{\text{Out}} (r,\ell,\vec{m}) =-\frac{\overbar{\multj}^E(\ell,\vec{m})}{r^{\ell+d-2}}\ ,\quad \
\overline{E}_{r}^{\text{Out}}(r,\ell,\vec{m}) =(\ell+d-2)\frac{\overbar{\multj}^E(\ell,\vec{m})}{r^{\ell+d-1}}\ .
\end{split}
\end{equation}
The radiative parts work out to be 
\begin{equation}\label{eq:RadFieldsE}
\begin{split}
\Phi_{E}^{\text{Rad}} &=\frac{1}{\ell}\frac{(-i\w )^{\nu-\frac{1}{2}}}{(2\nu-2)!!} \frac{e^{i\w r}}{r^{\frac{3-d}{2}}}\multj^E(\w,\ell,\vec{m}) \ ,\\
\Erf^{\text{Rad}} & =(\ell+d-2)\frac{(-i\w)^{\nu-\frac{1}{2}}}{(2\nu-2)!!} \frac{e^{i\w r}}{r^{\frac{d+1}{2}}}\ \multj^E(\w,\ell,\vec{m}) ,\\
\Esf^{\text{Rad}} &=-\frac{1}{\ell}\frac{(-i\w)^{\nu+\frac{1}{2}}}{(2\nu-2)!!} \frac{e^{i\w r}}{r^{\frac{d-3}{2}}} \multj^E(\w,\ell,\vec{m}),\\
\Hsf^{\text{Rad}} &=-\frac{1}{\ell}\frac{(-i\w)^{\nu+\frac{1}{2}}}{(2\nu-2)!!} \frac{e^{i\w r}}{r^{\frac{d-3}{2}}} \multj^E(\w,\ell,\vec{m}) ,
\end{split}
\end{equation}
where we have used Eq.\eqref{eq:GEasymp}. Note the faster fall-off of the radial electric field: this is consistent with the expectation that, at large $r$, EM fields can be thought of as transverse plane waves travelling outwards radially.

We now turn to the formulation in terms of cartesian STF tensors. We convert the orthonormal electric moments to STF electric moments by Eq.\eqref{eq:jEtoQE} generalised to arbiitrary frequency, i.e.,
\begin{equation}\label{eq:jEtoQE2}
\begin{split}
\frac{1}{\ell!}\ [{}^E\mathcal{Q}(\w)]_{<i_1 i_2\ldots i_\ell>}\ x^{i_1}\ldots x^{i_\ell}&\equiv \nn_{d,\ell-1}|\mathbb{S}^{d-1}|\sum_{\vec{m}} \multj^E(\w,\ell,\vec{m})\  r^{\ell}\mathscr{Y}_{\ell\vec{m}}(\hat{r})\ .
\end{split}
\end{equation}
The cartesian form of the scalar/vector potential can be obtained by substituting $\{\Esf^{\text{Out}},\Hsf^{\text{Out}}\}$ into Eq.\eqref{eq:VexpE}, and converting everything to cartesian coordinates. We end up with
\begin{equation}
\begin{split}
\gaugeVf_t^{\text{Out}}(\vec{r},\w)
&= -\sum_{\ell} \frac{G_{_E}^{\text{Out}}(r,\w,\ell)}{\ell! \nn_{d,\ell-1}|\mathbb{S}^{d-1}|}{}^E\mathcal{Q}_{<i_1 i_2\ldots i_\ell>}\frac{x^{i_1}\ldots x^{i_\ell}}{ r^{\ell+d-2}}\\
&\qquad +\omega^2\sum_{\ell>0}\frac{1}{\ell}\frac{G_{_E}^{\text{Out}}(r,\w,\ell-1)}{\ell! \nn_{d,\ell-2}|\mathbb{S}^{d-1}|}\ {}^E\mathcal{Q}_{<i_1 i_2\ldots i_\ell>}\frac{x^{i_1}\ldots x^{i_\ell}}{ r^{\ell+d-3}}\ ,\\
\gaugeVf_k^{\text{Out}}(\vec{r},\w) &=- i\w\sum_{\ell>0} \frac{1}{\ell}\frac{G_{_E}^{\text{Out}}(r,\w,\ell)}{\ell! \nn_{d,\ell-1}|\mathbb{S}^{d-1}|}{}^E\mathcal{Q}_{<i_1 i_2\ldots i_\ell>}\frac{x^k x^{i_1}\ldots x^{i_\ell}}{ r^{\ell+d-2}}\ .
\end{split}
\end{equation}
The field strengths are given by 
\footnote{Here we use the notation $\text{Anti}_{jk}[T_{jk}]\equiv T_{jk}-T_{kj}$ for the anti-symmetrisation operator.}
\begin{equation}\label{eq:EMcartGenDE}
\begin{split}
\FCf_{kt}^{\text{Out}}(\vec{r},\w) &=\sum_{\ell} \frac{G_{_E}^{\text{Out}}(r,\w,\ell)}{\ell! \nn_{d,\ell-1}|\mathbb{S}^{d-1}|}{}^E\mathcal{Q}_{<i_1 i_2\ldots i_\ell>}  [(2\ell+d-2)x^k x^{i_\ell}-r^2\ell\delta^{ki_\ell}]\frac{ x^{i_1}\ldots x^{i_{\ell-1}}}{ r^{\ell+d}}\\
&\qquad -\w^2 \sum_{\ell} \frac{ G_{_E}^{\text{Out}}(r,\w,\ell-1)}{\ell! \nn_{d,\ell-2}|\mathbb{S}^{d-1}|}{}^E\mathcal{Q}_{<i_1 i_2\ldots i_\ell>}[x^k x^{i_\ell}-r^2\delta^{ki_\ell}]\frac{x^{i_1}\ldots x^{i_{\ell-1}}}{ r^{\ell+d-1}}\ ,\\
\FCf_{jk}^{\text{Out}}(\vec{r},\w)&= \text{Anti}_{jk}\sum_{\ell}(-i\w) \frac{G_{_E}^{\text{Out}}(r,\w,\ell)}{\ell! \nn_{d,\ell-1}|\mathbb{S}^{d-1}|}{}^E\mathcal{Q}_{<ji_1 i_2\ldots i_{\ell-1}>}\frac{ x^k x^{i_1}\ldots x^{i_{\ell-1}}}{ r^{\ell+d-2}}\ . 
\end{split}
\end{equation}
These expressions can be derived by converting the spherical components to cartesian components or by directly differentiating the potentials (using Eq.\eqref{eq:GEraisingC} when necessary).
Using Eq.\eqref{eq:GEasymp}, the large $r$ asymptotics of the potentials work out to be
\begin{equation}
\begin{split}
\gaugeVf_t^{\text{Rad}}(\vec{r},\w)
&=  -\frac{e^{i\w r}}{(d-2)!!|\mathbb{S}^{d-1}| r^{\frac{d-3}{2}} }\sum_{\ell>0}\frac{1}{\ell}\frac{(-i\w)^{\nu+\frac{1}{2}}}{\ell!}\ {}^E\mathcal{Q}_{<i_1 i_2\ldots i_\ell>} n^{i_1}\ldots n^{i_\ell}\ ,\\
\gaugeVf_k^{\text{Rad}}(\vec{r},\w) &=\frac{e^{i\w r}}{(d-2)!!|\mathbb{S}^{d-1}| r^{\frac{d-3}{2}} }\sum_{\ell>0} \frac{1}{\ell}\frac{(-i\w)^{\nu+\frac{1}{2}}}{\ell! }{}^E\mathcal{Q}_{<i_1 i_2\ldots i_\ell>} n^k n^{i_1}\ldots n^{i_\ell}\ ,
\end{split}
\end{equation}
where we have used the notation $n^i\equiv \frac{x^i}{r}$. The fall-off here is slower than what might have been naively expected, e.g., in $d=3$, the potentials tend to a non-zero angle-dependent constant as $r\to\infty$ instead of becoming zero. The corresponding field strengths, however, have the correct asymptotic fall-off, viz.,
\begin{equation}
\begin{split}
\FCf_{kt}^{\text{Rad}}(\vec{r},\w) &=-\frac{e^{i\w r}}{(d-2)!!|\mathbb{S}^{d-1}| r^{\frac{d-1}{2}} }\sum_{\ell} \frac{(-i\w)^{\nu+\frac{1}{2}}}{\ell! }{}^E\mathcal{Q}_{<i_1 i_2\ldots i_\ell>}[\delta^{ki_\ell}-n^k n^{i_\ell}]n^{i_1}\ldots n^{i_{\ell-1}}\ ,\\
\FCf_{jk}^{\text{Rad}}(\vec{r},\w)&= \frac{e^{i\w r}}{(d-2)!!|\mathbb{S}^{d-1}| r^{\frac{d-1}{2}} }\times\text{Anti}_{jk}\Bigl\{n^k\sum_{\ell}\frac{(-i\w)^{\nu+\frac{1}{2}}}{\ell!}{}^E\mathcal{Q}_{<ji_1 i_2\ldots i_{\ell-1}>}  n^{i_1}\ldots n^{i_{\ell-1}}\Bigr\}\ . 
\end{split}
\end{equation}
This suggests that there should be a (large) gauge transformation that brings the potentials to the naively expected fall-offs. Such a gauge transformation can in fact be presented explicitly. Consider the large gauge transformation that removes all the $G_{_E}^{\text{Out}}(r,\w,\ell-1)$ terms from the scalar potential. The new potentials are then given by
\begin{equation}\label{eq:EMcartGenDEnew}
\begin{split}
\gaugeVf_t^{\text{Out,New}}(\vec{r},\w)
&= -\sum_{\ell} \frac{G_{_E}^{\text{Out}}(r,\w,\ell)}{\ell! \nn_{d,\ell-1}|\mathbb{S}^{d-1}|}{}^E\mathcal{Q}_{<i_1 i_2\ldots i_\ell>}\frac{x^{i_1}\ldots x^{i_\ell}}{ r^{\ell+d-2}}\ ,\\
\gaugeVf_k^{\text{Out,New}}(\vec{r},\w) &=- i\w\sum_{\ell>0} \frac{1}{\ell}\frac{G_{_E}^{\text{Out}}(r,\w,\ell)}{\ell! \nn_{d,\ell-1}|\mathbb{S}^{d-1}|}{}^E\mathcal{Q}_{<i_1 i_2\ldots i_\ell>}\frac{x^k x^{i_1}\ldots x^{i_\ell}}{ r^{\ell+d-2}}\\
&\qquad+\partial_k\Bigl\{-i\w\sum_{\ell>0}\frac{1}{\ell}\frac{G_{_E}^{\text{Out}}(r,\w,\ell-1)}{\ell! \nn_{d,\ell-2}|\mathbb{S}^{d-1}|}\ {}^E\mathcal{Q}_{<i_1 i_2\ldots i_\ell>}\frac{x^{i_1}\ldots x^{i_\ell}}{ r^{\ell+d-3}}\Bigr\}\\
&=- i\w\sum_{\ell>0}\frac{G_{_E}^{\text{Out}}(r,\w,\ell-1)}{\ell! \nn_{d,\ell-2}|\mathbb{S}^{d-1}|}\ {}^E\mathcal{Q}_{<ki_1 i_2\ldots i_{\ell-1}>}\frac{x^{i_1}\ldots x^{i_{\ell-1}}}{ r^{\ell+d-3}}
\ .
\end{split}
\end{equation}
In the last step, we have evaluated the cartesian derivative using Eq.\eqref{eq:GEraisingC}.
We can work out the large $r$ behaviour for these new potentials using Eq.\eqref{eq:GEasymp}:
\begin{equation}\label{eq:RadEflat}
\begin{split}
\gaugeVf_t^{\text{Rad,New}}(\vec{r},\w)
&=  -\frac{e^{i\w r}}{(d-2)!!|\mathbb{S}^{d-1}| r^{\frac{d-1}{2}} }\sum_{\ell}\frac{(-i\w)^{\nu-\frac{1}{2}}}{\ell!}\ {}^E\mathcal{Q}_{<i_1 i_2\ldots i_\ell>} n^{i_1}\ldots n^{i_\ell}\ ,\\
\gaugeVf_k^{\text{Rad,New}}(\vec{r},\w) &=\frac{e^{i\w r}}{(d-2)!!|\mathbb{S}^{d-1}| r^{\frac{d-1}{2}} }\sum_{\ell>0} \frac{(-i\w)^{\nu-\frac{1}{2}}}{\ell! }\ {}^E\mathcal{Q}_{<ki_1 i_2\ldots i_{\ell-1}>}  n^{i_1}\ldots n^{i_{\ell-1}}\ .
\end{split}
\end{equation}
These agree with what is expected. Further, potentials in this new gauge also have a nice repeated STF derivative representation (see Eq.\eqref{eq:dddGE}):
\begin{equation}
\begin{split}
\gaugeVf_t^{\text{Out,New}}(\vec{r},\w)
&=-\sum_{\ell} \frac{(-)^\ell}{\ell!}
\partial^{i_1}\partial^{i_2}\ldots \partial^{i_\ell}\Bigl\{\frac{{}^E\mathcal{Q}_{<i_1 i_2\ldots i_\ell>}G_{_E}^{\text{Out}}(r,\w,\ell=0)}{(d-2)|\mathbb{S}^{d-1}| r^{d-2}}\Bigr\}\ ,\\
\gaugeVf_k^{\text{Out,New}}(\vec{r},\w) 
&=-i\w\sum_{\ell>0} \frac{(-)^{\ell-1}}{\ell!}
\partial^{i_1}\partial^{i_2}\ldots \partial^{i_{\ell-1}}\Bigl\{\frac{{}^E\mathcal{Q}_{<ki_1 i_2\ldots i_{\ell-1}>}G_{_E}^{\text{Out}}(r,\w,\ell=0)}{(d-2)|\mathbb{S}^{d-1}| r^{d-2}}\Bigr\}
\ .
\end{split}
\end{equation}

The gauge transformation which gives the correct fall-off can also be performed on the full solution in the spherical coordinates. The new potentials are given by 
\begin{equation}
\begin{split}
\gaugeV^{\text{New}}_t(\vec{r},t) &\equiv \sum_{\ell\vec{m}} \int_\w e^{-i\w t}[\Esf(r,\w,\ell,\vec{m})-i\w\Lambda_s(r,\w,\ell,\vec{m})]\mathscr{Y}_{\ell\vec{m}}(\hat{r})\ ,\\
\gaugeV^{\text{New}}_r(\vec{r},t)
&\equiv -\sum_{\ell\vec{m}}\int_\w e^{-i\w t}[\Hsf(r,\w,\ell,\vec{m})-\partial_r\Lambda_s(r,\w,\ell,\vec{m})]\mathscr{Y}_{\ell\vec{m}}(\hat{r})\ ,\\
 \gaugeV^{\text{New}}_I(\vec{r},t) &\equiv \sum_{\ell\vec{m}}\int_\w e^{-i\w t}\ \Lambda_s(r,\w,\ell,\vec{m}) \mathscr{D}_I\mathscr{Y}_{\ell\vec{m}}(\hat{r})\ ,
\end{split}
\end{equation}
where the gauge transformation function is 
\begin{equation}
\begin{split}
\Lambda_s(r,\w,\ell,\vec{m})&=
\frac{1}{i\w r^{\ell+d-3}}\frac{\partial}{\partial r}\Bigl\{ r^\ell\Phi_{E}(r,\w,\ell,\vec{m})\Bigr\}\ .
\end{split}
\end{equation}
We can simplify the new potential using Eq.\eqref{eq:PhiEHelm} to get
\begin{equation}\label{eq:VexpEnew}
\begin{split}
\gaugeV^{\text{New}}_t(\vec{r},t) &\equiv -\sum_{\ell\vec{m}} \int_\w e^{-i\w t}\left[\frac{\ell}{r^{d-2}}\Phi_{E}(r,\w,\ell,\vec{m})+J_1(r,\w,\ell,\vec{m})\right]\mathscr{Y}_{\ell\vec{m}}(\hat{r})\ ,\\
\gaugeV^{\text{New}}_r(\vec{r},t)
&\equiv \sum_{\ell\vec{m}}\int_\w \frac{e^{-i\w t}}{i\w} \left[\frac{\ell}{r^{\ell+d-2}}\frac{\partial}{\partial r}\Bigl\{ r^\ell\Phi_{E}(r,\w,\ell,\vec{m})\Bigr\}-J_1(r,\w,\ell,\vec{m})+\partial_rJ_2(r,\w,\ell,\vec{m})\right]\mathscr{Y}_{\ell\vec{m}}(\hat{r})\ ,\\
\gaugeV^{\text{New}}_I(\vec{r},t) &\equiv \sum_{\ell\vec{m}}\int_\w \frac{e^{-i\w t}}{i\w}\ \frac{1}{r^{\ell+d-3}}\frac{\partial}{\partial r}\Bigl\{ r^\ell\Phi_{E}(r,\w,\ell,\vec{m})\Bigr\} \mathscr{D}_I\mathscr{Y}_{\ell\vec{m}}(\hat{r})\ .
\end{split}
\end{equation}

\subsection{\texorpdfstring{$\ell=0$}{l=0} and electrostatic limit}\label{appss:Elimit}
In our discussion of electric multipole radiation till now, we have avoided discussing the static limit, i.e., the limit as $\w\to 0$. We should expect the static limit to recover our results on static poloidal currents/charges (see \S\S\ref{appss:staticsE}). Unfortunately, this is not easy to see right away from the expressions we have seen till now. The key complication is the charge conservation that relates \emph{time-derivative}  of charge density to the currents. In practice, this means that various factors of $\w$ might be introduced or removed using charge conservation. Some care is thus required in \emph{how} we take the $\w\to 0$ limit.

Another related complication is the applicability of our expressions to the $\ell=0$ electric multipole. Many formulae in our discussion of electric multipole radiation have $\frac{1}{\ell}$ factors, and just setting $\ell=0$ does not work. The issue here is again charge conservation: the $\ell=0$ electric multipole is just the total electric charge, and it cannot have any time variation. In the frequency domain, this means that $\ell=0$ mode always comes with a delta function $\delta(\w)$, which should then be dealt with some care. The aim of this subsection is twofold: first, we will rewrite the expressions for $\{\Erf,\Esf,\Hsf\}$ in the last subsection in terms of (Fourier transforms of) charge and current densities. Next goal is to describe the $\ell=0$ and $\w\to 0 $ limits of such expressions.

In the last subsection, we parameterised the charge/current densities as  (See Eq.\eqref{eq:JfromP} and Eq.\eqref{eq:PfJ1J2})
\begin{align}
\begin{split}
    \bar{J}^t (\vec{r},t) &=-\frac{1}{r^{d-1}}\frac{\partial}{\partial r}[r^{d-1}\bar{P}^r(\vec{r},t)]-\mathscr{D}_I\bar{P}^I(\vec{r},t)  \ ,\\
    \bar{J}^r (\vec{r},t) &= \partial_t \bar{P}^r(\vec{r},t)\ ,\quad
    \bar{J}^I (\vec{r},t) =\partial_t \bar{P}^I(\vec{r},t)\ ,
\end{split}
\end{align}
with the electric polarisation fields expanded as
\begin{equation}
\begin{split}
\bar{P}^r(\vec{r},t) &\equiv \sum_{\ell\vec{m}} \int_\w e^{-i\w t}J_1(r,\w,\ell,\vec{m})\mathscr{Y}_{\ell\vec{m}}(\hat{r})\ ,\\
r^2\gamma_{_{IJ}} \bar{P}^J(\vec{r},t)
&\equiv \sum_{\ell\vec{m}}\int_\w e^{-i\w t}J_2(r,\w,\ell,\vec{m})\mathscr{D}_I\mathscr{Y}_{\ell\vec{m}}(\hat{r})\ .
\end{split}
\end{equation}
These equations are equivalent to parameterising the  \emph{Fourier transforms} of charge and current densities as
\begin{equation}
\begin{split}
J^t(\vec{r},\w) &=\sum_{\ell\vec{m}} \Bigl\{-\frac{1}{r^{d-1}}\frac{\partial}{\partial r}[r^{d-1}J_1(r,\w,\ell,\vec{m})]+\frac{\ell(\ell+d-2)}{r^2}J_2(r,\w,\ell,\vec{m})\Bigr\} \mathscr{Y}_{\ell\vec{m}}(\hat{r}) \ ,\\
J^r(\vec{r},\w) &= -i\w\sum_{\ell\vec{m}} J_1(r,\w,\ell,\vec{m})\mathscr{Y}_{\ell\vec{m}}(\hat{r})\ ,\\
r^2\gamma_{_{IK}}J^K(\vec{r},\w)
&= -i\w\sum_{\ell\vec{m}}J_2(r,\w,\ell,\vec{m})\mathscr{D}_I\mathscr{Y}_{\ell\vec{m}}(\hat{r})\ .
\end{split}
\end{equation}
For what follows, it is crucial to note that the SSH sum in the last equation has no $\ell=0$ contribution. Another observation is that the $\ell=0$ component of $J_2(r,\w,\ell,\vec{m})$ never enters these expressions. Without loss of generality, we can thus set $J_2(r,\w,\ell,\vec{m})|_{\ell=0}=0$ and assume that $J^I(\vec{r},\w)$ has no spherically symmetric component.

We will now turn to writing the EM fields in terms of charge/current densities instead of $\{J_1,J_2\}$. To this end, we use the above relations to derive  the following identity
\begin{equation}
\begin{split}
&i\w r^2 J^r(\vec{r},\w) -\partial_r[r^2J^t(\vec{r},\w)]\\
&=\sum_{\ell\vec{m}} r^{3-d}\mathscr{Y}_{\ell\vec{m}}(\hat{r}) \Bigl\{\frac{1}{r^{3-d}}\frac{\partial}{\partial r}\left(r^{3-d}\frac{\partial}{\partial r}\right)+\w^2 -\frac{\ell(\ell+d-2)}{r^2}\Bigr\}[r^{d-1}J_1(r,\w,\ell,\vec{m})]  \ ,\\
&\quad +\sum_{\ell\vec{m}} \ell(\ell+d-2)[J_1(r,\w,\ell,\vec{m})-\partial_r J_2(r,\w,\ell,\vec{m})] \mathscr{Y}_{\ell\vec{m}}(\hat{r}) \ .
\end{split}
\end{equation}
The combination here is chosen such that the differential operator in the first line of RHS
is the one defining the electric Green function $\mathbb{G}_{E}$ (See Eq.\eqref{eq:GEHelm}). In the last line of RHS, we recognise the $\{J_1,J_2\}$ source for the electric Debye field $\Phi_{E}$ in Eq.\eqref{eq:PhiEformGE} . Using these facts, we can then write 
\begin{equation}
\begin{split}
&\frac{1}{r^{d-1}}\int_0^\infty dr_0\int_{\hat{r}_0\in\mathbb{S}^{d-1}} \mathscr{Y}^\ast_{\ell\vec{m}}(\hat{r}_0) \ 
\mathbb{G}_E(r,r_0;\w,\ell) \left\{
i\w r_0^2 J^r(\vec{r}_0,\w) -\partial_{r_0}[r_0^2J^t(\vec{r}_0,\w)]\right\}\\
&=-J_1(r,\w,\ell,\vec{m})  +\frac{\ell(\ell+d-2)}{r^{d-1}}\Phi_{E}(r,\w,\ell,\vec{m})\ .
\end{split}
\end{equation}
We recognise here the combination that defines $E_r$ (c.f. Eq.\eqref{eq:EMfJ1J2}), i.e.,
\begin{equation}\label{eq:ErJ}
\begin{split}
\Erf(r,\w,\ell,\vec{m})&=\frac{1}{r^{d-1}}\int_0^\infty dr_0\int_{\hat{r}_0\in\mathbb{S}^{d-1}} \mathscr{Y}^\ast_{\ell\vec{m}}(\hat{r}_0) \ 
\mathbb{G}_E(r,r_0;\w,\ell) \Bigl\{
i\w r_0^2 J^r(\vec{r}_0,\w) -\partial_{r_0}[r_0^2J^t(\vec{r}_0,\w)]\Bigr\}\ .
\end{split}
\end{equation}
One corollary is a formula for electric multipole moment directly in terms of charges/currents: we use Eq.\eqref{eq:GE0F1} to evaluate $\mathbb{G}_E$ outside the sources and compare the result against Eq.\eqref{eq:EoutjE}. This yields
\begin{equation}\label{eq:jEGen}
\begin{split}
\multj^E(\w,\ell,\vec{m}) &\equiv\frac{1}{2\nu(\ell+d-2)}\int_0^\infty dr_0\int_{\hat{r}_0\in\mathbb{S}^{d-1}} \  r_0^{\nu+\frac{d}{2}-1}\mathscr{Y}^\ast_{\ell\vec{m}}(\hat{r}_0)\ {}_0F_1\left[1+\nu,-\frac{\w^2r_0^2}{4}\right]\\
&\qquad\times
\Bigl\{i\w r_0^2 J^r(\vec{r}_0,\w) -\partial_{r_0}[r_0^2J^t(\vec{r}_0,\w)]\Bigr\}\ .
\end{split}
\end{equation}
This generalises our earlier definition in Eq.\eqref{eq:JEfromJ1J2} to general source distributions. As a bonus, we now have an expression where both $\ell=0$ and $\w\to 0$ limits can be taken and be seen to give a non-zero electric moment, as expected. In fact, the static limit coincides with the electrostatics definition in \S\S\ref{appss:staticsE} as can be seen from 
\begin{equation}
\begin{split}
\multj^E(\w=0,\ell,\vec{m}) &=-\frac{1}{2\nu(\ell+d-2)}\int_0^\infty dr_0\int_{\hat{r}_0\in\mathbb{S}^{d-1}} \  r_0^{\ell+d-2}\mathscr{Y}^\ast_{\ell\vec{m}}(\hat{r}_0)\ \partial_{r_0}[r_0^2J^t(\vec{r}_0,\w)]\\
&=\frac{1}{2\nu}\int_0^\infty dr_0\int_{\hat{r}_0\in\mathbb{S}^{d-1}} \  r_0^{\ell+d-1}\mathscr{Y}^\ast_{\ell\vec{m}}(\hat{r}_0)\ J^t(\vec{r}_0,\w=0)\ .
\end{split}
\end{equation}
The last line follows via integration by parts. The above expression for electric multipole moment can also be converted into an STF moment via
\begin{equation}
\begin{split}
\frac{1}{\ell!}&\ {}^E\mathcal{Q}(\w)_{<i_1 i_2\ldots i_\ell>}x^{i_1}\ldots x^{i_\ell}\\
&\equiv \nn_{d,\ell-1}|\mathbb{S}^{d-1}|\sum_{\vec{m}} \multj^E(\w,\ell,\vec{m})\  r^{\ell}\mathscr{Y}_{\ell\vec{m}}(\hat{r})\\
&=\frac{1}{\ell+d-2}\int_{\vec{r}_0} \  \Pi^S(\vec{r}|\vec{r}_0)\ {}_0F_1\left[1+\nu,-\frac{\w^2r_0^2}{4}\right]
\Bigl\{i\w r_0 J^r(\vec{r}_0,\w) -\frac{1}{r_0}\partial_{r_0}[r_0^2J^t(\vec{r}_0,\w)]\Bigr\}\ .
\end{split}
\end{equation}
Stripping off the $x^i$'s on both sides, we get the STF electric multipole tensor as
\begin{equation}\label{eq:QEgen}
\begin{split}
{}^E\mathcal{Q}(\w)_{<i_1 i_2\ldots i_\ell>}
&=\frac{(\Pi^S)^{<i_1\ldots i_\ell>}_{<j_1\ldots j_\ell>}}{\ell+d-2}\int_{\vec{r}} \ x^{j_1}\ldots x^{j_\ell} \ {}_0F_1\left[1+\nu,-\frac{\w^2r^2}{4}\right]
\Bigl\{i\w r J^r(\vec{r},\w) -\frac{1}{r}\partial_{r}[r^2J^t(\vec{r},\w)]\Bigr\}\ .
\end{split}
\end{equation}
We now turn to how the magnetostatics of the poloidal currents is recovered from our expressions. We will do this by relating the  current combination that sources the poloidal magnetic field (Eq.\eqref{eq:HsbStatics}) to the source of the electric Debye field, viz.,
\begin{equation}
\begin{split}
\int_{\hat{r}\in\mathbb{S}^{d-1}}\mathscr{Y}_{\ell\vec{m}}^\ast(\hat{r})&\Bigl\{J^r(\vec{r},\w)+\frac{1}{\ell(\ell+d-2)}\partial_{r}\left[r^2\mathscr{D}_IJ^I(\vec{r},\w)\right]\Bigr\}\\
&=-i\w[J_1(r,\w,\ell,\vec{m})-\partial_r J_2(r,\w,\ell,\vec{m})]\ .
\end{split}
\end{equation}
This allows us to rewrite $\Hsf$ in Eq.\eqref{eq:EMfJ1J2} in the form
\begin{equation}\label{eq:HsJ}
\begin{split}
\Hsf 
&=-\frac{1}{r^{d-3}} \int_0^\infty dr_0 \int_{\hat{r}_0\in\mathbb{S}^{d-1}}\mathscr{Y}_{\ell\vec{m}}^\ast(\hat{r}_0)\ \mathbb{G}_E(r,r_0;\w,\ell) \\
&\qquad\times\Bigl\{J^r(\vec{r}_0,\w)+\frac{1}{\ell(\ell+d-2)}\partial_{r_0}\left[r_0^2\mathscr{D}_IJ^I(\vec{r}_0,\w)\right]\Bigr\} \ .
\end{split}
\end{equation}
This form then has a straightforward static limit where it reduces to Eq.\eqref{eq:HsbStatics}. If we account for the fact that $J^I$ has no $\ell=0$ component, the above expression also has a finite $\ell=0$ limit. The outside fields in Eq.\eqref{eq:EoutjE} also work out provided
\begin{equation}\label{eq:jEDot}
\begin{split}
0&=i\w\ \multj^E(\w,\ell,\vec{m}) + \frac{\ell}{2\nu}\int_0^\infty dr_0 \int_{\hat{r}_0\in\mathbb{S}^{d-1}}\mathscr{Y}_{\ell\vec{m}}^\ast(\hat{r}_0)\ r_0^{\nu+\frac{d}{2}-1}{}_0F_1\left[1+\nu,-\frac{\w^2r_{0}^2}{4}\right] \\
&\qquad\times\Bigl\{J^r(\vec{r}_0,\w)+\frac{1}{\ell(\ell+d-2)}\partial_{r_0}\left[r_0^2\mathscr{D}_IJ^I(\vec{r}_0,\w)\right]\Bigr\} \ .
\end{split}
\end{equation}
This seems to give a new expression for the electric multipole moment that is different from Eq.\eqref{eq:jEGen}. But,  using the   conservation equation (in the frequency domain)
\begin{equation}\begin{split}
i\w J^t=\frac{1}{r^{d-1}}\frac{\partial}{\partial r}(r^{d-1}J^r)+\mathscr{D}_IJ^I\ ,
\end{split}\end{equation}
the difference between the two $\multj^E$ definitions can be shown to be proportional to  \begin{equation}
\begin{split}
&\int_0^\infty dr_0 \int_{\hat{r}_0\in\mathbb{S}^{d-1}}\mathscr{Y}_{\ell\vec{m}}^\ast(\hat{r}_0)\ r_0^{\nu+\frac{d}{2}-1}{}_0F_1\left[1+\nu,-\frac{\w^2r_{0}^2}{4}\right] \\
&\qquad\times r_0^{3-d}\Bigl\{-\frac{1}{r_0^{3-d}}\frac{\partial}{\partial r_0}\left(r_0^{3-d}\frac{\partial}{\partial r_0}\right)-\w^2 +\frac{\ell(\ell+d-2)}{r_0^2}\Bigr\}[r_0^{d-1}J^r(\vec{r}_0,\w)] \ .
\end{split}
\end{equation}
This expression is zero since the derivative operator in the second line can be shifted onto its homogeneous solution in the first line via integration by parts. Finally, we can get an expression for $\Esf$ in terms of charge and current densities by using 
\begin{equation}
\begin{split}
\partial_r\Esf &= \Erf+ i\w \Hsf\\
&=\int_0^\infty dr_0\int_{\hat{r}_0\in\mathbb{S}^{d-1}} \mathscr{Y}^\ast_{\ell\vec{m}}(\hat{r}_0) \ 
\mathbb{G}_E(r,r_0;\w,\ell) \\
&\qquad\times\Bigl\{
i\w \frac{r_0^2-r^2}{r^{d-1}} J^r(\vec{r}_0,\w) -\partial_{r_0}\left[\frac{r_0^2}{r^{d-1}}J^t(\vec{r}_0,\w)+i\w \frac{r_0^2}{r^{d-3}} \frac{\mathscr{D}_IJ^I(\vec{r}_0,\w)}{\ell(\ell+d-2)}\right]\Bigr\}\ .
\end{split}
\end{equation}
The last line follows from   Eqs.\eqref{eq:ErJ} and \eqref{eq:HsJ}. As $r\to \infty$ the field  $\Esf\to \Hsf$ (see Eq.\eqref{eq:RadFieldsE}), and we can integrate the above equation to obtain
\begin{equation}
\begin{split}
\Esf 
&=\Hsf(r\to\infty)+\int_\infty^r dr_1\int_0^\infty dr_0\int_{\hat{r}_0\in\mathbb{S}^{d-1}} \mathscr{Y}^\ast_{\ell\vec{m}}(\hat{r}_0) \ 
\mathbb{G}_E(r_1,r_0;\w,\ell) \\
&\qquad\times\Bigl\{
i\w \frac{r_0^2-r_1^2}{r_1^{d-1}} J^r(\vec{r}_0,\w) -\partial_{r_0}\left[\frac{r_0^2}{r_1^{d-1}}J^t(\vec{r}_0,\w)+i\w \frac{r_0^2}{r_1^{d-3}} \frac{\mathscr{D}_IJ^I(\vec{r}_0,\w)}{\ell(\ell+d-2)}\right]\Bigr\} .
\end{split}
\end{equation}
In this form, we can easily take $\w\to 0$ as well as $\ell=0$ limits (provided we remember that $J^I$ has no $\ell=0$ component). Note that, for $d>3$, we have $\Hsf(r\to\infty)=0$, and we can drop the first term entirely.

\section{Multipole expansion in flat space II}
\label{app:FlatEMII}

\subsection{Radiative power loss (Larmor's formula)}
Before moving to the influence phase, it is instructive to generalise the textbook description of radiative power loss in EM  to arbitrary dimensions. The power carried away by the  radiation can be computed using  the EM energy-momentum tensor:
\begin{equation}
    {}^\text{(EM)}\overbar{T}^{\mu\nu}=\FC^{\mu\alpha}\FC_{\alpha}{}^\nu-\frac{1}{4}\eta^{\mu\nu}\FC_{\alpha\beta}\FC^{\alpha\beta}\ .
\end{equation}
The energy flux through a sphere at radial coordinate $r$ enclosing the origin is then the sphere integral of $T_t^r$ (i.e., the radial component of the Poynting vector in the frequency domain). We are interested in the energy flux at large $r$, which can be computed using radiative fields given in Eqs.\eqref{eq:RadFieldsE} and \eqref{eq:RadFieldsB}. We obtain 
\begin{equation}\label{eq:LarmorFlat}\begin{split}
     \mathcal{P}(\w)&\equiv\lim_{r\to\infty}\int_{\mathbb{S}^{d-1}_r} r^{d-1}\  
 {}^\text{(EM)}T_t{}^r(r,\w,\Omega) \\
     &=  \lim_{r\to\infty} r^{d-3}\left[\sum_{\ell\vec{m}}\ell(\ell+d-2)\Esf^\ast\Hsf+\sum_{\alpha \ell\vec{m}}\Evf^*\Hvf\right]\\
    &=  \sum_{\ell\vec{m}}\frac{\ell+d-2}{\ell}\frac{\w^{2\nu+1}}{[(2\nu-2)!!]^2}|\multj^E(\w,\ell,\vec{m})|^2+\sum_{\alpha \ell\vec{m}}\frac{\w^{2\nu+1}}{[(2\nu-2)!!]^2}|\multj^B(\w,\alpha,\ell,\vec{m})|^2  \ ,
\end{split}\end{equation}
where we have used the orthonormality of scalar/vector spherical harmonics to perform the sphere integrals. The sum over $\ell$ ranges from $\ell=1$ to $\ell=\infty$, since the monopole moment at $\ell=0$ (the total electric charge), is always time-independent and does not result in radiation. 

When the number of spatial dimensions $d$ is odd, the number $\nu\equiv\ell+\frac{d}{2}-1$ is a half-integer, and the power loss $\mathcal{P}(\w)$ is an even function of $\w$. This means that the power loss is always non-negative and is invariant under time reversal, i.e., time-reversing the charges/currents still results in an irreversible loss of energy into radiation.

The situation is qualitatively different when $d$ is even. We remind the reader that the radiative fields of Eqs.\eqref{eq:RadFieldsE} and \eqref{eq:RadFieldsB} as well as the power loss formula Eq.\eqref{eq:LarmorFlat} are still valid with an appropriate definition of double factorials. The main difference now is that the power loss above can be reversed by time-reversing the charges/currents. \emph{Such a reversible change in energy can then be absorbed into a redefinition of energy.} Physically, when $d$ is even, the radiation lingers on around the source, and its back-reaction serves to \emph{renormalise} the source properties without any dissipative effects.  We will see below that Eq.\eqref{eq:LarmorFlat} should really be interpreted as a beta function in classical EM.

Before proceeding, we would like to recast Eq.\eqref{eq:LarmorFlat}  in terms of cartesian multipole tensors. Using Eq.\eqref{eq:jEGen}, we have
\begin{equation}\begin{split}
&\sum_{\vec{m}}|\multj^E(\w,\ell,\vec{m})|^2= \frac{1}{[2\nu(\ell+d-2)]^2}\\
&\times \int_0^\infty dr_1\int_{\hat{r}_1\in\mathbb{S}^{d-1}} \  r_1^{\nu+\frac{d}{2}-1}{}_0F_1\left[1+\nu,-\frac{\w^2r_1^2}{4}\right]\int_0^\infty dr_2\int_{\hat{r}_2\in\mathbb{S}^{d-1}} \  r_2^{\nu+\frac{d}{2}-1}{}_0F_1\left[1+\nu,-\frac{\w^2r_2^2}{4}\right]\\
&\qquad\times
\Bigl\{i\w r_1^2 J^r(\vec{r}_1,\w) -\partial_{r_1}[r_1^2J^t(\vec{r}_1,\w)]\Bigr\}^\ast
\Bigl\{i\w r_2^2 J^r(\vec{r}_2,\w) -\partial_{r_2}[r_2^2J^t(\vec{r}_2,\w)]\Bigr\}\\
&\qquad\qquad\times\sum_{\vec{m}} \mathscr{Y}_{\ell\vec{m}}(\hat{r}_1)\ \mathscr{Y}^\ast_{\ell\vec{m}}(\hat{r}_2)\  \ .
\end{split}\end{equation}
The sum appearing in the last line can be performed by invoking the SSH addition theorem (Eq.\eqref{eq:SSHaddnThm}), and the answer factorised via symmetry/idempotence of the SSH projector:
\begin{equation}\begin{split}
\sum_{\vec{m}} \mathscr{Y}_{\ell\vec{m}}(\hat{r}_1)\ \mathscr{Y}^\ast_{\ell\vec{m}}(\hat{r}_2)&=\frac{1}{\nn_{d,\ell}|\mathbb{S}^{d-1}|}\ \Pi^S(\hat{r}_1|\hat{r}_2)_{d,\ell}\\
&=\frac{1}{\ell !\nn_{d,\ell}|\mathbb{S}^{d-1}|}\  (\Pi^S)^{<i_1\ldots i_\ell>}_{<j_1\ldots j_\ell>}\hat{r}_1^{j_1}\ldots \hat{r}_1^{j_\ell}\ (\Pi^S)^{<i_1\ldots i_\ell>}_{<k_1\ldots k_\ell>}\hat{r}_2^{k_1}\ldots \hat{r}_2^{k_\ell}\ .
\end{split}\end{equation}
In the next step, we use the definition of STF electric moment in Eq.\eqref{eq:QEgen} to write
\begin{equation}\begin{split}
&\sum_{\vec{m}}|\multj^E(\w,\ell,\vec{m})|^2= \frac{1}{(2\nu)^2\nn_{d,\ell}|\mathbb{S}^{d-1}|} \frac{1}{\ell!}\ 
\left[{}^E\mathcal{Q}(\w)_{<i_1\ldots i_\ell>}\right]^\ast\ \left[{}^E\mathcal{Q}(\w)_{<i_1\ldots i_\ell>}\right]\ .
\end{split}\end{equation}

A similar derivation can be given for the magnetic moment. From Eq.\eqref{eq:jBflat}, we have
\begin{equation}\begin{split}
\sum_{\alpha\vec{m}}|\multj^B(\w,\alpha,\ell,\vec{m})|^2&= \frac{1}{(2\nu)^2}\int_{\vec{r}_1} \  r_1^{\nu-\frac{d}{2}+2}{}_0F_1\left[1+\nu,-\frac{\w^2r_1^2}{4}\right]\int_{\vec{r}_2} \  r_2^{\nu-\frac{d}{2}+2}{}_0F_1\left[1+\nu,-\frac{\w^2r_2^2}{4}\right]\\
&\qquad\times
\Bigl\{J^I(\vec{r}_1,\w)\Bigr\}^\ast \Bigl\{J^J(\vec{r}_1,\w)\Bigr\}
 \sum_{\alpha\vec{m}} \mathbb{V}_I^{\alpha\ell\vec{m}}(\hat{r}_1)\ \mathbb{V}_J^{\alpha\ell\vec{m}\ast}(\hat{r}_2)\  \ .
\end{split}\end{equation}
We can relate the spherical components of the currents to 
cartesian ones by writing $J^I(\vec{r},\w)=J^i(\vec{r},\w)\frac{\partial \vartheta^I}{\partial x^i}$.
Applying the VSH addition theorem  as well as the symmetry/idempotence of the VSH projector, we get
\begin{equation}\begin{split}
\frac{\partial \vartheta_1^I}{\partial x_1^i}\frac{\partial \vartheta_2^J}{\partial x_2^j}&\sum_{\alpha\vec{m}} \mathbb{V}_I^{\alpha\ell\vec{m}}(\hat{r}_1)\ \mathbb{V}_J^{\alpha\ell\vec{m}\ast}(\hat{r}_2)\\
&=\frac{1}{\nn_{d,\ell}|\mathbb{S}^{d-1}|r_1 r_2}\ \Pi^V_{ij}(\hat{r}_1|\hat{r}_2)\\
&=\frac{1}{\ell !\nn_{d,\ell}|\mathbb{S}^{d-1}|r_1 r_2}\  (\Pi^V)^{k<k_1\ldots k_\ell>}_{i<i_1\ldots i_\ell>}\hat{r}_1^{i_1}\ldots \hat{r}_1^{i_\ell}\ (\Pi^V)_{j<j_1\ldots j_\ell>}^{k<k_1\ldots k_\ell>}\hat{r}_2^{j_1}\ldots \hat{r}_2^{j_\ell}\ .
\end{split}\end{equation}
We can then use the definition of ${}^B\mathcal{Q}$ given in Eq.\eqref{eq:QBgen} to write
\begin{equation}\begin{split}
&\sum_{\alpha\vec{m}}|\multj^B(\w,\alpha,\ell,\vec{m})|^2= \frac{1}{(2\nu)^2\nn_{d,\ell}|\mathbb{S}^{d-1}|} \frac{1}{\ell!}\ 
\left[{}^B\mathcal{Q}(\w)_{k<k_1\ldots k_\ell>}\right]^\ast\ \left[{}^B\mathcal{Q}(\w)_{k<k_1\ldots k_\ell>}\right]\ .
\end{split}\end{equation}
Putting these results together and using the explicit formula for $\nn_{d,\ell}$, we can rewrite the power loss formula in Eq.\eqref{eq:LarmorFlat} entirely in terms of STF moments:
\begin{equation}\label{eq:LarmorFlatSTF}\begin{split}
\mathcal{P}(\w)&=\frac{1}{(d-2)!!|\mathbb{S}^{d-1}|}\sum_{\ell=1}^\infty \frac{\ell+d-2}{\ell}\frac{\w^{2\ell+d-1}}{(2\ell+d-2)!!}\ \frac{1}{\ell!}\ 
\left[{}^E\mathcal{Q}(\w)_{<i_1\ldots i_\ell>}\right]^\ast\ \left[{}^E\mathcal{Q}(\w)_{<i_1\ldots i_\ell>}\right]  \\
&\quad +\frac{1}{(d-2)!!|\mathbb{S}^{d-1}|}\sum_{\ell=1}^\infty \frac{\w^{2\ell+d-1}}{(2\ell+d-2)!!}\ \frac{1}{\ell!}\ 
\left[{}^B\mathcal{Q}(\w)_{i<i_1\ldots i_\ell>}\right]^\ast\ \left[{}^
B\mathcal{Q}(\w)_{i<i_1\ldots i_\ell>}\right]  \ .
\end{split}\end{equation}

For odd $d$, the case where there is a dissipative power loss, the above  expression can be rewritten in the following suggestive form: 
\begin{equation}\begin{split}
\int_{-\infty}^\infty \frac{d\w}{2\pi}\mathcal{P}(\w)&=\int_0^\infty \w\ \frac{\w^{d-1}d\w}{(2\pi)^d 2\w}\sum_{\ell=1}^\infty \frac{\ell+d-2}{\ell}\frac{\nn_{d,\ell}|\mathbb{S}^{d-1}|}{\ell!}\ 
\left[\w^\ell\ {}^E\mathcal{Q}(\w)_{<i_1\ldots i_\ell>}\right]^\ast\ \left[\w^\ell\ {}^E\mathcal{Q}(\w)_{<i_1\ldots i_\ell>}\right]  \\
&\quad +\int_0^\infty \w\ \frac{\w^{d-1}d\w}{(2\pi)^d 2\w}\sum_{\ell=1}^\infty \frac{\nn_{d,\ell}|\mathbb{S}^{d-1}|}{\ell!}\ 
\left[\w^\ell\ {}^B\mathcal{Q}(\w)_{i<i_1\ldots i_\ell>}\right]^\ast\ \left[\w^\ell\ {}^
B\mathcal{Q}(\w)_{i<i_1\ldots i_\ell>}\right]  \ .
\end{split}\end{equation}
We recognise in front the Lorentz-invariant phase-space integral for a photon of energy $\w$, as well as another factor of $\w$, indicating that we are computing its energy. Since this is power loss, the remaining factor should be interpreted as the production rate of photon by a given multipole moment. This can be made even more explicit if we recognise each term in the sum as  the inner product on the sphere for SSHs and VSHs: See
Eq.\eqref{Eq:SSHintSTF} and Eq.\eqref{Eq:VSHintSTF}. We then have the following integral representations:
\begin{equation}\begin{split}
\int_{\hat{k}\in\mathbb{S}^{d-1}}&
\left[\frac{1}{\ell!}{}^E\mathcal{Q}(\w)_{<i_1\ldots i_\ell>}\hat{k}^{i_1}\ldots\hat{k}^{i_\ell}\right]^\ast \left[\frac{1}{\ell!}\ {}^E\mathcal{Q}(\w)_{<j_1\ldots j_\ell>}\hat{k}^{j_1}\ldots\hat{k}^{j_\ell}\right]\\
&=\frac{\nn_{d,\ell}|\mathbb{S}^{d-1}|}{\ell!}\ 
\left[ {}^E\mathcal{Q}(\w)_{<i_1\ldots i_\ell>}\right]^\ast\ \left[{}^E\mathcal{Q}(\w)_{<i_1\ldots i_\ell>}\right]\ ,  \\
\int_{\hat{k}\in\mathbb{S}^{d-1}}&
\left[\frac{1}{\ell!}{}^E\mathcal{Q}(\w)_{<pi_1\ldots i_{\ell-1}>}\hat{k}^{i_1}\ldots\hat{k}^{i_{\ell-1}}\right]^\ast \left[\frac{1}{\ell!}\ {}^E\mathcal{Q}(\w)_{<pj_1\ldots j_{\ell-1}>}\hat{k}^{j_1}\ldots\hat{k}^{j_{\ell-1}}\right]\\
&=\frac{1}{\ell^2}\frac{\nn_{d,\ell-1}|\mathbb{S}^{d-1}|}{(\ell-1)!}\ 
\left[ {}^E\mathcal{Q}(\w)_{<i_1\ldots i_\ell>}\right]^\ast\ \left[{}^E\mathcal{Q}(\w)_{<i_1\ldots i_\ell>}\right]\ ,  \\
\int_{\hat{k}\in\mathbb{S}^{d-1}}&
\left[\frac{1}{\ell!}{}^B\mathcal{Q}(\w)_{p<i_1\ldots i_\ell>}\hat{k}^{i_1}\ldots\hat{k}^{i_\ell}\right]^\ast \left[\frac{1}{\ell!}\ {}^B\mathcal{Q}(\w)_{p<j_1\ldots j_\ell>}\hat{k}^{j_1}\ldots\hat{k}^{j_\ell}\right]\\
&=\frac{\nn_{d,\ell}|\mathbb{S}^{d-1}|}{\ell!}\ 
\left[ {}^B\mathcal{Q}(\w)_{i<i_1\ldots i_\ell>}\right]^\ast\ \left[{}^B\mathcal{Q}(\w)_{i<i_1\ldots i_\ell>}\right]\ .
\end{split}\end{equation}
These, along with the identity
\begin{equation}
\frac{1}{\ell^2}\frac{\nn_{d,\ell-1}}{(\ell-1)!}-\frac{\nn_{d,\ell}}{\ell!}=\frac{\ell+d-2}{\ell}\frac{\nn_{d,\ell}}{\ell!}
\end{equation}
allows us to write the power loss as the Lorentz-invariant phase-space integral for the  with momentum $\vec{k}$ and energy $\w_k\equiv |\vec{k}|$, i.e.,
\begin{equation}\begin{split}
\int_{-\infty}^\infty \frac{d\w}{2\pi}\mathcal{P}(\w)&=\int \frac{d^dk}{(2\pi)^d 2\w_k}\times \w_k (\delta^{pq}-\hat{k}^p \hat{k}^q) \\
&\qquad \times \Biggl\{\sum_{\ell=1}^\infty\left[\frac{\w_k}{\ell!}{}^E\mathcal{Q}(\w_k)_{<pi_1\ldots i_{\ell-1}>}k^{i_1}\ldots k^{i_{\ell-1}}\right]^\ast \left[\frac{\w_k}{\ell!}\ {}^E\mathcal{Q}(\w_k)_{<qj_1\ldots j_{\ell-1}>}k^{j_1}\ldots k^{j_{\ell-1}}\right]  \\
&\qquad \qquad+\sum_{\ell=1}^\infty\left[\frac{1}{\ell!}{}^B\mathcal{Q}(\w_k)_{p<i_1\ldots i_{\ell}>}k^{i_1}\ldots k^{i_{\ell}}\right]^\ast \left[\frac{1}{\ell!}\ {}^B\mathcal{Q}(\w_k)_{q<j_1\ldots j_{\ell}>}k^{j_1}\ldots k^{j_{\ell}}\right] \Biggr\}  \ .
\end{split}\end{equation}
Here we have used the fact that ${}^B\mathcal{Q}_{p<i_1\ldots i_{\ell}>}\hat{k}^p k^{i_1}\ldots k^{i_{\ell}}=0$ due to the transversality. The factor  $(\delta^{pq}-\hat{k}^p \hat{k}^q)$ is the polarisation sum, summing over all transverse polarisations of the photon. The power loss written above corresponds to the following photon emission amplitudes by the multipoles:
\begin{equation}\begin{split}
&i(-i)^{\nu-\frac{1}{2}}\frac{ \varepsilon^{p\ast}(\vec{k})}{\ell!}{}^E\mathcal{Q}(\w_k)_{<pi_1\ldots i_{\ell-1}>}\w_k k^{i_1}\ldots k^{i_{\ell-1}}\ ,  \qquad
i(-i)^{\nu-\frac{1}{2}}\frac{\varepsilon^{p\ast}(\vec{k})}{\ell!}{}^B\mathcal{Q}(\w_k)_{p<i_1\ldots i_{\ell}>}k^{i_1}\ldots k^{i_{\ell}} \ .
\end{split}\end{equation}
Here $\varepsilon^{p}(\vec{k})$ is the polarisation for the photon with momentum $\vec{k}$ and energy $\w_k\equiv |\vec{k}|$, and we have fixed the overall phase by comparing against radiative fields in Eq.\eqref{eq:RadEflat} and Eq.\eqref{eq:RadBflat}.

\subsection{EM influence phase}
We will now turn to the description of radiation reaction in flat spacetime. Our goal here is to get some sort of effective action that captures the effect of radiation on charge/current sources. Since radiation carries away energy in some cases (not always: see below), what we need is an action that can describe dissipation. This implies that the correct language here is that of \emph{influence phase} ala Feynman-Vernon\cite{FEYNMAN1963118}, i.e., an action that doubles the system degrees of freedom to allow us to describe the evolution of density matrices, its decoherence and dissipation into the environment. As described by Feynman-Vernon, the influence phase is computed by doing a path integral over the doubled environment (here the EM fields) coupled to a doubled system(here the charges/currents), with a specific in-in boundary condition on the environment fields. As emphasised by us in \cite{Loganayagam:2023pfb}, such in-in boundary conditions are naturally implemented on the dS-SK geometry built by connecting two static patches at the future horizon. In the subsequent appendices, we will show that this works also for electromagnetism in dS. Coming back to the current topic of flat space EM, there is no such simple geometric construction: the in-in boundary conditions have to be imposed by hand.\footnote{Strictly speaking, flat spacetime EM path integral also has a variety of infrared subtleties. We will ignore them in what follows.  } For $d=3$, the reader can find such an analysis in \cite{Barone:1991zz,Anastopoulos:1997wh,Breuer:1999vm}. Given that we will be presenting a detailed derivation of the influence phase in the dS case, we will be content here with a brief sketch.

We remind the reader that the current density can be parametrised in terms of two functions such that it solves the conservation equations in the following manner:
\begin{equation}
\begin{split}
J^t(\vec{r},\w) &=\sum_{\ell\vec{m}} \Bigl\{-\frac{1}{r^{d-1}}\frac{\partial}{\partial r}[r^{d-1}J_1(r,\w,\ell,\vec{m})]+\frac{\ell(\ell+d-2)}{r^2}J_2(r,\w,\ell,\vec{m})\Bigr\} \mathscr{Y}_{\ell\vec{m}}(\hat{r}) \ ,\\
J^r(\vec{r},\w) &= -i\w\sum_{\ell\vec{m}} J_1(r,\w,\ell,\vec{m})\mathscr{Y}_{\ell\vec{m}}(\hat{r})\ ,\\
r^2\gamma_{_{IK}}J^K(\vec{r},\w)
&= -i\w\sum_{\ell\vec{m}}J_2(r,\w,\ell,\vec{m})\mathscr{D}_I\mathscr{Y}_{\ell\vec{m}}(\hat{r})+r^2\gamma_{_{IK}}\sum_{\alpha\ell\vec{m}}J_V(r,\w,\alpha,\ell,\vec{m})\mathbb{V}^K_{\alpha\ell\vec{m}}\ .
\end{split}
\end{equation}
We will now consider two copies of these currents($J_L$ and $J_R$), which can be independently specified for all cases except in the $\ell=0$ case. 

\begin{equation}
\begin{split}
&\int_\w \int dr\int dr_0 \\
&\Bigl\{\sum_{\ell\vec{m}}\ell(\ell+d-2)\left[J_1(r,\w,\ell,\vec{m})-\partial_{r} J_2(r,\w,\ell,\vec{m})\right]^*_D \mathbb{G}_E(r,r_0,\w,\ell)\left[ J_1(r_0,\w,\ell,\vec{m})-\partial_{r_0} J_2(r_0,\w,\ell,\vec{m})\right]_A\Bigr. \\
&\qquad\Bigl.\qquad+\sum_{\alpha\ell\vec{m}}[J_V(r,\w,\alpha,\ell,\vec{m})]^*_D \mathbb{G}_B(r,r_0,\w,\ell) J_V(r_0,\w,\alpha,\ell,\vec{m})_A \Bigr\}\
\end{split}
\end{equation}

For odd values of $d$, $\cot(\nu\pi)$ is zero, and the renormalised action is given by:
\begin{equation}
\begin{split}
\sum_{\ell\vec{m}}\ell(\ell+d-2)&\int_\w  \frac{\pi i}{2\Gamma(\nu+1)^2}\left(\frac{\w}{2}\right)^{2\nu} \\
& \times \int dr_0\  r_0^{\nu+\frac{d}{2}-1}\ {}_0F_1\left[1+\nu,-\frac{\w^2r_0^2}{4}\right] 
 \left[\partial_{r_0} J_2(r_0,\w)- J_1(r_0,\w)\right]^*_D \\ &\quad \times \int dr\  r^{\nu+\frac{d}{2}-1}\ {}_0F_1\left[1+\nu,-\frac{\w^2r^2}{4}\right]\left[\partial_r J_2(r,\w)- J_1(r,\w)\right]_A
\end{split}
\end{equation}

This action for even values of $d$ gets multiplied by a factor of $(1+i\cot(\nu\pi))$ and hence diverges. Further counterterms are needed for regularisation. These counterterms can be computed by expanding the action around $\nu=n\in \mathbb{Z}$. Consider:
\begin{equation}\label{eq:evenDimRegFlat}
(1+i\cot \pi \nu)\frac{2\pi i}{ \Gamma(\nu)^2} \left(\frac{\w }{2H}\right)^{2\nu}=\frac{1}{ \Gamma(n)^2} \left(\frac{\w }{2 H}\right)^{2n}\left\{\frac{2}{\nu-n}-4\psi^{(0)}(n) +\ln \left(\frac{\w}{2H}\right)^4+O(\nu-n)  \right\}\ . 
\end{equation}
where $H$ is the renormalisation scale. Following the same modified minimal subtraction scheme proposed in \cite{Loganayagam:2023pfb}, we will counterterm the first two terms in the RHS. Using this scheme, the RR action becomes:

\begin{equation}
\begin{split}
\sum_{\ell\vec{m}}\ell(\ell+d-2)&\int_\w  \frac{1}{4\Gamma(\nu+1)^2}\left(\frac{\w}{2}\right)^{2\nu}\ln{\left(\frac{\w^4}{H^4}\right)} \\
& \times \int dr_0\  r_0^{\nu+\frac{d}{2}-1}\ {}_0F_1\left[1+\nu,-\frac{\w^2r_0^2}{4}\right] 
  \left[\partial_{r_0} J_2(r_0,\w)- J_1(r_0,\w)\right]^*_D \\ &\quad \times \int dr\  r^{\nu+\frac{d}{2}-1}\ {}_0F_1\left[1+\nu,-\frac{\w^2r^2}{4}\right]\left[\partial_r J_2(r,\w)- J_1(r,\w)\right]_A
\end{split}
\end{equation}

A similar argument gives us the corresponding action for the magnetic parity action. We have:
\begin{equation}
\begin{split}
S^{\text{vector, bare}}_{RR}=\sum_{\ell\vec{m}}\int_\w \int r^{d-1}dr\int r_0^{d-1}dr_0 & J_{VD}^*(r_0,
w,\ell,\vec{m})J_{VA}(r_0,
w,\ell,\vec{m}) \mathbb{G}_B(r,r_0;\ell,\vec{m})
\end{split}
\end{equation}
For odd values of $d$, the dissipative part of the action is given by:
\begin{equation}
\begin{split}
\sum_{\ell\vec{m}}&\int_\w  \frac{\pi i}{2\Gamma(\nu+1)^2}\left(\frac{\w}{2}\right)^{2\nu} \\
& \times \int dr_0\  r_0^{\nu+\frac{d}{2}+1}\ {}_0F_1\left[1+\nu,-\frac{\w^2r_0^2}{4}\right] 
 J_{VD}^*(r_0,
w,\ell,\vec{m}) \\ &\quad \times \int dr\  r^{\nu+\frac{d}{2}+1}\ {}_0F_1\left[1+\nu,-\frac{\w^2r^2}{4}\right] J_{VA}^*(r_0,
w,\ell,\vec{m})
\end{split}
\end{equation}
Counterterming away the extra divergence for the case of even $d$ in the same way as for the electric sector, we obtain the action:
\begin{equation}
\begin{split}
\sum_{\ell\vec{m}}\ell(\ell+d-2)&\int_\w  \frac{1}{4\Gamma(\nu+1)^2}\left(\frac{\w}{2}\right)^{2\nu}\ln{\left(\frac{\w^4}{H^4}\right)}  \\
& \times \int dr_0\  r_0^{\nu+\frac{d}{2}+1}\ {}_0F_1\left[1+\nu,-\frac{\w^2r_0^2}{4}\right] 
 J_{VD}^*(r_0,
w,\ell,\vec{m}) \\ &\quad \times \int dr\  r^{\nu+\frac{d}{2}+1}\ {}_0F_1\left[1+\nu,-\frac{\w^2r^2}{4}\right] J_{VA}^*(r_0,
w,\ell,\vec{m})
\end{split}
\end{equation}

Given this reduced boundary action, we can write down an SK action for the EM radiation reaction as:
\begin{equation}\label{eq:FlatEMinfluence}
S_{\text{RR}}=\sum_{\ell\vec{m}}\frac{\ell+d-2}{\ell}\int_\w \frac{2\pi i}{\Gamma(\nu)^2}\left(\frac{\w}{2}\right)^{2\nu}\left\{\mathcal{J}^{D*}_E\mathcal{J}^A_E   +\frac{\ell}{\ell+d-2} \mathcal{J}^{D*}_B\mathcal{J}^A_B\right\}
\end{equation}
where we define:
\begin{equation}\label{eq:jsFlatdef}
	\begin{split}
\multj_A^E(\w,\ell,\vec{m}) & \equiv  \frac{\ell}{2\nu}\int dr\ \Bigg[J_2^A\ \partial_{r}\left\{r^{\nu+\frac{d}{2}-1}\ {}_0F_1\left[1+\nu,-\frac{\w^2r^2}{4}\right]\right\} 
\\ & \qquad\qquad\qquad\qquad\qquad\qquad
+J^A_1\  r^{\nu+\frac{d}{2}-1}\ {}_0F_1\left[1+\nu,-\frac{\w^2r^2}{4}\right]\Bigg]
\ ,\\
\multj_D^E(\w,\ell,\vec{m}) & \equiv  \frac{\ell}{2\nu}\int dr\ \Bigg[J_2^D\ \partial_{r}\left\{r^{\nu+\frac{d}{2}-1}\ {}_0F_1\left[1+\nu,-\frac{\w^2r^2}{4}\right]\right\} 
\\ & \qquad\qquad\qquad\qquad\qquad\qquad
+J_1^D\  r^{\nu+\frac{d}{2}-1}\ {}_0F_1\left[1+\nu,-\frac{\w^2r^2}{4}\right]\Bigg]
\ , \\
\multj_A^B(\w,\alpha,\ell,\vec{m}) & \equiv \frac{1}{2\nu}\int dr \Bigg\{J_V^A(\alpha,\ell,\vec{m}) r^{\nu+\frac{d}{2}+1}{}_0F_1\left[1+\nu,-\frac{\w^2r^2}{4}\right]\Bigg\} \ , \\
\multj_D^B(\w,\alpha,\ell,\vec{m}) & \equiv \frac{1}{2\nu}\int dr \Bigg\{J_V^D(\alpha,\ell,\vec{m}) r^{\nu+\frac{d}{2}+1}{}_0F_1\left[1+\nu,-\frac{\w^2r^2}{4}\right]\Bigg\} 
\end{split}
\end{equation}
Later, we will see that the influence phase obtained in de Sitter reduces to Eq.\eqref{eq:FlatEMinfluence} in the flat space limit. 

To compute the radiation reaction of a single particle, it is much more convenient to work with STF scalar harmonics. Corresponding to appendix A3 of \cite{Loganayagam:2023pfb}, one can rewrite the RR action in terms of STF moments $\mathcal{Q}^{<i_1\ldots i_\ell>}(\w)$ as:
\begin{equation}
	\begin{split}
{}^E\mathcal{Q}_{A}^{<i_1\ldots i_\ell>}(\w) & \equiv  \frac{1}{(\ell+d-2)}(\Pi^S)^{<i_1i_2\ldots i_\ell>}_{ <j_1j_2\ldots j_\ell>}\int d^{d}x\  x^{j_1}x^{j_2}\ldots x^{j_\ell} 
\\ & \qquad
\times \Bigg[\frac{1}{r^{\ell+d-3}}\partial_{r}\left\{r^{\ell+d-2}\ {}_0F_1\left[1+\nu,-\frac{r^2\w^2}{4}\right]\right\}J^t_A(\vec{r},\w) 
\\ & \qquad\qquad\qquad\qquad\qquad\qquad +i\w r \ {}_0F_1\left[1+\nu,-\frac{r^2\w^2}{4}\right]J^r_A(\vec{r},\w)\Bigg]\ ,
\end{split}
\end{equation}
\begin{equation}
	\begin{split}
{}^E\mathcal{Q}_{D}^{<i_1\ldots i_\ell>}(\w) & \equiv  \frac{1}{(\ell+d-2)}(\Pi^S)^{<i_1i_2\ldots i_\ell>}_{ <j_1j_2\ldots j_\ell>}\int d^{d}x\  x^{j_1}x^{j_2}\ldots x^{j_\ell} 
\\ & \qquad
\times \Bigg[\frac{1}{r^{\ell+d-3}}\partial_{r}\left\{r^{\ell+d-2}\ {}_0F_1\left[1+\nu,-\frac{r^2\w^2}{4}\right]\right\}J^t_D(\vec{r},\w) 
\\ & \qquad\qquad\qquad\qquad\qquad\qquad +i\w r \ {}_0F_1\left[1+\nu,-\frac{r^2\w^2}{4}\right]J^r_D(\vec{r},\w)\Bigg]\ .
\end{split}
\end{equation}
In terms of these STF multipole moments, we can write the RR action as follows:
\begin{equation}
S^E_{\text{RR}}=\sum_{\ell}\frac{\ell+d-2}{\ell}\int\frac{d\w}{2\pi} \frac{\pi i}{2 \Gamma(\nu+1)^2} \left(\frac{\w }{2}\right)^{2\nu} \frac{1}{\nn_{d,\ell}|\mathbb{S}^{d-1}|} \frac{1}{\ell!}{}^E\mathcal{Q}_{D}^{\ast<i_1i_2\ldots i_\ell>}{}^E\mathcal{Q}^{A}_{ <i_1i_2\ldots i_\ell>}\ .
\end{equation}
The magnetic multipole moment can be written in terms of the vector STF projector:
\begin{equation}
\begin{split}
    {}^B\mathcal{Q}_A^{i<i_1\ldots i_\ell>}\equiv (\Pi^V)^{i<i_1i_2\ldots i_\ell>}_{j<j_1j_2\ldots j_\ell>}\int d^{d}x \  x^{j_1}x^{j_2}\ldots x^{j_\ell}\ {}_0F_1\left[1+\nu,-\frac{r^2\w^2}{4}\right] J^j_A   \\
    {}^B\mathcal{Q}_D^{i<i_1\ldots i_\ell>}\equiv (\Pi^V)^{i<i_1i_2\ldots i_\ell>}_{j<j_1j_2\ldots j_\ell>}\int d^{d}x \ x^{j_1}x^{j_2}\ldots x^{j_\ell}\ {}_0F_1\left[1+\nu,-\frac{r^2\w^2}{4}\right] J^j_D  
\end{split}
\end{equation} 
Using the vector addition theorem \eqref{eq:VSHadd}, we can show that the magnetic part of the influence phase can be written in terms of the STF multipole moments as:
\begin{equation}
    S^B_{\text{RR}}=\sum_{\ell}\int\frac{d\w}{2\pi} \frac{\pi i}{ 2\Gamma(\nu+1)^2} \left(\frac{\w }{2}\right)^{2\nu} \frac{1}{\nn_{d,\ell}|\mathbb{S}^{d-1}|} \frac{1}{\ell!}{}^B\mathcal{Q}_{D,STF}^{\ast i<i_1i_2\ldots i_\ell>}\ {}^B\mathcal{Q}^{A,STF}_{ i<i_1i_2\ldots i_\ell>}\ .
\end{equation}

\newpage
\subsection{Multipole expansion in \texorpdfstring{$d=3$}{d=3} }
In $d=3$, we can trade the magnetic field 2-form for an axial vector field $\bar{B}_i\equiv\frac{1}{2}\varepsilon_{ijk}\FC_{jk}$. It is more convenient to deal with vectors than tensors, and the electric-magnetic duality is easier to see. We will describe here how the multipole expansion and radiation reaction formulae can be recast to make EM duality manifest.

At the level of orthonormal spherical harmonics, in $d=3$, all vector spherical harmonics can be replaced with the toroidal operator acting on scalar spherical harmonics, viz.,
\begin{equation}
\begin{split}
\mathbb{V}_{I}^{1\ell m}(\hat{r})=\frac{1}{\sqrt{\ell(\ell+1)}}\varepsilon_{IJ}\mathscr{D}^J\mathscr{Y}_{\ell m}(\hat{r})\ ,
  \end{split}
\end{equation}
where the spherical indices are raised using the unit sphere metric.
However, if we do this, we must contend with the irrational factors of $\sqrt{\ell(\ell+1)}$ everywhere in the spherical harmonic expansions. We will instead use the following strategy, motivated primarily by EM duality. We use a rescaled basis of VSHs\footnote{In what follows, it is useful to remember that\begin{equation}
\begin{split}
d\vartheta^I\mathbb{U}_{I}^{\ell m}(\hat{r})=\frac{1}{\ell}\left(\frac{d\vartheta}{\sin\vartheta}\ \frac{\partial}{\partial \varphi}-\sin\vartheta\ d\varphi\ \frac{\partial}{\partial \vartheta}\right){Y}_{\ell m}(\hat{r})=\overrightarrow{dr}\cdot\left\{-\frac{\hat{e}_r}{\ell}\times\vec{\nabla}{Y}_{\ell m}(\hat{r})\right\}\ .
\end{split}
\end{equation}
The last expression gives a cartesian form for our rescaled VSH.}
\begin{equation}
\begin{split}
\mathbb{U}_{I}^{\ell m}(\hat{r})\equiv\sqrt{\frac{\ell+1}{\ell}}\mathbb{V}_{I}^{1\ell m}(\hat{r})=\frac{1}{\ell}\varepsilon_{_{IJ}}\mathscr{D}^J\mathscr{Y}_{\ell m}(\hat{r})\ .
\end{split}
\end{equation}
Similarly, we scale the components appearing in the vector spherical harmonic expansions. We will add a $\vee$ symbol to all scaled components to avoid confusion. This rescaling factor should also be included in orthonormality, addition theorem, etc. A rough thumb rule is to  replace all occurrences of $\mathbb{V}_{I}$s in our formulae with $\mathbb{U}_{I}$s, but replace $\mathbb{V}_{I}^\ast$ with $\frac{\ell}{\ell+1}\mathbb{U}_{I}^\ast$ (apart from adding $\vee$ symbol to the components). We will see an example of this below.

\subsubsection*{Toroidal magnetostatics in $d=3$}
Let us first see how this works in the simpler setting of magnetostatics: the generalisation to the full dynamical situation is straightforward.

In terms of the rescaled VSH, the magnetostatic expansion in Eq.\eqref{eq:MstatVSHexp} becomes
\begin{equation}
\begin{split}
\gaugeV_r &=0\ ,\quad \gaugeV_I\equiv \sum_{\ell m} \overline{\Phi}^\vee_{_B}(r,\ell,m)\mathbb{U}_{I}^{\ell m}(\hat{r})\ ,\\
\FC_{rI}&\equiv \sum_{\ell m} \overline{H}^\vee_v (r,\alpha,\ell,\vec{m})\mathbb{U}_{I}^{\ell m}(\hat{r}) =\varepsilon_{_{IJ}}\sum_{\ell m}\frac{1}{\ell}\overline{H}_v^{\vee} (r,\ell,m)\mathscr{D}^J\mathscr{Y}_{\ell m}(\hat{r})\ ,\\
\FC_{IJ} &\equiv \sum_{\ell\vec{m}\alpha} \overline{H}^\vee_{vv}(r,\ell, m)\mathscr{D}_{[I}\mathbb{U}_{J]}^{\ell m}(\hat{r}) =\varepsilon_{_{IJ}}\sum_{\ell m}(\ell+1)\overline{H}_{vv}^{\vee} (r,\ell,m)\mathscr{Y}_{\ell m}(\hat{r})\ ,
\end{split}
\end{equation}
where, in the last step, we have used the identity
\begin{equation}
\begin{split}
\mathscr{D}_{[I}\mathbb{U}_{J]}^{\ell m}(\hat{r})=\varepsilon_{_{IJ}}\varepsilon^{MN}\mathscr{D}_{M}\mathbb{U}_{N}^{\ell m}=(\ell+1)\mathscr{Y}_{\ell m}(\hat{r})\ \varepsilon_{_{IJ}}\ .
\end{split}
\end{equation}
The vector magnetic field is obtained by stripping off the $\varepsilon_{_{rIJ}}=r^2\varepsilon_{_{IJ}}$ factors in $\FC_{ij}$ and then lowering the indices.The spherical/radial components of the vector magnetic field are
\begin{equation}
\begin{split}
 \overline{B}_I&=r^2\gamma_{IJ}\times \frac{1}{r^2}\sum_{\ell m}\frac{1}{\ell}\overline{H}_v^{\vee} (r,\ell,m)\mathscr{D}^J\mathscr{Y}_{\ell m}(\hat{r})=\sum_{\ell m}\frac{1}{\ell}\overline{H}_v^{\vee} (r,\ell,m)\mathscr{D}_I\mathscr{Y}_{\ell m}(\hat{r})\ ,\\
 \overline{B}_r&=\sum_{\ell m}\frac{\ell+1}{r^2}\overline{H}_{vv}^{\vee} (r,\ell,m)\mathscr{Y}_{\ell m}(\hat{r})\ .
\end{split}
\end{equation}
The expression for the Debye field in Eq.\eqref{eq:PhiBStatics} becomes
\begin{equation}
\begin{split}\label{eq:phiBcheck}
\overline{\Phi}_{_B}^\vee(r,\ell,m)&\equiv \frac{\ell}{\ell+1}
\int_{\vec{r}_0}\mathbb{G}_B(r,r_0;\ell)\mathbb{U}_J^{\ell m\ast}(\hat{r}_0)\bar{J}^{J}(\vec{r}_0)=\sqrt{\frac{\ell}{\ell+1}}\overline{\Phi}_{_B}(r,\alpha=1,\ell,m)\ .
\end{split}
\end{equation}
In the first step, there is a rescaling pre-factor $\frac{\ell}{\ell+1}$ which multiplies $\mathbb{U}_J^{\ell m\ast}$ in accordance with the thumb rule mentioned above. Similar factors appear in the definition of the magnetic moment:
\begin{equation}
\begin{split}\label{eq:JBcheck}
\overbar{\multj}^{\vee B}(\ell,m) &\equiv\frac{1}{2\ell+1}\frac{\ell}{\ell+1}\int_{\vec{r}_0}r_0^{\ell+1}
\mathbb{U}_J^{\ell m\ast}(\hat{r}_0)\bar{J}^{J}(\vec{r}_0)=\sqrt{\frac{\ell}{\ell+1}}\overbar{\multj}^{B}(\alpha=1,\ell,m)\ .
\end{split}
\end{equation}
The magnetic Debye field and the magnetic field components outside the sources take the form 
\begin{equation}\label{eq:Phi Bcheck}
\begin{split}
\overline{\Phi}_{_B}^{\vee,\text{Out}}(r,\ell,m)&=\frac{\overbar{\multj}^{\vee B}(\ell,m)}{r^{\ell}}\ ,\\
\frac{1}{\ell}\overline{H}_v^{\vee\text{Out}} (r,\ell,m) &=-\frac{\overbar{\multj}^{\vee B}(\ell,m)}{r^{\ell+1}}\ ,\quad \
\frac{\ell+1}{r^2}\overline{H}_{vv}^{\vee\text{Out}}(r,\ell,m) =(\ell+1)\frac{\overbar{\multj}^{\vee B}(\ell,m)}{r^{\ell+2}}\ .
\end{split}
\end{equation}
This should be compared against the electrostatic expressions written in terms of electric multipole moments:
\begin{equation}
\begin{split}
\overline{\Phi}_{_E}^{\vee,\text{Out}}(r,\ell,\vec{m})&=\frac{\overbar{\multj}^{E}(\ell,\vec{m})}{r^{\ell}}\ ,\\
 \overline{E}_s^{\text{Out}} (r,\ell,\vec{m}) &=-\frac{\overbar{\multj}^E(\ell,\vec{m})}{r^{\ell+d-2}}\ ,\quad \
\overline{E}_{r}^{\text{Out}}(r,\ell,\vec{m}) =(\ell+d-2)\frac{\overbar{\multj}^E(\ell,\vec{m})}{r^{\ell+d-1}}\ .
\end{split}
\end{equation}
We see that the forms of the outside electrostatic vs magnetostatic fields agree in $d=3$. In this normalisation,  EM duality acts by mapping electric to magnetic fields and $\overbar{\multj}^E$ to $\overbar{\multj}^{\vee B}$.
As to the cartesian moments, in analogy with electrostatic multipole tensors in Eq.\eqref{eq:jEtoQE}, we define\footnote{We use the result that $\nn_{d,\ell-1}|\mathbb{S}^{d-1}|=\frac{4\pi}{(2\ell-1)!!}$ for $d=3$.}
\begin{equation}
\begin{split}
\frac{1}{\ell!}\ {}^B\overbar{\mathcal{Q}}^\vee_{<i_1 i_2\ldots i_\ell>}x^{i_1}\ldots x^{i_\ell}&\equiv \frac{4\pi}{(2\ell-1)!!}\sum_{m}\overbar{\multj}^{\vee B}(\ell,m) r^{\ell}\mathscr{Y}_{\ell m}(\hat{r})\\
&=\frac{1}{\ell+1}\int_{\vec{r}_0}[\vec{r}_0\times\vec{\bar{J}}(\vec{r}_0)]\cdot\vec{\nabla}_0\Pi^S(\vec{r}|\vec{r}_0)_{d=3,\ell}\ .
\end{split}
\end{equation}
In the second line, we have used the expression for $\overbar{\multj}^{\vee B}$ in Eq.\eqref{eq:JBcheck}, the explicit form of $\mathbb{U}_I$, as well as the SSH addition theorem. The integral appearing here can be evaluated as
\begin{equation}
\begin{split}
\int_{\vec{r}_0}[\vec{r}_0\times\vec{\bar{J}}(\vec{r}_0)]\cdot\vec{\nabla}_0\Pi^S(\vec{\kappa}|\vec{r}_0)&=\frac{\ell}{\ell !}\kappa^{i_1}\ldots \kappa^{i_\ell}(\Pi^S)^{<i_1 i_2\ldots i_\ell>}_{<j_1 j_2\ldots j_\ell>}\int_{\vec{r}_0}[\vec{r}_0\times\vec{\bar{J}}(\vec{r}_0)]^{j_1}x_0^{j_2}\ldots x_0^{j_\ell}\ ,
\end{split}
\end{equation}
thus yielding a direct cartesian definition 
\begin{equation}
\begin{split}\label{eq:Qbcheck}
{}^B\overbar{\mathcal{Q}}^\vee_{<i_1i_2\ldots i_\ell>}&\equiv \frac{\ell}{\ell+1}
\int_{\vec{r}_0}[\vec{r}_0\times\vec{\bar{J}}(\vec{r}_0)]^{<i_1}x_0^{i_2}\ldots x_0^{i_\ell>}\ .
\end{split}
\end{equation}
By construction, the normalisation here is fixed to ensure that EM duality maps the STF tensors  ${}^E\overbar{\mathcal{Q}}$ to ${}^B\overbar{\mathcal{Q}}^\vee$. The outside vector potential/magnetic field can be written in terms of this STF magnetic moment as
\begin{equation}
\begin{split}
\gaugeV_k^{\text{Out}} &=- \varepsilon_{kij}x^i\sum_{\ell} {}^B\overbar{\mathcal{Q}}^\vee_{<ji_1\ldots i_{\ell-1}>} 
\frac{x^{i_1}\ldots x^{i_{\ell-1}}}{ 4\pi r^{2\ell+1}}\frac{(2\ell-1)!!}{\ell!}\\
&= -\varepsilon_{kij}\sum_{\ell} \frac{(-)^\ell}{\ell!}
\partial^i\partial^{i_1}\partial^{i_2}\ldots \partial^{i_{\ell-1}}\Bigl\{\frac{{}^B\overbar{\mathcal{Q}}^\vee_{<ji_1\ldots i_{\ell-1}>} }{4\pi r}\Bigr\}\ ,\\
\overline{B}_j^{\text{Out}}&= \sum_{\ell} {}^B\overbar{\mathcal{Q}}^\vee_{<i_1 i_2\ldots i_\ell>}[(2\ell+1)x^j x^{i_\ell}-r^2\ell\delta^{ji_\ell}]\frac{ x^{i_1}\ldots x^{i_{\ell-1}}}{4\pi r^{2\ell+3}} \frac{(2\ell-1)!!}{\ell!}\ .
\end{split}
\end{equation}
As expected from EM duality, the magnetic field here has the same form as the outside electrostatic field in Eq.\eqref{eq:EstaicSTF}. The description in terms of the magnetic moment ${}^B\overbar{\mathcal{Q}}^\vee$ can be related to ones in terms of  ${}^B\overbar{\mathcal{Q}}$ as follows: first, we begin by rewriting the VSH projector as
\begin{equation}\label{eq:3dPiVPiS}
\begin{split}
\Pi^V_{ij}(\vec{\kappa}|\vec{r})|_{d=3}=
\frac{[\vec{\kappa}\times\vec{\nabla}_\kappa]_i[\vec{r}\times\vec{\nabla}]_j}{\ell(\ell+1)}\Pi^S(\vec{\kappa}|\vec{r})|_{d=3}\ .
\end{split}
\end{equation}
This follows from the fact that, in $d=3$, all the VSHs can be obtained by applying a single toroidal operator $[\vec{r}\times\vec{\nabla}]$ on SSHs. To get orthonormal VSHs, we need to divide by a factor $\sqrt{\ell(\ell+1)}$. This means that the VSH addition theorem in $d=3$ can be obtained by sandwiching the SSH addition theorem between two toroidal operators\cite{freeden1998constructive,freeden2013special,freeden2022spherical}. The relation between the two addition theorems then yields the above relation between the projectors.  

In the next step, we write
\begin{equation}
\begin{split}
\frac{1}{\ell !}{}^B\overbar{\mathcal{Q}}_{i<i_1 i_2\ldots i_\ell>}\kappa^{i_1}\ldots \kappa^{i_\ell}&\equiv\int_{\vec{r}_0}\Pi^V_{ij}(\vec{\kappa}|\vec{r}_0)\bar{J}^j(\vec{r}_0)\\
&=-\frac{1}{\ell}
[\vec{\kappa}\times\vec{\nabla}_\kappa]_i\frac{1}{\ell+1}\int_{\vec{r}_0}[\vec{r}_0\times\vec{\bar{J}}(\vec{r}_0)]\cdot\vec{\nabla}_0\Pi^S(\vec{\kappa}|\vec{r}_0)\ .
\end{split}
\end{equation}
In RHS, we recognise the integral that defines the ${}^B\overbar{\mathcal{Q}}^\vee$. Stripping off the $\kappa^i$'s, we get the relation between the two kinds of moments as 
\begin{equation}\label{eq:QbcToQb}
\begin{split}
{}^B\overbar{\mathcal{Q}}_{i<i_1 i_2\ldots i_\ell>}
&=
\frac{1}{\ell}\sum_{p=1}^\ell{}^B\overbar{\mathcal{Q}}^\vee_{<ji_1\ldots \underline{i_p}\ldots i_\ell>}\ \varepsilon_{iji_p}\ .
\end{split}
\end{equation}
Here, the underlining on the $i_p$ index indicates that it should be dropped, i.e., the indices of ${}^B\overbar{\mathcal{Q}}^\vee$ are $ji_1i_2\ldots i_{p-1}i_{p+1}\ldots i_\ell$. We can  invert this relation by rewriting Eq.\eqref{eq:3dPiVPiS} as
\begin{equation}
\begin{split}
[\vec{\kappa}\times\vec{\nabla}_\kappa]_i\Pi^V_{ij}(\vec{\kappa}|\vec{r})|_{d=3}=
-[\vec{r}\times\vec{\nabla}]_j\Pi^S(\vec{\kappa}|\vec{r})|_{d=3}\ .
\end{split}
\end{equation}
This is equivalent to the tensorial relation
\begin{equation}\begin{split}
   \sum_{p=1}^\ell \varepsilon_{ii_pn}   (\Pi^V)^{i<n i_1\ldots \underline{i_p}\ldots i_{\ell}>}_{j<j_1 j_2\ldots j_\ell>} 
&= - \sum_{p=1}^\ell\varepsilon_{jj_pn}   (\Pi^S)^{<i_1 i_2\ldots i_{\ell}>}_{<n j_1\ldots\underline{j_p} \ldots j_\ell>}\ .
 \end{split}\end{equation}
Multiplying by $x_0^{j_1}x_0^{j_2}\ldots x_0^{j_\ell} \bar{J}^n(\vec{r}_0)$ and integrating, we get  
\begin{equation}\label{eq:QbToQbc}
\begin{split}
{}^B\overbar{\mathcal{Q}}^\vee_{<i_1 i_2\ldots i_\ell>}
&=
-\frac{1}{\ell+1}\sum_{p=1}^\ell{}^B\overbar{\mathcal{Q}}_{i<ji_1\ldots \underline{i_p}\ldots i_\ell>}\ \varepsilon_{iji_p}\ .
\end{split}
\end{equation}
The relations Eq.\eqref{eq:QbcToQb} and Eq.\eqref{eq:QbToQbc} show that same information is contained in the STF tensors  ${}^B\overline{\mathcal{Q}}$ and 
${}^B\overline{\mathcal{Q}}^\vee$.

\subsubsection*{Multipole radiation in $d=3$}
We will now move to describing the full radiative fields in $d=3$ outside the sources. The decomposition of EM fields in terms of spherical harmonics has the form
\begin{equation}
\begin{split}
E_I(\vec{r},\w)&=\sum_{\ell m}\Esf (r,\w,\ell,m)\mathscr{D}_I\mathscr{Y}_{\ell m}(\hat{r})+\sum_{\ell m}\Evf^{\vee} (r,\w,\ell,m)\mathbb{U}_I^{\ell m}(\hat{r})\ ,\\
E_r(\vec{r},\w)&=\sum_{\ell m}\Erf (r,\w,\ell,m)\mathscr{Y}_{\ell m}(\hat{r})\ ,\\
B_I(\vec{r},\w)&=\sum_{\ell m}\frac{1}{\ell}\Hvf^{\vee} (r,\w,\ell,m)\mathscr{D}_I\mathscr{Y}_{\ell m}(\hat{r})-\sum_{\ell m}\ell\ \Hsf (r,\w,\ell,m)\mathbb{U}_I^{\ell m}(\hat{r})\ ,\\
B_r(\vec{r},\w)&=\sum_{\ell m}\frac{\ell+1}{r^2}H_{vv}^{\vee} (r,\w,\ell,m)\mathscr{Y}_{\ell m}(\hat{r})\ .
 \end{split}
\end{equation}
We can write this succinctly as 
\begin{equation}
\begin{split}
 \vec{E}(\vec{r},\w) 
    &= \sum_{\ell m}\left[\Erf (r,\w,\ell,m)\mathscr{Y}_{\ell m}\hat{e}_r+ \Esf(r,\w,\ell,m)\
    \vec{\nabla}\mathscr{Y}_{\ell m}
+\Evf^{\vee}(r,\w,\ell,m)\ \vec{\mathbb{U}}_{\ell m}\right]\ ,\\
\vec{B}(\vec{r},\w) &= \sum_{\ell m}\left[\frac{\ell+1}{r^2}H_{vv}^{\vee} (r,\w,\ell,m)\mathscr{Y}_{\ell m}\hat{e}_r+\frac{1}{\ell} H_v^{\vee}(r,\w,\ell,m)\ \vec{\nabla}\mathscr{Y}_{\ell m}-\ell\Hsf(r,\w,\ell,m)\ \vec{\mathbb{U}}_{\ell m}\right] \ , \\
\end{split}
\end{equation}
where we have introduced 
\begin{equation}
\vec{\mathbb{U}}_{\ell m}\equiv -\frac{1}{\ell}\hat{e}_r\times\vec{\nabla}\mathscr{Y}_{\ell m}\ ,\quad  \vec{\nabla}\mathscr{Y}_{\ell m}=\ell\ \hat{e}_r\times \vec{\mathbb{U}}_{\ell m}\ .  
\end{equation}
With this understanding, much of our description of 
EM fields in this section continue to hold true. Explicit expressions for outside fields in terms of multipole moments can be written down using Eq.\eqref{eq:HoutjB} and Eq.\eqref{eq:EoutjE}. We have
\begin{equation}\label{eq:EMout3dSph}
\begin{split}
 \vec{E}^{\text{Out}}(\vec{r},\w) 
    &= \sum_{\ell m}G_{_E}^{\text{Out}}(r,\w,\ell)\ \multj^E(\w,\ell,m)\left[\frac{\ell+1}{r^2}\mathscr{Y}_{\ell m}\hat{e}_r-\frac{1}{r} \
    \vec{\nabla}\mathscr{Y}_{\ell m}\right]\\
    &+\sum_{\ell m}\frac{\w^2}{\ell(2\ell-1)}G_{_E}^{\text{Out}}(r,\w,\ell-1)\ \multj^E(\w,\ell,m)\vec{\nabla}\mathscr{Y}_{\ell m}\\
    &+i\w\ \sum_{\ell m} G_{_B}^{\text{Out}}(r,\w,\ell)\ \multj^{\vee B}(\w,\ell,m)\ \vec{\mathbb{U}}_{\ell m}\ ,\\
\vec{B}^{\text{Out}}(\vec{r},\w)    &= \sum_{\ell m}G_{_B}^{\text{Out}}(r,\w,\ell)\ \multj^{\vee B}(\w,\ell,m)\left[\frac{\ell+1}{r^2}\mathscr{Y}_{\ell m}\hat{e}_r-\frac{1}{r} \
    \vec{\nabla}\mathscr{Y}_{\ell m}\right]\\
&+\sum_{\ell m}\frac{\w^2}{\ell(2\ell-1)}G_{_B}^{\text{Out}}(r,\w,\ell-1)\ \multj^{\vee B}(\w,\ell,m)\vec{\nabla}\mathscr{Y}_{\ell m}\\
    &-i\w\ \sum_{\ell m} G_{_E}^{\text{Out}}(r,\w,\ell)\ \multj^E(\w,\ell,m)\ \vec{\mathbb{U}}_{\ell m}\ .
\end{split}
\end{equation}
In $d=3$, the electric outgoing fields are the same as the magnetic ones, i.e.,
\begin{align}
\begin{split}
G_{_E}^{\text{Out}}(r,\w,\ell)=G_{_B}^{\text{Out}}(r,\w,\ell)\equiv r^{-\ell}{}_0F_1\left[\frac{1}{2}-\ell,-\frac{\w^2r^2}{4}\right]= \frac{\theta_{\ell}(-i\w r)}{(2\ell-1)!!} \frac{e^{i\w r}}{r^{\ell}}\ ,
\end{split}
\end{align}
where the $\theta_\ell(z)$ are the  reverse Bessel polynomials (see  Table \ref{tab:SmallTheta} for explicit expressions). We remind the reader that the above function is essentially the outgoing, spherical Hankel function up to a frequency-dependent normalisation factor:
\begin{equation}
{}_0F_1\left[\frac{1}{2}-\ell,-\frac{\w^2r^2}{4}\right]= \frac{i(\w r)^{\ell}}{(2\ell-1)!!}h^{(1)}_\ell(\w r)\ .  
\end{equation}

The above expressions for outgoing EM waves are consistent with EM duality, which maps
\begin{equation}\label{eq:EMduality}
(\vec{E},\multj^E)\mapsto (\vec{B},\multj^{\vee B})\ ,\quad  (\vec{B},\multj^{\vee B})\mapsto (-\vec{E},-\multj^E)\ .       
\end{equation}
The explicit expressions for the spherical multipole moments are given by
\begin{equation}\label{eq:jEJBch3d}
\begin{split}
\multj^E(\w,\ell,\vec{m}) &\equiv\frac{1}{(2\ell+1)(\ell+1)}\int_{\vec{r}\in\mathbb{R}^{3}} \  r^{\ell-1}\mathscr{Y}^\ast_{\ell m}(\hat{r})\ {}_0F_1\left[\ell+\frac{3}{2},-\frac{\w^2r^2}{4}\right]\\
&\qquad\times
\Bigl\{i\w r^2 J^r(\vec{r},\w) -\partial_{r}[r^2J^t(\vec{r},\w)]\Bigr\}\ ,\\
\overbar{\multj}^{\vee B}(\w,\ell,m) &\equiv\frac{1}{2\ell+1}\frac{\ell}{\ell+1}\int_{\vec{r}\in \mathbb{R}^3}r^{\ell+1}
\mathbb{U}_I^{\ell m\ast}(\hat{r}){}_0F_1\left[\ell+\frac{3}{2},-\frac{\w^2r^2}{4}\right]J^{I}(\vec{r},\w)\\
    &= \frac{1}{(2\ell+1)(\ell+1)}\int_{\vec{r}_0\in \mathbb{R}^3}\ r^\ell{}_0F_1\left[\ell+\frac{3}{2},-\frac{\w^2r^2}{4}\right][\vec{r}\times \vec{J}(\vec{r},\w)]\cdot \vec{\nabla} \Yellma(\hat{r})\ .
\end{split}
\end{equation}
Here, we have specialised the electric moment of Eq.\eqref{eq:jEGen} to $d=3$, and we have generalised the magnetic moment in Eq.\eqref{eq:JBcheck} to dynamical situations. The ${}_0F_1$ is the time-delay smearing function and is essentially the  spherical Bessel function up to a normalisation:
\begin{equation}\label{eq:0F13d}
  {}_0F_1\left[\ell+\frac{3}{2},-\frac{\w^2r^2}{4}\right]=\frac{(2\ell+1)!!}{(\w r)^\ell}j_{\ell}(\w r)=\frac{(2\ell+1)!!}{2^\ell\ell!}\int_{-1}^{1}(1-z^2)^\ell\ e^{\pm i\w z r}\ .  
\end{equation}

At large $r$, far away from sources, the EM fields given in Eq.\eqref{eq:EMout3dSph} become
\begin{equation}\label{eq:RadFields3d}
\begin{split}
 \vec{E}_{\text{Rad}}(\vec{r},\w) 
    &= -e^{i\w r}\sum_{\ell m}\frac{(-i\w)^{\ell+1}}{\ell(2\ell-1)!!} \left[ \multj^{E}(\w,\ell,m)\
    \vec{\nabla}\mathscr{Y}_{\ell m}
-\multj^{\vee B}(\w,\ell,m)\ 
\hat{e}_r\times\vec{\nabla}\mathscr{Y}_{\ell m}\right]\ ,\\
\vec{B}_{\text{Rad}}(\vec{r},\w) &= -e^{i\w r}\sum_{\ell m}\frac{(-i\w)^{\ell+1}}{\ell(2\ell-1)!!} \left[ \multj^{E}(\w,\ell,m)\ \hat{e}_r\times\vec{\nabla}\mathscr{Y}_{\ell m}\ + \multj^{\vee B}(\w,\ell,m)\ \vec{\nabla}\mathscr{Y}_{\ell m}\right] \ .
\end{split}
\end{equation}
Here the sum runs over $\ell\geq 1$, and we recognise in these formulae the transverse, radially outgoing  EM waves with $\vec{B}_{\text{Rad}}=\hat{e}_r\times \vec{E}_{\text{Rad}}$.
The cartesian forms of these equations can be obtained by using the definitions
\begin{equation}
\begin{split}
\sum_{m}\multj^{E}(\w,\ell,m) \mathscr{Y}_{\ell m}(\hat{r})\equiv \frac{(2\ell-1)!!}{\ell!}\ {}^E\mathcal{Q}(\w)_{<i_1 i_2\ldots i_\ell>}\frac{x^{i_1}\ldots x^{i_\ell}}{4\pi r^\ell}\ ,\\
\sum_{m}\multj^{\vee B}(\w,\ell,m) \mathscr{Y}_{\ell m}(\hat{r})\equiv \frac{(2\ell-1)!!}{\ell!}\ {}^B\mathcal{Q}^\vee(\w)_{<i_1 i_2\ldots i_\ell>}\frac{x^{i_1}\ldots x^{i_\ell}}{4\pi r^\ell}\ .
\end{split}
\end{equation}
These generalise the relation between spherical and cartesian STF moments beyond statics. The explicit expressions for the cartesian moments are given by
\begin{equation}\label{eq:QEQBch3d}
\begin{split}
{}^E\mathcal{Q}(\w)_{<i_1 i_2\ldots i_\ell>}&\equiv\frac{1}{\ell+1}\int_{\vec{r}\in\mathbb{R}^{3}} \  x^{<i_1}x^{i_2}\ldots x^{i_\ell>} \ {}_0F_1\left[\ell+\frac{3}{2},-\frac{\w^2r^2}{4}\right]\\
&\qquad\qquad\times
\Bigl\{i\w r J^r(\vec{r},\w) -\frac{1}{r}\partial_{r}[r^2J^t(\vec{r},\w)]\Bigr\}\ ,\\
{}^B\mathcal{Q}^\vee(\w)_{<i_1 i_2\ldots i_\ell>}
    &= \frac{\ell}{\ell+1}\int_{\vec{r}_0\in \mathbb{R}^3}\ {}_0F_1\left[\ell+\frac{3}{2},-\frac{\w^2r^2}{4}\right][\vec{r}\times \vec{J}(\vec{r},\w)]^{<i_1}x^{i_2}\ldots x^{i_\ell>}\ .
\end{split}
\end{equation}
It is then straightforward to see that Eq.\eqref{eq:QbcToQb} and Eq.\eqref{eq:QbToQbc} also generalise to time-dependent situations.

The cartesian components of the EM fields can be worked out by directly substituting the above definitions into Eq.\eqref{eq:EMout3dSph}. Alternately, we can take the cartesian Electric fields in general dimensions (i.e. the sum of electric fields in Eq.\eqref{eq:EMcartGenDE} and Eq.\eqref{eq:EMcartGenDE}), and specialise to $d=3$. In either case, once the electric field has been figured out, the magnetic field expressions follow from EM duality. The EM fields outside the sources evaluate to
\begin{equation}
\begin{split}
E_k^{\text{Out}}(\vec{r},\w) &=e^{i\w r}\sum_{\ell}\frac{\theta_{\ell}(-i\w r)}{\ell! }{}^E\mathcal{Q}(\w)_{<i_1 i_2\ldots i_\ell>}  [(2\ell+1)n^k n^{i_\ell}-\ell\delta^{ki_\ell}]\frac{ n^{i_1}\ldots n^{i_{\ell-1}}}{4\pi r^{\ell+2} }\\
&\qquad +\w^2 e^{i\w r} \sum_{\ell} \frac{\theta_{\ell-1}(-i\w r) }{\ell! }{}^E\mathcal{Q}(\w)_{<i_1 i_2\ldots i_\ell>}[\delta^{ki_\ell}-n^k n^{i_\ell}]\frac{n^{i_1}\ldots n^{i_{\ell-1}}}{4\pi r^{\ell}}\\
&\qquad +i\w e^{i\w r} \varepsilon_{kji_1}  \sum_{\ell} \frac{\theta_{\ell}(-i\w r)}{\ell!}{}^B\mathcal{Q}^\vee(\w)_{<ji_2\ldots i_{\ell}>}\ 
\frac{n^{i_1}\ldots n^{i_\ell}}{4\pi r^{\ell+1}},\\
B_k^{\text{Out}}(\vec{r},\w) &=e^{i\w r}\sum_{\ell}\frac{\theta_{\ell}(-i\w r)}{\ell! }{}^B\mathcal{Q}^\vee(\w)_{<i_1 i_2\ldots i_\ell>}  [(2\ell+1)n^k n^{i_\ell}-\ell\delta^{ki_\ell}]\frac{ n^{i_1}\ldots n^{i_{\ell-1}}}{4\pi r^{\ell+2} }\\
&\qquad +\w^2 e^{i\w r} \sum_{\ell} \frac{\theta_{\ell-1}(-i\w r) }{\ell! }{}^B\mathcal{Q}^\vee(\w)_{<i_1 i_2\ldots i_\ell>}[\delta^{ki_\ell}-n^k n^{i_\ell}]\frac{n^{i_1}\ldots n^{i_{\ell-1}}}{4\pi r^{\ell}}\\
&\qquad -i\w e^{i\w r} \varepsilon_{kji_1}  \sum_{\ell} \frac{\theta_{\ell}(-i\w r)}{\ell!}{}^E\mathcal{Q}(\w)_{<ji_2\ldots i_{\ell}>}\ 
\frac{n^{i_1}\ldots n^{i_\ell}}{4\pi r^{\ell+1}}\ . 
\end{split}
\end{equation}
Here we have used a shorthand $n^i\equiv \frac{x^i}{r}$. The reader might recognise the form appearing in the first line of the fields from the EM fields outside static multipoles. Only the second and the third lines in each field contribute to the radiative part. These are again manifestly duality invariant if we also take 
\begin{equation}
{}^E\mathcal{Q}\mapsto {}^B\mathcal{Q}^\vee\ ,\quad  {}^B\mathcal{Q}^\vee\mapsto -{}^E\mathcal{Q}\ .       
\end{equation}
With some relabelling of indices, the radiative EM fields can be cast into the following form
\begin{equation}
\begin{split}
E_k^{\text{Rad}}(\vec{r},\w) &=-
 \frac{e^{i\w r}}{4\pi r} \sum_{\ell} \frac{(-i\w )^{\ell+1} }{\ell! }\\
 &\quad\times\Biggl\{{}^E\mathcal{Q}(\w)_{<i_1 i_2\ldots i_\ell>}[\delta^{ki_1}-n^k n^{i_1}]+ \varepsilon_{kji_1}  {}^B\mathcal{Q}^\vee(\w)_{<ji_2\ldots i_{\ell}>}\ 
n^{i_1}\Biggr\}n^{i_2}\ldots n^{i_{\ell}}\\
B_k^{\text{Rad}}(\vec{r},\w) &= -\frac{e^{i\w r}}{4\pi r} \sum_{\ell} \frac{(-i\w )^{\ell+1} }{\ell! }\\
 &\quad\times\Biggl\{{}^B\mathcal{Q}^\vee(\w)_{<i_1 i_2\ldots i_\ell>}[\delta^{ki_1}-n^k n^{i_1}]- \varepsilon_{kji_1}  {}^E\mathcal{Q}(\w)_{<ji_2\ldots i_{\ell}>}\ 
n^{i_1}\Biggr\}n^{i_2}\ldots n^{i_{\ell}} . 
\end{split}
\end{equation}

The scalar and vector potentials corresponding to the above EM fields can be written down by choosing a gauge. Combining Eqs.\eqref{eq:EMcartGenDB} and \eqref{eq:EMcartGenDEnew}, we can write
\begin{equation}
\begin{split}
\gaugeVf_t^{\text{Out,New}}(\vec{r},\w)
&= -\sum_{\ell} \frac{G_{_E}^{\text{Out}}(r,\w,\ell)}{\ell! \nn_{d,\ell-1}|\mathbb{S}^{d-1}|}{}^E\mathcal{Q}(\w)_{<i_1 i_2\ldots i_\ell>}\frac{x^{i_1}\ldots x^{i_\ell}}{ r^{\ell+d-2}}\ ,\\
\gaugeVf_k^{\text{Out,New}}(\vec{r},\w)
&=- i\w\sum_{\ell>0}\frac{G_{_E}^{\text{Out}}(r,\w,\ell-1)}{\ell! \nn_{d,\ell-2}|\mathbb{S}^{d-1}|}\ {}^E\mathcal{Q}(\w)_{<ki_1 i_2\ldots i_{\ell-1}>}\frac{x^{i_1}\ldots x^{i_{\ell-1}}}{ r^{\ell+d-3}}\\
&\qquad+\sum_{\ell>0} \frac{G_{_B}^{\text{Out}}(r,\w,\ell)}{\ell! \nn_{d,\ell-1}|\mathbb{S}^{d-1}|}{}^B\mathcal{Q}(\w)_{k<i_1 i_2\ldots i_\ell>}\frac{x^{i_1}\ldots x^{i_\ell}}{ 4\pi r^{\ell+1}}
\ .
\end{split}
\end{equation}
Specialising to $d=3$ and rewriting the magnetic moment part accordingly, we obtain
\begin{equation}
\begin{split}
\gaugeVf_t^{\text{Out,New}}(\vec{r},\w)
&= -e^{i\w r}\sum_{\ell} \frac{\theta_{\ell}(-i\w r) }{\ell! }{}^E\mathcal{Q}(\w)_{<i_1 i_2\ldots i_\ell>}\frac{n^{i_1}\ldots n^{i_\ell}}{ 4\pi r^{\ell+1}}\ ,\\
\gaugeVf_k^{\text{Out,New}}(\vec{r},\w)
&=- i\w e^{i\w r}\sum_{\ell>0}\frac{\theta_{\ell-1}(-i\w r) }{\ell! }\ {}^E\mathcal{Q}(\w)_{<ki_1 i_2\ldots i_{\ell-1}>}\frac{n^{i_1}\ldots n^{i_{\ell-1}}}{4\pi r^{\ell}}\\
&\qquad+e^{i\w r}\varepsilon_{kji_1}\sum_{\ell>0} \frac{\theta_{\ell}(-i\w r) }{\ell! }{}^B\mathcal{Q}^\vee(\w)_{<j i_2\ldots i_\ell>}\frac{n^{i_1}\ldots n^{i_\ell}}{ r^{\ell+1}}
\ .
\end{split}
\end{equation}
These outgoing solutions and their time-reversed counterparts are useful in quantising electrodynamics in spherical coordinates. Far away from the sources, we get the radiative fields
\begin{equation}
\begin{split}
\gaugeVf_t^{\text{Rad,New}}(\vec{r},\w)
&= -\frac{e^{i\w r}}{4\pi r}\sum_{\ell} \frac{(-i\w)^\ell }{\ell! }{}^E\mathcal{Q}(\w)_{<i_1 i_2\ldots i_\ell>}n^{i_1}\ldots n^{i_\ell}\ ,\\
\gaugeVf_k^{\text{Rad,New}}(\vec{r},\w)
&= \frac{e^{i\w r}}{4\pi r}\sum_{\ell>0}\frac{(-i\w )^\ell }{\ell! }\Biggl\{
{}^E\mathcal{Q}(\w)_{<ki_1 i_2\ldots i_{\ell-1}>}
+\varepsilon_{kji_\ell}n_{i_\ell}{}^B\mathcal{Q}^\vee(\w)_{<j i_1\ldots i_{\ell-1}>}\Biggr\}n^{i_1}\ldots n^{i_{\ell-1}}
\ .
\end{split}
\end{equation}

\subsubsection*{Power loss in $d=3$}
We will now show how the power loss formulae in general dimension can be cast into familiar forms when $d=3$. The main point is to rewrite the magnetic multipole power loss to make the EM duality manifest. In terms of spherical multipoles, this is easy:  we start with the power radiated in terms of our multipole moments is
\begin{equation}
    \mathcal{P}(\w)=\sum_{\ell\vec{m}}\frac{2\pi}{\Gamma(\nu)^2}\frac{\w^{2\nu+1}}{2^{2\nu}}\frac{\ell+d-2}{\ell}|\multj^E(\w,\ell,\vec{m})|^2+\sum_{\alpha\ell\vec{m}}\frac{2\pi}{\Gamma(\nu)^2}\frac{\w^{2\nu+1}}{2^{2\nu}}|\multj^B(\w,\alpha,\ell,\vec{m})|^2,
\end{equation}
and rewrite it in $d=3$ using Eq.\eqref{eq:JBcheck} to get
\begin{equation}
    \mathcal{P}(\w)_{d=3}=\sum_{\ell m}\frac{\w^{2\ell+2}}{[(2\ell-1)!!]^2}\frac{\ell+1}{\ell}\left[|\multj^E(\w,\ell,m)|^2+|\multj^{\vee B}(\w,\ell,m)|^2\right]\ .
\end{equation}
This expression is now manifestly invariant under EM duality.

Power loss in terms of STF moments requires a bit more work: the $d=3$ version of Eq.\eqref{eq:LarmorFlatSTF}  is 
\begin{equation}
\begin{split}\label{eq:STFpower3d}
\mathcal{P}(\w)_{d=3}&=\frac{1}{4\pi}\sum_{\ell=1}^\infty \frac{\ell+1}{\ell}\frac{\w^{2\ell+2}}{(2\ell+1)!!}\ \frac{1}{\ell!}\ 
\left[{}^E\mathcal{Q}(\w)_{<i_1\ldots i_\ell>}\right]^\ast\ \left[{}^E\mathcal{Q}(\w)_{<i_1\ldots i_\ell>}\right]  \\
&\quad +\frac{1}{4\pi}\sum_{\ell=1}^\infty \frac{\w^{2\ell+2}}{(2\ell+1)!!}\ \frac{1}{\ell!}\ 
\left[{}^B\mathcal{Q}(\w)_{k<i_1\ldots i_\ell>}\right]^\ast\ \left[{}^
B\mathcal{Q}(\w)_{k<i_1\ldots i_\ell>}\right]  \\
\end{split}\end{equation}
We want to rewrite this formula in terms of ${}^B\mathcal{Q}^\vee$ tensor. Using Eq.\eqref{eq:QbcToQb}, we can write 
\begin{equation}
\begin{split}
    &\frac{1}{\ell!}[{}^B\mathcal{Q}_{k<i_1i_2\ldots i_\ell>}]^\ast[{}^B\mathcal{Q}^{k<i_1i_2\ldots i_\ell>}]|_{d=3} \\
        &=\frac{1}{\ell!}\left[\frac{1}{\ell}\sum_{p=1}^\ell\varepsilon_{k q i_p}{}^B\mathcal{Q}^{\vee <qi_1i_2\dots i_{\underline{p}}\ldots i_{\ell-1}>}\right]^\ast\left[\frac{1}{\ell}\sum_{p'=1}^\ell\varepsilon_{k q' i_{p'}}{}^B\mathcal{Q}^\vee_{<q'i_1i_2\dots i_{\underline{p'}}\ldots i_{\ell-1}>}\right]\ .
\end{split}
\end{equation}
In this product of sums made of $\ell^2$ terms, we get $\ell$ terms with $p=p'$ and $\ell(\ell-1)$ terms with $p\neq p'$. Here every $p\neq p'$ term evaluates to 
\begin{equation}\begin{split}
\varepsilon_{kq_1i_1}&\varepsilon_{kq_2i_2}[{}^B\mathcal{Q}^{\vee<q_1i_2i_3\dots  i_{\ell}>}]^\ast[{}^B\mathcal{Q}^\vee_{<q_2i_1i_3\dots  i_{\ell}>}]\\
&=(\delta_{q_1 q_2}\delta_{i_1i_2}-\delta_{q_1i_2}\delta_{q_2i_1})[{}^B\mathcal{Q}^{\vee<q_1i_2i_3\dots  i_{\ell}>}]^\ast[{}^B\mathcal{Q}^\vee_{<q_2i_1i_3\dots  i_{\ell}>}]\\
&=[{}^B\mathcal{Q}^{\vee<i_1i_2i_3\dots  i_{\ell}>}][{}^B\mathcal{Q}^\vee_{<i_1i_2i_3\dots  i_{\ell}>}]\ .
\end{split}\end{equation}
Here, the second term evaluates to zero because of the trace-free condition on ${}^B\mathcal{Q}^\vee$.
Every $p=p'$ term evaluates to 
\begin{equation}\begin{split}
\varepsilon_{kq_1i_1}\varepsilon_{kq_2i_1}[{}^B\mathcal{Q}^{\vee<q_1i_2i_3\dots  i_{\ell}>}]^\ast[{}^B\mathcal{Q}^\vee_{<q_2i_2i_3\dots  i_{\ell}>}]
&=2\ [{}^B\mathcal{Q}^{\vee<i_1i_2i_3\dots  i_{\ell}>}]^\ast[{}^B\mathcal{Q}^\vee_{<i_1i_2i_3\dots  i_{\ell}>}]\ .
\end{split}\end{equation}
Adding up all the terms yields then a factor of $\ell(\ell-1)+2\ell=\ell(\ell+1)$. Thus, we have proved that 
\begin{equation}
\begin{split}
    \frac{1}{\ell!}[{}^B\mathcal{Q}_{k<i_1i_2\ldots i_\ell>}]^\ast[{}^B\mathcal{Q}^{k<i_1i_2\ldots i_\ell>}]|_{d=3} 
        &=\frac{\ell+1}{\ell}\times\frac{1}{\ell!}[{}^B\mathcal{Q}^{\vee<i_1i_2i_3\dots  i_{\ell}>}]^\ast[{}^B\mathcal{Q}^\vee_{<i_1i_2i_3\dots  i_{\ell}>}]\ .
\end{split}
\end{equation}
This identity can be used to recast the power loss formula in Eq.\eqref{eq:STFpower3d} into a duality invariant form:
\begin{equation}
\begin{split}\label{eq:STFpower3dalt}
\mathcal{P}(\w)_{d=3}&=\frac{1}{4\pi}\sum_{\ell=1}^\infty \frac{\ell+1}{\ell}\frac{\w^{2\ell+2}}{(2\ell+1)!!}\\
&\quad\times\frac{1}{\ell!}\ \Biggl\{  
\left[{}^E\mathcal{Q}_{<i_1\ldots i_\ell>}\right]^\ast\ \left[{}^E\mathcal{Q}_{<i_1\ldots i_\ell>}\right] +[{}^B\mathcal{Q}^{\vee<i_1i_2i_3\dots  i_{\ell}>}]^\ast[{}^B\mathcal{Q}^\vee_{<i_1i_2i_3\dots  i_{\ell}>}]\Biggr\} \ .
\end{split}\end{equation}

The first few terms here correspond to electric/magnetic dipole/quadrupole moments, i.e., if we set
\begin{equation}\begin{split}
{}^E\mathcal{Q}_{i_1}&\equiv d^E_{i_1} \ ,\quad {}^E\mathcal{Q}_{<i_1 i_2>}\equiv q^E_{<i_1 i_2>} \ ,\quad
{}^B\mathcal{Q}^\vee_{ki_1}\equiv d^B_{i_1} \ ,\quad {}^B\mathcal{Q}^\vee_{<i_1 i_2>}=q^B_{<i_1i_2>} \ .
\end{split}\end{equation}
The power loss formula then takes the form
\begin{equation}\begin{split}
\mathcal{P}(\w)_{d=3}&=\frac{\w^4}{6\pi}\sum_{i}|d^E_{i}|^2+ \frac{\w^6}{80\pi}\sum_{ij}|q^E_{<ij>}|^2+\ldots\\
&\quad +\frac{\w^4}{6\pi}\sum_{i}|d^B_{i}|^2+ \frac{\w^6}{80\pi}\sum_{ij}|q^B_{<ij>}|^2+\ldots
\end{split}\end{equation}
This agrees with the Larmor formula quoted in standard textbooks.\footnote{For example, Jackson defines his cartesian STF electric quadrupole moment to be three times our quadrupole moment, and his definition of power loss in the frequency domain is half of ours due to Fourier transform conventions. So, in his textbook, he gives $2\times 6\pi=12\pi$ as the denominator for dipole power loss and $2\times 3^2 \times 80\pi=1440\pi$ as the denominator for quadrupole power loss.}

\subsection{Comparison of normalisations}\label{sec:MultNflat}
In our discussion of multipole expansions, we have chosen our multipole moment definitions uniformly across statics and radiation, and we have tried to use the simplest normalisations consistent with electric-magnetic duality. This is consistent with modern treatments of gravitational multipole expansions based on STF tensors. Unfortunately, our notations differ from popular textbooks on electromagnetism where  \emph{radiative} multipole moments are treated very differently from \emph{static} multipole moments. This fact complicates the comparison of our expression with those available in standard EM textbooks like that of Jackson\cite{Jackson:1998nia} and Zangwill\cite{zangwill2013modern}.  Our normalisations also differ slightly from papers on STF multipole expansion \cite{Damour:1990gj,Ross:2012fc,Amalberti:2023ohj}. Given this bewildering array of existent normalisations for EM multipoles, we will conclude this appendix by providing the necessary dictionary for translation to the notations in such texts and papers. There are no new results here, and a reader disinterested in notational fine print may safely skip what follows.

Our expressions can be converted to static moments appearing in textbooks via
\begin{equation}
\begin{split}
    \bar{\mathcal{J}}^E(\ell,m)&=\frac{1}{2\ell+1}q^{\text{Jackson}}_{\ell m}=\frac{1}{4\pi}A^{\text{Zangwill}}_{\ell m}=\frac{Q^{(e)\text{LBP}}_{\ell m}}{\sqrt{4\pi(2\ell+1)}}\ , \\
      \overbar{\multj}^{\vee B}(\ell,m) &= \ell\frac{\mu^{\text{Jackson}}_{\ell m}}{\sqrt{4\pi(2\ell+1)}}= \frac{1}{2\ell+1}M^{\text{Zangwill}}_{\ell m} = \frac{Q^{(m)\text{LBP}}_{\ell m}}{\sqrt{4\pi(2\ell+1)}}\ .
\end{split}
\end{equation}
These factors can be figured out by comparing our static moments (i.e., Eqs.\eqref{eq:jEstat} and \eqref{eq:JBcheck}) against the definitions given in these texts.\footnote{The static moments of Jackson\cite{Jackson:1998nia} are defined in JEq(4.3) and Problem(5.8b),  that of  Zangwill\cite{zangwill2013modern} are defined in ZEq.(4.87) and ZEq.(11.66), and of  Lifshitz-Berestetskii-Pitaevskii\cite{Berestetskii:1982qgu} are given in LBPEq.(46.7) and LBPEq(47.3).} As indicated in the footnote, Jackson does not define a magnetostatic multipole moment in his text. Instead, in Problem(5.8b) involving axisymmetric currents of the form $\vec{J}=J_\varphi(r,\theta)\hat{\varphi}$, Jackson defines
\begin{equation}
    \mu^{\text{Jackson}}_{\ell,m=0}=-\frac{1}{\ell(\ell+1)}\int_{\mathbb{R}^3} r^\ell P^1_\ell(\cos\vartheta) \bar{J}_\varphi(r,\theta)=\frac{1}{\ell+1}\sqrt{\frac{4\pi}{2\ell+1}}\int_{\mathbb{R}^3} r^{\ell+1}\vec{\mathbb{U}}^\ast_{\ell, m=0}(\hat{r})\cdot \bar{J}(\vec{r})\ ,
\end{equation}
where we have used the fact that 
\begin{equation}
\vec{r}\times\vec{\nabla}\mathscr{Y}_{\ell,m=0}= \hat{e}_{\varphi}\sqrt{\frac{2\ell+1}{4\pi}}P^{1}_\ell(\cos\vartheta)\ .
\end{equation}
We have generalised Jackson's definition to general $m$ to determine the relative normalisation quoted above. With this dictionary, we have checked that the multipole expansions for static fields given in these texts agree with our expressions.

Moving on to radiative multipole moments, the relative normalisations are given by\footnote{This follows from comparing Zangwill's ZEq.(20.224),(20.225), as well as Jackson's JEq.( 9.167), (9.168) against our Eq.\eqref{eq:jEJBch3d}.} 
\begin{equation}\label{eq:RadFJZ}
\begin{split}
\Lambda_{\ell m}^{E,\text{Zangwill}}(\w)&=\frac{1}{\sqrt{\ell(\ell+1)}} \  a_E^{\text{Jackson}}(\w,\ell,m)=-i\frac{\w^{\ell+2}}{\ell(2\ell-1)!!} \ \multj^{E}(\w,\ell,m)\ ,\\
\Lambda_{\ell m}^{M,\text{Zangwill}}(\w)&=\frac{1}{\sqrt{\ell(\ell+1)}}   a_M^{\text{Jackson}}(\w,\ell,m)=i\frac{\w^{\ell+2}}{\ell(2\ell-1)!!} \multj^{\vee B}(\w,\ell,m)\ .
\end{split}
\end{equation}
Our normalisations are closer to that of Campbell-Macek-Morgan\cite{Campbell:1977jf} with
\begin{equation}
\begin{split}
    Q^{\text{CMM}}_{\ell m}(\w)=(2\ell+1)\multj^E(\w,\ell,m)\ , \qquad
    M^{\text{CMM}}_{\ell m}(\w)=(2\ell+1)\multj^{\vee B}(\w,\ell,m) \ .
\end{split}
\end{equation}
With this dictionary, we have checked that the multipole expansions for radiative fields as well as power loss given in these texts agree with our expressions.\footnote{Note that power loss formulae in Jackson's chapter\S 9 are half of ours due to differences in Fourier transform conventions.}

We now turn to STF multipole moments. In general $d$, the authors Amalberti-Larrouturou-Yang (ALY) \cite{Amalberti:2023ohj} define STF electric and magnetic moments.  Their definitions are related to ours by
\begin{equation}
\begin{split}
{}^E\mathcal{Q}_{<i_1 i_2\ldots i_\ell>}&=I_{<i_1i_2\dots i_\ell>}^{\text{ALY}}\ ,\quad
{}^B\mathcal{Q}_{i<i_1 i_2\ldots i_\ell>}=\frac{2\ell}{\ell+1}J_{i<i_1i_2\dots i_\ell>}^{\text{ALY}}
\ .
\end{split}
\end{equation}
These normalisations can be fixed by comparing the Fourier transform of ALYEq(3.17) against our definitions in Eq.\eqref{eq:QEgen} and Eq.\eqref{eq:QBgen}. For the vector projector, ALY seem to use Eq.\eqref{eq:PiVAntiTF}, but their formula seems to omit the final symmetrisation. The conversion rule quoted above assumes that such a symmetrisation is implicit in their expressions.\footnote{The reader should note that  ALY's anti-symmetric projection involves an additional factor of half compared to our conventions here.}

The rest of the references dealing with STF moments are specific to $d=3$: Damour-Iyer\cite{Damour:1990gj} and Ross\cite{Ross:2012fc}. We claim that  the relative normalisation factors are
 \begin{equation}
\begin{split}
{}^E\mathcal{Q}_{<i_1 i_2\ldots i_\ell>}&= Q_{<i_1 i_2\ldots i_{\ell}>}^{\text{Damour-Iyer}}=I^{<i_1i_2\dots i_\ell>}_{\text{Ross}}\ ,\\
{}^B\mathcal{Q}^\vee_{<i_1 i_2\ldots i_\ell>}&= \frac{\ell}{\ell+1}M_{<i_1 i_2\ldots i_{\ell}>}^{\text{Damour-Iyer}}=\frac{\ell}{\ell+1}J^{<i_1i_2\dots i_\ell>}_{\text{Ross}}
\ .
\end{split}
\end{equation}
Here, we have converted the time-domain expressions of \cite{Damour:1990gj,Ross:2012fc} into the frequency domain. The Damour-Iyer definitions in  DIeq(4.18) of \cite{Damour:1990gj} can be  Fourier transformed to frequency domain as
\begin{equation}
\begin{split}
   Q^\text{Damour-Iyer}_{<i_1 i_2\ldots i_{\ell}>} &=  \frac{1}{\ell+1}\int_{\mathbb{R}^3}\  \  x^{<i_1}x^{i_2}\ldots x^{i_{\ell}>} \int\limits_{-1}^{1} dz \ \frac{(2\ell+1)!!}{2^{\ell+1}\ell!}(1-z^2)^{\ell}e^{-i\w r z} \\
   &\qquad \times \left[(\ell+1-i\w rz)J^t(\vec{r},\w)+i\w J^r(\vec{r},\w)\right] \ , \\
   M^\text{Damour-Iyer}_{<i_1 i_2\ldots i_{\ell}>} &= \frac{1}{\ell}  \int_{\mathbb{R}^3}\ \int\limits_{-1}^{1} dz \ \frac{(2\ell+1)!!}{2^{\ell+1}\ell!}(1-z^2)^{\ell}e^{-i\w r z} \  x^{<i_1}x^{i_2}\ldots x^{i_{\ell}>}  (\vec{r}\times\vec{\nabla})\cdot \vec{J}(\vec{r},\w) \ .
\end{split}
\end{equation}
To relate it to Eq.\eqref{eq:QEQBch3d},  we perform an integration by parts: 
\begin{equation}
\begin{split}
   Q^\text{Damour-Iyer}_{<i_1 i_2\ldots i_{\ell}>} &=  \frac{1}{\ell+1}\int_{\mathbb{R}^3}\  \  x^{<i_1}x^{i_2}\ldots x^{i_{\ell}>} \int\limits_{-1}^{1} dz \ \frac{(2\ell+1)!!}{2^{\ell+1}\ell!}(1-z^2)^{\ell}e^{-i\w r z} \\
   &\qquad \times \Bigl\{i\w r J^r(\vec{r},\w) -\frac{1}{r}\partial_{r}[r^2J^t(\vec{r},\w)]\Bigr\} \ , \\
   M^\text{Damour-Iyer}_{<i_1 i_2\ldots i_{\ell}>} &=  \int_{\mathbb{R}^3}\ \int\limits_{-1}^{1} dz \ \frac{(2\ell+1)!!}{2^{\ell+1}\ell!}(1-z^2)^{\ell}e^{-i\w r z} \  [\vec{r}\times \vec{J}(\vec{r},\w)]^{<i_1}x^{i_2}\ldots x^{i_\ell>} \ .
\end{split}
\end{equation}
Invoking Eq.\eqref{eq:0F13d}, we then obtain the normalisations claimed above.

As for Ross\cite{Ross:2012fc}, his magnetic moment expressions are directly of the form Eq.\eqref{eq:0F13d}. After converting to the frequency domain, his electric  moment definition is
\begin{equation}
\begin{split}
& I(\w)^{<i_1i_2\dots i_\ell>}_{\text{Ross}}  = \int_{\mathbb{R}^3}\  \Bigg\{ {}_1F_2\left[\frac{\ell}{2}\ ; \frac{\ell}{2}+1\ , \ell+\frac{3}{2}\ ; -\frac{\w^2r^2}{4} \right] J^t (\vec{r},\w)\ x^{<i_1}x^{i_2}\dots x^{i_\ell>} \\
&  -i\w \frac{\ell}{(\ell+1)(\ell+2)}{}_1F_2\left[\frac{\ell}{2}+1\ ; \frac{\ell}{2}+2\ , \ell+\frac{3}{2}\ ; -\frac{\w^2r^2}{4} \right] \\
&\qquad \times \left\{r^2 J^{<i_\ell}(\vec{r},\w)\ x^{<i_1}x^{i_2}\dots x^{i_{\ell-1}>}-x_kJ^k(\vec{r},\w)\ x^{<i_1}x^{i_2}\dots x^{i_\ell>}\right\}\Bigg\}\ .
\end{split}
\end{equation}
From charge conservation, we have the following identity:
\begin{equation}\begin{split}
 &-i\w\int_{\mathbb{R}^3} \ J^0(\vec{r},\w)\  r^{2p}x^{<i_1}x^{i_2}\dots x^{i_{\ell}>}=-\int_{\mathbb{R}^3} \ \partial_kJ^k(\vec{r},\w)\  r^{2p}x^{<i_1}x^{i_2}\dots x^{i_{\ell}>}\\
 &=\int_{\mathbb{R}^3} r^{2p-2}\left[\ell\ J^{<i_\ell}(\vec{r},\w)\ x^{<i_1}x^{i_2}\dots x^{i_{\ell-1}>}+2p\  x_kJ^k(\vec{r},\w) \ x^{<i_1}x^{i_2}\dots x^{i_\ell>}\right]
\end{split}\end{equation}
We use this to write
\begin{equation}
\begin{split}
& I(\w)^{<i_1i_2\dots i_\ell>}_{\text{Ross}}  = \int_{\mathbb{R}^3}\  \Bigg\{ {}_1F_2\left[\frac{\ell}{2}\ ; \frac{\ell}{2}+1\ , \ell+\frac{3}{2}\ ; -\frac{\w^2r^2}{4} \right] J^t (\vec{r},\w)\ x^{<i_1}x^{i_2}\dots x^{i_\ell>} \\
    & - \frac{\w^2 r^2}{(\ell+1)(\ell+2)}{}_1F_2\left[\frac{\ell}{2}+1\ ; \frac{\ell}{2}+2\ , \ell+\frac{3}{2}\ ;- \frac{\w^2r^2}{4} \right]   J^t(\vec{r},\w)\ x^{<i_1}x^{i_2}\dots x^{i_\ell>} \\ 
    & -\frac{1}{(\ell+1)}\Big\{ {}_1F_2\left[\frac{\ell}{2}+1\ ; \frac{\ell}{2}+2\ , \ell+\frac{3}{2}\ ; -\frac{\w^2r^2}{4} \right]  \\ &  \qquad -\frac{\w^2 r^2}{(\ell+4)(2\ell+3)} {}_1F_2 \left[ \frac{\ell}{2}+2\ ; \frac{\ell}{2}+3\ , \ell+\frac{5}{2}\ ; -\frac{\w^2r^2}{4} \right] \Big\} x_kJ^k(\vec{r},\w) \ x^{<i_1}x^{i_2}\dots x^{i_\ell>}\Bigg\}\ . 
\end{split}
\end{equation}
The above expression can then  be further simplified to 
\begin{equation}
\begin{split}
    I(\w)^{<i_1i_2\dots i_\ell>}_{\text{Ross}} = & \frac{1}{\ell+1}\int_{\mathbb{R}^3}\  \Bigg\{\frac{1}{r^{\ell}} \partial_r\left(r^{\ell+1} {}_0F_1\left[\ell+\frac{3}{2},- \frac{\w^2r^2}{4} \right]\right) J^t(\vec{r},\w)\ x^{<i_1}x^{i_2}\dots x^{i_\ell>} \\
    &\qquad\qquad\qquad +i\w {}_0F_1\left[\ell+\frac{3}{2}, -\frac{\w^2r^2}{4} \right]  x_kJ^k(\vec{r},\w) \ x^{<i_1}x^{i_2}\dots x^{i_\ell>}\Bigg\}\ .
\end{split}
\end{equation}
After an integration by parts, this matches our definition. We have also checked that the power loss formula and the radiative fields given by Ross agrees with our expressions.

\subsection*{Dimensional analysis}

This section provides a list of scaling dimensions of various physical quantities defined in our analysis for the reader's convenience. This allows for easy checks on the dimensional compatibility of all the equations. We work in units where the speed of light($c$), the permittivity of vacuum($\epsilon_0$) and the reduced Planck's constant($\hbar$) are set to one. This leaves us a single scaling dimension, which we pick to be mass $[M]$. 

\begin{table}
\centering
\resizebox{\columnwidth}{!}{%
\begin{tabular}{|c||c|c||c|c|}
\hline
\textbf{Quantity}        & \textbf{Quantity}        & \textbf{Mass Dimension}    &\textbf{Quantity}        & \textbf{Mass Dimension}                              \\ \hline\hline
Debye potential & $\left[\Phi_E\right]$    & $-\frac{d-3}{2}-1$                              &
$\left[\Phi_B\right]$    & $\frac{d-3}{2}-1$                               \\ \hline\hline
EM field strength & $\left[\Erf\right]$       & $\frac{d-3}{2}+1$                               &
$\left[\Bvvf\right]$      & $\frac{d-3}{2}-1$                               \\ \hline
&$\left[\Esf\right]$       & $\frac{d-3}{2}$                              &
$\left[\Hsf\right]$       & $\frac{d-3}{2}$                               \\ \hline
&$\left[\Evf\right]$       & $\frac{d-3}{2}$                              &
$\left[\Hvf\right]$       & $\frac{d-3}{2}$ 
\\ \hline\hline
Gauge potential & $\left[\gaugeVf_t\right]$       & $\frac{d-3}{2}$                               &
$\left[\gaugeVf_r\right]$       & $\frac{d-3}{2}$                               \\ \hline
& & & $\left[\gaugeVf_I\right]$       & $\frac{d-3}{2}-1$                              
\\ \hline\hline
Charge/current density & $\left[J^t\right]$       & $\frac{d-3}{2}+2$                               &
$\left[J^r\right]$       & $\frac{d-3}{2}+2$                               \\ \hline
& & & $\left[J^I\right]$       & $\frac{d-3}{2}+3$                               \\ \hline\hline
Spherical multipole moments & $\left[\multj_E\right]$  & $-\frac{d-3}{2}-(\ell+1)$             &
$\left[\multj_B\right]$  & $-\frac{d-3}{2}-(\ell+1)$             \\ \hline\hline
STF multipole moments & $\left[{}^E\mathcal{Q}_{<i_1 i_2\ldots i_\ell>}\right]$  & $-\frac{d-3}{2}-(\ell+1)$             &
$\left[{}^B\mathcal{Q}_{k<i_1 i_2\ldots i_\ell>}\right]$  & $-\frac{d-3}{2}-(\ell+1)$             \\ \hline\hline
Outgoing waves & $\left[G_E^{\text{Out}}\right]$       & $\ell $               &
$\left[G_B^{\text{Out}}\right]$       & $(d-3)+\ell  $        \\ \hline
EM Green fns. & $\left[\mathbb{G}_E\right]$ & $-(d-3)-1$                                      &
$\left[\mathbb{G}_B\right]$ & $(d - 3)-1$                                      \\ \hline
\end{tabular}}
\caption{Mass dimensions of various quantities in the frequency domain. The electric quantities appear in the second and the third column, whereas the magnetic counterparts appear in the third and the fourth. The time domain versions are denoted by a tilde over the symbol, and their mass dimensions are 1 more than the dimensions quoted above.}
\end{table}

For our de Sitter analysis, we will also work with the Hubble's constant, $H$, set to 1. To check the dimensional consistency of equations in the following sections, one should restore the $H$'s, after which they can be checked against the above table. 

\section{EM point source in dS}\label{app:EMptsource}

In this section, we will describe the EM fields of a comoving point source in  dS$_{d+1}$. The discussion here is straightforward and is provided mainly to establish our conventions about multipole moments in dS. We will also frame our discussion in a way that the similarities to holographic renormalisation\cite{Witten:1998qj,Balasubramanian:1999re,Skenderis:2002wp} are obvious. Our discussion here also closely parallels the discussion of EM fields around planar AdS black holes
\cite{Ghosh:2020lel} but with crucial differences. Some other authors have also studied electromagnetic fields in de Sitter in various settings, such as \cite{Allen:1985wd,Ishibashi:2004wx,Higuchi:2008fu,Montaquila:2010at,AtulBhatkar:2021sdr,Bini:2009nes,Bhatkar:2022qhz}.

Consider then a classical point particle placed at the south pole of global de Sitter, at the centre of the southern static patch. We will characterise this point particle by its electric and magnetic multipole moments, which we take to be time-dependent. We will give a precise definition of these moments below for particles in dS, using how EM fields behave as we approach the particle. Our goal would be to characterise the radiative loss suffered by the particle in terms of these moments.

Our definition of near-zone multipole moments is strongly guided by the principle that they should simply extend the corresponding flat spacetime definitions. This fact should be contrasted with \emph{far-zone} multipole moments defined, say using asymptotic behaviour at future time-like infinity. As we shall see, given the different asymptotics of dS, such far-zone moments do not have any simple relation to their flat-space counterparts\cite{Ashtekar:2015lla}.

Let $\FCf_{\mu\nu}$ denote the EM field strength due to the particle in the frequency domain corresponding to outgoing time $u$. In the outgoing coordinates of the static patch, we will expand each component of this field strength into appropriate scalar/vector spherical harmonics as follows:
\begin{equation}
\begin{split}
    \FCf_{ru}(r,\w,\Omega)&\equiv \sum_{\ell\vec{m}}\Er(r,\w,\ell,\vec{m})\Yellm(\Omega) =  \FCf^{ur}(r,\w,\Omega)\ , \\
   \FCf_{rI}(r,\w,\Omega) &\equiv  \sum_{\ell\vec{m}} \Brs(r,\w,\ell,\vec{m}) \mathscr{D}_{I}\Yellm(\Omega)  + \sum_{\alpha\ell\vec{m}}\Brv(r,\w,\alpha,\ell,\vec{m})  \VSH_I(\Omega) = r^2\gamma_{IJ}\FCf^{Ju}(r,\w,\Omega) \ , \\
    \FCf_{Iu}(r,\w,\Omega) &= \sum_{\ell\vec{m}}\Es(r,\w,\ell,\vec{m}) \mathscr{D}_I \Yellm(\Omega) + \sum_{\alpha\ell\vec{m}} \Ev(r,\w,\alpha,\ell,\vec{m})   \VSH_I(\Omega)\ , \\
    \FCf_{IJ}(r,\w,\Omega) &\equiv \sum_{\alpha\ell\vec{m}} \Bvv(r,\w,\alpha,\ell,\vec{m}) \left[\mathscr{D}_I\VSH_J(\Omega)-\mathscr{D}_J \VSH_I(\Omega)\right] = r^4 \gamma_{IK} \gamma_{JL} \FCf^{KL}(r,\w,\alpha,\ell,\vec{m})\ .
\end{split}
\end{equation}
Here $\Omega$ denotes the angular co-ordinates on the sphere $\mathbb{S}^{d-1}$, indices $I,J,K$ denote vector indices on the sphere, $\gamma_{IJ}$ denotes the unit-sphere metric and $\mathscr{D}_{I}$ is its associated covariant derivative. We have also indicated the relation to the tensor $\FCf^{\mu\nu}$ with raised indices for future convenience. Another useful combination is 
\begin{equation}
\begin{split}
    r^2\gamma_{IJ}\FCf^{rJ}(r,\w,\Omega) &= (1-r^2)\FCf_{rI}(r,\w,\Omega) +\FCf_{Iu}(r,\w,\Omega)\\
    &\equiv \sum_{\ell\vec{m}}\Hs(r,\w,\ell,\vec{m}) \mathscr{D}_{I} \Yellm(\Omega) + \sum_{\alpha\ell\vec{m}} \Hv(r,\w,\alpha,\ell,\vec{m}) \VSH_I(\Omega)\ ,
\end{split}
\end{equation}
where we have defined 
\begin{equation}
\begin{split}
    \Hs &\equiv (1-r^2) \Brs + \Es \ ,\quad 
    \Hv \equiv (1-r^2) \Brv + \Ev\ .
\end{split}
\end{equation}
Our definitions here closely parallel the spherical harmonic expansion in flat space described in Appendix 
\ref{app:FlatEMI}. We use calligraphic letters here to encode the fact that the above quantities are defined in the frequency domain conjugate to $u$, which differ, in the flat limit, from frequency domain expressions conjugate to  $t$  by an additional factor of $e^{i\w r}$. We use subscripts $s$ and $v$ to denote sphere vector indices in the scalar/vector sector, respectively. The time-reversal covariant combinations are $\{\Es,\Ev,\Er\}$ and $\{\Hs,\Hv,\Bvv\}$, with the intrinsic time-reversal parity being even and odd for the two sets, respectively.

Given the spherical harmonic decomposition given above, we can recast the Maxwell equations in terms of the above components. The source-free Maxwell equations (or equivalently the Bianchi identity of electromagnetism $\partial_{[\mu}\FC_{\nu\lambda]}=0$) take the following form:
\begin{equation}\label{Eq:dF}
\begin{split}
\Ev &= i\w\ \Bvv \ , \quad
    \Brv = \partial_r \Bvv \ ,\quad
    \Er = \partial_r \Es - i\w\ \Brs\ ,
\end{split}
\end{equation}
or an equivalent time-reversal covariant form
\begin{equation}\label{Eq:dFcov}
\begin{split}
    \Ev &= i\w\ \Bvv \ , \quad
    \Hv = \Dp \Bvv \ ,\quad
    (1-r^2)\Er =\Dp \Es - i\w\ \Hs\ ,
\end{split}
\end{equation}
where $\Dp\equiv(1-r^2)\partial_r+i\w$. As usual, these equations can be solved by introduction of the vector potential in some gauge. While we will indeed chose a gauge eventually, it is more illuminating to first analyse this system via gauge invariant observables (i.e., EM fields).

The sourced Maxwell equations outside the particle take the form $\partial_{\mu}\left[\sqrt{-g}\ \FC{}^{\mu\nu}\right]=0$, which when written out in terms of the components become
\begin{equation}\label{Eq:dFstPt}
\begin{split}
       - i\w\ \Er + \frac{\ell(\ell+d-2)}{r^2}\Hs &=0 \qquad \text{(r-Eqn)}\ ,\\
    \frac{1}{r^{d-1}}\partial_r \left[r^{d-1}\Er\right] + \frac{\ell(\ell+d-2)}{r^2}\Brs &=0 \qquad \text{(u-Eqn)}\ ,\\
    \frac{1}{r^{d-3}}\partial_r \left[r^{d-3}\Hs\right] + i\w\  \Brs &=0 \qquad \text{(I-Eqn scalar)}\ ,\\
    \frac{1}{r^{d-3}}\Dp\left[r^{d-3}\Hv\right]-i\w\ \Ev-\frac{(\ell+1)(\ell+d-3)}{r^2}\Bvv&=0 \qquad \text{(I-Eqn vector)}\ .
\end{split}
\end{equation}
Here, we have indicated the equation coming from each component. As is well-known, these set of equations are not all independent: the $r$-equation above is the Gauss constraint for radial evolution, which is preserved by the next two equations with radial derivatives. Alternately,  the $u$-equation above is the Gauss constraint for $u$ evolution, which is preserved by the first/third equations with time derivatives. Since our goal is to examine how the near zone data is given in terms of multipole moments of the particle, we will take the radial evolution perspective.

In terms of time reversal covariant quantities, we can recast the above equations in the form
\begin{equation}\label{Eq:dFstPtcov}
\begin{split}
 \frac{\ell(\ell+d-2)}{r^2}\Hs &=i\w\ \Er \ ,\\
    \frac{1}{r^{d-1}}\Dp\left[r^{d-1}\Er\right] &= \frac{\ell(\ell+d-2)}{r^2}\Es  \ ,\\
    \frac{1}{r^{d-3}}\Dp \left[r^{d-3}\Hs\right] &=i\w\ \Es\ , \\
    \frac{1}{r^{d-3}}\Dp\left[r^{d-3}\Hv\right]&=i\w\  \Ev+\frac{(\ell+1)(\ell+d-3)}{r^2}\Bvv\ .
\end{split}
\end{equation}
To conclude, the set of seven equations in Eqs.\eqref{Eq:dFcov}
and \eqref{Eq:dFstPtcov} define the Maxwell system in dS$_{d+1}$.

Examining them as a set of coupled radial ODEs for the six quantities $\{\Es,\Ev,\Er,\Hs,\Hv,\Bvv\}$, we note the following structure:  we can first solve the two equations with no radial derivatives to express  $\{\Ev,\Hs\}$  in terms of rest four quantities. Once this is done, one of the radial equations involving $\Dp \left[r^{d-3}\Hs\right]$ is also solved for `free'. Among the seven equations, we are then left with four first-order radial evolution equations (one each for $\{\Es,\Er,\Hv,\Bvv\}$). Thus, given the near zone values of these four fields, one can radially evolve the above set of equations to get the EM fields far away from the particle.  

Among all possible near-zone data, it is intuitively clear that roughly half would give rise to outgoing EM waves, whereas the other half would result in incoming EM waves.\footnote{We say roughly since this argument ignores time-independent solutions, which are neither incoming nor outgoing. But such zero frequency solutions are a set of measure zero in the space of all solutions of Maxwell equations and can hence be justifiably ignored in this counting.} More generally, if we wanted to also impose boundary conditions far away (or in the case of dS-SK on the near zone of the other branch), the correct thing to do is to constrain only two out of the four quantities. Motivated by the above heuristic argument, we will seek EM fields that satisfy the following boundary conditions
at $r=0$:
\begin{equation}\begin{split}
     \mathcal{J}^E(\w,\ell,\vec{m}) &\equiv  -\lim_{r\to 0}\ r^{\ell+d-2}\Es(r,\w,\ell,\vec{m}) \ ,\\
    \mathcal{J}^B(\w,\alpha,\ell,\vec{m}) &\equiv \lim_{r\to 0}\ r^{\ell+d-3}\Bvv(r,\w,\alpha,\ell,\vec{m}) \ .  
\end{split}\end{equation}
Here $\mathcal{J}^{E,B}$ are the electric/magnetic multipole moments of the particle, and we have chosen here the appropriate powers of $r$ to agree with the flat space definitions in Eqs.\eqref{eq:jBflat} and \eqref{eq:JEfromJ1J2}. We then expect the near-zone behaviour of the form
\begin{equation}
\begin{split}\label{eq:nearzoneBCs}
\Er &\sim (\ell+d-2)\frac{\mathcal{J}^E}{r^{\ell+d-1}} \ ,\
\Es \sim -\frac{\mathcal{J}^E}{r^{\ell+d-2}} \ , \
\Ev \sim \frac{i\w\mathcal{J}^B}{r^{\ell+d-3}}\ ,  \\
\Hs &\sim \frac{i\w \mathcal{J}^E}{\ell r^{\ell+d-3}} \ , \ \Hv \sim-(\ell+d-3) \frac{\mathcal{J}^B}{r^{\ell+d-2}}\ ,\ \Bvv \sim \frac{\mathcal{J}^B}{r^{\ell+d-3}}\ .
%,\\ \Brs &\sim \frac{\mathcal{J}^E}{r^{\ell+d-2}}\ ,\
%\Brv \sim  -(\ell+d-3) \frac{\mathcal{J}^B}{r^{\ell+d-2}}\ . \\
\end{split}
\end{equation}
Here, the terms without $i\w$ correspond exactly to the standard static multipole solutions in flat spacetime. The $i\w$ terms represent the leading quasi-static correction in flat spacetime: the component $\Ev$ is the induced EMF due to changing magnetic moment, whereas 
the component $\Hs$ is the magnetic field due to Maxwell's displacement current. This fact 
gives a physical justification of why we treat the pair $\{\Ev,\Hs\}$ as being derived from the other four: at a given order in quasi-static expansion, these components can be obtained from the other four at one less order.

For a given multipole moment  $\mathcal{J}^{E,B}$, we should then arrange the sub-leading near zone behaviour of the other two fields $\{\Er,\Hv\}$ to get an outgoing wave solution: the powers that need to be controlled happen to be $1/r^{\ell-1}$ terms in $\Er$ and $1/r^{\ell}$ terms in $\Hs$. In fact, as we shall show below, the near zone behaviour of these two other fields, appropriately renormalised, actually encodes the radiation reaction on the particle corresponding to these multipole moments. More precisely, we will show that 
\begin{equation}\begin{split}
    \lim_{r\to 0}r^{1-\ell}\left[\Er+\text{(counter-term proportional to $\Es$)}\right] &=\text{Radiation reaction on $\mathcal{J}^E$}\ ,\\
     \lim_{r\to 0}r^{-\ell}\left[\Hv+\text{(counter-term proportional to $\Bvv$)}\right] &=\text{Radiation reaction on $\mathcal{J}^B$}\ . 
\end{split}\end{equation}
This statement relates the sub-dominant behaviour in the near zone to radiation reaction.

To make these statements precise, it is convenient to express the $\ell\neq 0$ EM field strengths in terms of  electric/magnetic \emph{Hertz-Debye scalars}, i.e., we write
\begin{equation}\label{eq:HDemdSC}
\begin{split}
\Er(r,\w,\ell,\vec{m}) &= \frac{\ell(\ell+d-2)}{r^{d-1}}\Phi_E(r,\w,\ell,\vec{m}) \ ,\\
\Es(r,\w,\ell,\vec{m}) = \frac{1}{r^{d-3}}\Dp \Phi_E(r,\w,\ell,\vec{m}) \ &, \
\Ev(r,\w,\alpha,\ell,\vec{m}) = i\w \Phi_B(r,\w,\alpha,\ell,\vec{m})\ , \\
\Hs(r,\w,\ell,\vec{m}) = \frac{i\w }{r^{d-3}}\Phi_E(r,\w,\ell,\vec{m}) \ &, \ \Hv(r,\w,\alpha,\ell,\vec{m})  = \Dp \Phi_B(r,\w,\alpha,\ell,\vec{m}) \ ,\\ \Bvv(r,\w,\alpha,\ell,\vec{m})  &= \Phi_B(r,\w,\alpha,\ell,\vec{m}) \ 
,\\ \Brs(r,\w,\ell,\vec{m}) = -\frac{1}{r^{d-3}}\partial_r \Phi_E(r,\w,\ell,\vec{m})\ &,\
\Brv(r,\w,\alpha,\ell,\vec{m})  = \partial_r \Phi_B(r,\w,\alpha,\ell,\vec{m}) \ . \\
\end{split}
\end{equation}
The reader is encouraged to check that this form automatically satisfies the Maxwell equations in Eqs.\eqref{Eq:dFcov} and \eqref{Eq:dFstPtcov}, provided the Hertz-Debye scalar fields satisfy the following radial ODEs:
\begin{equation}\label{eq:PhiEqdSEM}\begin{split}
\frac{1}{r^{3-d}}D_+\left[r^{3-d}D_+\Phi_E\right]&+\omega^2\Phi_E-(1-r^2)\frac{\ell(\ell+d-2)}{r^2}\Phi_E=0\ ,\\
\frac{1}{r^{d-3}}D_+\left[r^{d-3}D_+\Phi_B\right]&+\omega^2\Phi_B-(1-r^2)\frac{(\ell+1)(\ell+d-3)}{r^2}\Phi_B=0\ .
\end{split}\end{equation}
In terms of the scalar fields, the Maxwell system reduces to a set of decoupled radial ODEs. An alternate route to get the same results is to write down the gauge potentials $\gaugeVf_{\mu}$ in the \emph{Hertz-Debye gauge}, i.e., we take
\begin{equation}\label{eq:HDemdSV}
\begin{split}
    \gaugeVf_u(r,\w,\Omega) &= r^{3-d}\sum_{\ell\vec{m}} D_{+}\Phi_E(r,\w,\ell,\vec{m})\ \Yellm(\Omega) \ , \\
    \gaugeVf_r(r,\w,\Omega) &= r^{3-d}\sum_{\ell\vec{m}} \partial_r\Phi_E(r,\w,\ell,\vec{m}) \ \Yellm(\Omega) \ , \\
    \gaugeVf_I(r,\w,\Omega) &= \sum_{\alpha\ell\vec{m}} \Phi_B(r,\w,\alpha,\ell,\vec{m}) \ \mathbb{V}_{I}^{\alpha\ell\vec{m}}(\Omega)\ .
\end{split}
\end{equation}
In this gauge, we essentially set all the electric sector gauge fields to be normal to the sphere directions. The magnetic sector is gauge invariant due to the divergencelessness of the $ \mathbb{V}^{\alpha\ell\vec{m}}$. It can then be checked that the field strengths derived from these potentials turn out to be the expressions in Eq.\eqref{eq:HDemdSC}.

The radial ODEs for the Debye potentials can be solved with appropriate boundary conditions inherited from the corresponding boundary conditions on the field strengths defined in \eqref{eq:nearzoneBCs}, which are:

\begin{equation}
\begin{split}
   - \lim_{r\to 0}r^{\ell+1}\Dp\Phi_E(r,\w,\ell,\vec{m})&= \mathcal{J}^E(\w,\ell,\vec{m}) \ , \\ 
    \lim_{r\to 0}r^{\ell+d-2}\Phi_B(r,\w,\alpha,\ell,\vec{m})&= \mathcal{J}^B(\w,\alpha,\ell,\vec{m})\ .
\end{split}
\end{equation}
We can now solve the fields in terms of bulk to boundary propagators for the two scalars. These propagators are special cases of the propagators we found for the generic class of scalars in part I\cite{Loganayagam:2023pfb}. We will review some of the properties of these propagators that we need in our analysis here, but we refer the reader to part I for a much more extensive discussion. One writes the Debye potentials as:
\begin{equation}
\begin{split}
    \Phi_E(r,\w,\ell,\vec{m}) &=\frac{1}{\ell}\Gout_E(r,\w,\ell)\mathcal{J}^E(\w,\ell,\vec{m}) \ ,\\ 
    \Phi_B(r,\w,\alpha,\ell,\vec{m}) &=\Gout_B(r,\w,\ell)\mathcal{J}^B(\w,\alpha,\ell,\vec{m})\ .
\end{split}
\end{equation}
The extra factor of $\ell$ is a convenient normalisation for the $\Phi_E$ Debye scalar because the scalar satisfies a Neumann boundary condition at the origin.  

The boundary to bulk propagators can be written explicitly in terms of Gauss hypergeometric functions given as follows:
\begin{equation}
	\begin{split}
  &G^{\text{Out}}_E(r,\w,\ell)=\frac{r^{\ell+d-2}(1+r)^{-i\w}}{\Gamma(1-i\w)\Gamma\left(\ell+\frac{d}{2}-1 \right)}
		\\
		&\quad\times \Gamma\left(\frac{\ell+2-i\w}{2}\right)\Gamma\left(\frac{\ell+d-2-i\w}{2}\right)\ {}_2F_1\left[\frac{\ell+2-i\w}{2},\frac{\ell+d-2-i\w}{2};1-i\w;1-r^2\right]\ ,\\
   	&G^{\text{Out}}_B(r,\w,\ell)=\frac{r^{\ell+1}(1+r)^{-i\w}}{\Gamma(1-i\w)\Gamma\left(\ell+\frac{d}{2}-1 \right)}
		\\
		&\quad\times \Gamma\left(\frac{\ell+1-i\w}{2}\right)\Gamma\left(\frac{\ell+d-1-i\w}{2}\right)\ {}_2F_1\left[\frac{\ell+1-i\w}{2},\frac{\ell+d-1-i\w}{2};1-i\w;1-r^2\right]\ .
	\end{split}
\end{equation}
The above hypergeometric functions, for odd values of $d$, are polynomials in $r$ that generalise the reverse Bessel polynomials one obtains in the study of outgoing radiation in $3+1$ dimensional spacetime\footnote{In appendix \ref{app:FlatEMII}, we have generalised the reverse Bessel polynomials to arbitrary even dimensional spacetimes.}. Later in this appendix, we show how the $\Gout_{E/B}$ gives Hubble corrections to the reverse Bessel polynomials in flat spacetimes with odd values of $d$. Much like the case in the corresponding flat spacetimes, for even values of $d$, the functions do not reduce to a polynomial form in $r$. 

The hypergeometric functions are defined by a series expansion about the points where the last argument goes to zero. This tells us that the above retarded boundary-to-bulk propagators have a nice expansion at the horizon $r=1$. On the other hand, to obtain their behaviour at $r=0$, the following equivalent form is instructive: 
\begin{equation}\label{eq:GoutII}
	\begin{split}
		G^{\text{Out}}_E&=r^{-\ell}(1+r)^{-i\w}
		\\
		&\times\Bigl\{ {}_2F_1\left[\frac{-\ell-i\w}{2},\frac{4-d-\ell-i\w}{2};2-\frac{d}{2}-\ell;r^2\right]\Bigr.\\
		&\Bigl.\quad\qquad -(1+i\cot\nu \pi)\Khout_E(\w,\nu)\frac{r^{2\nu}}{2\nu}\ {}_2F_1\left[\frac{2+\ell -i\w}{2},\frac{d+\ell-2-i\w}{2};\frac{d}{2}+\ell;r^2\right] \Bigr\}\ , \\
        G^{\text{Out}}_B&=r^{3-d-\ell}(1+r)^{-i\w}
		\\
		&\times\Bigl\{ {}_2F_1\left[\frac{1-\ell-i\w}{2},\frac{3-d-\ell-i\w}{2};2-\frac{d}{2}-\ell;r^2\right]\Bigr.\\
		&\Bigl.\quad\qquad -(1+i\cot\nu \pi)\Khout_B(\w,\nu)\frac{r^{2\nu}}{2\nu}\ {}_2F_1\left[\frac{1+\ell -i\w}{2},\frac{d+\ell-1-i\w}{2};\frac{d}{2}+\ell;r^2\right] \Bigr\}\ .
	\end{split}
\end{equation}
Here, $\nu=\ell+\frac{d}{2}-1$ as defined in the previous sections. The functions $\Khout$ are given by:
\begin{equation}\label{eq:KhOut}
	\begin{split}
    \Khout_E(\w,\nu) 
   &=-e^{i\nu\pi}\frac{2\pi i}{\Gamma(\nu)^2} \frac{\Gamma\left(\frac{3-\frac{d}{2}+\nu-i\w}{2}\right)\Gamma\left(\frac{-1+\frac{d}{2}+\nu-i\w}{2}\right)}{\Gamma\left(\frac{3-\frac{d}{2}-\nu-i\w}{2}\right)\Gamma\left(\frac{-1+\frac{d}{2}-\nu-i\w}{2}\right)} \ ,\\
   \Khout_B(\w,\nu) 
   &=-e^{i\nu\pi}\frac{2\pi i}{\Gamma(\nu)^2} \frac{\Gamma\left(\frac{2-\frac{d}{2}+\nu-i\w}{2}\right)\Gamma\left(\frac{\frac{d}{2}+\nu-i\w}{2}\right)}{\Gamma\left(\frac{2-\frac{d}{2}-\nu-i\w}{2}\right)\Gamma\left(\frac{\frac{d}{2}-\nu-i\w}{2}\right)} \ .
	\end{split}
\end{equation}
When $d$ is odd, $\nu$ takes half-integer values and the above expression is well-defined. For even dimensional spacetimes, the above expressions for the propagators should be treated as a limit as $\nu$ takes the desired integer value. This will play a role when we obtain renormalised boundary correlators on the worldline next.

Now that we have the fields satisfying the prescribed boundary conditions, we want to understand the \emph{self-force} on the multipole moments due to the radiation, i.e. we would like to ask how the fields cause the multipole moments to dissipate energy. We claim that the radiation reaction is encoded in the boundary behaviour of \emph{renormalised} components of the electric and magnetic fields $\Er$ and $\Hv$. These field components exert a radial force on a spherical shell current density. One can think of the point source as the zero radius limit of such a shell. In this limit, the radial force naively diverges, but if one focuses on purely dissipative parts of the force, they are regular. One would like to remove the conservative pieces containing all the divergences by systematically subtracting them from the $\Er$ and $\Hv$. This is accomplished by subtracting from the fields, terms proportional to $\Es$ and $\Bvv$ as follows:
\begin{equation}\begin{split}\label{eq:EMfieldct}
    -\lim_{r\to 0}r^{-\ell}\left[r\Er+\frac{\ell(\ell+d-2)}{\ct_{3-d}}\Es\right]&=\frac{\ell+d-2}{\ell}(1+i\cot\pi\nu)\Khout_E\  \mathcal{J}^E\ ,\\
    -\lim_{r\to 0}r^{-\ell}\left[\Hv+\frac{\ct_{d-3}}{r}\Bvv\right]&=(1+i\cot\pi\nu)\Khout_B\  \mathcal{J}^B
\end{split}\end{equation}
where the $\ct$'s are :
\begin{equation}
	\begin{split}\label{eq:EMcts}
		\frac{\ct_{d-3}}{1-r^2}&\equiv
		-r\frac{d}{dr}\ln\left\{ r^{3-d-\ell} (1-r^2)^{-\frac{i\w}{2}} {}_2F_1\left[\frac{1-\ell-i\w}{2},\frac{3-d-\ell-i\w}{2};2-\frac{d}{2}-\ell;r^2\right] \right\}\ , \\
        \frac{\ct_{3-d}}{1-r^2}&\equiv
		-r\frac{d}{dr}\ln\left\{ r^{-\ell} (1-r^2)^{-\frac{i\w}{2}} {}_2F_1\left[\frac{-\ell-i\w}{2},\frac{4-d-\ell-i\w}{2};2-\frac{d}{2}-\ell;r^2\right] \right\} \ .
	\end{split}
\end{equation}
The $\ct$'s are special cases of the counterterm $\ct_\nn$ that we obtained in part I for the generic class of scalars. The essential difference here is in the case of the electric sector, where the Debye scalar $\Phi_E$ satisfies a Neumann boundary condition. We review this counterterming procedure for the same generic class of scalars discussed in part I, now satisfying Neumann boundary conditions, in appendix \ref{app:NeuDesSca}. 

The radiation reaction kernel is encoded in the boundary values of the renormalised fields which we will call $\Kout_{E/B}$. For odd values of $d$(half-integer values of $\nu$), they are given by:
\begin{equation}\label{eq:KOutOdd}
\begin{split}
	\Kout_E|_{\text{Odd d}} &= (1+i\cot\pi\nu)\Khout_E|_{\text{Odd d}}
   =-e^{i\nu\pi}\frac{2\pi i}{\Gamma(\nu)^2} \frac{\Gamma\left(\frac{3-\frac{d}{2}+\nu-i\w}{2}\right)\Gamma\left(\frac{-1+\frac{d}{2}+\nu-i\w}{2}\right)}{\Gamma\left(\frac{3-\frac{d}{2}-\nu-i\w}{2}\right)\Gamma\left(\frac{-1+\frac{d}{2}-\nu-i\w}{2}\right)}\ , \\
   \Kout_B|_{\text{Odd d}} &= (1+i\cot\pi\nu)\Khout_B|_{\text{Odd d}}
   =-e^{i\nu\pi}\frac{2\pi i}{\Gamma(\nu)^2} \frac{\Gamma\left(\frac{2-\frac{d}{2}+\nu-i\w}{2}\right)\Gamma\left(\frac{\frac{d}{2}+\nu-i\w}{2}\right)}{\Gamma\left(\frac{2-\frac{d}{2}-\nu-i\w}{2}\right)\Gamma\left(\frac{\frac{d}{2}-\nu-i\w}{2}\right)}\ .
	\end{split}
\end{equation}
The table \ref{tab:KoutMark} lists explicit expressions for this function in even-dimensional spacetimes up to quadrupole. The $\Kout_{E/B}$ for odd values of $d$ are polynomials which signify the markovian nature of the electromagnetic radiation reaction. This is the same `boundary two-point function' whose poles are used to analyse the quasinormal mode spectrum of the static patch\cite{Natario:2004jd,Lopez-Ortega:2006aal,Anninos:2011af}.
\begin{table}[H]
	\centering
		\caption{$\frac{\kO}{-i\w}$ for Electromagnetic radiation}
		\setlength{\extrarowheight}{2pt}
	\begin{tabular}{|c|c|c|c|}
		\hline
		Magnetic&$\ell=0$&$\ell=1$&$\ell=2$\\[0.5ex]\hline
		$d=3$& $1$&$ \w^2+1 $ & $\frac{ \w
			^4}{9}+\frac{5 \w^2}{9}+\frac{4 }{9}$  \\[0.5ex] $d=5$ & $
		\w^2+4$ & $\frac{\w^4}{9}+\frac{10
			\w^2}{9}+1 $ & $ \frac{\w
			^6}{225}+\frac{7 \w^4}{75}+\frac{28 \w
			^2}{75}+\frac{64 }{225} $\\[0.5ex]
		$d=7$&$\frac{\w^4}{9}+\frac{20 \w^2}{9}+\frac{64 }{9} $ &$\frac{\w^6}{225}+\frac{7 \w ^4}{45}+\frac{259 \w^2}{225}+1$ &$ \frac{ \w^8}{11025}+\frac{19 \w^6}{3675}+\frac{8 \w ^4}{105}+\frac{3088 \w^2}{11025}+\frac{256
			}{1225} $\\[0.5ex] \hline
		\hline
		Electric &$\ell=0$&$\ell=1$&$\ell=2$\\[0.5ex]\hline
		$d=3$& $1$&$ \w^2+1 $ & $\frac{ \w
			^4}{9}+\frac{5 \w^2}{9}+\frac{4 }{9}$  \\ $d=5$ & $\w^2+1$ & $\frac{\w^4}{9}+\frac{5 \w ^2}{9}+\frac{4 }{9} $ & $\frac{ \w
			^6}{225}+\frac{14 \w^4}{225}+\frac{49 \w ^2}{225}+\frac{4 }{25} $\\
		$d=7$&$\frac{\w^4}{9}+\frac{10 \w^2}{9}+1 $ &$\frac{\w^6}{225}+\frac{7 \w
			^4}{75}+\frac{28 \w^2}{75}+\frac{64
		}{225}$ &$\frac{ \w^8}{11025}+\frac{13 \w^6}{3675}+\frac{19 \w^4}{525}+\frac{1261  \w
		^2}{11025}+\frac{4 }{49} $\\[0.5ex] \hline
	\end{tabular}\label{tab:KoutMark}
\end{table}

On the other hand, for even values of $d$, the $\cot\pi\nu$ diverges and one needs additional counterterms to obtain the correct radiation reaction kernel. These can be obtained by adding the following counterterm action to our electromagnetic action:
\begin{equation}\label{eq:PhiActCtEven}
	\begin{split}
		S_{ct,\text{Even}}&= \sum_{\ell\vec{m}}\frac{1}{\nu-n}\int\frac{d\w}{2\pi} \Bigg[r^{d-4+2n}\Delta\left(n,\frac{d}{2}-1,\w\right)\Phi_B^* \Phi_B|_{r_c} \\ &\qquad \qquad\qquad +\ell(\ell+d-2)r^{2-d+2n}\Delta\left(n,\frac{d}{2}-2,\w\right)| \Dp\Phi_E|^2|_{r_c} \Bigg]\ ,
	\end{split}
\end{equation}
where $n=\ell+\frac{d}{2}-1$ 
and,
\begin{equation}\label{eq:ctEvenH}
\begin{split}
   \Delta(n,\mu,\w)&\equiv
  \frac{(-)^n}{\Gamma(n)^2} \frac{\Gamma\left(\frac{1+n-\mu-i\w}{2}\right)\Gamma\left(\frac{1+n+\mu-i\w}{2}\right)}{\Gamma\left(\frac{1-n+\mu-i\w}{2}\right)\Gamma\left(\frac{1-n-\mu-i\w}{2}\right)}=\frac{1}{\Gamma(n)^2}\prod\limits_{k=1}^{n}\left[\frac{\w^2}{4}+\frac{1}{4}(\mu-n+2k-1)^2\right]\\
  &=\Delta^\ast(n,\mu,\w)\ .
\end{split}
\end{equation} 

With this counterterm, we obtain the following form of the radiation reaction kernel for even-dimensional spacetimes:
\begin{equation}\label{eq:KoutEven}
	\begin{split}
		\Kout_E|_\text{Even $d$} &=\Delta_\nn\left(\nu,\frac{d}{2}-2,\w\right)\left[\psi^{(0)}\left(\frac{3-\frac{d}{2}+\nu-i\w}{2}\right)+\psi^{(0)}\left(\frac{-1+\frac{d}{2}\nu-i\w}{2}\right)\right.\\ &\left.+\psi^{(0)}\left(\frac{3-\frac{d}{2}-\nu-i\w}{2}\right)+\psi^{(0)}\left(\frac{-1+\frac{d}{2}-\nu-i\w}{2}\right)-4\psi^{(0)}(\nu)\right]\ , \\
        \Kout_B|_\text{Even $d$} &=\Delta_\nn\left(\nu,\frac{d}{2}-1,\w\right)\left[\psi^{(0)}\left(\frac{2-\frac{d}{2}+\nu-i\w}{2}\right)+\psi^{(0)}\left(\frac{\frac{d}{2}\nu-i\w}{2}\right)\right.\\ &\left.+\psi^{(0)}\left(\frac{2-\frac{d}{2}-\nu-i\w}{2}\right)+\psi^{(0)}\left(\frac{\frac{d}{2}-\nu-i\w}{2}\right)-4\psi^{(0)}(\nu)\right]\ .
	\end{split}
\end{equation}
Unlike the case for odd values of $d$, we see that these functions do not reduce to polynomials in $\w$, which is expected from the non-markovian nature of radiation reaction in odd spacetime dimensions\cite{Harte:2018iim}.

In the limit where the Hubble constant is small, i.e. we are looking at sources moving much more rapidly compared to the cosmological time scales, the radiation reaction kernel reduces to its flat space analogues, which have the same behaviour for both magnetic and electric sectors\cite{Birnholtz:2013ffa,Birnholtz:2013nta}:
\begin{align}
\begin{split}
\Kout_{E/B} \approx \left\{\begin{array}{cc} \frac{2\pi i}{\Gamma(\nu)^2}\left(\frac{\w}{2}\right)^{2\nu} & \text{for $d$  odd ,} \\
\frac{1}{ \Gamma(\nu)^2} \left(\frac{\w }{2}\right)^{2\nu} \ln \left(\frac{\w^4}{H^4}\right) & \text{for $d$  even .} \\
\end{array}\right.
\end{split}
\end{align}

The procedure we just illustrated provides us with a dS version of the famous Son-Starinets prescription\cite{Son:2002sd} in AdS/CFT: in this prescription,  the field's value ( $\gaugeV_\mu$ in our case) is fixed at the boundary and an outgoing boundary condition is imposed at the horizon. This corresponds to imposing a Dirichlet boundary condition on the $\Phi_B$, whereas the $\Phi_E$ satisfies a Neumann boundary condition at $r=0$. Then one takes the boundary limit of the conjugate field $\mathcal{C}^{r\mu}$ and renormalises it to obtain the radiation reaction kernel $K^{\text{Out}}$. 

Given this procedure of obtaining the radiation reaction, we will now justify it as an on-shell action computation on the dS-SK geometry described in detail in part I\cite{Loganayagam:2023pfb}. This is the de Sitter version of the real-time GKPW prescription: We specify boundary data as the observer's multipole moments, and the dS-SK saddle computes the effective action of the observer's dynamics. 

We will begin by reviewing some useful geometric details required in our analysis here. We take the de Sitter static patch and complexify the radial coordinate. Then, we consider a hypersurface defined by a contour in the complex $r$ plane. The following \emph{mock tortoise coordinate} is useful to make this notion precise:
\begin{equation}
		\zeta(r)=\frac{1}{i\pi}\int\limits^{0-i\epsilon}_r\frac{dr'}{1-r'^2}=\frac{1}{2\pi i}\ln\left(\frac{1-r}{1+r}\right)\ .\label{eq:zeta-def}
\end{equation}
This integral has logarithmic branch points at $r=\pm 1$, and we pick the branch cut to run from $r=-1$ to $r=1$. If we start from $r=1+i\epsilon$ and go around the branch cut to $r=1-i\epsilon$, the $\zeta$ coordinate is normalised to go from $1$ to $0$ in its real part. We define $\zeta=0$ as the left boundary and $\zeta=1$ as the right boundary. Given this geometry, we will now turn to the question of obtaining the electromagnetic fields on it.  

We need ingoing counterparts of the outgoing propagators to define the correct boundary-to-bulk propagator on the dS-SK geometry. We can obtain the ingoing propagator simply by time-reversing the outgoing one:
\begin{equation}\label{eq:GinI}
	\Gin_{E/B}(r,\w,\ell) =e^{-2\pi\w\zeta}\Gouts_{E/B}(r,\w,\ell)\ .
\end{equation}
The boundary-to-bulk propagators on the dS-SK geometry then turn out to be:
\begin{equation}\label{eq:gLgRdef}
	\begin{split}
		g^{E/B}_L(r,\w,\ell)&\equiv n_\w\Bigl[\Gout_{E/B}(r,\w,\ell)-e^{2\pi\w(1-\zeta)}\Gouts_{E/B}(r,\w,\ell)\Bigr]\ ,\\
		g^{E/B}_R(r,\w,\ell)&\equiv (1+n_\w)\Bigl[\Gout_{E/B}(r,\w,\ell)-e^{-2\pi\w\zeta}\Gouts_{E/B}(r,\w,\ell)\Bigr]\ .
 	\end{split}
\end{equation}
where $n_\w =\frac{1}{e^{2\pi\w}-1}$, is the Bose-Einstein factor. Essentially, the $g_L$ connects the sources on the left boundary to the fields on the dS-SK geometry, whereas $g_R$ does the same for the sources on the right boundary. These boundary-to-bulk propagators are hence defined so as to satisfy the following boundary conditions:
\begin{equation}\label{eq:gLRbc}
	\begin{split}
		\lim_{\zeta\to 0}r^{\ell}g^E_L =-1\ ,&\quad \lim_{\zeta\to 0}r^{\ell}g^E_R =0\ ,\\
	\lim_{\zeta\to 1}r^{\ell}g^E_L =0\ ,&\quad \lim_{\zeta\to 1}r^{\ell}g^E_R =1,\\
    	\lim_{\zeta\to 0}r^{d-3+\ell}g^B_L =-1\ ,&\quad \lim_{\zeta\to 0}r^{d-3+\ell}g^B_R =0\ ,\\
	\lim_{\zeta\to 1}r^{d-3+\ell}g^B_L =0\ ,&\quad \lim_{\zeta\to 1}r^{d-3+\ell}g^B_R =1.
	\end{split}
\end{equation}
Hence, the $g_{L/R}$ is defined to be regular on the right/left boundary and to have a source singularity on the other boundary\footnote{In part I, we give a more detailed analysis of these boundary-to-bulk propagators, along with explicit expressions, which the reader can refer to for further details.}. 

Using these boundary-to-bulk propagators, we can write down the solutions for $\Phi_E$ and $\Phi_B$ as:
\begin{equation}
\begin{split}
    \Phi_E(\zeta,\w,\ell,\vec{m}) &=\frac{1}{\ell}\left\{g_R^E(\zeta,\w,\ell) \multj^E_R(\w,\ell,\vec{m})-g_L^E(\zeta,\w,\ell) \multj^E_L(\w,\ell,\vec{m})\right\}, \\
    \Phi_B(\zeta,\w,\alpha,\ell,\vec{m}) &= g_R^B(\zeta,\w,\ell) \multj^B_R(\w,\alpha,\ell,\vec{m})-g_L^B(\zeta,\w,\ell) \multj^B_L(\w,\alpha,\ell,\vec{m})\ ;
\end{split}
\end{equation}
where $\ \multj^{E/B}_L$ and $\ \multj_{R}^{E/B}$ are the electric/magnetic multipole moments on the left and the right boundary, respectively. The $\Phi_E$ satisfies Neumann boundary conditions, whereas $\Phi_B$ satisfies Dirichlet boundary conditions at both the $R$ and $L$ boundaries of the dS-SK geometry. Correspondingly, we can also write the conjugate fields to $\pi_{E/B}$ in the following manner:
\begin{equation}
\begin{split}
    \pi_E(\zeta,\w,\ell,\vec{m}) &=\frac{1}{\ell}\left\{\pi_L^E(\zeta,\w,\ell) \multj^E_L(\w,\ell,\vec{m})-\pi_R^E(\zeta,\w,\ell) \multj^E_R(\w,\ell,\vec{m})\right\}, \\
    \pi_B(\zeta,\w,\alpha,\ell,\vec{m}) &= \pi_L^B(\zeta,\w,\ell) \multj^B_L(\w,\alpha,\ell,\vec{m})-\pi_R^B(r,\w,\ell) \multj^B_R(\w,\alpha,\ell,\vec{m})
\end{split}    
\end{equation}
where we have defined $\pi_{L/R}^{E/B}=\Dp g_{L/R}^{E/B}$. These fields are fixed such that they satisfy the following boundary conditions:
\begin{equation}
\begin{split}
    \lim_{\zeta\to 0}r^{\ell+d-2}\pi_E = \multj^E_L\ ,& \qquad \lim_{\zeta\to 1}r^{\ell+d-2}\pi_E = \multj^E_R \ ,\\
    \lim_{\zeta\to 0}r^{\ell+d-3}\Phi_B = \multj^B_L\ ,& \qquad \lim_{\zeta\to 1}r^{\ell+d-3}\Phi_B = \multj^B_R\ .
\end{split}
\end{equation}

Given the solutions with appropriate boundary conditions, one can now substitute them into the action to obtain the boundary Schwinger-Keldysh action:  
\begin{equation}
\begin{split}
 S&=   -\frac{1}{4}\int_{dSSK} d^{d+1}x\sqrt{-g} \FC^{\mu\nu}\FC_{\mu\nu}+S_{ct}\\
 &=\frac{1}{2}\int_{dSSK}d^{d+1}x\left[\gaugeV_\nu\partial_{\mu}\left(\sqrt{-g}\FC^{\mu\nu}\right)-\partial_\mu\left(\sqrt{-g}\FC^{\mu\nu} \gaugeV_\nu\right)\right] +S_{ct}\\ 
 &\underset{\text{(On-Shell)}}{=} \left[-\frac{1}{2}\int r^{d-1} dt \ d\Omega_{d-1} \FC^{r\mu}\gaugeV_\mu\right]^{r_c-i\epsilon}_{r_c+i\epsilon} +S_{ct}\ .
\end{split}
\end{equation}
Here, the first term on the second line can be set to zero using the equations of motion, and the second one evaluates to a boundary term. We can compute this boundary term using the dS-SK solutions:
\begin{equation}
\begin{split}
    \lim_{\zeta\to 1}r^{d-1}\FCf^{ru}_{\text{ren}}\gaugeVf_u^* &= \lim_{\zeta\to 1} \frac{1}{r^{\ell-1}}\FCf^{ru}_{\text{ren}}\times\lim_{\zeta\to 1}r^{\ell+d-2}\gaugeVf^*_u \\ &=-\frac{(\ell+d-2)}{\ell}\multj^{E\ast}_R\left\{K^E_{LR}\multj^E_R-K^E_{LL}\multj^E_L\right\} \ ,\\
    \lim_{\zeta\to 0}r^{d-1}\FCf^{ru}_{\text{ren}}\gaugeVf_u^* &= \lim_{\zeta\to 0} \frac{1}{r^{\ell-1}}\FCf^{ru}_{\text{ren}}\times\lim_{\zeta\to 0}r^{\ell+d-2}\gaugeVf^*_u \\ &=-\frac{(\ell+d-2)}{\ell}\multj^{E\ast}_L\left\{K^E_{RR}\multj^E_R-K^E_{RL}\multj^E_L\right\} \ ,\\
    \lim_{\zeta\to 1}r^{d-1}\FCf^{rI}_{\text{ren}}\gaugeVf_I^* &= \lim_{\zeta\to 1} \frac{1}{r^{\ell-2}}\FCf^{rI}_{\text{ren}}\times\lim_{\zeta\to 1}r^{\ell+d-3}\gaugeVf^*_I \\ &=-\multj^{E\ast}_R\left\{K^B_{LR}\multj^B_R-K^B_{LL}\multj^B_L\right\} \ ,\\
    \lim_{\zeta\to 0}r^{d-1}\FCf^{ru}_{\text{ren}}\gaugeVf_u^* &= \lim_{\zeta\to 0} \frac{1}{r^{\ell-2}}\FCf^{ru}_{\text{ren}}\times\lim_{\zeta\to 0}r^{\ell+d-3}\gaugeVf^*_u \\ &=-\multj^{B\ast}_L\left\{K^B_{RR}\multj^B_R-K^B_{RL}\multj^B_L\right\} \ .
\end{split}
\end{equation}
We have defined these combinations of the wordline two-point functions:
\begin{equation}\label{eq:Ksk}
	\begin{split}
		K^{E/B}_{LL}\equiv n_\w \Kout_{E/B}-(1+n_\w)\Kouts_{E/B}\ ,&\quad K^{E/B}_{LR}\equiv(1+n_\w)\Bigl( \Kout_{E/B}-\Kouts_{E/B}\Bigr)\ ,\\
		K^{E/B}_{RL}\equiv n_\w\Bigl( \Kout_{E/B}-\Kouts_{E/B}\Bigr)\ ,&\quad K^{E/B}_{RR}\equiv(1+n_\w) \Kout_{E/B}-n_\w\Kouts_{E/B}\ .
	\end{split}
\end{equation}
Given the above expressions on the boundaries of dS-SK, we can write down the on-shell action as follows:
\begin{equation}\label{eq:SCIPptapp}
\begin{split}
	\SCIP&=-\sum_{\alpha\ell\vec{m}}\int\frac{d\w}{2\pi} \Kout_B(\w,\ell)\  \multj_D^{B*}\ \Bigl[\multj^B_A+\left(n_\w+\frac{1}{2}\right)\multj_D^B\Bigr] \\
    &\quad -\sum_{\ell\vec{m}}\frac{(\ell+d-2)}{\ell}\int\frac{d\w}{2\pi} \Kout_E(\w,\ell)\  \multj_D^{E*}\ \Bigl[\multj_A^E+\left(n_\w+\frac{1}{2}\right)\multj_D^E\Bigr]\ ,
\end{split}
\end{equation}
where we have defined the average and difference combinations of the source multipole moments:
\begin{equation}
\multj^{E/B}_A\equiv \frac{1}{2}\multj^{E/B}_R+\frac{1}{2}\multj^{E/B}_L\ ,\qquad \multj^{E/B}_D\equiv\multj^{E/B}_R-\multj^{E/B}_L \ .
\end{equation}
This is otherwise known as the Keldysh basis, which is convenient for extracting the physics from these expressions. The average-difference terms capture the dissipative piece: they encode the physics, as we will see in the next section, of the Abraham-Lorentz-Dirac force in dS. The difference-difference term encodes the Hawking fluctuations. This can be seen through a Hubbard-Stratonovich transformation of the difference moments, which will induce a noise field whose fluctuations are controlled by the Hawking temperature\footnote{See part I\cite{Loganayagam:2023pfb} for a proper derivation of the fluctuating field in the long time limit.}. The thermality of the correlators is encoded in the fact that the fluctuations are proportional to the dissipation, as can be seen from our action.

Lastly, we will show how the $S_{ct}$ can be written gauge-invariantly. The counterterm lagrangian can be written in a gauge-invariant manner in the following way:
\begin{equation}
\begin{split}\label{eq:GaugeInvCT}
S_{ct}&=\left[\frac{1}{2}\int r^{d-1} dt \ d\Omega_{d-1} \frac{r}{\ct_{\nn=3-d}}\left\{(\mathscr{D}^{I}\FC_{uI})\gaugeV_u\right\}\right.\\
&\left.-\frac{1}{2}\int r^{d-1} dt \ d\Omega_{d-1} \left\{\frac{\ct_{\nn=d-3}-\frac{r^2\partial_u^2}{\ct_{\nn=3-d}}}{(\ell+1)(\ell+d-3)}r\mathscr{D}_J\FC^{IJ}+\frac{r}{\ct_{\nn=3-d}}\partial_u\FC_u^{\ I}\right\}\gaugeV_I\right]^{r_c-i\epsilon}_{r_c+i\epsilon}\\
&=\left[\frac{1}{2}\int r^{d-1} dt \ d\Omega_{d-1} \frac{r}{\ct_{\nn=3-d}}\left\{(\mathscr{D}^{I}\FC_{uI})\gaugeV_u-(\partial_u\FC_u^{\ I}) \gaugeV_I+\frac{r^2}{(\ell+1)(\ell+d-3)}(\partial_u^2\mathscr{D}_J\FC^{IJ})\gaugeV_I\right\}\right.\\
&\left.-\frac{1}{2}\int r^{d-1} dt \ d\Omega_{d-1} \frac{r\ct_{\nn=d-3}}{(\ell+1)(\ell+d-3)}(\mathscr{D}_J\FC^{IJ})\gaugeV_I\right]^{r_c-i\epsilon}_{r_c+i\epsilon}\\
&=\left[\frac{1}{2}\int r^{d-1} dt \ d\Omega_{d-1} \frac{r}{\ct_{\nn=3-d}}\left\{\FC_{uI}\FC_u^{\ I}-\frac{1}{2}\frac{r^2}{(\ell+1)(\ell+d-3)}(\partial_u\FC^{IJ})(\partial_u\FC_{IJ})\right\}\right.\\
&\left.-\frac{1}{4}\int r^{d-1} dt \ d\Omega_{d-1} \frac{r\ct_{\nn=d-3}}{(\ell+1)(\ell+d-3)}\FC^{IJ}\FC_{IJ}\right]^{r_c-i\epsilon}_{r_c+i\epsilon}\ .
\end{split}
\end{equation}
As we can see, the counterterm action is local in time and gauge invariant. This concludes our analysis of constructing a regularised effective action for a point source observer.

\subsection{Energy flux through the horizon}
The Electromagnetic stress tensor, in our notation, is given by:
\begin{equation}
    \overline{T}_{\text{EM}}^{\mu\nu}=\FC^{\mu\alpha}\FC^\nu_{\ \alpha}-\frac{1}{4}g^{\mu\nu}\FC_{\alpha\beta}\FC^{\alpha\beta}\ .
\end{equation}
We want to calculate the electromagnetic energy flux that exits through the horizon. This is encoded in the $T_u^r$ component of the stress tensor, which we integrate over the sphere to obtain the total flux:
\begin{equation}
\begin{split}\label{eq:dSEMflux}
     \int_{\mathbb{S}^{d-1}_r} r^{d-1}(\overline{T}_\text{EM})_u^{\ r} &= -\sum_{\ell\vec{m}}\int_\w r^{d-3}\left[\sum_{\alpha}\Ev^*\Hv+\ell(\ell+d-2)\Hs^*\Es\right] \\
     &=\sum_{\ell\vec{m}}\int_\w  i\w \left[\sum_{\alpha}r^{d-3}\  \Phi_B^* D_+\Phi_B+  r^{3-d}\ \ell(\ell+d-2)  \Phi_E^* D_+\Phi_E\right]\ . 
\end{split}
\end{equation}
Here, we have expressed the stress tensor in terms of the Debye scalars.
%whereas at $\mathscr{I}^+$,
%\begin{equation}
%    \lim_{r\to \infty}r^{\nn+1-2\mu}\gs D_+ \gO=\frac{2 \Gamma (1+\nu)^2}{\Gamma(\mu)^2 ( \nn-2 \mu+2)}\frac{ \Gamma \left(\frac{\mu -\nu -i \omega
%   +1}{2}\right) \Gamma \left(\frac{\mu -\nu +i \omega
 %  +1}{2}\right)}{ \Gamma
  % \left(\frac{ \nu-\mu -i \omega +1}{2}\right) \Gamma
   %\left(\frac{ \nu -\mu+i \omega +1}{2} \right)} 
%\end{equation}
In \cite{Ashtekar:2015lla}, the authors compute using covariant phase space formalism, the flux through $\mathscr{I}^+$ given by:
\begin{equation}
    \int_{\mathscr{I}^+} dV_{\mathscr{I}^+} \ g^{\mu\nu}\left[\overbar{\mathcal{C}}_{u\mu}\overbar{\mathcal{C}}_{u\nu}-(1-r^2)\overbar{\mathcal{C}}_{u\mu}\overbar{\mathcal{C}}_{r\nu}\right]\ ,
\end{equation}
which matches our expression for the flux through a constant $r$ slice.

The boundary to bulk retarded Green's functions in de Sitter static patch for the Debye scalars are:
\begin{equation}\label{eq:GoutI}
\begin{split}
	\Gout_E(r,\w,\ell) &=H^{2\nu}r^{\ell+d-2}(1+H r)^{-\frac{i\w}{H}} \frac{\Gamma\left(\frac{\ell+2-\frac{i\w}{H}}{2}\right)\Gamma\left(\frac{\ell+d-2-\frac{i\w}{H}}{2}\right)}{\Gamma(1-\frac{i\w}{H})\Gamma\left(\ell+\frac{d}{2}-1 \right)} \\
	&\qquad\times \ {}_2F_1\left[\frac{\ell+2-\frac{i\w}{H}}{2},\frac{\ell+d-2-\frac{i\w}{H}}{2};1-\frac{i\w}{H};1-r^2H^2\right]\ , \\
	\Gout_B(r,\w,\ell) &=H^{2\nu}r^{\ell+1}(1+Hr)^{-\frac{i\w}{H}} \frac{\Gamma\left(\frac{\ell+1-\frac{i\w}{H}}{2}\right)\Gamma\left(\frac{\ell+d-1-\frac{i\w}{H}}{2}\right)}{\Gamma(1-\frac{i\w}{H})\Gamma\left(\ell+\frac{d}{2}-1 \right)} \\
	&\qquad\times \ {}_2F_1\left[\frac{\ell+1-\frac{i\w}{H}}{2},\frac{\ell+d-1-\frac{i\w}{H}}{2};1-\frac{i\w}{H};1-H^2r^2\right]\ .
\end{split}
\end{equation}
This form of writing the propagators allows us to easily read off the horizon behaviour as the hypergeometric function goes to 1. We have:
\begin{equation}
\begin{split}
    \lim_{r \to H^{-1}}\Gout_E &= 2^{-\frac{i\w}{H}}H^{\ell}\frac{\Gamma\left(\frac{\ell+2-\frac{i\w}{H}}{2}\right)\Gamma\left(\frac{\ell+d-2-\frac{i\w}{H}}{2}\right)}{\Gamma(1-\frac{i\w}{H})\Gamma\left(\ell+\frac{d}{2}-1 \right)} \equiv H^{3-d}\mathfrak{f}_E(\w)\ ,\\
    \lim_{r \to H^{-1}}\Gout_B &= 2^{-\frac{i\w}{H}}H^{\ell+d-3}\frac{\Gamma\left(\frac{\ell+1-\frac{i\w}{H}}{2}\right)\Gamma\left(\frac{\ell+d-1-\frac{i\w}{H}}{2}\right)}{\Gamma(1-\frac{i\w}{H})\Gamma\left(\ell+\frac{d}{2}-1 \right)}\equiv\mathfrak{f}_B(\w)\ .
\end{split}
\end{equation}
where we have defined the $\mathfrak{f}_E$ to have the same mass dimension as $\mathfrak{f}_B$.
Similarly, we can also obtain the following:
\begin{equation}
\begin{split}
    \lim_{r \to H^{-1}}\Dp\Gout_E &= i\w H^{3-d}\ \mathfrak{f}_E\ ,\\
    \lim_{r \to H^{-1}}\Dp\Gout_B &= i\w \ \mathfrak{f}_B\ .
\end{split}
\end{equation}
 To find the flux of outgoing radiation, we simply substitute the outgoing solutions evaluated at the horizon. Substituting this relation back into \eqref{eq:dSEMflux} evaluated at the horizon, we have:
\begin{equation}
\begin{split}
     \lim_{r\to H^{-1}}\int_{\mathbb{S}^{d-1}_r} r^{d-1}(\overline{T}_\text{EM})_u^{\ r} &=-\sum_{\ell\vec{m}}\int_\w  i\w \left[\sum_{\alpha}\Kout_B |\multj^B|^2 +  \frac{\ell+d-2}{\ell}  \Kout_E|\multj^E|^2\right]\ . 
\end{split}
\end{equation}
From the above expressions, we can obtain the behaviour of the electromagnetic fields at the horizon:
\begin{equation}
\begin{split}\label{eq:dShorizonFields}
 \lim_{r \to H^{-1}}\Er =(\ell+d-2) H^{2}\ \mathfrak{f}_E\ \mathcal{J}^E \ ,\qquad 
 \lim_{r \to H^{-1}}\Es &= i\w \ \mathfrak{f}_E \frac{\mathcal{J}^E}{\ell}\ ,   \qquad
\lim_{r \to H^{-1}}\Hs = i\w \ \mathfrak{f}_E\frac{\mathcal{J}^E}{\ell}\ , \\ 
\lim_{r \to H^{-1}}\Bvv =\mathfrak{f}_B\ \mathcal{J}^B\ ,\qquad 
 \lim_{r \to H^{-1}}\Ev &=i\w \ \mathfrak{f}_B\ \mathcal{J}^B\ ,  \qquad
\lim_{r \to H^{-1}}\Hv =i\w\ \mathfrak{f}_B\ \mathcal{J}^B\ .
\end{split}
\end{equation}
We can see that these expressions reproduce the flat space expressions as $H\to 0$ with $r$ set to $\frac{1}{H}$. To see this, we give the flat space limits of the $\mathfrak{f}_{E/B}$(for odd values of $d$):
\begin{equation}
\begin{split}
    \lim_{H\to 0}H^{\frac{3-d}{2}}\mathfrak{f}_E=\lim_{H\to 0}H^{\frac{3-d}{2}}\mathfrak{f}_B=\frac{(-i\w)^{\nu-\frac{1}{2}}}{(2\nu-2)!!} \ .
\end{split}
\end{equation}
For generic values of $d$ we have:
\begin{equation}
\begin{split}
    \lim_{H\to 0}H^{\frac{3-d}{2}}\mathfrak{f}_E=\lim_{H\to 0}H^{\frac{3-d}{2}}\mathfrak{f}_B=\frac{\sqrt{\pi}}{\Gamma\left(\nu\right)}\left(-\frac{i\w}{2}\right)^{\nu-\frac{1}{2}} \ .
\end{split}
\end{equation}

\subsection{dS-Bessel Polynomials}\label{app:dSBesselPoly}

In this section, we will generalise the flat space reverse Bessel polynomials\cite[{\href{https://dlmf.nist.gov/18.34}{18.34}}]{NIST:DLMF}, obtained in the study of outgoing radiation in $3+1$ dimensions, to a generic class of scalar fields in all even-dimensional de Sitter spacetimes. In part I, we introduced a system of `designer scalars' that are governed by the action:
\begin{equation}\label{eq:ActPhiMain}
		S=-\frac{1}{2}\int d^{d+1}x \sqrt{-g}\ r^{\nn+1-d}\left\{(\partial\Phi_{\nn})^2+\frac{\Phi_{\nn}^2}{4r^2}\left[(d+\nn-3)(d-\nn-1)-r^2\left(4\mu^2-(\nn+1)^2\right)\right]\right\}\ .
\end{equation}
where the $\nn$ and $\mu$ parametrise the various scalar fields. The centrifugal and mass terms are chosen in the action such that for appropriate values of $\nn$ and $\mu$, one can obtain various scalar fields that are obtained in the study of massive Klein-Gordon fields, electromagnetism, and linearised gravity. 

In even dimensional de Sitter spacetimes, for massless fields, i.e., when $4\mu^2=(\nn+1)^2$, one finds that the outgoing boundary to bulk propagator $\gO$  can be written in polynomials in $\w r$ and $rH$. These polynomials are the de Sitter analogues of the reverse Bessel polynomials generalised to $d+1$ dimensions and reproduce them in the zero curvature limit. To make this explicit, we will write the designer scalar EOM as follows:
\begin{equation}
    \frac{1}{r^{1-2\nu}}\Dp\left(r^{1-2\nu}\Dp\psi_\nn\right)+\w^2\psi_\nn+H^2(1-r^2H^2)\left[\mu^2-(\nu-1)^2\right]\psi_\nn=0\ .
\end{equation}
where we have scaled the designer scalar with a power of $r$ for convenience:
\begin{equation}
    \Phi_\nn(r,\w,\ell)=r^{\frac{1-\nn}{2}-\nu}\psi_\nn(r,\w,\ell)\ .
\end{equation}
We remind the reader that $\nu=\ell+\frac{d}{2}-1$.
One can write the solution for this equation in the following form:
\begin{equation}
    \psi_\nn= \sum_{n=0}^{\infty}(r^2H^2)^n\frac{\left(\frac{1-\mu-\nu}{2}\right)_n\left(\frac{1+\mu-\nu}{2}\right)_n}{\Gamma(n+1)\left(1-\nu\right)_n}\Theta_{\nu-\frac{1}{2}-n}(z)\ .
\end{equation}
Here, 
\begin{equation}
    (a)_n=\prod\limits_{k=0}^{k=n-1}(a-k)\ ,
\end{equation} is the falling factorial and $\Theta_{\nu-\frac{1}{2}}$ satisfies the $\mu$ independent equation of motion:
\begin{equation}
    \frac{1}{r^{1-2\nu}}\Dp\left(r^{1-2\nu}\Dp\Theta_{\nu-\frac{1}{2}}\right)+\w^2\Theta_{\nu-\frac{1}{2}}=0\ .
\end{equation}
Although this differential equation can be solved in terms of Hypegeometric functions, that form is not particularly illuminating to extract out the polynomial nature of the propagator. Instead, we will express the solutions in a Hubble expansion, which makes the polynomial nature explicit.

\begin{table}
	\centering
		\caption{$\Theta_{\nu-\frac{1}{2}}$ for various values of $\nu$ ($H=1$)}
		\setlength{\extrarowheight}{2pt}
	\begin{tabular}{|c|c|}
		\hline\hline
		$\nu $&$\Theta_{\nu-\frac{1}{2}}$\\[0.5ex]\hline \hline
		$\frac{1}{2}$&  $1$  \\[0.5ex]\hline$\frac{3}{2}$  & $1+z$\\[0.5ex]\hline
		$\frac{5}{2}$&$3+3z+z^2+r^2z$\\[0.5ex]\hline
  $\frac{7}{2}$&$15+15z+6z^2+z^3+r^2z\left(5+3z\right)+3r^4z$\\[0.5ex]\hline
  $\frac{9}{2}$&$105+105z+45z^2+10z^3+z^4+r^2z\left(35+26z+6z^2\right)+r^4z\left(21+15z\right)+15r^6z$\\[0.5ex]\hline
  $\frac{11}{2}$&\makecell{$945+945z+420z^2+105z^3+15z^4+z^5+r^2z\left(315+255z+80z^2+10z^3\right)$\\$+r^4z\left(189+170z+45z^2\right)+r^6z\left(135+105z\right)+105r^8z$}\\[0.5ex]\hline
  $\frac{13}{2}$&\makecell{$10395+10395z+4725z^2+1260z^3+210z^4+21z^5+z^6$\\$+r^2z\left(3465+2940z+1050z^2+190z^3+15z^4\right)$\\$+r^4z\left(2079+2059z+750z^2+105z^3\right)+r^6z\left(1485+1470z+420z^2\right)$\\$+r^8z\left(1155+945z\right)+945r^{10}z$ }\\[0.5ex]\hline
  \end{tabular}\label{tab:BigTheta}
\end{table}

Taking $z\equiv-i\w r$, we can write the Hubble expanded solutions as:
\begin{equation}\begin{split}
&\Theta_{\ell}(z,Hr)=\sum_{k=0}^\ell\frac{z^{\ell-k}}{2^k k!}\frac{(\ell+k)!}{(\ell-k)!}\\
&\quad+\frac{H^2 r^2}{2!}\sum_{k=0}^{\ell-1}(\ell-1-k)\frac{z^{\ell-1-k}}{2^k k!}\frac{(\ell-1+k)!}{(\ell-1-k)!}\Bigl\{\ell+\frac{1}{3}k\Bigr\}\\
&\quad +\frac{H^4 r^4}{4!}\sum_{k=0}^{\ell-2}(\ell-2-k)\frac{z^{\ell-2-k}}{2^k k!}\frac{(\ell-2+k)!}{(\ell-2-k)!}\Bigl\{\ell (\ell - 1) (3 \ell + 3 - k)-\frac{1}{15}k(k-1)(25\ell+5k-3)\Bigr\}\\
&\quad +\frac{H^6 r^6}{6!}\sum_{k=0}^{\ell-3}(\ell-3-k)\frac{z^{\ell-3-k}}{2^k k!}\frac{(\ell-3+k)!}{(\ell-3-k)!}\\
&\qquad\times\Bigl\{15\ell(\ell^2-1)(\ell^2-4)+k\ell (\ell-1) [(42 + 25 \ell -15 \ell^2)  - (12 + 10 \ell) k] \Bigr.\\ 
&\qquad\Bigl.\qquad+ \frac{1}{63} k (k-1) (k-2) [-2 - 441 \ell + 350 \ell^2 + 
    7 (-9 + 35 \ell) k + 35 k^2]\Bigr\}\\
&\quad+ O(H^8)\ .
\end{split}\end{equation}

We give explicit expressions for the retarded boundary to bulk propagators in tables \ref{tab:GoutM1}, \ref{tab:GoutM2}, \ref{tab:GoutE1} and \ref{tab:GoutE2}.
\begin{table}[H]
	\centering
		\caption{$r^{\nu+\frac{d-4}{2}}\Gout$ for magnetic Debye potentials ($z=-i\w r$, $H=1$). }
		\setlength{\extrarowheight}{2pt}
	\begin{tabular}{|c|c|c|}
		\hline
		$\mu=\frac{d}{2}-1$&$\ell=0$&$\ell=1$\\[0.5ex]\hline
		$d=3$& $1$&$ 1+z $  \\[0.5ex] $d=5$ & $
		1+z+r^2$ & $1+\frac{z^2}{3}+r^2+\frac{r^2z}{3}$ \\[0.5ex]
		$d=7$&$1+z+\frac{z^2}{3}+r^2\left(z+\frac{2}{3}\right)+r^4 $ &$1+z+\frac{z^2}{15}(z+6)+\frac{r^2z}{15}(3z+5)+\frac{r^4z}{5}$ \\[0.5ex] \hline
  \end{tabular}\label{tab:GoutM1}
\end{table}
\begin{table}[H]
	\centering
		\caption{$r^{\nu+\frac{d-4}{2}}\Gout$ for magnetic Debye potentials ($z=-i\w r$, $H=1$)}
		\setlength{\extrarowheight}{2pt}
	\begin{tabular}{|c|c|}
		\hline
		$\mu=\frac{d}{2}-1$&$\ell=2$\\[0.5ex]\hline
		$d=3$&  $1+z+\frac{z^2}{3}-\frac{r^2}{3}$  \\[0.5ex] $d=5$  & $1+z+\frac{2z^2}{5}+\frac{z^3}{15}-\frac{r^2}{15}\left(6+z-z^2\right)-\frac{r^4}{15}$\\[0.5ex]
		$d=7$&$1+z+\frac{3z^2}{7}+\frac{2z^3}{21}+\frac{z^4}{105}-\frac{r^2}{105}\left(45+10z+8z^2+3z^3\right)-\frac{r^4}{35}\left(3+z-z^2\right)-\frac{r^6}{35}$\\[0.5ex] \hline
  \end{tabular}\label{tab:GoutM2}
\end{table}
\begin{table}[H]
	\centering
		\caption{$r^{\nu+\frac{2-d}{2}}\Gout$ for electric Debye potentials ($z=-i\w r$, $H=1$). }
		\setlength{\extrarowheight}{2pt}
	\begin{tabular}{|c|c|c|}
		\hline
		$\mu=\frac{d}{2}-2$&$\ell=0$&$\ell=1$\\[0.5ex]\hline
		$d=3$& $1$&$ 1+z $ \\ $d=5$ & $1+z$ & $1+z+\frac{z^2}{3}-\frac{r^2}{3}$ \\
		$d=7$&$1+z+\frac{z^2}{3}+\frac{r^2z}{3}$ &$1+z+\frac{z^2}{5}+\frac{2z^3}{15}-\frac{r^2}{15}(6+z-z^2)-\frac{r^4}{15}$ \\[0.5ex] \hline
	\end{tabular}\label{tab:GoutE1}
\end{table}
\begin{table}[H]
	\centering
		\caption{$r^{\nu+\frac{2-d}{2}}\Gout$ for electric Debye potentials ($z=-i\w r$, $H=1$). }
		\setlength{\extrarowheight}{2pt}
	\begin{tabular}{|c|c|}
		\hline
		$\mu=\frac{d}{2}-2$&$\ell=2$\\[0.5ex]\hline
		$d=3$& $1+z+\frac{z^2}{3}-\frac{r^2}{3}$ \\ $d=5$  & $1+z+\frac{2z^2}{5}+\frac{z^3}{15}-r^2(\frac{3}{5}+\frac{4z}{15})$\\
		$d=7$ &$1+z+\frac{3z^2}{7}+\frac{2z^3}{21}+\frac{z^4}{105}-\frac{r^2}{105}(75+40z+4z^2-z^3)-\frac{4r^4z}{105}$\\[0.5ex] \hline
	\end{tabular}\label{tab:GoutE2}
\end{table}

\subsection{Regularisation for Neumann scalars}\label{app:NeuDesSca}
This section gives a procedure for obtaining the boundary 2-point function $\kO$ for the designer scalar with the scalar satisfying Neumann boundary conditions at the $r=0$ boundary. The action \eqref{eq:ActPhiMain} describes the dynamics of the designer scalar field with two parameters $\nn$ and $\mu$. For the specific case of the electric Debye potential, for which we use the results of this section, these parameters take the values $\nn=d-3$ and $\mu= \frac{d}{2}-2$.

The conjugate field for radial evolution for $\phN$ is given by $\pi_{_\nn}=-r^\nn\Dp\phN$, where we have defined $\Dp\equiv(1-r^2)\partial_r+i\w$ and $\Dm\equiv (1-r^2)\partial_r-i\w$. For a Neumann boundary condition, we will fix the value of $\pi_{_\nn}$ to some source multipole moment. The boundary 2-pt function is then specified by the behaviour of the $\phN$ at $r\to 0$. Naively, this limit yields a divergence similar to the divergence of the Coulomb field of a point charge at $r=0$. We would like to regulate this divergence by the usual QFT technique of adding appropriate counterterms to our action. To this end, let us first look at the divergent behaviour of the $\phN$ as we take the limit $r\to 0$. 

We require $\phN$ to satisfy outgoing boundary conditions at the horizon, which is equivalent to demanding analyticity at $r=1$ in the outgoing Eddington-Finkelstein coordinates. The Neumann boundary condition at $r=0$ is imposed as:
\begin{equation}
    \lim_{r\to 0}r^{\nu+\frac{\nn+1}{2}}\left\{-r^\nn\Dp\phN\right\}=\mathcal{J}_{\ell\vec{m}}(\w)\ .
\end{equation}
Given these boundary condition, $\phN$ can be written as:
\begin{equation}
    \phN=\frac{\gO}{\nu+\frac{\nn-1}{2}} \mathcal{J}_{\ell\vec{m}}\ ,
\end{equation}
where $\gO$ is given by:
\begin{equation}\label{eq:Goutdes}
	\begin{split}
		\gO &=r^{-\nu-\frac{1}{2}(\nn-1)}(1+r)^{-i\w}
		\\
		&\times\Bigl\{ {}_2F_1\left[\frac{1-\nu+\mu-i\w}{2},\frac{1-\nu-\mu-i\w}{2};1-\nu;r^2\right]\Bigr.\\
		&\Bigl.\quad\qquad -(1+i\cot\nu \pi)\khO\frac{r^{2\nu}}{2\nu}\ {}_2F_1\left[\frac{1+\nu-
			\mu-i\w}{2},\frac{1+\nu+\mu-i\w}{2};1+\nu;r^2\right] \Bigr\}\ .
	\end{split}
\end{equation}
As one can see from the above formula, $\phN$ has a term with leading behaviour of $r^{-\nu-\frac{\nn-1}{2}}$ as $r\to0$ that diverges and needs to be countertermed away.  

The renormalised field ($\phNren$) is given by:
\begin{equation}
    \phNren=\phN +\frac{r}{\ct_\nn}\Dp\phN\ ,
\end{equation}
where $\ct_\nn$ is the same function that appears in the counterterming of the conjugate field in the corresponding Dirichlet problem (see  part I). In particular,
\begin{equation}\label{eq:ctnn}
	\begin{split}
		\frac{\ct_\nn}{1-r^2}&\equiv
		-r\frac{d}{dr}\ln\left\{ r^{-\nu -\frac{1}{2}
			(\nn-1)} (1-r^2)^{-\frac{i\w}{2}} {}_2F_1\left[\frac{1-\nu+\mu-i\w}{2},\frac{1-\nu-\mu-i\w}{2};1-\nu;r^2\right] \right\}\ .
	\end{split}
\end{equation}
In part I, we showed that $\ct_\nn$ is an even function in $\w$ and has a well-behaved small $r$ expansion. Let us see how $\frac{1}{\ct_\nn}$ behaves at small $r$:
\begin{equation}
\frac{1}{\ct_\nn}=\frac{2}{2\nu+\nn-1}\left\{1+r^2-\frac{(\nu-1)^2-\mu^2-\w^2}{(\nu-1)(2\nu+\nn-1)}r^2+\dots \right\}\ .
\end{equation}
Even in this case, the counterterm is local in time, which can be verified by further expanding the above function. Given this definition of the renormalised field, its boundary behaviour becomes:
\begin{equation}
    \lim_{r\to 0}r^{-\nu+\frac{\nn-1}{2}}\phNren=-\frac{1+i\cot\nu \pi}{\left(\nu+\frac{\nn-1}{2}\right)^2}\khO
\end{equation}

Renormalising the $\phN$ in this manner is equivalent to adding the following counterterm to the action:
\begin{equation}\label{Eq:PhiActCT}
	\begin{split}
		S_{ct}&= \frac{1}{2}\sum_{\ell\vec{m}}\int\frac{d\w}{2\pi} r^{\nn+1} \frac{1}{\ct_\nn(r,\w,\ell,\vec{m})} (\Dp\phN)^* \Dp\phN|_{\text{Bnd}}\ .
	\end{split}
\end{equation}
For the case of even $d$, one needs an additional counterterm to make the action finite. This counterterm is the same as the one required in the Dirichlet case:
\begin{equation}\label{Eq:PhiActCTnu}
	\begin{split}
		S_{ct,\text{Even}}&= \sum_{\ell\vec{m}}\frac{1}{\nu-n}\int\frac{d\w}{2\pi} r^{\nn-1+2n}\Delta_\nn(n,\mu,\w)\phN^* \phN|_{\text{Bnd}}\ .
	\end{split}
\end{equation}

\section{Extended EM sources in dS and radiation reaction}
In this section, we want to describe in detail the results about extended EM sources in de Sitter that were alluded to in the main text. Our goal here is twofold: first, we want to describe radiative multipole expansion in de Sitter with correct normalisations for multipole moments, etc., which in the $H\to 0$ limit reproduces the flat space analysis. The second goal is to compute the analogue of ALD force in de Sitter (they also have a
$H\to 0$ limit). One main difference to our flat spacetime analysis is the following: the analysis in this section uses retarded time $u$ instead of the Schwarzschild time $t$. 

\subsection*{Magnetic Multipole Radiation}
We will begin by describing the magnetic multipole radiation due to toroidal currents in dS$_{d+1}$. These currents are identically conserved; hence, conservation equations play no role in this sector, making the analysis conceptually simpler. One begins with decomposing the currents in terms of vector spherical harmonics on the sphere. 
\begin{equation}
\begin{split}\label{eq:SrcParMag}
    \bar{J}^u(r,u,\hat{r})= \bar{J}^r(r,u,\hat{r})= 0\ , \quad
    \bar{J}^I(r,u,\hat{r}) &=\sum_{\alpha \ell \vec{m}}\int_\w J_V^{\alpha \ell \vec{m}}(r,\w) \mathbb{V}^I_{\alpha \ell \vec{m}}(\hat{r}) \ .
\end{split}
\end{equation}
As in the flat space as well as the point source analysis, we will use a convenient parameterization of the gauge field in terms of the magnetic Debye scalar:
\begin{equation}
\begin{split}
    \gaugeV_u(r,u,\hat{r}) = \gaugeV_r(r,u,\hat{r}) = 0\ ,\quad
    \gaugeV_I(r,u,\hat{r}) = \sum_{\alpha,\ell,\vec{m}}\int_\w e^{-i\w u} \Phi_B(r,\w,\alpha,\ell,\vec{m}) \ \mathbb{V}_{I}^{\alpha\ell\vec{m}}(\hat{r})\ .
\end{split}
\end{equation}
We will remind the reader that due to the orthogonality of the VSH with $\mathscr{D}_I\Yellm$, the electromagnetic fields due to toroidal currents completely decouple from those due to charge densities and poloidal currents. Hence, we can proceed with independently analysing the effects of toroidal current distributions.

Given the above parametrisation of the sources and the gauge fields, the electromagnetic field equations reduce to the following inhomogeneous equation for the magnetic Debye potential: 
\begin{equation}
\begin{split}
        \frac{1}{r^{d-3}}D_+\left[r^{d-3}D_+\Phi_B\right]+\omega^2\Phi_B-&\frac{(\ell+1)(\ell+d-3)(1-r^2)}{r^2}\Phi_B +
	r^{2} (1-r^2)J_V^{\alpha \ell \vec{m}}= 0 \ .
\end{split}
\end{equation}
We will construct a Green function for solving the above inhomogeneous differential equation, such that it satisfies the following equation:
\begin{equation}
\begin{split}
        \frac{1}{r^{d-3}}D_+\left[r^{d-3}D_+\mathbb{G}_B(r,r_0;\ell)\right]+\omega^2\mathbb{G}_B(r,r_0;\ell)-&\frac{(\ell+1)(\ell+d-3)(1-r^2)}{r^2}\mathbb{G}_B(r,r_0;\ell) \\ &\qquad+
	 (1-r^2)\frac{\delta(r-r_0)}{r^{d-3}}= 0 \ .
\end{split}
\end{equation}
One interprets this green function as the field generated by a single shell of unit toroidal current placed at $r=r_0$. The appropriate boundary conditions for this problem are that the field should be outgoing at the horizon, i.e. at $r=1$ and that it should be finite at $r=0$. The field is also required to satisfy the correct jump condition at the sphere obtained by integrating the above equation about $r=r_0$:
\begin{equation}
    r^{d-3}\Dp\mathbb{G}_B\Big|^{r_0-}_{r_0+}=1\ .
\end{equation}

We have already analysed the homogeneous solution to this equation that satisfies the outgoing boundary condition: $\Gout_B$, which dictates the field outside the sphere. We also need a homogeneous solution normalisable at the origin to obtain the appropriate Green function for the inhomogeneous solution. This normalisable solution is given by:  
\begin{equation}
\begin{split}
    \left(\frac{1-r}{1+r}\right)^{\frac{i\w}{2}}\Xi^B_n(r,\w,\ell)&\equiv\frac{1}{2\ell+d-2}r^{\ell+1}(1+r)^{-i\w} {}_2F_1\left[\frac{\ell+1-i\w}{2},\frac{\ell+d-1-i\w}{2};\ell+\frac{d}{2};r^2\right] \ .
\end{split}
\end{equation}
The function $\Xi^B_n(r,\w,\ell)$ is the corresponding normalisable solution in the Schwarzschild time $t$\cite{Anninos:2011af}. $\Xi^B_n(r,\w,\ell)$ reproduces the flat space normalisable solution (the Bessel function) \eqref{eq:normBflat} in the $H\to0$ limit(see appendix \ref{app:nearflatexp}). Given these solutions to the homogeneous equation satisfying appropriate boundary conditions, we can construct the Green function for the above inhomogeneous equation such that the solution takes the form:
\begin{equation}
    \Phi_B(r,\w,\alpha,\ell,\vec{m}) =\int dr_0 r_0^{d-1}\mathbb{G}_B(r,r_0;\w,\ell) J_V^{\alpha \ell \vec{m}}(r_0,\w)
\end{equation}
Imposing these boundary conditions and the appropriate jump condition at $r=r_0$, we can write the form of the Green function as follows:
\begin{equation}
    \mathbb{G}_B(r,r_0;\w,\ell)=\frac{1}{W_B(r_0,\w,\ell)}\left(\frac{1-r_<}{1+r_<}\right)^{\frac{i\w }{2}}\Xi^B_n(r_<,\w,\ell)\Gout_B(r_>,\w,\ell)\ ,
\end{equation}
where the Wronskian $W_B(r_0,\w,\ell)$ is given by:
\begin{equation}
    W_B(r_0,\w,\ell)=\left(\frac{1-r_0}{1+r_0}\right)^{i\w}
\end{equation}
and
\begin{equation}
\begin{split}
r_>\equiv \text{Max}(r,r_0)\ ,\quad	
r_<\equiv \text{Min}(r,r_0)\ .
\end{split}
\end{equation}

We have now solved for the Debye potential $\Phi_B$ given a poloidal current distribution. Given the potential, we can now use \eqref{eq:HDemdSC} to obtain the solutions for the fields:
\begin{equation}
\begin{split}
    \Ev &= i\w\  \Phi_B =\int_{\vec{r}_0}i\w \ \mathbb{G}_B(r,r_0;\w,\ell) \VSH_I(\hat{r}_0) J^I(\vec{r}_0,\w)\ , \\
    \Hv &= \Dp \Phi_B =\int_{\vec{r}_0}\Dp\mathbb{G}_B(r,r_0;\w,\ell) \VSH_I(\hat{r}_0) J^I(\vec{r}_0,\w)\ , \\
    \Bvv &= \Phi_B=\int_{\vec{r}_0}\mathbb{G}_B(r,r_0;\w,\ell) \VSH_I(\hat{r}_0) J^I(\vec{r}_0,\w)\ .
\end{split}
\end{equation}
These equations are analogous to the equations \eqref{eq:MagneticFieldsFlat} in flat space. As pointed out in the flat space analysis, the time-dependent toroidal currents give rise not only to a magnetic field but also to an induced electric field.  Given the expressions of the fields, we can now describe the fields outside the sources. We identify the magnetic multipole moment to be:
\begin{equation}
\begin{split}
    \multj^B(\w,\alpha,\ell,\vec{m}) &= \int_{\vec{r}_0}\left(\frac{1-r_0}{1+r_0}\right)^{-\frac{i\w }{2}}\Xi^B_n(r_0,\w,\ell)\VSH_I(\hat{r}_0) J^I(\vec{r}_0,\w) \\
    &= \frac{1}{2\ell+d-2}\int_{\vec{r}_0} r_0^{\ell+1}(1-r_0)^{-i\w}\\
    &\quad \times{}_2F_1\left[\frac{\ell+1-i\w}{2},\frac{\ell+d-1-i\w}{2};\ell+\frac{d}{2};r_0^2\right] \VSH_I(\hat{r}_0) J^I(\vec{r}_0,\w) \ .
\end{split}
\end{equation}
We can rewrite the multipole moment in Schwarzschild time to obtain a formula that can be compared directly with the corresponding flat space expression\eqref{eq:jBflat}. For this, we need to convert our expressions in the Fourier transform of the outgoing time to expressions in terms of Schwarzschild time $t$. This can be achieved by first recognising that:
\begin{equation}
\begin{split}
    J^I(\vec{r},\w)=
    \int du\ e^{i\w u}\bar{ J}^I(\vec{r},t) =\int dt\ e^{i\w t}\left(\frac{1-r}{1+r}\right)^{\frac{i\w}{2}}\bar{ J}^I(\vec{r},t)\ .
\end{split}
\end{equation}
This allows us to rewrite our expression for $\multj_B$ as:
\begin{equation}
    \multj^B(\w,\alpha,\ell,\vec{m})=\int dt\ e^{i\w t}\int_{\vec{r}_0}\Xi^B_n(r_0,i\partial_t,\ell)\VSH_I(\hat{r}_0) J^I(\vec{r}_0,t)
\end{equation}
\subsection*{Electric Multipole Radiation}

Charge density and poloidal currents source the `electric' multipole radiation. Unlike toroidal currents, which are conserved identically and hence carry no constraint coming from conservation equations, poloidal currents are interlinked with the temporal change of charge density. One can rewrite the current density in a form where the conservation equation is manifest:
\begin{equation}
\begin{split}\label{eq:SrcParElec}
    \bar{J}^u(r,u,\hat{r}) &= \sum_{\ell \vec{m}}\int_\w \left[\frac{\ell(\ell+d-2)}{r^2}J_2(r,\w,\ell,\vec{m})-\frac{1}{r^{d-1}}\partial_r\left\{r^{d-1}J_1(r,\w,\ell,\vec{m})\right\}\right] \mathscr{Y}_{\ell\vec{m}}(\hat{r})\ ,\\   
    \bar{J}^r(r,u,\hat{r}) &=  -\sum_{\ell \vec{m}}\int_\w i\w J_1(r,\w,\ell,\vec{m}) \mathscr{Y}_{\ell\vec{m}}(\hat{r})\ , \\ 
    \bar{J}^I(r,u,\hat{r}) &= -\sum_{\ell \vec{m}}\int_\w \frac{i\w}{r^2} J_2(r,\w,\ell,\vec{m})\mathscr{D}^I\mathscr{Y}_{\ell\vec{m}}(\hat{r}) \ .
\end{split}
\end{equation}
The electric parity multipole radiation has a gauge redundancy in its description tied to the above conservation equation. Due to such a gauge redundancy, one cannot impose the same gauge conditions as we did for free electromagnetic fields in the presence of extended sources. In particular, the gauge field parametrisation, in terms of the Debye scalar, requires modification with additional source-local terms to satisfy the sourced Maxwell equations. The gauge field parametrisation then becomes:
\begin{equation}
\begin{split}
    \gaugeV_u(r,u,\hat{r}) &= \sum_{\ell\vec{m}}\int_\w e^{-i\w u} \left[r^{3-d}D_{+}\Phi_E(r,\w,\ell,\vec{m}) - (1-r^2)J_2\right] \Yellm(\hat{r}) \ , \\
    \gaugeV_r(r,u,\hat{r}) &= \sum_{\ell\vec{m}}\int_\w e^{-i\w u} \left[r^{3-d}\partial_r\Phi_E(r,\w,\ell,\vec{m}) - J_2\right] \Yellm(\hat{r}) \ , \\
    \gaugeV_I(r,u,\hat{r}) &= \sum_{\alpha,\ell,\vec{m}}\int_\w e^{-i\w u} \Phi_B(r,\w,\ell,\vec{m}) \ \mathbb{V}_{I}^{\alpha\ell\vec{m}}(\hat{r})\ .
\end{split}
\end{equation}
One can check the consistency of this gauge choice by plugging it into the Maxwell equations, which are satisfied contingent on the fact that the Debye scalar is a solution to second-order inhomogeneous differential equations:  
\begin{equation}
\begin{split}
	\frac{1}{r^{3-d}}D_+\left[r^{3-d}D_+\Phi_E\right]+\omega^2\Phi_E-&\frac{\ell(\ell+d-2)(1-r^2)}{r^2}\Phi_E\\-
	r^{d-3}&\left[\Dp\left[(1-r^2)J_2\right]-(1-r^2)J_1\right]= 0\ .
\end{split}
\end{equation}
The particular combination of the source parameters $J_1$ and $J_2$ that appears on the RHS of the EOM of $\Phi_E$ can be considered the `radiative' source combination. This combination is solely responsible for the radiative energy loss.

Given the above parametrisation of the gauge field, the electromagnetic field strengths in the presence of sources can be computed to give:
\begin{equation}
\begin{split}
    \Er &= \frac{\ell(\ell+d-2)}{r^{d-1}}\Phi_E-J_1\ ,\\
    \Es &=\frac{1}{r^{d-3}}D_{+}\Phi_E - (1-r^2)J_2 \ ,\\
    \Hs &=\frac{i\w}{r^{d-3}}\Phi_E \ .
\end{split}
\end{equation}

One can think of the above formulae alternatively as being obtained from the gauge field parametrisation in terms of the electromagnetic fields: 
\begin{equation}
\begin{split}
\gaugeV_u(r,u,\hat{r}) &= \sum_{\ell\vec{m}} \int_\w e^{-i\w u}\Es(r,\w,\ell,\vec{m})\mathscr{Y}_{\ell\vec{m}}(\hat{r})\ ,\\
\gaugeV_r(r,u,\hat{r})
&= \sum_{\ell\vec{m}}\int_\w \frac{e^{-i\w u}}{1-r^2}\left[\Es(r,\w,\ell,\vec{m})-\Hs(r,\w,\ell,\vec{m})\right]\mathscr{Y}_{\ell\vec{m}}(\hat{r}) \ , \\
    \gaugeV_I(r,u,\hat{r}) &= 0 .
\end{split}
\end{equation}

Given the inhomogeneous equations of motion for $\Phi_E$, one can solve for the electromagnetic fields in de Sitter by finding the corresponding Green's functions. The natural boundary condition to impose in de Sitter is the outgoing boundary condition at the horizon. The outgoing Green's functions can be written as: 
\begin{equation}
\begin{split}
\mathbb{G}_E(r,r_0;\w,\ell)&=\frac{1}{W(r_0,\w,\ell)}\left(\frac{1-r_<}{1+r_<}\right)^{\frac{i\w }{2}}\Xi^E_n(r_<,\w,\ell)\Gout_E(r_>,\w,\ell)\ .
\end{split}
\end{equation}
where the $\Gout_E$ is the boundary-to-bulk outgoing propagator defined in the previous section and $\Xi^E_n$ is the \emph{normalisable} mode:
\begin{equation}
\begin{split}
\Xi^E_n(r,\w,\ell)&\equiv\frac{1}{2\ell+d-2}r^{\ell+d-2}(1-r^2)^{-\frac{i\w}{2}}
\\
  &\qquad\times {}_2F_1\left[\frac{\ell+2-i\w}{2},\frac{\ell+d-2-i\w}{2};\ell+\frac{d}{2};r^2\right] \ .
\end{split}
\end{equation}
For odd values of $d$, the above expressions are well-defined, but for even values of $d$, one should evaluate the expressions as a limiting case.
With these Green's functions, we can write the Debye scalars as:
\begin{equation}
\begin{split}
    \Phi_E(r,\w,\ell,\vec{m}) &=\int dr_0 \mathbb{G}_E(r,r_0;\w,\ell)\left[J_1(r_0,\w,\ell,\vec{m})-\frac{1}{1-r^2}\Dp^0\left\{(1-r^2)J_2(r_0,\w,\ell,\vec{m})\right\}\right] \ .
\end{split}
\end{equation}
The above solutions in terms of the source parameters $J_1$ and $J_2$ can be rewritten in terms of the source currents to give:
\begin{equation}
\begin{split}
    \Phi_E= &- \frac{1}{i\w}\int dr_0 \int d\Omega_{d-1} \Bigg[\Yellm^*J^r\mathbb{G}_E(r,r_0;\w,\ell)-\frac{1}{\ell(\ell+d-2)}r_0^2J^I\mathscr{D}_I\Yellm^*\Dm^0 \mathbb{G}_E(r,r_0;\w,\ell)\Bigg]  \\
    \Hs  = &- \frac{1}{r^{d-3}}\int dr_0 \int d\Omega_{d-1} \Bigg[\Yellm^*J^r\mathbb{G}_E(r,r_0;\w,\ell)-\frac{1}{\ell(\ell+d-2)}r_0^2J^I\mathscr{D}_I\Yellm^*\Dm^0 \mathbb{G}_E(r,r_0;\w,\ell)\Bigg] 
    \\
    \Er  = & \frac{1}{r^{d-1}}\int dr_0 \int d\Omega_{d-1} \Yellm^*r_0^2\Bigg[J^t\Dm^0\mathbb{G}_E(r,r_0;\w,\ell) + i\w  J^r\mathbb{G}_E(r,r_0;\w,\ell)\Bigg]\ .
\end{split}
\end{equation}

The expression for $\Es$ can be obtained from the above by integrating the first Bianchi as we did in flat space. The first Bianchi in EF coordinates in dS is:
\begin{equation}
    \Dp \Es=(1-r^2)\Er+i\w \Hs\ .
\end{equation}
To integrate this equation, we will rewrite it as:
\begin{equation}
\begin{split}
    \partial_r\left(e^{i\w r_\ast}\Es\right) &= e^{i\w r_\ast}\left[\Er+\frac{i\w}{(1-r^2)}\Hs\right]\ \\
    &= e^{i\w r_\ast}\int_0^\infty dr_0\int_{\hat{r}_0\in\mathbb{S}^{d-1}} \mathscr{Y}^\ast_{\ell\vec{m}}(\hat{r}_0) \ 
\mathbb{G}_E(r,r_0;\w,\ell) \Bigl\{
i\w \frac{r_0^2-\frac{r^2}{1-r^2}}{r^{d-1}} J^r(\vec{r}_0,\w)\\
&\qquad -\frac{1}{1-r_0^2}\Dp^0\left[\frac{(1-r_0^2)r_0^2}{r^{d-1}}J^t(\vec{r}_0,\w)+i\w \frac{(1-r_0^2)r_0^2}{r^{d-3}} \frac{\mathscr{D}_IJ^I(\vec{r}_0,\w)}{\ell(\ell+d-2)}\right]\Bigr\}\ ,
  \end{split}
\end{equation}
where $r_\ast$ is the tortoise coordinate. We can now integrate this equation to write an expression for $\Es$ as follows:
\begin{equation}
 \begin{split}
    \Es &=e^{-i\w r_\ast}\int dr_1 e^{i\w r_{1\ast}}\left[\Er(r_1)+\frac{i\w}{(1-r_1^2)}\Hs(r_1)\right]\ \\
    &= e^{-i\w r_\ast}\int dr_1e^{i\w r_{1\ast}}\int_0^\infty dr_0\int_{\hat{r}_0\in\mathbb{S}^{d-1}} \mathscr{Y}^\ast_{\ell\vec{m}}(\hat{r}_0) \ 
\mathbb{G}_E(r_1,r_0;\w,\ell) \Bigl\{
i\w \frac{r_0^2-\frac{r_1^2}{1-r_1^2}}{r_1^{d-1}} J^r(\vec{r}_0,\w)\\
&\qquad -\frac{1}{1-r_0^2}\Dp^0\left[\frac{(1-r_0^2)r_0^2}{r_1^{d-1}}J^t(\vec{r}_0,\w)+i\w \frac{(1-r_0^2)r_0^2}{r_1^{d-3}} \frac{\mathscr{D}_IJ^I(\vec{r}_0,\w)}{\ell(\ell+d-2)}\right]\Bigr\}\ .
  \end{split}   
\end{equation}

\subsection*{Action reduction}
We must evaluate the on-shell action on the dS-SK geometry to obtain the effective action describing the extended observer. Although the explicit derivation specific to the dS-SK geometry is treated in the next section, we want to simplify the sourced Maxwell action in this section after imposing the equations of motion to bring it to a simpler form. We begin with the Maxwell action:
\begin{equation}
    S_{EM}= -\int d^{d+1}x \ \left[\frac{1}{4}\bar{\mathcal{C}}_{\mu\nu}\bar{\mathcal{C}}^{\mu\nu}-\bar{\mathcal{V}}_\mu J^\mu\right]\ .
\end{equation}
We perform, as before, an expansion in the spherical harmonics and a Fourier transform on the outgoing time to obtain, in terms of the field strengths:
\begin{equation}
\begin{split}
    S_{EM}= -\frac{1}{2}\sum_{\ell\vec{m}}\int_\w \int dr \frac{r^{d-1}}{1-r^2}\Bigg[& (1-r^2)|\Er|^2-\frac{\ell(\ell+d-2)}{r^2}\left\{|\Hs|^2-|\Es|^2\right\}+2i\w\Hs J_1^*\\
    &\qquad +\Es\left\{\frac{1}{r^{d-1}}\Dm(r^{d-1}J_1^*)-\frac{\ell(\ell+d-2)}{r^2}(1-r^2)J_2^*\right\}\Bigg]\ .
\end{split}
\end{equation}

Using integration by parts and the $\Phi_E$ EOM:
\begin{equation}
\begin{split}
    S_{\text{on-shell}}=\frac{1}{2}\sum_{\ell\vec{m}}\int\frac{d\w}{2\pi} \Bigg[\partial_r\left\{-r^2(\Dp\Phi_E)^*J_1+r^2(1-r^2)\ell(\ell+d-2)\Phi_E^* J_2+r^{d+1}J_1J_2^*\right\}\\
    +\ell(\ell+d-2) \Phi_E^* \left\{\frac{1}{1-r^2}\Dp\left[(1-r^2)J_2\right]+ J_1\right\}\\
    -r^{d-1}|J_1|^2-r^{d+1}\ell(\ell+d-2)(1-r^2)|J_2|^2\Bigg]\ .
\end{split}
\end{equation}

To obtain the parameters $J_1$ and $J_2$ given a current density, we invert the equations in \eqref{eq:SrcParElec} to obtain:
\begin{equation}
\begin{split}\label{eq:J1J2}
    J_1(r,\w,\ell,\vec{m}) & = \int d\Omega_{d-1}\Yellm^*(\hat{r}) \frac{J^r(r,\w,\hat{r})}{-i\w}\ ,\\
    J_2(r,\w,\ell,\vec{m}) & = -\int d\Omega_{d-1}\Yellm^*(\hat{r})\frac{r^2}{-i\w \ell(\ell+d-2)}\mathscr{D}_IJ^I(r,\w,\hat{r})\\
    & = \int d\Omega_{d-1}\Yellm^*(\hat{r}) \frac{1}{\ell(\ell+d-2)}\frac{1}{1-r^2}\left[\frac{1}{-i\w r^{d-3}}\Dp\left(r^{d-1}J^r\right)- r^2 J_u\right]\ .
\end{split}
\end{equation}
Here we have rewritten the $\mathscr{D}_IJ^I$ term using the conservation equation:
\begin{equation}
    \mathscr{D}_IJ^I=-\frac{1}{1-r^2}\left[i\w J_u+\frac{1}{r^{d-1}}\Dp\left(r^{d-1}J^r\right)\right] \ .
\end{equation}

\subsection{Radiative multipole moments on dS-SK}
\label{app:dSSKaction}
We will use the result of the previous section and identify the average/difference electromagnetic radiative multipole moments on the dS-SK geometry. The equation of motion for $\Phi_E$ shows that the $J_2$ acts like a `Neumann' source, i.e., the $\Phi_E$ is sourced by a derivative operator acting on $J_2$. On the other hand, $J_1$ acts like a `Dirichlet' source. For the `Neumann' case, we identify $\rho$ with $\frac{J_2}{r^{3-d}}$. Consider the contribution to the `$R$' electric multipole moment from the $J_2$ source takes the form:
\begin{equation}
\begin{split}
    & -\ell\int_R dr\  \left(\frac{1-r}{1+r}\right)^{-\frac{i\w}{2}}(1-r^2)\partial_r\Xi^E_n(r,\w,\ell)\ J_2(r,\w,\ell,\vec{m}) \\
    = & \frac{1}{\ell+d-2}\int_R dr\  \left(\frac{1-r}{1+r}\right)^{-\frac{i\w}{2}}\partial_r\Xi^E_n(r,\w,\ell)\int d\Omega_{d-1}\Yellm^*(\hat{r})\left[ r^2 J_u+\frac{1}{i\w r^{d-3}}\Dp\left(r^{d-1}J^r\right)\right]\ .
\end{split}
\end{equation}

We can use integration by parts on the $J^r$ term along with the following identity for $\Xi_n$:
\begin{equation}
\begin{split}
    \Dm&\left[\left(\frac{1-r}{1+r}\right)^{-\frac{i\w}{2}}r^{3-d}(1-r^2)\partial_r\Xi^E_n\right] \\&\qquad \qquad \qquad = -\left\{\w^2-\frac{\ell(\ell+d-2)}{r^2}(1-r^2)\right\}\left(\frac{1-r}{1+r}\right)^{-\frac{i\w}{2}}r^{3-d}\Xi^E_n \ .
\end{split}
\end{equation}
The full radiative multipole moment then becomes:
\begin{equation}
\begin{split}
    \multj^E_R(r,\w,\ell,\vec{m}) & = \frac{1}{2\nu(\ell+d-2)}\int_R d^dx \ \Yellm^*(\hat{r})\frac{1}{r^{d-3}}\left(\frac{1-r}{1+r}\right)^{-\frac{i\w}{2}}\\
    &\qquad\qquad\qquad \times \Bigg[ \partial_r\Xi^E_n(r,\w,\ell) J_u(r,\w,\hat{r})+i \w\ \Xi^E_n(r,\w,\ell) J^r(r,\w,\hat{r})\Bigg] \\
    \multj^E_L(r,\w,\ell,\vec{m}) & =  \frac{1}{2\nu(\ell+d-2)}\int_R d^dx \ \Yellm^*(\hat{r})\frac{1}{r^{d-3}}\left(\frac{1-r}{1+r}\right)^{-\frac{i\w}{2}}\\
    &\qquad\qquad\qquad \times \Bigg[ \partial_r\Xi^E_n(r,\w,\ell) J_u(r,\w,\hat{r})+i \w\ \Xi^E_n(r,\w,\ell) J^r(r,\w,\hat{r})\Bigg] \ .
\end{split}
\end{equation}
The computation of the magnetic multipole moments on the dS-SK geometry, on the other hand, is straightforward as there is no ambiguity due to conservation equations. We find:
\begin{equation}
\begin{split}
    \multj^B_R(\w,\alpha,\ell,\vec{m}) &= \int_R d^{d}r\left(\frac{1-r_0}{1+r_0}\right)^{-\frac{i\w }{2}}\Xi^B_n(r_0,\w,\ell)\VSH_I(\hat{r}_0) J^I(\vec{r}_0,\w)\ , \\
    \multj^B_L(\w,\alpha,\ell,\vec{m}) &= \int_L d^{d}r\left(\frac{1-r_0}{1+r_0}\right)^{-\frac{i\w }{2}}\Xi^B_n(r_0,\w,\ell)\VSH_I(\hat{r}_0) J^I(\vec{r}_0,\w)\ .   
\end{split}
\end{equation}

 Corresponding to these multipole moments, we can write down the STF moments, which are better suited for a post-newtonian expansion:
\begin{equation}
\begin{split}\label{eq:STFelectric}
{}^E\mathcal{Q}_{A,STF}^{i_1\ldots i_\ell}(\w) & \equiv  \frac{1}{(\ell+d-2)}(\Pi^S)^{<i_1i_2 \ldots i_\ell>}_{<j_1j_2\ldots j_\ell>}\int d^dr\  \hat{r}^{j_1}\hat{r}^{j_2}\ldots \hat{r}^{j_\ell}\ \\
&\times\frac{2\nu}{r^{d-3}}\Bigg[ \partial_r\Xi^E_n(r,\w,\ell) J_u(r,\w,\hat{r}) +i \w\ \Xi^E_n(r,\w,\ell) J^r(r,\w,\hat{r})\Bigg] \ ,\\
{}^E\mathcal{Q}_{D,STF}^{i_1\ldots i_\ell}(\w) &\equiv \frac{1}{(\ell+d-2)}(\Pi^S)^{<i_1i_2\ldots i_\ell>}_{ <j_1j_2\ldots j_\ell>}\int d^dr\  \hat{r}^{j_1}\hat{r}^{j_2}\ldots \hat{r}^{j_\ell}\ \\
&\times\frac{2\nu}{r^{d-3}} \left(\frac{1-r}{1+r}\right)^{-\frac{i\w}{2}}\Bigg[\partial_r\Xi^E_n(r,\w,\ell) J_u(r,\w,\hat{r})\\ & \qquad\qquad\qquad\qquad\qquad\qquad +i \w \Xi^E_n(r,\w,\ell) J^r(r,\w,\hat{r})\Bigg] \ .
\end{split}
\end{equation}
Similarly, the magnetic multipole moments on dS-SK can be written as:
\begin{equation}
\begin{split}\label{eq:STFmagnetic}
    {}^B\mathcal{Q}_A^{i<i_1\ldots i_\ell>}\equiv (\Pi^V)^{i<i_1i_2\ldots i_\ell>}_{j<j_1j_2\ldots j_\ell>}\int d^{d}x \ x^{j_1}x^{j_2}\ldots x^{j_\ell}\ \frac{2\nu}{r^{\ell+1}} \left(\frac{1-r}{1+r}\right)^{-\frac{i\w}{2}}\Xi^B_n(r,\w,\ell) J^j_A \ ,  \\
    {}^B\mathcal{Q}_D^{i<i_1\ldots i_\ell>}\equiv (\Pi^V)^{i<i_1i_2\ldots i_\ell>}_{j<j_1j_2\ldots j_\ell>}\int d^{d}x \ x^{j_1}x^{j_2}\ldots x^{j_\ell}\ \frac{2\nu}{r^{\ell+1}} \left(\frac{1-r}{1+r}\right)^{-\frac{i\w}{2}}\Xi^B_n(r,\w,\ell)J^j_D  
\end{split}
\end{equation} 
The dissipative part of the on-shell action in terms of these STF multipoles can then be written as:
\begin{equation}
\begin{split}\label{eq:dSSKSTFaction}
    S_{RR}^\text{Odd $d$}&=-\sum_{\ell}\int\frac{d\w}{2\pi}  \frac{1}{4\nu^2\nn_{d,\ell}|\mathbb{S}^{d-1}|} \frac{1}{\ell!}\Bigg[\Kout_E\frac{\ell+d-2}{\ell}{}^E\mathcal{Q}_{D,STF}^{\ast<i_1i_2\ldots i_\ell>}{}^E\mathcal{Q}^{A,STF}_{ <i_1i_2\ldots i_\ell>}\\ 
    & \qquad\qquad \qquad\qquad \qquad\qquad \qquad\qquad +\Kout_B\ {}^B\mathcal{Q}_{D,STF}^{\ast i<i_1i_2\ldots i_\ell>}{}^B\mathcal{Q}^{A,STF}_{ i<i_1i_2\ldots i_\ell>}\Bigg]\ .
\end{split}
\end{equation}

\subsection*{Near Flat Expansion}\label{app:nearflatexp}
In the next section, we will calculate the radiation reaction of a point particle moving along an arbitrary trajectory. This result is obtained in a Hubble expansion about flat spacetime up to order $H^4$ terms. To facilitate this calculation, we quote the Hubble expansions of some useful quantities in this section\footnote{The expressions quoted here are special cases of those derived in appendix D.1 of part I\cite{Loganayagam:2023pfb}. There, one can find a detailed derivation of these formulae.}. Since the radiation reaction force is well defined only in $dS_{d+1}$ for odd $d$, we will restrict our analysis to that particular case. 

The radiation reaction kernels $\Kout_{E/B}$ have the following expansions in the small $H$ limit:
\begin{equation}\label{eq:KoutBflat}
\begin{split}
\Kout_B|_\text{Odd $d$} &= \frac{2\pi i}{\Gamma(\nu)^2}\left(\frac{\w}{2}\right)^{2\nu} \Bigg[1+\left\{\nu^2+\frac{3 }{4}d(d-4)+2\right\}\frac{\nu}{3!!}\frac{H^2}{\w^2}+c_B\frac{\nu(\nu-1)}{5!!}\frac{H^4}{\w^4}+O\left(\frac{H^6}{\w^6}\right)\Bigg]\ ,
\end{split}
\end{equation}
as well as
\begin{equation}\label{eq:KoutEflat}
\begin{split}
&\Kout_E|_\text{Odd $d$} = \frac{2\pi i}{\Gamma(\nu)^2}\left(\frac{\w}{2}\right)^{2\nu} \Bigg[1+\left\{\nu^2+\frac{3 }{4}d(d-8)+11\right\}\frac{\nu}{3!!}\frac{H^2}{\w^2}+c_E\frac{\nu(\nu-1)}{5!!}\frac{H^4}{\w^4}+O\left(\frac{H^6}{\w^6}\right)\Bigg]\ .
\end{split}
\end{equation}
Here, we have defined the coefficients
\begin{equation}
\begin{split}
c_B&\equiv \frac{5\nu^4-4\nu^3+\left\{\frac{15d(d-4)+32}{2}\right\}\nu^2-\left\{15d(d-4)+44\right\}\nu+\frac{45}{16}d^2(d-4)^2-24}{2\times 3}\ ,\\
c_E&\equiv \frac{5\nu^4-4\nu^3+\left\{\frac{15d(d-8)+212}{2}\right\}\nu^2-\left\{15d(d-8)+224\right\}\nu+\frac{45}{16}d(d-8)\left\{d(d-8)+24\right\}+381}{2\times 3}\ .
\end{split}
\end{equation}
These expressions can be obtained using the Stirling approximation. In the flat limit, the combination appearing in the influence phase evaluates to
\begin{equation}
    \frac{\Kout|_\text{Odd $d$}}{4\nu^2 \nn_{d,\ell}}=\frac{\w^{2\ell+d-2}}{(d-2)!!(2\ell+d-2)!!}\ ,
\end{equation}
so that the above expressions can be used to give an explicit expression for the influence phase in a small $H$ expansion.

The smearing functions for the multipole moments $\Xi^{E/B}_n$ can also be expanded about small $H$ in the following manner:
\begin{equation}\label{eq:XiExp}
	\begin{split}
	\Xi^{E/B}_n =\sum_{k=0}^\infty \mathfrak{p}^{E/B}_{k}(\nu,H^2,\w^2)
		\mathfrak{B}^{E/B}_k\ ,
	\end{split}
\end{equation}
where,
\begin{equation}
	\begin{split}
\mathfrak{B}^E_k&\equiv\frac{r^{\nu+\frac{d}{2}-1+2k}}{2\nu(\nu+1)\ldots(\nu+k)}\  {}_0F_1\left[1+k+\nu,-\frac{\w^2r^2}{4}\right]=\frac{\Gamma(\nu)\ r^{\frac{d-2}{2}+k}}{2(\w/2)^{k+\nu}}J_{k+\nu}(\w r)\ , \\
\mathfrak{B}^B_k&\equiv\frac{r^{\nu-\frac{d}{2}+2+2k}}{2\nu(\nu+1)\ldots(\nu+k)}\  {}_0F_1\left[1+k+\nu,-\frac{\w^2r^2}{4}\right]=\frac{\Gamma(\nu)\ r^{2-\frac{d}{2}+k}}{2(\w/2)^{k+\nu}}J_{k+\nu}(\w r)\ ,
\end{split}
\end{equation}
and
\begin{equation}
	\begin{split}
		\mathfrak{p}^{E/B}_k
		&\equiv \frac{H^{2k}}{k!}	\sum_{m=0}^k (-)^m\binom{k}{m}\sum_{n=0}^m(-)^n\binom{m}{n} \sigma^{2k-2m}\frac{\Gamma(\alpha_{E/B}+m)\Gamma(1+\nu+m)}{\Gamma(\alpha_{E/B}+m-n)\Gamma(1+\nu+m-n)}\ \\
		&\quad \qquad\qquad \times\frac{\Gamma(\alpha_{E/B}+i\sigma+m-n) \Gamma(\alpha_{E/B}-i\sigma+m-n) }{\Gamma(\alpha_{E/B}+i\sigma) \Gamma(\alpha_{E/B}-i\sigma) }\ .
	\end{split}
\end{equation}
The electric vs magnetic parity smearing function only differs in the $\alpha$ parameter in the above formula: 
\begin{equation}
	\begin{split}
\alpha_E\equiv \frac{1}{2}(3-\frac{d}{2}+\nu)\ ,\quad \alpha_B\equiv \frac{1}{2}(2-\frac{d}{2}+\nu) \ , \quad \sigma= \frac{\w}{2H}\ .
	\end{split}
\end{equation}
This expansion was derived for the generic case of the designer scalar in part I, which we have used for the specific cases of $\{\nn=3-d,\mu=\frac{d}{2}-2\}$ for the electric and $\{\nn=d-3,\mu=\frac{d}{2}-1\}$ for the magnetic smearing function. 

\subsection{Non-relativistic expansion}\label{app:dSPN}
We will now derive the Abraham-Lorentz-Dirac force in arbitrary dimensions and find curvature corrections in dS spacetime. We will also find terms, up to cubic order in amplitude, contributing to the full radiation reaction(RR) force. We will follow the technique used in \cite{Loganayagam:2023pfb} with some crucial differences for a charged particle interacting with electromagnetic fields.

Let us start with a point source travelling along a worldline $x(\tau)$. We will evaluate the RR force in a non-relativistic approximation. We will also take the particle to move close to the south pole, i.e. $rH\ll 1$. The wavelength of the radiation is much larger than the `amplitude' of the trajectory about the south pole($\w r\ll 1$) but much smaller than the Hubble constant ($\w\gg H$). In \cite{Loganayagam:2023pfb}, we referred to these approximations as the post-newtonian(PN) approximations adapted to dS.

The 4-current density associated with a charged particle in dS is given by,
\begin{align}
\label{eq:ParticleSrc}
	\begin{split}
		\bar{J}^\mu(x')=\int \frac{dx^\mu}{d\tau}\delta^{d+1}(x(\tau)-x')d\tau = \frac{dx^\mu}{dt}\delta(\vec{x}-\vec{x}')\ 
	\end{split}
\end{align}

In the doubled dS-SK geometry, this source will also be doubled, i.e. given by two worldlines $x_L(\tau)$ and $x_R(\tau)$. Correspondingly, they will source the electromagnetic fields by current densities $\bar{J}_L$ and $\bar{J}_R$. The particle degrees of freedom, on which the effective action of radiation reaction is defined, are the positions of the two particles on either side of the geometry, as well as their time derivatives i.e. $\{x_L,x_R,\dot{x}_L,\dot{x}_R,\ddot{x}_L,\ddot{x}_R,\dots\}$. In the RR Lagrangian, we will retain only up to quartic terms in the $x$'s. The RR force is determined by the terms linear in the difference of their positions as well as time derivatives, i.e. $\{x_D,\dot{x}_D,\ddot{x}_D,\dots\}$. The terms cubic in $x_D$ give rise to non-thermal fluctuations, which will be discussed later. The fact that there are only linear and cubic $x_D$ terms in the Lagrangian follows from the fact that the action is odd under $R-L$ exchange.

Such an amplitude expansion of the lagrangian allows us to evaluate the forces and fluctuations in a straightforward way by Taylor expanding the lagrangian about the point where $x_D$ and its time derivatives are zero. As an illustration, consider a given function $\mathfrak{f}$ of position. Its corresponding average and difference are obtained as:
\begin{align}
	\frac{1}{2}\left[\mathfrak{f}\left(x_A+\frac{x_D}{2}\right)+\mathfrak{f}\left(x_A-\frac{x_D}{2}\right)\right]=\mathfrak{f}(x_A)+\frac{x_D^2}{4}\frac{\partial^2 \mathfrak{f}}{\partial x_A^2}+O(x_D^4)\ ,\\
	\mathfrak{f}\left(x_A+\frac{x_D}{2}\right)-\mathfrak{f}\left(x_A-\frac{x_D}{2}\right)=x_D\frac{\partial\,\mathfrak{f}}{\partial x_A}+\frac{x_D^3}{24}\frac{\partial^3\,\mathfrak{f}}{\partial x_A^3}+O(x_D^4)\ .
\end{align}
In general, we will need to expand functions which are not just functions of positions but also depend on the time derivatives of the positions, in which case one uses a multi-variable Madhava-Taylor expansion.

The electric sector on-shell action gives the following RR lagrangian: 
\begin{equation}
	\begin{split}
& |\mathbb{S}^{d-1}|  (d-2)!!\times(-1)^{\frac{d+1}{2}}  L_{E} \\ = & (d-1)[x_i]_D\mathbb{D}_1[x^i]_A-\frac{d}{4}\left[x_ix_j-\frac{x^2}{d}\delta_{ij}\right]_D\mathbb{D}_2\left[x^i x^j-\frac{x^2}{d}\delta_{ij}\right]_A\\
		&+\left\{\frac{1}{2}(x_i)_D\mathbb{D}_{1}^{X}[x^ix^2]_A+\frac{1}{2}(x^2x_i)_D\mathbb{D}_{1}^{X}[x^i]_A\right\}-\left\{(x_i)_D\mathbb{D}_{1}^{V}\partial_t[(\vec{x}\cdot\vec{v})x^i]_A+((\vec{x}\cdot\vec{v})x^i)_D\mathbb{D}_{1}^{V}\partial_t[x^i]_A\ \right\}\ .
	\end{split}
\end{equation}
Only the dipole and quadrupole terms contribute to this order. The differential operators used in the above expressions $\{\mathbb{D}_1,\mathbb{D}_2,\mathbb{D}_1^X,\mathbb{D}_1^V\}$ are given explicitly in table \ref{tab:RRdiffOp}. The number on the differential operator indicates which multipole contributes to that particular term whereas the subscripts signify the structure on which this operator acts. Similar to the scalar case, the magnetic sector action gives the following RR Lagrangian:
\begin{equation}
	\begin{split}
|\mathbb{S}^{d-1}|(d-2)!!\times(-1)^{\frac{d+1}{2}}L_{B}=\frac{1}{4}(x_iv_j-x_jv_i)_D\mathbb{D}_1^\mathbb{V}[x^iv^j-x^jv^i]_A\ .
	\end{split}
\end{equation}
In this case, only the dipole contributes to the quartic lagrangian, and we only have one differential operator. The full lagrangian is just a sum of these two contributions.

Given the lagrangian, we can use integration by parts to rewrite it in a way that one can read off the RR force:
\begin{equation}
L=\frac{(-1)^{\frac{d-1}{2}}}{|\mathbb{S}^{d-1}|(d-2)!!}\left[f_i(x_A)x^i_D+\frac{1}{4}N_i(x_D)x^i_A\right]\ .
\end{equation}
Here, $f^i$ are the Euler-Lagrange derivatives of the terms linear in $x_D$ with respect to $x^i_D$. Similarly, $N^i$ are the Euler-Lagrange derivatives of the terms linear in $x_A$ with respect to $x^i_A$. The $f^i$'s can be written as:
\begin{equation}
\begin{split}
f^i=&-(d-1)\mathbb{D}_1[x^i]+\frac{d}{2}x_j\mathbb{D}_2[x^i x^j]-\frac{x^i}{2}\mathbb{D}_2[x^2]\\
&-\left\{\frac{1}{2}\mathbb{D}_{1}^{X}[x^ix^2]+\frac{1}{2}x^2\mathbb{D}_{1}^{X}[x^i]+x^i x^j \mathbb{D}_{1}^{X}[x_j]\right\} \\ &+\left\{\mathbb{D}_{1}^{V}\partial_t\left(x^i(x_jv^j)\right)+(x_jv^j)\mathbb{D}_{1}^{V}[v^i]+(x^jv^i)\mathbb{D}_{1}^{V}[v_j]-\partial_t\left(x^i x^j \mathbb{D}_{1}^{V}[v_j]\ \right)\right\}\\
&+v_j\mathbb{D}^{\mathbb{V}}_1[x^jv^i-x^iv^j]+\frac{1}{2}x_j\mathbb{D}^{\mathbb{V}}_1[x^ja^i-x^ia^j]\ .
\end{split}
\end{equation}

\begin{table}[H]
\resizebox{\columnwidth}{!}{
	\begin{tabular}{||c||c||}
		\hline\hline
		Symbol & $f^\mu_d$\\\hline & \\	
		$\mathbb{D}_1$ & $\frac{\partial_t^d}{d!!}-\frac{H^2}{3!}(d^2-6d+11)\frac{\partial_t^{d-2}}{(d-2)!!} + \frac{H^4}{5!}\frac{(d-1)(d-3)}{3}(5d^2-48d+127)\frac{\partial_t^{d-4}}{(d-4)!!}$\\ 	 & $\qquad$  \\ 
    \hline & \\	
	$\mathbb{D}_2$ & $\frac{\partial_t^{d+2}}{(d+2)!!}-\frac{H^2}{3!}(d^2-5d+12)\frac{\partial_t^{d}}{d!!}+\frac{H^4}{5!}\frac{(d-1)(d-2)}{3}(5d^2-43d+132)\frac{\partial_t^{d-2}}{(d-2)!!}$ \\ & $\qquad$\\
		\hline & \\		  	
	$\mathbb{D}_1^X$ &  $(d+1)\frac{\partial_t^{d+2}}{(d+2)!!}-\frac{H^2}{3!}(d-3)(d^2-4d+1)\frac{\partial_t^{d}}{d!!}+\frac{H^4}{5!}\frac{(d-1)}{3}(5d^4-78d^3+420d^2-946d+711)\frac{\partial_t^{d-2}}{(d-2)!!}$ \\		 & $\qquad$  \\ 
    \hline & \\
	$\mathbb{D}_1^V$ & $\frac{\partial_t^d}{d!!}-\frac{H^2}{3!}(d^2-6d+11)\frac{\partial_t^{d-2}}{(d-2)!!} + \frac{H^4}{5!}\frac{(d-1)(d-3)}{3}(5d^2-48d+127)\frac{\partial_t^{d-4}}{(d-4)!!}$\\ 	 & $\qquad$ \\ 	
	\hline & \\	
 $\mathbb{D}^\mathbb{V}_1$ & $\frac{\partial_t^d}{d!!}-\frac{H^2}{3!}(d-1)(d-2)\frac{\partial_t^{d-2}}{(d-2)!!} + \frac{H^4}{5!}\frac{(d-1)(d-3)(d-4)(5d+2)}{3}\frac{\partial_t^{d-4}}{(d-4)!!}$ \\ & $\qquad$ \\
\hline \hline
	\end{tabular}}
\caption{The differential operators that appear in de Sitter electromagnetic radiation reaction (for $d$ odd).}
\label{tab:RRdiffOp}
\end{table}	

\subsection{dS covariantisation}\label{app:covRR}
The $f^i$'s can be obtained from the following de Sitter covariant vectors:
 \begin{equation}
\begin{split}
f^\mu_3 &\equiv \frac{P^{\mu\nu}}{3!!}\left\{-2a_\nu^{(1)}\right\}\ ,\\
f^\mu_5&\equiv \frac{ P^{\mu\nu}}{5!!}\left\{-4a_\nu^{(3)}+10\ (a\cdot a)\  a_\nu^{(1)}+30\ (a\cdot a^{(1)})\ a_\nu\right\}-H^2\frac{ P^{\mu\nu}}{5!!}\left\{16a_\nu^{(1)}\right\}\ ,\\
f^\mu_7&\equiv \frac{P^{\mu\nu}}{7!!}\left\{-6a_\nu^{(5)}+42\ (a\cdot a)\  a_\nu^{(3)}+210\ (a\cdot a^{(1)})\ a_\nu^{(2)}+224\ (a\cdot a^{(2)})\ a_\nu^{(1)}+\frac{574}{3}\ (a^{(1)}\cdot a^{(1)})\  a_\nu^{(1)}\right.\\
&\left.\qquad\qquad +126\ (a\cdot a^{(3)})\ a_\nu+280\ (a^{(1)}\cdot a^{(2)})\ a_\nu+O(a^5)\right\}\\
&\quad+H^2\frac{P^{\mu\nu}}{7!!}\left\{120 a_\nu^{(3)}-342\ (a\cdot a)\  a_\nu^{(1)}-978\ (a\cdot a^{(1)})\ a_\nu\right\} -H^4\frac{P^{\mu\nu}}{7!!}\left\{384a^{(1)}_\nu\right\}\ ,\\
\end{split}
\end{equation}
 \begin{equation}
\begin{split}
f^\mu_9&\equiv \frac{P^{\mu\nu}}{9!!}\left\{-8 a_\nu^{(7)}+132\ (a\cdot a)\  a_\nu^{(5)}+924\ (a\cdot a^{(1)})\ a_\nu^{(4)}+1512\ (a\cdot a^{(2)})\ a_\nu^{(3)}\right.\\
&\left.\qquad+1470\ (a\cdot a^{(3)})\ a_\nu^{(2)}+888\ (a\cdot a^{(4)})\ a_\nu^{(1)}+324\ (a\cdot a^{(5)})\ a_\nu\right.\\
&\left.\qquad+1344\ (a^{(1)}\cdot a^{(1)})\  a_\nu^{(3)}+3570\ (a^{(1)}\cdot a^{(2)})\  a_\nu^{(2)}+2640\ (a^{(1)}\cdot a^{(3)})\  a_\nu^{(1)}\right.\\
&\left.\qquad+1092\ (a^{(1)}\cdot a^{(4)})\ a_\nu+1830\ (a^{(2)}\cdot a^{(2)})\  a_\nu^{(1)}+1890\ (a^{(2)}\cdot a^{(3)})\  a_\nu+O(a^5)\right\}\\
&-H^2\frac{P^{\mu\nu}}{9!!} \left\{448 a_\nu^{(5)}+3488\ (a\cdot a)\  a_\nu^{(3)}+17240\ (a\cdot a^{(1)})\ a_\nu^{(2)}+18132\ (a\cdot a^{(2)})\ a_\nu^{(1)}\right.\\
&\left.\qquad+9944\ (a\cdot a^{(3)})\ a_\nu+15424\ (a^{(1)}\cdot a^{(1)})\  a_\nu^{(1)}+21980\ (a^{(1)}\cdot a^{(2)})\ a_\nu+O(a^5)\right\}\\
&+H^4\frac{P^{\mu\nu}}{9!!}\left\{-784a_\nu^{(3)}+19286\ (a\cdot a)\  a_\nu^{(1)}+53770\ (a\cdot a^{(1)})\ a_\nu\right\}+O(H^6)\ ,\\
\end{split}
\end{equation}
 \begin{equation}
\begin{split}
f^\mu_{11}&\equiv \frac{P^{\mu\nu}}{11!!}\left\{-10a_\nu^{(9)}+330\ (a\cdot a)\  a_\nu^{(7)}+2970\ (a\cdot a^{(1)})\ a_\nu^{(6)}+6600\ (a\cdot a^{(2)})\ a_\nu^{(5)}\right.\\
&\left.\qquad\qquad\quad+9240\ (a\cdot a^{(3)})\ a_\nu^{(4)}+8712\ (a\cdot a^{(4)})\ a_\nu^{(3)}\right.\\
&\left.\qquad\qquad\quad+5610\ (a\cdot a^{(5)})\ a_\nu^{(2)}+2420\ (a\cdot a^{(6)})\ a_\nu^{(1)}+660\ (a\cdot a^{(7)})\ a_\nu\right.\\
&\left.\qquad\qquad+5940\ (a^{(1)}\cdot a^{(1)})\  a_\nu^{(5)}+23100\ (a^{(1)}\cdot a^{(2)})\  a_\nu^{(4)}+27324\ (a^{(1)}\cdot a^{(3)})\  a_\nu^{(3)}\right.\\
&\left.\qquad\qquad\quad+20790\ (a^{(1)}\cdot a^{(4)})\  a_\nu^{(2)}+10120\ (a^{(1)}\cdot a^{(5)})\  a_\nu^{(1)}+2970\ (a^{(1)}\cdot a^{(6)})\ a_\nu\right.\\
&\left.\qquad\qquad+19140\ (a^{(2)}\cdot a^{(2)})\  a_\nu^{(3)}+36960\ (a^{(2)}\cdot a^{(3)})\  a_\nu^{(2)}+21560\ (a^{(2)}\cdot a^{(4)})\  a_\nu^{(1)}\right.\\
&\left.\qquad\qquad +7260\ (a^{(2)}\cdot a^{(5)})\  a_\nu+13706\ (a^{(3)}\cdot a^{(3)})\  a_\nu^{(1)}+11088\ (a^{(3)}\cdot a^{(4)})\  a_\nu\right\}\\
&+H^2\frac{P^{\mu\nu}}{11!}\left\{-1200 a_\nu^{(7)}+21460\ (a\cdot a)\  a_\nu^{(5)}+149660\ (a\cdot a^{(1)})\ a_\nu^{(4)}+244048\ (a\cdot a^{(2)})\ a_\nu^{(3)}\right.\\
&\left.\qquad\qquad\quad+236030\ (a\cdot a^{(3)})\ a_\nu^{(2)}+141308\ (a\cdot a^{(4)})\ a_\nu^{(1)}+50660\ (a\cdot a^{(5)})\ a_\nu\right.\\
&\left.\qquad\qquad+216186 \ (a^{(1)}\cdot a^{(1)})\  a_\nu^{(3)}+570000\ (a^{(1)}\cdot a^{(2)})\  a_\nu^{(2)}+417254\ (a^{(1)}\cdot a^{(3)})\  a_\nu^{(1)}\right.\\
&\left.\qquad\qquad +169692\ (a^{(1)}\cdot a^{(4)})\ a_\nu+288650\ (a^{(2)}\cdot a^{(2)})\  a_\nu^{(1)}+292932\ (a^{(2)}\cdot a^{(3)})\  a_\nu+O(a^5)\right\}\ \\
&-H^4\frac{P^{\mu\nu}}{11!!} \left\{43680 a_\nu^{(5)}+363860\ (a\cdot a)\  a_\nu^{(3)}+1786540\ (a\cdot a^{(1)})\ a_\nu^{(2)}+1864072\ (a\cdot a^{(2)})\ a_\nu^{(1)}\right.\\
&\left.\qquad\qquad +1006750\ (a\cdot a^{(3)})\ a_\nu+1581920\ (a^{(1)}\cdot a^{(1)})\  a_\nu^{(1)}+2218698\ (a^{(1)}\cdot a^{(2)})\ a_\nu+O(a^5)\right\}\\
&+O(H^6)\ .
\end{split}
\end{equation}

Here $v^\mu=\frac{dx^\mu}{d\tau}$ is the proper velocity of the particle computed using dS metric, $a^\mu\equiv \frac{D^2x^\mu}{D\tau^2}$ is its proper acceleration and $P^{\mu\nu}\equiv g^{\mu\nu}+v^\mu v^\nu$ is the transverse projector to the worldline. We use  $a_\mu^{(k)}\equiv \frac{D^ka^\mu}{D\tau^k}$ to denote the proper-time derivatives of the acceleration. All the spacetime dot products are computed using the dS metric.

The problem of flat space electromagnetic radiation reaction in 3+1 dimensions has been discussed in textbooks of classical electrodynamics (see for reference \cite{Jackson:1998nia,wald2022advanced}). The corresponding RR force in higher dimensions has been treated in many works \cite{Kosyakov:1999np, Galtsov:2007zz, Mironov:2007mv, Mironov:2007nk, Birnholtz:2013ffa, Birnholtz:2013nta}. In \cite{Birnholtz:2013nta}, the authors compute the electromagnetic radiation reaction action in arbitrary dimensions. The flat space limit of our action matches the one they obtained. We match their post-newtonian expansion of the RR force with ours for the cases $d=3$ and $d=5$ given in their paper.
We also match the flat limit of our covariant expressions to those given in previous works. Our expressions match the flat space results from \cite{Birnholtz:2013nta} and \cite{Galtsov:2007zz} for $d=3$ and $d=5$. Galakhov\cite{Galakhov:2007my} gives covariant expressions up to $d=7$, which matches ours up to signs of certain terms. We disagree with the curved space results of \cite{Galtsov:2007zz} at the $H^2$ order and higher, even though we match the flat space result. The source of this disagreement is unclear due to the very different nature of our derivations.   

\section{Neumann designer scalar on the dS-SK Geometry}\label{app:NeuBlktoBlk}
In paper I, we analysed extended sources in de Sitter coupled to designer scalars through Dirichlet boundary conditions. In this section, we will solve for the designer scalar fields in the presence of extended sources but obeying Neumann boundary conditions. This will be relevant to our study of the electric Debye scalar. 

For a generic bulk source $\rho_\nn$, the Neumann boundary condition arises from a coupling of $\rho_\nn$ to the conjugate field $\pi_\nn=-r^\nn\Dp\phN$, where we have defined $\Dp\equiv(1-r^2)\partial_r+i\w$ and $\Dm\equiv (1-r^2)\partial_r-i\w$.
\begin{equation}\label{Eq:phiActs}
	\begin{split}
		S&= -\frac{1}{2}\sum_{\ell\vec{m}}\int\frac{d\w}{2\pi}\oint  \frac{r^\nn dr}{1-r^2}\Bigl[(\Dp \phN)^* \Dp \phN-\w^2\phN^* \phN\Bigr.\\
		&\Bigl.\qquad\qquad\qquad\qquad\qquad -\frac{1-r^2}{4r^2}\Bigl\{(\nn-1)^2-4\nu^2+[4\mu^2-(\nn+1)^2]r^2\Bigr\} \phN^* \phN\Bigr]\\
		&\qquad + \sum_{\ell\vec{m}}\int\frac{d\w}{2\pi}\oint dr\ \piN^*\varrho_{_\nn} +S_{ct}[\varrho_{_\nn}]\ .
	\end{split}
\end{equation}
Here $\oint$ denotes the integral over the complex radial contour of dS-SK geometry.
The inhomogeneous equation of motion satisfied by such a Neumann scalar is given by:
 \begin{equation}\label{Eq:phiRadODEs}
 	\begin{split}
 		&\frac{1}{r^\nn}\Dp [r^\nn \Dp \phN] +\w^2\phN
 		+\frac{1-r^2}{4r^2}\Bigl\{(\nn-1)^2-4\nu^2+[4\mu^2-(\nn+1)^2]r^2\Bigr\}\phN \\ & \qquad\qquad\qquad +\frac{1}{r^\nn}\Dp\left[r^\nn(1-r^2)\varrho_\nn(\zeta,\w,\ell,\vec{m})\right]=0\ .
    \end{split}
\end{equation}  
Notice that the equation is the same as the inhomogeneous equation of motion for the Dirichlet scalar with the inhomogeneous term replaced with $\frac{1}{r^\nn}\Dp\left[r^\nn(1-r^2)\varrho_\nn(\zeta,\w,\ell,\vec{m})\right]$. In such a case, we can use the same bulk-to-bulk Green function derived in \cite{Loganayagam:2023pfb}\footnote{See appendix C.3 of the reference for a detailed derivation. The corresponding bulk-to-bulk two-point functions in the case of black holes can be found in \cite{Loganayagam:2024mnj}.}, convolved with the appropriate Neumann source, to write the solution. In particular:
 \begin{equation}\label{eq:dSSKBlkN}
	\begin{split}
		\phN(\zeta,\w,\ell,\vec{m})&=\oint   dr_0\ \frac{\mathbb{G}(\zeta|\zeta_0,\w,\ell)}{1-r_0^2}\Dp^0\left[r_0^\nn(1-r_0^2)\varrho_{_\nn}(\zeta_0,\w,\ell,\vec{m})\right]\ .
	\end{split}
\end{equation}
The bulk-to-bulk Green's function $\mathbb{G}$ is given by:
\begin{equation}\label{eq:BlkBlkdS}
	\begin{split}
	 \mathbb{G}(\zeta|\zeta_0,\w,\ell)&=\frac{1}{W_{LR}(\zeta_0,\w,\ell)} g_R(\zeta_\succ,\w,\ell)g_L(\zeta_\prec,\w,\ell)\\
  &\equiv\frac{1}{W_{LR}(\zeta_0,\w,\ell)}\begin{cases}
	 	 g_R(\zeta,\w,\ell)g_L(\zeta_0,\w,\ell) &\quad \text{if}\; \zeta\succ\zeta_0\\
         g_L(\zeta,\w,\ell)g_R(\zeta_0,\w,\ell) &\quad \text{if}\; \zeta\prec\zeta_0
	 \end{cases}	\ .
	\end{split}
\end{equation}
where $\succ$ and $\prec$ respectively mean `\emph{succeeds}' and `\emph{precedes}' on the dS-SK contour. In \eqref{eq:dSSKBlkN}, one can use integration by parts to rewrite it in the more conventional definition of the bulk-to-bulk Green function:
\begin{equation}
\begin{split}
	\phN(\zeta,\w,\ell,\vec{m})&=-\oint   dr_0\ r_0^\nn\Dm^0\mathbb{G}(\zeta|\zeta_0,\w,\ell)\varrho_{_\nn}(\zeta_0,\w,\ell,\vec{m})\ .
	\end{split}    
\end{equation}
If we now repackage this new Green function into a `Neumann' Green function defined by:
\begin{equation}
\begin{split}
	\phN(\zeta,\w,\ell,\vec{m})&=\oint   dr_0\ \widetilde{\mathbb{G}}(\zeta|\zeta_0,\w,\ell)\varrho_{_\nn}(\zeta_0,\w,\ell,\vec{m})\ ,
	\end{split}    
\end{equation}
we find the expression for the Neumann Green function in terms of bulk to boundary propagators as:
\begin{equation}
	\begin{split}
	 \widetilde{\mathbb{G}}(\zeta|\zeta_0,\w,\ell)&=-r_0^{\nn}\Dm^0\left[\frac{1}{W_{LR}(\zeta_0,\w,\ell)} g_R(\zeta_\succ,\w,\ell)g_L(\zeta_\prec,\w,\ell)\right]\\
  &=\frac{1}{W_{LR}(\zeta_0,\w,\ell)}\begin{cases}
	 	 g_R(\zeta,\w,\ell)\pi_L(\zeta_0,\w,\ell) &\quad \text{if}\; \zeta\succ\zeta_0\\
         g_L(\zeta,\w,\ell)\pi_R(\zeta_0,\w,\ell) &\quad \text{if}\; \zeta\prec\zeta_0
	 \end{cases}	\ .
	\end{split}
\end{equation}
In going from the first line of the equation to the second, we have used the fact that for any function $\mathfrak{f}(r)$:
\begin{equation}
    \Dm\left[\frac{\mathfrak{f}(r)}{W_{LR}}\right]=\frac{\Dp\mathfrak{f}(r)}{W_{LR}}.
\end{equation}
This Neumann Green function then solves the following differential equation:
\begin{equation}
	\begin{split}
		&\frac{1}{r^\nn}\Dp [r^\nn \Dp \widetilde{\mathbb{G}}] +\w^2\widetilde{\mathbb{G}}\\
		&\qquad+\frac{1-r^2}{4r^2}\Bigl\{(\nn-1)^2-4\nu^2+[4\mu^2-(\nn+1)^2]r^2\Bigr\}\widetilde{\mathbb{G}}+\frac{1}{r^\nn}\Dp\left[r^\nn(1-r^2)\delta_c(r-r_0)\right]=0\ .
	\end{split}
\end{equation}

\subsection*{Spherical shell influence phase with Neumann boundary conditions}
In this section, we will derive the influence phase of an extended source for scalar fields that obey Neumann boundary conditions. We will discretize the extended source into a set of spherical shells, centered at the origin, on both left and right static patches. The discontinuity in the field is given by the surface density $\sigma_i^{R}$ for the sphere placed at $\zeta=1+\zeta_i$ on the right patch and by the surface density $\sigma_i^{L}$ for the sphere placed at $\zeta=\zeta_i$ on the left patch. One can think of such a source in terms of the bulk source $\rho$ defined in the previous section as being given by:
\begin{equation}
	\begin{split}
r^\nn\varrho_{_\nn}(\zeta,\w,\ell,\vec{m})=\sum_i\sigma^R_i(\w,\ell,\vec{m})\ \delta_c(\zeta|1+\zeta_i)-\sum_i\sigma^L_i(\w,\ell,\vec{m})\ \delta_c(\zeta|\zeta_i)\ .
	\end{split}
\end{equation}
The crucial difference from the Dirichlet case is that when this source enters the inhomogeneous differential equation for $\varphi$, it is acted upon by a $\Dp$ operator. 
The solution is given by:
\begin{equation}
\begin{split}
& \phN = \\
  & \sum_i\frac{1}{W_{LR}(\zeta_i,\w,\ell)}
	\begin{cases}
        e^{2\pi\w}g_L(\zeta,\w,\ell)\ \Bigl[\pi_L(1+\zeta_i,\w,\ell)\ \sigma^L_i-\pi_R(1+\zeta_i,\w,\ell)\ \sigma^R_i\Bigr]&\quad \text{if}\; \zeta\prec 1+\zeta_i\ ,\\ & \\
	\quad \pi_R(\zeta_i,\w,\ell)\ \Bigl[g_L(\zeta,\w,\ell)\ \sigma^L_i-g_R(\zeta,\w,\ell)\ \sigma^R_i\Bigr] &\quad \text{if}\; 1+\zeta_i\prec\zeta\prec \zeta_i\	, \\  & \\
  	\quad g_R(\zeta,\w,\ell)\ \Bigl[\pi_L(\zeta_i,\w,\ell)\ \sigma^L_i-\pi_R(\zeta_i,\w,\ell)\ \sigma^R_i\Bigr] &\quad \text{if}\;  \zeta\succ \zeta_i\ .
	\end{cases}		
\end{split}
\end{equation}

We can substitute this solution into the action to obtain the effective action in terms of the surface charge densities of the shells. This yields the following: 

\begin{equation}
	\begin{split}
		S|_{\textbf{On-shell}}&= \frac{1}{2} \sum_{\ell\vec{m}}\int\frac{d\w}{2\pi}\oint  r^\nn dr\ \varrho_{_\nn}^*\Dp\phN|_{\textbf{On-shell}}\\
  &=\frac{1}{2} \sum_{i,j,\ell,\vec{m}}\int\frac{d\w}{2\pi} \frac{\pi_R(\zeta_i,\w,\ell,\vec{m})}{W_{LR}(\zeta_i,\w,\ell,\vec{m})}\Bigl\{ \sigma^{R\ast}_j \ \Bigl[\pi_R(1+\zeta_j,\w,\ell,\vec{m})\sigma^R_i-\pi_L(1+\zeta_j,\w,\ell,\vec{m})\sigma^L_i\Bigr]\Bigr.\\
  &\qquad \Bigl.\qquad -\sigma^{L\ast}_j\ \Bigl[\pi_R(\zeta_j,\w,\ell,\vec{m})\sigma^R_i-\pi_L(\zeta_j,\w,\ell,\vec{m})\sigma^L_i\Bigr] \Bigr\} \ .
	\end{split}
\end{equation}
We can now use explicit expressions for the bulk to boundary propagators and rewrite the action in a convenient form. We make use of the following equations: 
\begin{equation}
	\begin{split}   
W_{LR}(\zeta_i,\w,\ell) &=- (1+n_\w)\left(\frac{1-r_i}{1+r_i}\right)^{i\w} [\kO-\kI]\ ,\\  
\frac{\pi_R(\zeta_i,\w,\ell)}{W_{LR}(\zeta_i,\w,\ell)} &=\left(\frac{1-r_i}{1+r_i}\right)^{-\frac{i\w}{2}} (1-r_i^2)\partial_{r_i} \Xi_{n}(r_i,\w,\ell)\ ,\\
\pi_L(\zeta_i,\w,\ell)&= -\left(\frac{1-r_i}{1+r_i}\right)^{\frac{i\w}{2}}(1-r_i^2)\partial_{r_i}\Biggl\{\Xi_{nn}(r_i,\w,\ell) +\left[n_\w \kO-(1+n_\w)\kI  \right]\Xi_{n}(r_i,\w,\ell)\Biggr\}\ ,\\
\pi_L(1+\zeta_i,\w,\ell)&= -n_\w\left(\frac{1-r_i}{1+r_i}\right)^{\frac{i\w}{2}}\left[ \kO-\kI  \right](1-r_i^2)\partial_{r_i}\Xi_{n}(r_i,\w,\ell)\ ,\\
\pi_R(\zeta_i,\w,\ell)&= -(1+n_\w)\left(\frac{1-r_i}{1+r_i}\right)^{\frac{i\w}{2}}\left[\kO-\kI \right] (1-r_i^2)\partial_{r_i}\Xi_{n}(r_i,\w,\ell)\ ,\\
\pi_R(1+\zeta_i,\w,\ell)&= \left(\frac{1-r_i}{1+r_i}\right)^{\frac{i\w}{2}}(1-r_i^2)\partial_{r_i}\Biggl\{\Xi_{nn}(r_i,\w,\ell)  -\left[(1+n_\w)\kO-n_\w \kI \right]\Xi_{n}(r_i,\w,\ell)\Biggr\}\ .
\end{split}
\end{equation}
Substituting these expressions in the above action yields the following:
\begin{equation}
	\begin{split}
		S|_{\textbf{On-shell}}&= \frac{1}{2} \sum_{\ell\vec{m}}\int\frac{d\w}{2\pi}\oint  r^\nn dr\ \varrho_{_\nn}^*\Dp\phN|_{\textbf{On-shell}}\\
  &=\frac{1}{2} \sum_{i,j,\ell,\vec{m}}\int\frac{d\w}{2\pi} \left(\frac{1-r_i}{1+r_i}\right)^{-\frac{i\w}{2}} \left(\frac{1-r_j}{1+r_j}\right)^{\frac{i\w}{2}}(1-r_i^2)\partial_{r_i} \Xi_{n}(r_i,\w,\ell) \\ &\Bigl\{ \sigma^{R\ast}_j \ \Bigl[ (1-r_j^2)\partial_{r_j}\Biggl\{\Xi_{nn}(r_j,\w,\ell)  -\left[(1+n_\w)\kO-n_\w \kI \right]\Xi_{n}(r_j,\w,\ell)\Biggr\}\sigma^R_i\\ & +n_\w\left[ \kO-\kI  \right] (1-r_j^2)\partial_{r_j}\Xi_{n}(r_j,\w,\ell)\sigma^L_i\Bigr]\Bigr.\\
  & \Bigl.-\sigma^{L\ast}_j\ \Bigl[ -(1+n_\w)\left[\kO-\kI \right] (1-r_j^2)\partial_{r_j}\Xi_{n}(r_j,\w,\ell)\sigma^R_i\\ &  + (1-r_j^2)\partial_{r_j}\Biggl\{\Xi_{nn}(r_j,\w,\ell) +\left[n_\w \kO-(1+n_\w)\kI  \right]\Xi_{n}(r_j,\w,\ell)\Biggr\}\sigma^L_i\Bigr] \Bigr\} \ .
	\end{split}
\end{equation}
Simplifying the above expression and rewriting the terms in specific combinations of the sources, we obtain
\begin{equation}\label{Eq:shellDoubleSum}
	\begin{split}
		&S|_{\textbf{On-shell}}
  =\frac{1}{2}\sum_{ij\ell}\int\frac{d\w}{2\pi}\left(\frac{1-r_i}{1+r_i}\right)^{-\frac{i\w}{2}}\left(\frac{1-r_j}{1+r_j}\right)^{\frac{i\w}{2}}\\
  &\quad \times\Bigl\{ 
 (1-r_i^2)\partial_{r_i} \Xi_{n}(r_i,\w,\ell)(1-r_j^2)\partial_{r_j}\Xi_{nn}(r_j,\w,\ell)\ [\sigma^{R\ast}_j\sigma^R_i-\sigma^{L\ast}_j\sigma^L_i]\Bigr.\\
  &\quad\qquad-(1-r_i^2)\partial_{r_i} \Xi_{n}(r_i,\w,\ell) (1-r_j^2)\partial_{r_j}\Xi_{n}(r_j,\w,\ell)\ \kO(\sigma^R_j-\sigma^L_j)^\ast[(1+n_\w)\sigma^R_i-n_\w\sigma^L_i]\\
  &\quad\qquad\Bigl.-(1-r_i^2)\partial_{r_i} \Xi_{n}(r_i,\w,\ell) (1-r_j^2)\partial_{r_j}\Xi_{n}(r_j,\w,\ell)\ \kI(\sigma^R_i-\sigma^L_i)[(1+n_{-\w})\sigma^{R\ast}_j-n_{-\w}\sigma^{L\ast}_j]\Bigr\}\ .
 	\end{split}
\end{equation}
The last two lines of the above expression are related by relabelling $\w\to -\w$. These two terms can be interpreted physically by defining the radiative multipole moments:
\begin{equation}\label{eq:ShellMultDef}
\begin{split}
	\multj_R(\w,\ell,\vec{m}) &\equiv \sum_i \left(\frac{1-r_i}{1+r_i}\right)^{-\frac{i\w}{2}}(1-r_i^2)\partial_{r_i} \Xi_n(r_i,\w,\ell)\ \sigma_i^R \\ &\equiv \int_R dr\  r^\nn (1-r^2)\partial_r\Xi_n(r,\w,\ell)\ \left(\frac{1-r}{1+r}\right)^{-\frac{i\w}{2}}\varrho_{_\nn}(\zeta,\w,\ell,\vec{m})  \ ,\\
 \multj_L(\w,\ell,\vec{m}) &\equiv \sum_i \left(\frac{1-r_i}{1+r_i}\right)^{-\frac{i\w}{2}} (1-r_i^2)\partial_{r_i}\Xi_n(r_i,\w,\ell)\ \sigma_i^L \\ &\equiv -\int_L dr\ r^\nn(1-r^2)\partial_r\Xi_n(r,\w,\ell)\ \left(\frac{1-r}{1+r}\right)^{-\frac{i\w}{2}}\varrho_{_\nn}(\zeta,\w,\ell,\vec{m}) \  .
\end{split}
\end{equation}
This now allows us to recast the last two lines of the action into the cosmological influence phase we obtained for the point source:
\begin{equation}
    \label{eq:SCIPpt2}
	\begin{split}
		\SCIP^\text{Pt}
  &\equiv  -\sum_{\ell\vec{m}}\int\frac{d\w}{2\pi} \kO(\multj_R-\multj_L)^\ast[(1+n_\w)\multj_R-n_\w\multj_L]\\ & =-\sum_{\ell\vec{m}}\int\frac{d\w}{2\pi} \kO\  \multj_D^*\ \Bigl[\multj_A+\left(n_\w+\frac{1}{2}\right)\multj_D\Bigr]\ .
	\end{split}
\end{equation}
where we have defined the average and difference multipole moments as
\begin{equation}\begin{split}
\multj_A(\w,\ell,\vec{m})&\equiv\frac{1}{2}[\multj_R(\w,\ell,\vec{m})+\multj_L(\w,\ell,\vec{m})]\ ,\\
 \multj_D(\w,\ell,\vec{m})&\equiv \multj_R(\w,\ell,\vec{m})-\multj_L(\w,\ell,\vec{m}) \ .    
\end{split}\end{equation}

\bibliographystyle{JHEP}
\bibliography{biblio}
\end{document}